\documentclass[showpacs,preprintnumbers,pre]{revtex41}

\usepackage{amsmath,amssymb,graphicx,subfigure,color,psfrag}
\usepackage{graphics,graphicx}

\newcommand{\veps}{\varepsilon}
\newcommand{\average}[1]{\left\langle #1 \right\rangle}
\newcommand{\dd}{\mbox{d}}

\begin{document}
%\begin{frontmatter}
\title{
%Long-range interacting systems: statistical mechanics and dynamics of solvable models\\or\\
Statistical mechanics and dynamics of solvable models with
long-range interactions}

\author{Alessandro Campa$^1$, Thierry Dauxois$^2$, Stefano Ruffo$^3$}

%\address
\affiliation {1. Complex Systems and Theoretical Physics Unit,
Health and Technology Department, Istituto Superiore di Sanit\`a,
and INFN Roma1, Gruppo Collegato Sanit\`a, Viale Regina Elena 299,
00161 Roma, Italy\\
2. Universit\'e de Lyon, Laboratoire de Physique de l'\'Ecole
Normale Sup\'erieure de Lyon, CNRS, 46 all\'{e}e d'Italie, 69364
Lyon cedex 07, France\\
3. Dipartimento di Energetica ``S. Stecco" and CSDC, Universit{\`a}
di Firenze, INFN, Via S. Marta, 3 I-50139, Firenze, Italy}

\date{\today}

\begin{abstract}
For systems with long-range interactions, the two-body potential
decays at large distances as $V(r)\sim 1/r^\alpha$, with $\alpha\leq
d$, where $d$ is the space dimension. Examples are: gravitational
systems, two-dimensional hydrodynamics, two-dimensional elasticity,
charged and dipolar systems. Although such systems can be made
extensive, they are intrinsically {\em non additive}: the sum of the
energies of macroscopic subsystems is not equal to the energy of the
whole system. Moreover, the space of accessible macroscopic
thermodynamic parameters might be {\em non convex}. The violation of
these two basic properties of the thermodynamics of short-range
systems is at the origin of {\em ensemble inequivalence}. In turn, this
inequivalence implies that specific heat can be negative in the
microcanonical ensemble, temperature jumps can appear at
microcanonical first order phase transitions. The lack of convexity
allows us to easily spot regions of parameters space where ergodicity may be
broken. Historically, negative specific heat had been found for
gravitational systems and was thought to be a specific property of a
system for which the existence of standard equilibrium statistical
mechanics itself was doubted. Realizing that such properties may be
present for a wider class of systems has renewed the interest in
long-range interactions. Here, we present a comprehensive review of
the recent advances on the statistical mechanics and
out-of-equilibrium dynamics of solvable systems with long-range
interactions. The core of the review consists in the detailed
presentation of the concept of ensemble inequivalence, as exemplified
by the exact solution, in the microcanonical and canonical ensembles,
of mean-field type models. Remarkably, the entropy of all these models
can be obtained using the method of large deviations. Long-range
interacting systems display an extremely slow relaxation towards
thermodynamic equilibrium and, what is more striking, the convergence
towards {\em quasi-stationary states}. The understanding of such
unusual relaxation process is obtained by the introduction of an
appropriate kinetic theory based on the Vlasov equation. A statistical
approach, founded on a variational principle introduced by Lynden-Bell, is
shown to explain qualitatively and quantitatively some features of
quasi-stationary states. Generalizations to models with both short and
long-range interactions, and to models with weakly decaying
interactions, show the robustness of the effects obtained for
mean-field models.
\end{abstract}

\maketitle

%\end{frontmatter}

\tableofcontents

\section{Introduction}

A wide range of problems in physics concerns systems with long-range
interactions. However, their statistical and dynamical properties
are much less understood than those of short-range systems
\cite{leshouches,Assisi}. One finds examples of long-range
interacting systems in astrophysics
\cite{Padmanabhan90,Chavanisreview2006}, plasma physics
\cite{Escande}, hydrodynamics \cite{Robert90,Miller90}, atomic
physics \cite{Courteille}, nuclear physics \cite{chomazdd}. This
ubiquitous presence in different physics disciplines alone would
itself justify the need for a general and interdisciplinary
understanding of the physical and mathematical problems raised by
long-range interacting systems.

In this review, we are interested in systems with a large number $N$
of degrees of freedom, for which a statistical physics approach is
mandatory, independently of the specific features of the
interactions. Therefore, we will discuss in the following
equilibrium and out--of--equilibrium properties of long-range
systems relying on the tools of statistical mechanics \cite{Huang}.

Let us define which is the property of the interaction that makes it
short or long-ranged. We consider systems where the interaction
potential is given by the sum, over pairs of the elementary
constituents, of a two-body translationally invariant potential. For
sufficiently large distances $r$, the absolute value of the two-body
potential is bounded by $r^{-\alpha}$. If the positive power
$\alpha$ is larger than the dimension $d$ of the space where the
system is embedded, $\alpha>d$, we define the system to be
short-range. Otherwise, if $\alpha \le d$, the system is long-range.
The reason for this definition is that, in the large $N$ limit all
the mathematical and physical differences between short and
long-range systems can be traced back to this property of the
interaction potential. We should remark that this definition of the
range of the interaction does not coincide with others, where the
range is instead defined by a characteristic length appearing in the
interaction potential. In this latter definition, any interaction
decaying as a power law at large distances, thus without
characteristic length, is considered as long range. However, for the
physical and mathematical problems found in the statistical
mechanics of many-body systems, the definition used throughout this
review is more appropriate, since, as we have stressed, it is
related to the interaction property that determines the behavior of
such systems for large $N$.

Our purpose will be to illustrate, especially through the use of
simple models, the peculiar properties of long-range systems, and
the tools and techniques that are employed to describe them. To give
a flavour of the issues that will be considered, we would like to
begin with a very simple description of the main problems that one
faces in the study of these systems.

The aim of equilibrium statistical mechanics is to derive the
thermodynamic properties of a macroscopic system from microscopic
interactions \cite{Huang}. The connection between micro and macro is
realized through the introduction of statistical ensembles.
Different thermodynamic potentials describe situations in which
different thermodynamic parameters are used in the characterization
of the system, and, in the aforementioned connection with the
microscopic interactions, different statistical ensembles are
related to different thermodynamic potentials. However, it is usually
stated (and experimentally verified for many physical systems) that,
as far as macroscopic averages are concerned, i.e. in the
thermodynamic limit ($N\rightarrow \infty$, $V \to \infty$ with
$N/V=const.$), the predictions of statistical mechanics do not
depend on the chosen ensemble. {\it Ensemble equivalence} is related
to the fact that, given a sufficient number of macroscopic
thermodynamic parameters (two in a one-component system), the others
are fixed in the large volume limit, apart from vanishingly small
relative fluctuations.

An important feature of long-range systems is that {\it ensembles
can be inequivalent}
\cite{Thirringdd,Thirring,KiesslingLebowitz97,BMR,Chavanisreview2006}, and therefore
one of the main issues in the statistical mechanics of these systems
is a careful examination of the relations between the different
ensembles, in particular of the conditions that determine their
equivalence or inequivalence. We emphasize that ensemble
inequivalence is not merely a mathematical drawback, but it is the
cause, as it will be shown in this review, of physical properties of
these systems that could be experimentally verified. Probably, one
of the most striking features of long-range systems is the
possibility to display {\it negative specific heat} in the
microcanonical ensemble \cite{Emden,eddigton,Antonov, Lyndenwood68,
Thirringdd, Thirring}. Specific heat is always positive in the
canonical ensemble, independently of the nature of the interactions,
since it is given by the expectation value of a positive quantity.
It turns out that microcanonical equilibrim contains all
informations about canonical equilibrium, while the converse is
wrong in case of ensemble inequivalence \cite{Dieter,Touchette2003}.
The discrepancy between the two ensembles extends to other
observables related to the response of the system to a change in a
thermodynamic parameter: a concrete example will be given discussing
magnetic susceptibility.

As far as out--of--equilibrium dynamical properties are concerned,
many-body long-range systems again show peculiar behaviors. The
approach to equilibrium of short-range systems is usually
characterized by the time scales that govern the equations of motion
of the elementary constituents \cite{Balian}. For systems without
disorder, these time scales are typically small when compared to the
observational time scales. Sometimes these systems can be trapped in
metastable states that last for long times. These states are local
extrema of thermodynamic potentials and, in practical cases, their
realization requires a very careful preparation of the system (e.g.
undercooled liquids and superheated solids). If perturbed, the
system rapidly converges towards the equilibrium state.
It can also happen that the relaxation time depends on the volume
if hydrodynamics modes are present or in coarsening processes.

For long-range systems, dynamics can be extremely slow and the
approach to equilibrium can take a very long time, that increases
with the number $N$ of elementary constituents
\cite{binneytremaine}. This feature is induced by the long-range
nature of the interaction itself and is not a consequence of the
existence of a collective phenomenon. The state of the system during
this long transient is {\it quasi-stationary}
\cite{Lyndenbell67,ChavanisSommeriaRobert,lrr,yoshi}, since its very
slow time evolution allows us to define slowly varying macroscopic
observables, like for local equilibrium or quasi-static
transformations. It should be however remarked that quasi-stationary
states are not thermodynamic metastable states, since they do not
lie on local extrema of equilibrium thermodynamic potentials. The
explanation of their widespread presence should rely only upon the
dynamical properties of the systems. It must be stressed that the
nature of quasi-stationary states can be strongly dependent on the
initial condition, as it will become clear from the examples that we
will give. In addition, a variety of macroscopic structures can form
spontaneously in out--of--equilibrium conditions for isolated
systems: a fact that should not be a surprise given that already the
equilibrium states of long-range systems are usually inhomogeneous.
In short-range systems, macroscopic structures can arise as an
effect of an external forcing (Rayleigh-B\'enard convection,
Benjamin-Feir instability, Faraday waves) \cite{Fauve} or due to the
nonlinearity of the governing equations of motion (solitons,
breathers) \cite{dauxoispeyrard}, but are usually strongly selected
by the specific dynamical properties and by the geometrical
conditions. All this shows the great richness of the dynamics of
long-range systems.

Summarizing, a satisfying theoretical framework concerning the
behaviour of system with long-range interactions should necessarily
address the following aspects:
\begin{itemize}
\item For what concerns equilibrium properties, the determination of the
physical conditions that determine equivalence or inequivalence of
the statistical ensembles and, for the latter case, the relation
between macrostates in the different statistical ensembles.
\item As for the out--of--equilibrium features, the development of a consistent
kinetic theory able to explain the formation of quasi-stationary
states and their final relaxation to equilibrium.
\end{itemize}
Although different aspects of systems with long-range interactions
have been studied in the past in specific scientific communities,
notably astrophysics and plasma physics, this has not constituted a
seed for more general theoretical studies. In the last decade or so,
it has become progressively clear that the ubiquitous presence of
long-range forces needs an approach that integrates different
methodologies \cite{leshouches,Assisi}. This has induced a
widespread interest in long-range systems throughout numerous
research groups. The successive development has led to a better
understanding of both the equilibrium and out--of--equilibrium
properties of such systems. Time is ripe to summarize what is known
on firm basis about the equilibrium statistical mechanics of systems
with long-range interactions and to describe the preliminaries of a
theory of non equilibrium.

In this review, we have chosen to present the main problems, tools
and solutions by discussing paradigmatic examples, which are simple
and general enough to be useful also for specific applications.
Therefore, we will mostly emphasize:
\begin{itemize}
\item the equilibrium statistical mechanics solutions of
simple mean-field toy models, which is a first step for
understanding phase diagrams of more complex long-range systems;
\item the basic ingredients of a kinetic theory of a model able
to catch the essential properties of out--of--equilibrium dynamics.
\item the analysis of the phase diagram of models with both short and long-range
interactions, or with forces that weakly decay in space.
\end{itemize}

The structure of the review is the following. In
section~\ref{Theproblemofadditivity}, we introduce the subject
presenting the definition of long-range interactions and discussing
the non additivity property and its consequences. In
section~\ref{physicalexamples}, we briefly review the relevant
physical systems with long-range interactions: gravitational
systems, 2D hydrodynamics, 2D elasticity, charged systems, dipolar
systems and small systems. Section~\ref{equilibriumsec} constitutes
the core of the review. We present there the equilibrium properties
of several mean-field models for which one can analytically compute
both the free energy and the entropy. The solution is obtained by
different methods, including the powerful large deviation method,
which allows to solve models with continuous variables. Dynamics is
tackled in section~\ref{outofequilibrium}, where the main result
reviewed is the widespread presence of quasi-stationary states that
hinder relaxation to Boltzmann-Gibbs equilibrium. This phenomenon is
strongly supported by numerical simulations and can be studied using
kinetic equations explicitly devised for long-range interactions
(Vlasov, Lenard-Balescu). Few analytical results exist for non
mean-field models: we collect some of them in
section~\ref{perspectives}. The message is that the introduction of
short-range terms doesn't spoil the features of mean-field models
and that some weakly decaying interactions can be treated
rigorously. Finally, we draw conclusions and we discuss some
perspectives in section~\ref{conclusions}.

\section{The additivity property and the definition of long-range systems}
\label{Theproblemofadditivity}

\subsection{Extensivity vs. Additivity}
\label{additivity}

\begin{figure}[htpb]
\begin{center}
\includegraphics[width=.3\textwidth]{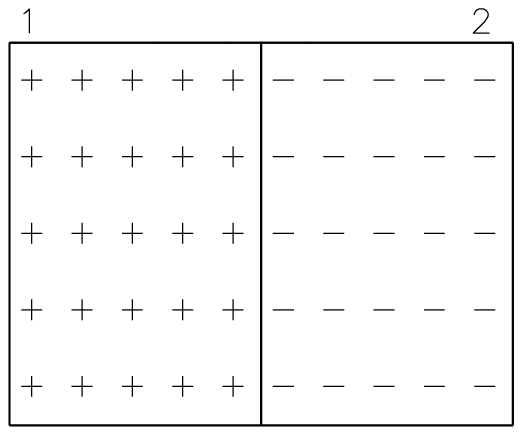}
\end{center}
\vskip -1truecm \caption{Schematic picture of a system separated in
two equal parts with $N/2$ spins up in domain 1 and $N/2$ spins down
in domain 2.} \label{additivite}
\end{figure}

In order to easily illustrate the issues of {\em extensivity} and
{\em additivity}, it is useful to consider a concrete example and
restrict ourselves to the energy as the thermodynamic extensive
variable. We employ a very simple model that is used in the study of
magnetic systems, namely the Curie-Weiss Hamiltonian
\begin{equation}
H_{CW}= -\frac{J}{2N}\sum_{i,j=1}^{N} S_i S_j =
-\frac{J}{2N}\left(\sum_{i=1}^N S_i\right)^2, \label{hamiladditi}
\end{equation}
where the spins $S_i=\pm 1$ are attached to sites labeled by
$i=1,\ldots,N$. In this example, the interaction does not decay at
all with the distance: indeed, each spin interacts with equal
strength with all the other spins. Such systems are usually referred
to as mean-field systems. With the $1/N$ prefactor in
(\ref{hamiladditi}), the total energy increases as $N$, then the
energy per spin converges to a finite value in the thermodynamic
limit, which is a physically reasonable requirement. We recall that
one can find rigorous definitions of the thermodynamic limit in
Ruelle's book \cite{ruelle}. Model (\ref{hamiladditi}) is {\em extensive}:
for a given intensive magnetization
\begin{equation}
m=\frac{\sum_i S_i}{N}=\frac{M}{N} \, ,\label{defmagnetization}
\end{equation}
where $M$ is the extensive magnetization, if one doubles the number
of spins the energy doubles. On the other hand,
Hamiltonian~(\ref{hamiladditi}) is not additive, in spite of the
presence of the regularizing factor $1/N$. Indeed, let us divide the
system, schematically pictured in Fig.~\ref{additivite}, in two
equal parts. In addition, let us consider the particular case in
which all spins in the left part are equal to $+1$, whereas all
spins in the right part are equal to $-1$. The energy of the two
parts is $E_1 = E_2 = -{JN}/{4}$. However, if one computes the total
energy of the system, one gets $E=0$. Since  $E \neq E_1+E_2$, such
a system is not additive, at least for this configuration. One could
easily generalize the argument to generic configurations. The
problem is not solved by increasing system size $N$, because the
``interaction" energy $E_{1,2}=E-E_1-E_2=JN/2$ increases with $N$.

{\it Extensivity} in the Curie-Weiss model is provided by the $1/N$
prefactor in Hamiltonian (\ref{hamiladditi}), which makes the energy
proportional to $N$. This is the so-called Kac prescription
\cite{KacUhlenbeck}. This energy rescaling guarantees a competition
between energy $E$ and entropy $S$, which is crucial for phase
transitions. Indeed, introducing the free energy $F=E-TS$, where $T$
is temperature, Kac prescription implies that the two competing
terms on the r.h.s. both scale as $N$, since temperature is
intensive and entropy scales like $N$. This latter scaling deserves
a further analysis which will be developed later on. Intuitively,
the fact that the interaction is long-range does not alter the
density of states, which usually grows factorially with $N$. An
alternative prescription would be to make temperature extensive $T
\to TN$, in such a way that energy and entropy term of the free
energy both scale as $N^2$. The two prescriptions give equivalent
physical consequences. In systems with kinetic energy, being
temperature the average kinetic energy per particle, rescaling
temperature corresponds to a renormalization of velocities, and
finally of the time scale.

After having defined the two distinct concepts of {\em extensivity}
and {\em additivity} with reference to a specific model, let us
clarify more generally these two notions.

Indeed, it is very important, as a first step, to give the general
definitions of {\em extensivity} and {\em additivity}, and to
clarify the distinction between these two concepts. It is convenient
to first consider the situation that one encounters in short-range
systems. We can imagine to divide a system at equilibrium in two
parts occupying equal volumes. Some thermodynamic variables of each
half of the system will be equal to the corresponding ones of the
total system, others will be halved. Temperature and pressure are
examples of the first kind of thermodynamic variables: they do not
depend on the size of the system and are called intensive variables.
Energy, entropy and free energy are variables of the second kind;
their value is proportional to system size, i.e. to the number of
elementary constituents (for given values of the intensive
variables), and they are called extensive variables. The property
that the size dependent thermodynamic variables are proportional to
system size is called {\em extensivity}, and systems with this
property are called extensive. The specific value of extensive
variables (e.g., the energy per unit particle, or per unit mass, or
per unit volume) give rise to new intensive quantities. Considering
the energy of a system, we see that it has also the property of {\em
additivity}, that consists in the following. Dividing the systems in
two macroscopic parts, the total energy $E$ will be equal to $E_1 +
E_2 + E_{int}$, with $E_i$ the energy of the $i$-th part, and
$E_{int}$ the interaction energy between the two parts. In the
thermodynamic limit the ratio $E_{int}/(E_1 + E_2)$ tends to zero;
therefore in this limit $E \approx E_1 +E_2$. This property is
called {\em additivity}; systems with this property (for the energy
as well as for other size dependent quantities) are called additive.
It is evident that extensivity and additivity are related. Indeed,
in the definition of extensivity just given, we could not have
concluded that the energy of each part of the system is half the
total energy, if the interaction energy $E_{int}$ would not be
negligible. Additivity implies extensivity (thus non extensivity
implies non additivity), but not the reverse, since the interaction
energy might scale with $N$, as in the Curie-Weiss model. This
comment applies more generally to all long-range systems. The
unusual properties of these systems derive from the lack of
additivity.

\subsection{Definition of long-range systems}
\label{decayint}

We have just shown that mean-field systems like the Curie-Weiss
model can be made extensive using Kac's trick. However, extensive
systems could be non additive. Here, we will discuss the case of
interactions that decay as a power law at large distances. We will
show that the energy $\varepsilon$ of a particle (excluding
self-energy) diverges if the potential does not decay sufficiently
fast, implying that the total energy grows superlinearly with volume
at constant density, which violates extensivity. These interactions
are called {\it long-range} or {\it non integrable}, just referring
to this divergence of the energy. Energy convergence can be restored
by an appropriate generalization of Kac's trick.

\begin{figure}[htpb]
\begin{center}
\includegraphics[width=.3\textwidth,height=.25\textwidth]{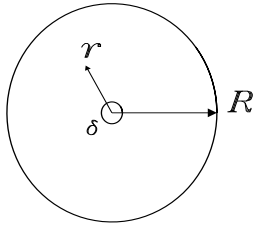}
\end{center}
\vskip-1truecm \caption[]{Schematic picture of the domain considered
for the evaluation of the energy $\varepsilon$ of a particle. It is
a spherical shell of outer radius $R$ and inner radius $\delta$.}
\label{sphere}
\end{figure}

Let us estimate the energy $\varepsilon$ by considering a given
particle placed at the center of a sphere of radius $R$ where the
other particles are homogeneously distributed. We will exclude the
contribution to $\varepsilon$ coming from the particles located in a
small neighborhood of radius $\delta$ (see Fig.~\ref{sphere}). This
is motivated by the necessity to regularize the divergence of the
potential at small distances, which has nothing to do with its
long-range nature.

If the other particles interact with the given one via a potential
that at large distances decays like ${1}/{r^\alpha}$, we obtain in
$d$--dimensions
\begin{equation}
\varepsilon = \int_\delta^R \dd^d  r \; \rho\frac{ J}{r^\alpha}=
\rho J \Omega_d \int_\delta^Rr^{d-1-\alpha} \dd r= \frac{\rho J
\Omega_d}{d-\alpha} \left[R^{d-\alpha}-\delta^{d-\alpha}\right],
\hskip 2truecm \mbox{if}\ \alpha \ne d~,
\end{equation}
where $\rho$ is the generic density (e.g. mass, charge), $J$ is the
coupling constant and $\Omega_d$ is the angular volume in dimension
$d$ ($2\pi$ in $d=2$, $4 \pi$ in $d=3$, etc.). When increasing the
radius $R$, the energy $\varepsilon$ remains finite if and only if
$\alpha>d$. This implies that the total energy $E$ will increase
linearly with the volume $V$, i.e. the system is extensive. Such
interactions are the usual short-range ones. On the contrary, if
$\alpha \le d$ the energy $\varepsilon$ grows with volume as
$V^{1-\alpha/d}$ (logarithmically in the marginal case $\alpha=d$).
This implies that the total energy $E$ will increase superlinearly,
$E \propto V^{2-\alpha/d}$, with volume. However, analogously to the
Kac's prescription, one can redefine the coupling constant $J \to J
V^{\alpha/d-1}$ and get a perfectly extensive system. Mean-field
models, like Hamiltonian~(\ref{hamiladditi}), correspond to the
value $\alpha=0$, since the interaction does not depend on the
distance, and one recovers the usual Kac's rescaling. Cases where
the energy grows superlinearly define the long-range nature of the
interaction. However, as we have shown for mean-field systems, the
fact that energy can be made extensive, does not imply that the
system is additive. Implications of the lack of additivity for
long-range systems will be discussed throughout the paper.
%in Section
%\ref{decayinginteraction}.

\subsection{Convexity in thermodynamic parameters}
\label{Convexity}

\begin{figure}[htpb]
\begin{center}
\includegraphics[width=.3\textwidth]{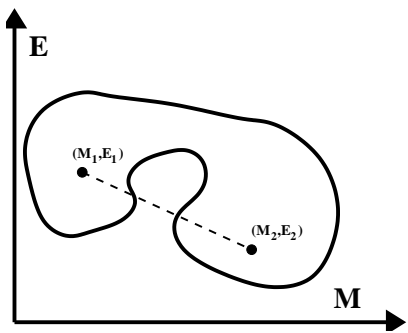}
\end{center}
\caption{The set of accessible macrostates in the $(M,E)$ space can
have a non-convex shape for systems with long-range interactions,
such that if $(M_1,E_1)$ and $(M_2,E_2)$ can be realized
macroscopically, this is not necessarily true for all the states
joining these two along the straight dashed line.}
\label{convexspace}
\end{figure}

Let us consider the space of the extensive thermodynamic parameters.
To be specific, as in Fig.~\ref{convexspace}, we consider the $(E,M)$
plane of energy and
magnetization (\ref{defmagnetization}). The attainable region in
this space is always convex when only short-range interactions are
present. This property is a direct consequence of additivity.
Consider two different subsystems with two different energies $E_1$
and $E_2$, and two different magnetization values $M_1$ and $M_2$.
Introducing a parameter $\lambda$ taking values between $0$ and $1$,
depending on the relative size of the subsystems, the system
obtained by combining the two subsystems has an energy $E= \lambda
E_1 +(1-\lambda)E_2$, and a magnetization $M=\lambda
M_1+(1-\lambda)M_2$. Any value of $\lambda$ between $0$ and $1$ is
realized thermodynamically, just varying the relative size of the
two subsystems. This is exactly the {\it convexity property} of the
attainable region of the space of thermodynamic parameters. In
particular, one can consider two subsystems with the same energy but
different magnetization. Varying the relative size of the two
subsystems, the combined system will have the same energy and any
possible magnetization between the two values of the two subsystems.
%Hence, the combined system is composed of two phases, i.e., the two
%subsystems with different magnetizations, $M_1$ and $M_2$, giving the average
%magnetization $M$.
%It is of course possible that the homogeneous
%state with the same magnetization $M$ has a higher entropy. In any
%case, there will be always be an equilibrium state with magnetization
%$M$.
It is important to stress that the convexity property is possible if
additivity is satisfied, since the interaction energy between the
two subsystems has been neglected. As already remarked, the
additivity property is generically valid for large enough and
short-range interacting systems. Moreover convexity implies that the
space of thermodynamic parameters is connected.

On the contrary, systems with long-range interactions are not
additive, and thus intermediate values of extensive parameters are
not necessarily accessible (see Fig.~\ref{convexspace}). This feature has
profound consequences
on the dynamics of systems with long-range interactions. Gaps may
open up in the space of extensive variables. Since the space of
thermodynamic parameters is no more connected, {\it ergodicity
breaking} might appear when considering continuous microcanonical
dynamics of such a system. We will discuss again this question in
detail in Sec.~\ref{generalizedHMF} by emphasizing simple examples.

\subsection{Lattice systems}
\label{systemtypes}

When defining the long-range nature of the interaction, care must be
taken of the specific nature of the microscopic variables. These
could be divided in two classes: the coordinates related to the
translational degrees of freedom (e.g. cartesian coordinates), and
those giving the internal state of each particle (e.g. spin
variables). Both could be either continuous or discrete. When
coordinates take fixed discrete values, one speaks of {\it lattice
systems}. As for the internal degrees of freedom the variables could
be discrete (e.g. spin systems) or continuous (e.g. systems of
rotators).

For systems with continuous translational degrees of freedom that do
not possess internal degrees of freedom, the potential energy can be
written in the general form
\begin{equation}
\label{potcont} U(\overrightarrow{r}_1,\dots,\overrightarrow{r}_N)=
\sum_{1\le i<j\le N}V(|\overrightarrow{r}_i -
\overrightarrow{r}_j|)\, ,
\end{equation}
where $(\overrightarrow{r}_1,\dots,\overrightarrow{r}_N)$ are
cartesian coordinates in $d$-dimensional space and we assume that
the translationally invariant pair potential $V$ depends only on the
modulus of the distance between two particles. Systems of
gravitational point masses or Coulomb point charges fall into this
category.

In a lattice system, for which each site $i$ of a $d$-dimensional
lattice, located by the position vector $\mathbf{r}_i$, hosts a
particle possessing one or more internal degrees of freedom,
collectively denoted by the vector ${\bf q}_i$ (the dimensionality
of this vector is independent of $d$), the potential energy can be
written as
\begin{equation}
\label{potlat} U({\bf q}_1,\dots,{\bf q}_N)=\sum_{1\le i < j \le
N}C_{ij}V({\bf q}_i,{\bf q}_j) + g \sum_{i=1}^N V_e({\bf q}_i)~,
\end{equation}
where the coupling constants $C_{ij}$ are translationally and
rotationally invariant (i.e., they depend only on
$|\overrightarrow{r}_i - \overrightarrow{r}_j|$). We allow also the
presence of an external field that couples to the particles via the
function $V_e({\bf q})$ with a strength $g$. As pointed out above,
the variables ${\bf q}_i$ may take continuous or discrete values.

We have already discussed in the previous Subsection that long-range
systems of type (\ref{potcont}) are those for which, at large
distance, $V(r) \sim r^{-\alpha}$ with $\alpha \le d$. Similarly,
long-range lattice systems (\ref{potlat}) can be characterized by a
slow decay of the coupling constants $C_{ij}$. If these latter
behave at large distance like $|\overrightarrow{r}_i -
\overrightarrow{r}_j|^{-\alpha}$ with $\alpha \le d$, the system is
long-range. The energy grows also in this case superlinearly with
the volume and one will need to introduce a Kac rescaling factor. At
variance with systems of type (\ref{potcont}) we need not worry
about the behavior at short distances, because the lattice
regularizes any possible divergence.

A kinetic energy term can be added to the potential energy one for
systems of type (\ref{potcont}) and for lattice systems
(\ref{potlat}) when the variables are continuous.

In this review, we will mostly concentrate our attention on lattice
systems. Since they cannot display any short distance singularity,
their thermodynamic and dynamical behaviour highlights the essential
features of long-range interactions. Therefore, these systems are
more suitable to present an overview of the main results and of the
tools used to deal with long-range interactions.

\subsection{Non additivity and the canonical ensemble}
\label{addcanon}

Let us recall that, in $d=3$, neglecting the dependence on the
number of particles $N$ and on the volume $V$, the microcanonical
partition function (proportional to the number of microstates with a
given energy $E$) is defined as
\begin{equation}
\Omega (E) \sim \int \dd^{3N} q\, \dd^{3N} p\ \delta (E-H(p,q)),
\label{mic}
\end{equation}
where $(p,q)$ are the phase space coordinates, $H$ is the
Hamiltonian and we forget for the moment about multiplicative
$N$-dependent factors and dimensional constants (for a more precise
definition see Section~\ref{equilibriumsec}). The entropy is defined
via the classical Boltzmann formula
\begin{equation}
S(E)=\ln \Omega(E)~,
\end{equation}
where we adopt units for which the Boltzmann constant $k_B$ is equal
to $1$.

The non additivity has strong consequences on the construction of
the canonical ensemble from the microcanonical. The reasoning
usually goes as follows. One considers an isolated  system with
energy $E$, that we divide into a ``small" part with energy $E_1$
(the subsystem of interest) and a ``large" part with energy $E_2$
which plays the role of the bath. The additivity of the energy
implies that the probability distribution that the ``small'' system
has an energy~$E_1$, let us call it $p(E_1)$, is given by
\begin{eqnarray}\label{hypprem}
p(E_1)&=& \int \Omega_2(E_2)\
\delta(E_1+E_2-E)\dd E_2\\
&=&\Omega_2(E-E_1)\; .
\label{hyp}
\end{eqnarray}
Using the entropy to express $\Omega_2$ and expanding the term
$S_2(E-E_1)$, one gets
\begin{eqnarray}
p(E_1)&=&\exp \left[S_2(E-E_1)\right]%\nonumber
\\
& \approx & \exp \left[S_2(E)-E_1\left.\frac{\partial
S_2}{\partial E}\right|_E + \cdots \right]%\nonumber
\\
& \propto &\Omega_2(E)\;e^{-\beta E_1}\; ,
\end{eqnarray}
where
\begin{equation}
\beta=\left.\frac{\partial S_2}{\partial E}\right|_E.
\end{equation}
One would end up with the usual canonical distribution for the
system of interest after performing the thermodynamic limit
\cite{ruelle}. Let us however remark that this derivation is valid
also before taking the thermodynamic limit. This has led many
authors to develop a thermodynamic formalism for finite systems
\cite{chomazdd}.

It is clear that additivity is crucial to justify the factorization
hypothesis implied in~(\ref{hypprem}), and hence the existence of
the canonical distribution. This hypothesis is clearly violated when
the system size is finite, because of a contribution to the entropy
coming from the surface that separates the two subsystems. However,
after performing the thermodynamic limit, this contribution becomes
negligible when the interactions are short-range. This is not true
for long-range interactions which are non additive also in the
thermodynamic limit. This has led the community to split into two
different attitudes. On one side are those who think that the
canonical ensemble cannot be appropriately defined for long-range
interactions and claim that all the analyses should be performed
using the microcanonical ensemble \cite{Dieter, grossdd}. Although
clearly appropriate for isolated system, this approach cannot
describe open systems. On the other side are those who stress that
the canonical ensemble could be still formally defined and used
\cite{kiesling89,Padmanabhan90, paddy,ChavanisHouches}. Among the
two sides are those who attempt operative definitions of a heat
bath. For instance, one has imagined that the heat bath is a
short-range system that interacts with the system of interest with
short-range interactions, so that the non additivity is due only to
the system \cite{BaldovinOrlandini,BaldovinOrlandinibis}. The
failure of the usual derivation of the canonical ensemble suggests
that non additive systems might have a very peculiar behavior if
they are in contact with a thermal reservoir (see
Refs.~\cite{PoschThirringpre2006,Lyndenbell_epl2008,
RamirezHernandezPRL, RamirezHernandezPRE,VelazquezCurilef} for
recent literature on this topic). A third position was recently
initiated by Bouchet and Barr\'e who proposed that when considering
systems with long-range interaction, the canonical ensemble does not
describe fluctuations of a small part of the whole system. However,
they argued~\cite{julienfreddyjstat,Villain} that it may describe
fluctuations of the whole system when coupled to a thermostat via a
negligibly small coupling. This interesting line of thought needs to
be pursued theoretically and confirmed by numerical simulations.

In the following Section, we will discuss some physical examples
before proceeding to the core of the review.

\section{Physical examples of long-range interacting systems}
\label{physicalexamples}

There are many systems in nature where particles interact with a
pair potential that decays at large distances as $V(r) \sim
r^{-\alpha}$ with $\alpha \le d$. Although these systems are deeply
studied in their own sake (e.g. gravitational many body systems,
Coulomb systems, systems of vortices, etc.), they rarely appear in
books of statistical mechanics. This is of course due to the
difficulty to deal with systems that are non extensive and non
additive. We give here a brief sketch of the physics involved in
some of these systems.

\begin{figure}[htpb]
\begin{center}
\includegraphics[width=.5\textwidth]{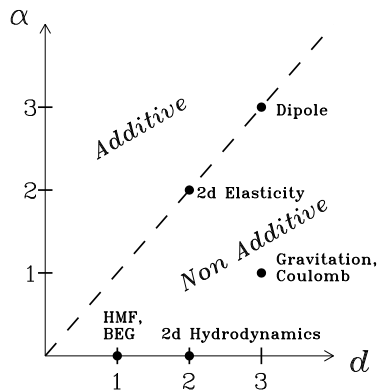}
\end{center}
\vskip -1.5truecm \caption[]{Location of some physical systems in
the plane where the abscissa is space dimension $d$ and the ordinate
is the exponent $\alpha$ characterizing the spatial decay at large
distances of the pair potential.} \label{diagramshortlongrange}
\end{figure}
In Fig.~\ref{diagramshortlongrange}, we draw in the $(\alpha,d)$
plane some of the physical systems with long-range interactions that
we will discuss afterwards.

\subsection{Gravitational systems}

Gravitational systems, which correspond to $\alpha=1$ in dimension
$d=3$, clearly belong to the category of long-range interacting
systems. The gravitational problem is particularly difficult
because, in addition to the non additivity due to the long-range
character of the interaction, one also needs a careful
regularization of the potential at short distances to avoid
collapse. To be more specific, let us consider the canonical
partition function of a system of $N$ self-gravitating particles of
the same mass $m$ moving inside a volume $V$
\begin{equation}
\label{cutoffgravit} Z_N=\frac{1}{(2 \pi \lambda^2)^{3N/2}N!}
\int_{V} \prod_{i=1}^N \dd\overrightarrow{r_i}\ \exp \left[-\beta
U(\overrightarrow{r}_1,\ldots,\overrightarrow{r}_N)\right],
\end{equation}
where
\begin{equation}
U(\overrightarrow{r}_1,\ldots,\overrightarrow{r}_N)=-{\cal
G}m^2\sum_{i<j}^N V(|\overrightarrow{r_i}-\overrightarrow{r_j}|)~,
\end{equation}
with
\begin{equation}
\label{unsurer} V(r)=\frac{1}{r}~,
\end{equation}
where $\beta$ is the inverse temperature, ${\cal G}$ the
gravitational constant,
 and $\lambda=\hbar(\beta/m)^{1/2}$ the De Broglie wavelength.
In the following, we will not introduce dimensional constants, i.e.
$\hbar$, in the definition of partition functions for a matter of
convenience (see e.g. Eq. (\ref{canon})).
\begin{figure}[htb]
\begin{center}
\includegraphics[width=.5\textwidth]{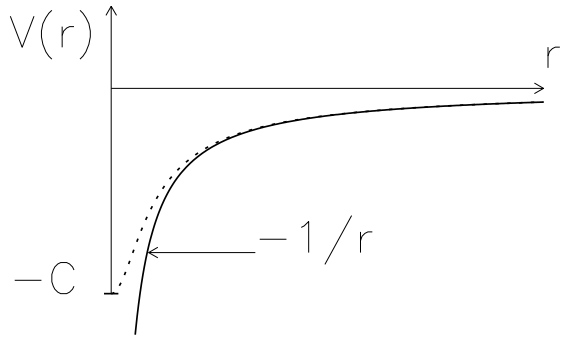}
\end{center}
\vskip -1.5truecm \caption[]{The gravitational potential as a
function of the distance $r$ is represented by the solid curve. The
dotted curve shows the short-distance regularized potential which
avoids gravitational collapse, $-C$ is the lower bound of the
potential.} \label{courtedistance}
\end{figure}
From the shape of the potential represented in
Fig.~\ref{courtedistance} by the solid line, one clearly sees that
$Z_N$ will diverge if at least two particles collapse towards the
same point. This difficulty arises because the potential is not
bounded from below as for the Lennard-Jones or Morse potential. In
quantum mechanics, the collapse of self-gravitating fermions
\cite{HertelThirringcommmathphys, Chavanisreview2006}, a system which is physically
relevant for dwarfs and neutron stars, is forbidden by the Pauli
exclusion principle, that introduces a natural effective small scale
cut-off. However, to avoid the use of quantum concepts
and stick to a classical model, the usual trick is to introduce an
{\it ad-hoc} cut-off \cite{binneytremaine}. One way of regularizing
the potential is shown in Fig.~\ref{courtedistance} by the dotted
line. One can imagine that the potential has a hard core, which
represents the particles' size. As a consequence, the inequality
$U(r)\ge -G m^2 C \equiv  -C' $ allows to easily determine a finite
upper bound of the configurational partition function
\begin{equation}
Z_N \leq  \frac{V^N}{(2 \pi \lambda^2)^{3N/2}N!} \exp [\beta C'
N(N-1)/2]~.
\end{equation}

As far as the microcanonical ensemble is concerned, a standard
argument tells us that the entropy of a self-gravitating system in a
finite volume might have convergence problems if the potential is
not regularized at short distances. Indeed, let us consider $N$
gravitating particles grouped together in a finite volume. A strong
decrease of the potential energy of a pair of particles $-{\cal G}
m^2/(|\overrightarrow{r_i}-\overrightarrow{r_j}|)$ is obtained when
$\overrightarrow{r_i}$ tends to $\overrightarrow{r_j}$. Since the
total energy is a conserved quantity, the kinetic energy would
correspondingly increase. This process leads to an increase of the
accessible phase volume in the direction of momentum, and hence to
an entropy increase. Since the process extends to the limit of zero
distance among the particles, it might induce a divergence of the
density of states, and then of Boltzmann entropy. However, one can
show that this does not happens for $N=2$ and a more subtle
derivation \cite{Padmanabhan90,Chabanol} reveals that the entropy integral
diverges only when $N \geq 3$. It's remarkable that this also
corresponds to the transition from an integrable ($N=2$) to a non
integrable ($N=3$) gravitational system. As a consequence of this
discussion, ``equilibrium" states can exist only in association with
local entropy maxima \cite{Antonov,Lyndenbell67}.

Besides these unusual properties of the canonical and microcanonical
partition functions, self-gravitating systems were historically the
first physical system for which ensemble inequivalence was
discovered through the phenomenon of negative specific heat. The
possibility of finding a negative specific heat in gravitational
systems was already emphasized by Emden~\cite{Emden} and
Eddington~\cite{eddigton}.
The divergence of the phase space volume for gravitational systems was
proved by Antonov~\cite{Antonov}, and the fact that this implies
negative specific heats was later stressed by
Lynden-Bell~\cite{Lyndenwood68}.
An early remark on the possibility
of having a negative specific heat can be found in the seminal
review paper on statistical mechanics by Maxwell~\cite{Maxwell}.
This was for long time considered a paradox, until
Thirring~\cite{Thirringdd,Thirring} finally clarified the
controversial point by showing that the paradox disappears if one
realizes that the microcanonical specific heat can be negative only
in the microcanonical ensemble. Therefore, one can attribute to
Thirring the discovery of ensemble inequivalence.

Let's rephrase Thirring's argument. In the canonical ensemble, the
mean value of the energy is computed from the partition function
(\ref{cutoffgravit}) as
\begin{equation}
\label{canen} \langle E\rangle=-\frac{\partial \ln Z }{\partial
\beta} \, .
\end{equation}
It is then straightforward to compute the heat capacity at constant
volume
\begin{equation}
\label{cvpos} { C_V}=\frac{\partial \langle E\rangle}{\partial T} =
\beta^2 \langle \left(E   -\langle E\rangle\right)^2\rangle{ >0}.
\end{equation}
Let's remind that $k_B=1$. This clearly shows that the canonical
specific heat is always positive. Notice also that this condition is
true for systems of any size, regardless of whether a proper
thermodynamic limit exists or not.

For self-gravitating systems at constant energy (i.e., in the
microcanonical ensemble) a simple physical argument which justifies
the presence of a negative specific heat has been given by
Lynden-Bell \cite{LyndenPhysA}. It is based on the virial theorem,
which, for the gravitational potential, states that
\begin{equation}
\label{virial} 2\langle K \rangle+\langle U \rangle=0,
\end{equation}
where $K$ and $U$ are the kinetic and potential energy,
respectively. Recalling that the total energy $E$ is constant
\begin{equation}
E=\langle K \rangle+\langle U \rangle=-\langle K \rangle,
\end{equation}
where in the second identity we have used the virial theorem
(\ref{virial}), and since the kinetic energy $K$ defines the
temperature, one gets
\begin{equation}
{C_V}=\frac{\partial E}{\partial T} \propto \frac{\partial
E}{\partial K} <0.
\end{equation}
Loosing its energy, the system becomes hotter.

There is a further difficulty in the case of gravitational
interactions: the system is open, i.e. without boundary, strictly
speaking. Therefore, the microcanonical partition function
(\ref{mic}) will diverge. This divergence is actually not peculiar
of self-gravitating systems since it would also occur for a perfect
gas. Although for gases it is natural to confine them in a box, this
is completely unjustified for gravitational systems. A way out from
this pitfall is to consider an expanding universe and reconsider the
problem in a wider context \cite{paddy}.

A detailed discussion on phase transitions in self-gravitating
systems, in both the canonical and microcanonical ensemble, is not the
aim of this review, and can be found in
Refs.~\cite{Padmanabhan90,Chavanisreview2006}.

\subsection{Two-dimensional hydrodynamics}

Two-dimensional incompressible hydrodynamics is another important
case where long-range interactions appear. Although high Reynolds
number flows have a very large number of degrees of freedom, one can
often identify structures in the flow. This suggests that one could
use a much smaller number of effective degrees of freedom to
characterize the flow. This remark is particularly valid for
two-dimensional flows, where the inverse energy cascade leads to the
irreversible formation of large coherent structures (e.g. vortices).
A system with a large number of degrees of freedom which can be
characterized by a small number of effective parameters, is
reminiscent of what happens in thermodynamics~\cite{Castaing}, where
a few macroscopic variables describe the behavior of systems
composed of many particles. Statistical mechanics for turbulence is
nowadays a very active field of research \cite{EyingSreenivasan},
initiated long ago by Lars Onsager~\cite{Onsager49}. We will discuss
in the following the long-range character of two-dimensional
hydrodynamics.

The velocity of a two-dimensional flow can be expressed in terms of
the stream function~$\psi (\overrightarrow{r})$, where
$\overrightarrow{r}=(x,y)$ is the coordinate on the plane
\begin{eqnarray}
v_x&=& +\frac{\partial \psi}{\partial y} \\
v_y&=& -\frac{\partial \psi}{\partial x}~.
\end{eqnarray}
The vorticity $\omega$ is related to the velocity field
\begin{equation}
\omega=\frac{\partial v_y}{\partial x}-\frac{\partial v_x}{\partial
y}
\end{equation}
and, hence, to the stream function by the Poisson equation
\begin{equation}
\omega=-\Delta\psi~. \label{Poisson}
\end{equation}
Using the Green's function
$G\left(\overrightarrow{r},\overrightarrow{r}'\right)$, one easily
finds the solution of the Poisson equation in a given domain $D$
\begin{equation}
   \psi( \overrightarrow{r})=\int_D \dd \overrightarrow{r}'\;
   \omega(\overrightarrow{r}')\;G\left(\overrightarrow{r},\overrightarrow{r}'\right),
\end{equation}
plus surface terms \cite{Alastueybook}. In an infinite domain,
\begin{equation}
G\left(\overrightarrow{r},\overrightarrow{r}'\right)=-\frac{1}{2
\pi} \ln |\overrightarrow{r}-\overrightarrow{r}'|.
\end{equation}
The energy is conserved for the Euler equation and is given by
\begin{eqnarray}
\label{energyvorticity}
H &=& \int_D \dd \overrightarrow{r} \frac{1}{2}(v_x^2+v_y^2)\\
  &=& \int_D \dd \overrightarrow{r} \frac{1}{2}\left(\nabla\psi\right)^2\\
  &=& \frac{1}{2}\int_D \dd \overrightarrow{r} \ \omega(\overrightarrow{r})
  \psi(\overrightarrow{r})\\
  &=&-\frac{1}{4 \pi} \int_D\! \int_D \dd\overrightarrow{r} \dd \overrightarrow{r}'\;
  \omega(\overrightarrow{r}')\omega(\overrightarrow{r})
  \ln|\overrightarrow{r}-\overrightarrow{r}'|~.
\end{eqnarray}
This emphasizes that one gets a logarithmic interaction between
vorticies at distant locations, which corresponds to a decay with
an effective exponent $\alpha=0$, well within the case of long-range
interactions (see Fig.~\ref{diagramshortlongrange}). For a finite
domain $D$ the Green's function contains additional surface
terms~\cite{Alastueybook}, which however gives no contribution to
the energy (\ref{energyvorticity}) if the velocity field is tangent
to the boundary of the domain (no outflow or inflow).

Another important conserved quantity is enstrophy, defined as
\begin{equation}
{\cal A} =\frac{1}{2}\int_D \dd \overrightarrow{r}
\left[\omega(\overrightarrow{r})\right]^2.\label{enstrophy}
\end{equation}

The long-range character of the interaction is even more evident if
one approximates the vorticity field by $N$ point vortices located
at $\overrightarrow{r}_i=(x_i,y_i)$, with a given
circulation~$\Gamma_i$
\begin{equation}
\omega(\overrightarrow{r})=\sum_{i=1}^N\Gamma_i\delta
(\overrightarrow{r}-\overrightarrow{r}_i)~.
\end{equation}
The energy of the system reads now
\begin{equation}
H=-\frac{1}{4 \pi}\sum_{i\neq
j}\Gamma_i\Gamma_j\ln|\overrightarrow{r}_i-\overrightarrow{r}_j|~,
\end{equation}
where we have dropped the self-energy term because, although
singular, it would not induce any motion~\cite{MarchioroPulvirenti}.
%The stream function is then
%\begin{equation}
%\psi( \overrightarrow{r})=\sum_{i=1}^N \Gamma_i \;
%G\left(\overrightarrow{r},\overrightarrow{r}_i \right).
%\end{equation}
Considering the two coordinates of the point vortex on the plane,
the equations of motion are
\begin{eqnarray}
\label{kirchoffa}
\Gamma_i\frac{\dd x_i}{\dd t}&=&+\frac{\partial H}{\partial y_i}\\
\Gamma_i\label{kirchoffb} \frac{\dd y_i}{\dd t}&=&-\frac{\partial
H}{\partial x_i}.
\end{eqnarray}

The phase-space volume contained inside the energy shell $H=E$ can
be written as
\begin{equation}
\Phi(E)=\int\prod_{i=1}^N\dd \overrightarrow{r}_i\,
\theta(E-H(\overrightarrow{r}_1,...,\overrightarrow{r}_N)),
\end{equation}
$\theta$ being the Heaviside step function. The total phase space
volume is $\Phi(\infty)=A^N$, where $A$ is the area of the domain
$D$. One immediately realizes that $\Phi(E)$ is a non-negative
increasing function of the energy $E$ with limits $\Phi(-\infty)=0$
and $\Phi(\infty)=A^N$. Therefore, its derivative, which is nothing
but the microcanonical partition function (\ref{mic}), is given by
\begin{equation}
\Omega(E)=\Phi'(E)=\int\prod_{i=1}^N\dd \overrightarrow{r}_i\,
\delta(E-H(\overrightarrow{r}_1,...,\overrightarrow{r}_N)),
\end{equation}
and is a non-negative function going to zero at both extremes
$\Omega(\pm \infty)=0$. Thus the function must achieve at least one
maximum at some finite value $E_m$ where $\Omega'(E_m)=0$. For
energies $E>E_m$, $\Omega'(E)$ will then be negative. Using the
entropy $S(E)=\ln\Omega(E)$, one thus gets that the inverse
temperature $\dd S/\dd E$ is negative for $E>E_m$. This argument for
the existence of negative temperatures was proposed by
Onsager~\cite{Onsager49} two years before the experiment on nuclear
spin systems by Purcell and Pound~\cite{PurcellPound51} reported the
presence of negative ``spin temperatures".

Onsager also pointed out that negative temperatures could lead to
the formation of large-scale vortices by clustering of smaller ones.
Although, as anticipated, the canonical distribution has to be used
with cautions for long-range interacting systems, the statistical
tendency of vortices of the same circulation sign to cluster in the
negative temperature regime can be justified using the canonical
distribution $\exp(-\beta H)$. Changing the sign of the inverse
temperatures $\beta$ would be equivalent to reversing the sign of
the interaction between vortices, making them repel (resp. attract)
if they are of opposite (resp. same) circulation sign.

After a long period in which Onsager's statistical theory was not
further explored, this domain of research has made an impressive
progress recently. One might in particular cite the work of Joyce
and Montgomery~\cite{Joyce73}, who have considered a system of
vortices with total zero circulation. Maximizing the entropy at
fixed energy, they obtain an equation for the stream function which
gives exact stable stationary solutions of the 2D Euler equations,
able to describe the macroscopic vortex formation proposed by
Onsager for negative temperatures. Later, Lundgren and Pointin~\cite{Lundgren}
studied the effect of far vorticity field on the motion of a single
vortex, showing that it produces a positive eddy viscosity term
leading to an increase of cluster size. Subsequently,
Robert~\cite{Robert90} and Miller~\cite{Miller90} have elaborated an
equilibrium statistical mechanical theory directly for the continuum
2D Euler equation. Nice prolongations along these lines are
Refs.~\cite{RobertSommeria91,MichelRobert94,ChavanisSommeriaRobert},
together with applications to the Great Red Spot of
Jupiter~\cite{Turkington, Bouchet01,Bouchet02,chavanise}.

The canonical ensemble can also be defined for both the Euler
equation in $2D$ and the Onsager point vortex model. It turns out
that in both cases, canonical and microcanical ensembles may be
inequivalent \cite{Caglioti_95,KiesslingLebowitz97, ellisdd, Ellis02}. In
particular for the Euler equation in the region of negative
temperature (where vorticity tends to accumulate at the center of
the domain) some hybrid states are shown to be realized in the
microcanonical ensemble but not in the canonical \cite{ellisdd}. As
far as the point vortex model is concerned, microcanonical stable
states, that are unstable in the canonical ensemble, are found for
specific geometries \cite{KiesslingLebowitz97}. In both cases,
microcanonical specific heat is negative. Reviews on 2D turbulence can
be found in Refs~\cite{chavanisf,chavanisg,chavanish,ChavanisHouches}.
Another related interesting case of ensemble inequivalence has been
recently reported in the context of physical
oceanography \cite{venaillebouchet}.

\subsection{Two-dimensional elasticity}

Let us discuss the planar stress and displacement fields around the
tip of a slit-like plane crack in an ideal Hookean continuum solid.
The classical approach to a linear elasticity problem of this sort
involves the search for a suitable ``stress function'' that
satisfies the so-called biharmonic equation
\begin{equation}
\nabla^2(\nabla^2\psi)=0
\end{equation}
where $\psi$ is the Airy stress function, which has to satisfy
appropriate boundary conditions. The deformation energy density is
then defined as $U \propto\sigma \epsilon$ where $\sigma$ is the
fracture stress field around the tip, whereas $\epsilon$ is the
deformation field. Considering a crack-width $a$ and using the exact
Muskhelishvili's solution~\cite{muskhelishvili}, one obtains the
elastic potential energy due to the crack
\begin{equation}
U\simeq\frac{\sigma_\infty^2(1-\nu)}{2E}\;\frac{a^2}{r^2},
\end{equation}
where $E$ is the Young modulus, $\sigma_\infty$ the stress field at
infinity, $\nu$ the Poisson coefficient and $r$ the distance to the
tip. The elasticity equation in the bulk of two-dimensional
materials leads therefore to a marginal case of long-range
interaction, since $U \sim 1/r^2$ in $d=2$. Looking at the important
engineering applications, the dynamics of this non conservative
system should be better studied: the difficulty lies again in the
long-range nature of the interaction. In addition, in such a two
dimensional material, the presence of several fractures could
exhibit a very interesting new type of screening effects.

\subsection{Charged systems}

The partition function of system of charges is basically the same as
the one for gravitational systems displayed in formula
(\ref{cutoffgravit}) with the potential given by
\begin{equation}
U(\overrightarrow{r}_1,\ldots,\overrightarrow{r}_N)= \frac{1}{4 \pi
\varepsilon_0} \sum_{i<j}^N e_i e_j
V(|\overrightarrow{r_i}-\overrightarrow{r_j}|)~,
\end{equation}
where $e_i=\pm e$ is the charge of the particle and $V$ is given by
formula (\ref{unsurer}). If the total charge is non zero, the excess
charge is expelled to the boundary of the domain and the bulk is
neutral. Hence, neglecting boundary contributions (which are non
extensive), one usually considers an infinite medium with total zero
charge \cite{LebowitzLieb69,LiebLebowitz72}. Similarly to the
gravitational case, the partition function would diverge if not
properly regularized at short distances. This is usually done by
supposing that additional forces are present at short distances,
either hard core of radius $\lambda$ or smoothed Coulombic
singularities of the type
\begin{equation}
V_{smooth} \sim \frac{1-\exp (-r/\lambda)}{r}.
\end{equation}
The regularized partition function allows one to perform the
thermodynamic limit and derive physical quantities, like the
pressure.

As far as the large distance behavior is concerned, several rigorous
results exist \cite{Martin} that prove, under appropriate hypotheses
(low density and high temperature), that the effective two-body
potential is Debye-H\"uckel screened
\begin{equation}
V_{eff} \propto \frac{\exp (-r/\ell_D)}{r},
\end{equation}
where $\ell_D=(\varepsilon_0/(2 n e^2 \beta))^{1/2}$ is the Debye
length, with $n$ the density. Among the hypotheses, the most
important one for its possible physical consequences is the one of
low density or high temperature. This suggests that all pathologies
related to ensemble inequivalence (e.g. negative specific heat),
that is a consequence of the long-range nature of the interaction,
will be indeed absent for charged systems.

An interesting different situation is the one of plasmas consisting
exclusively of single charged particles (pure electron or pure ion
plasmas) \cite{DubinOneil99}. Charged particles are confined by
external electric and magnetic fields. The thermodynamics of such
systems is not affected by short-distance effects because particles
repel each other. On the contrary, the behavior at large distances
is different from the one of globally neutral low density plasmas.
Nevertheless, by studying the correlation at equilibrium, a Debye
length emerges~\cite{DubinOneil99} (this behavior is also present in
self-gravitating systems). Experimentally, relaxation to thermal
equilibrium has been observed only in some specific conditions. The
system has been shown to relax into quasi-stationary states,
including minimum enstrophy states (see Eq.~(\ref{enstrophy})) and
vortex crystal states \cite{HuangDriscoll,Fine}. Indeed, in the
particular case of a pure electron plasma in a cylindrical container
with a strong magnetic field applied along the axis of the cylinder,
it can be shown that the electron motion in the plane perpendicular
to the magnetic field obeys equations that are the same as the
Onsager point vortex model (\ref{kirchoffa}) and (\ref{kirchoffb}).
Therefore, all what has been written above about two-dimensional
hydrodynamics applies to this system, including the existence of
negative specific heat which has been explicitly demonstrated in a
magnetically self-confined plasma torus~\cite{Kiessling03}.

\subsection{Dipolar systems}

Systems of electric and magnetic dipoles share many similarities,
but also have important differences (force in non uniform external
fields, the symmetry axial/polar of
the dipolar vector). However, for what concerns statistical
mechanics, the two systems are equivalent. Magnetic dipolar systems
are easier to realize in nature;  let us then concentrate on them.
The interaction energy between two magnetic dipoles is
\begin{equation}
E_{ij}=\frac{\mu_0}{4 \pi} \left[ \frac{\overrightarrow{\mu}_i \cdot
\overrightarrow{\mu}_j}{|\overrightarrow{r}_{ij}|^3} - \frac{ 3
(\overrightarrow{\mu}_i \cdot \overrightarrow{r}_{ij})
(\overrightarrow{\mu}_j \cdot \overrightarrow{r}_{ij})}
{|\overrightarrow{r}_{ij}|^5} \right]~, \label{dipolarenergy}
\end{equation}
where $\overrightarrow{\mu}_i$ is the magnetic
moment,$\overrightarrow{r}_{ij}= (\overrightarrow{r}_{j}-
\overrightarrow{r}_{i})$ is the distance between the two dipoles and
$\mu_0$ is the magnetic permeability of the vacuum. Dipolar
interaction energy is strongly anisotropic: on a lattice, magnetic
moments parallel to a bond interact ferromagnetically, while when
they are perpendicular to the bond they interact
antiferromagnetically. When magnetic moments are placed on a
triangular or a square lattice the interaction is {\it frustrated}.
Dipolar forces are long-range only in $d=3$, because the energy per
spin scales as $\int \dd^d r/r^3$. They are therefore {\it
marginally} long-range in $d=3$, while they are short-range in
$d=1,2$. It should be remarked that elasticity is instead marginally
long-range in $d=2$.

It can be shown in general that, for samples of ellipsoidal shape
and when the magnetization $\overrightarrow{M}=\sum_i
\overrightarrow{\mu}_{i} /V$ lies along the longest principal axis of the
ellipsoid, the energy per volume of a system of dipoles on a lattice
can be written as
\begin{equation}
\frac{H}{V} = \frac{1}{2V} \sum_{i,j} E_{ij} = E_0 + \frac{1}{2}
\mu_0 |M|^2 D~, \label{finaldipolarformula}
\end{equation}
where $E_0$ is an energy that depends on the crystal structure, and
$D$ is the so called demagnetizing factor, which is equal to $1/3$
for spherically shaped samples, tends to $D=0$ for needle shape
samples and to $D=1$ for disk shaped ones. The well-known {\it shape
dependence} of dipolar energy is hence accounted for by the highly
frustrating antiferromagnetic demagnetizing term in formula
(\ref{finaldipolarformula}) \cite{LandauLifshitz}. We will see
examples of energies of the form (\ref{finaldipolarformula}) in
subsection~\ref{splusl}, while in subsection \ref{ramaz}, we will
work out in detail a specific example corresponding to a realistic
system.

Using methods similar to those introduced to prove the existence of
a thermodynamic limit for short-range forces \cite{ruelle}, it can
be shown \cite{Griffiths68,Griffiths00} that a system of dipolar
spins posses a well defined bulk free energy, independent of sample
shape, only in the case of zero applied field. The key to the
existence of this thermodynamic limit is the reduction in
demagnetization energy when uniformly magnetized regions break into
ferromagnetically ordered domains \cite{Kittel51}. Technically, the
proof is performed by showing that under the hypothesis of zero field
the free energy is additive,
using invariance under time reversal of the dipole energy, which is
a consequence of its bilinearity in $\overrightarrow{\mu}_i$.

In the future, magnetic dipolar systems might constitute a field
where the results on the statistical mechanics of systems with
long-range will find a fruitful application. Besides the example
discussed in subsection \ref{ramaz}, holmium titanate materials
\cite{Bramwell01},  where dipolar interactions dominate over
Heisenberg exchange energy, deserve some attention. The possibility
to perform experiments with single domain needle shaped dipolar
materials has been also stressed \cite{Barbara}.

\subsection{Small systems}
\label{secsmallsyst}

As we have seen, the presence of long-range interactions causes the
lack of additivity in macroscopic systems. However, even if the
interactions are short-range, systems of a linear size comparable to
the range of the interaction are nonadditive. Then, we should expect
that some of the peculiar features found for macroscopic systems
with long-range interactions are present also in microscopic or
mesoscopic systems. Examples of this sort are: atomic clusters,
quantum fluids, large nuclei, dense hadronic matter.

As it will be shown in the following section, the study of phase
transitions is very important for the characterization of the
properties of macroscopic systems with long-range interactions,
especially in relation to the issue of ensemble inequivalence.
However,  phase transitions do occur also in atomic clusters
(liquid-gas and solid-liquid transitions), quantum fluids
(Bose-Einstein condensation or super-fluid transition), large nuclei
(liquid-gas transition), dense hadronic matter (formation of
quark-gluon plasma). Extensive studies have been devoted to phase
transitions in the thermodynamic limit. This limit introduces
simplifications in the analytical treatment, also in the case of
long-range interactions. On the contrary, a consistent theory of
phase transitions for small systems has not yet been developed.
Signatures of phase transitions in finite systems are, however,
often found both in numerical and laboratory experiments.

Gross has focused his attention on the theoretical treatment of
``Small'' systems~\cite{grossphysrep,Dieter,grossdd}. His approach
privileged the use of the microcanonical ensemble. Therefore, he
immediately realized the possibility that specific heat could be
negative and pointed out the feasibility of experiments with heavy
nuclei.

According to Chomaz and collaborators, the key point is to begin
with the general definition of entropy in the framework of
information theory (see Refs.~\cite{chonucphys1999,chopre2001,
choepj2006,chomazassisi,chomazdd,chopre2002}). In order to treat on
the same ground classical and quantum systems, Chomaz introduces the
density matrix
\begin{equation}
\label{densmatrsmall} \hat{D}=\sum_n \left| \Psi_n \right\rangle p_n
\left\langle \Psi_n \right| \, ,
 \end{equation}
where $\left| \Psi_n \right\rangle$ are the states of the system and
$p_n$ their probabilities. The entropy is thus defined by
\begin{equation}
S[\hat{D}]=-{\rm Tr}\hat{D}\ln \hat{D} \, . \label{entropysmallsys}
\end{equation}
Different Gibbs ensembles are obtained by maximizing $S[\hat{D}]$
with respect to the probabilities $p_n$ under some constraints. Each
set of constraints defines an ensemble. The different number and
functional forms of the constraints are related to the different
physical situations. Within this framework, the finiteness of small
systems is described by the introduction of specific constraints. It
has been shown that phase transitions in finite systems can be
equivalently signalled by three different effects. The first one is
the bimodality of the density of states as a function of
energy~\cite{LabastieWhetten}, with the distance between maxima
corresponding to different phases scaling as the number of
particles. The second one is a negative slope of the microcanonical
caloric curve, i.e. a negative specific heat~\cite{LyndenMrMME}. The
third one is the presence of anomalously large fluctuations in the
energy partition between potential energy and kinetic energy. When
the interactions are short-range, all these features give rise in
the thermodynamic limit to the usual phase transitions, with the
disappearance of the negative slope of the caloric curve. It is also
possible to find a further signature of a phase transition, that
makes a connection with Yang-Lee theory of phase transitions. In
this latter theory, all zeroes of the partition function lie in the
complex plane of the temperature for the canonical ensemble or of
the fugacity for the grand-canonical ensemble, with an imaginary
part different from zero as long as the system is finite. Phase
transitions in the infinite system are associated to the approach of
some of these zeros to the real axis, as system size increases. It
has been stressed that the way in which zeroes approach the real
axis may serve as a classification of phase transitions in finite
systems~\cite{chomazassisi}.

Some of the previously mentioned signatures of phase transitions in
finite systems have been also experimentally reported. We mention
here experiments on atomic clusters~\cite{haberland,gobet}, and
experiments on nuclear fragmentation~\cite{dagostino}. In all
experiments, the microcanonical caloric curve is compatible with the
presence of an energy range where specific heat is negative.

The first set of experiments is realized using atomic sodium
clusters Na$_{147}^+$ and hydrogen cluster ions
H$_3^+$(H$_2$)$_{m\le 14}$. In the first case, the negative specific
heat has been found in correspondence to a solid-liquid phase
transition, while in the second case in the vicinity of a liquid-gas
transition. Sodium clusters~\cite{haberland} are produced in a gas
aggregation source and then thermalized with Helium gas of
controlled temperature and selected to a single cluster size by a
first mass spectrometer. Energy of the clusters is then increased by
laser irradiation leading finally to evaporation. A second mass
spectrometer allows the reconstruction of the size distribution, and
correspondingly of their energies. Performing this experiment at
different temperatures of the Helium gas, a caloric curve is
constructed. The procedure assumes that, after leaving the source, a
microcanonical temperature can be assigned to the clusters.
Conceptually,  this is probably the most delicate point of the
experiment. A region of negative specific heat, corresponding to the
solid-liquid transition, has been reported.

In the second set of experiments, performed with hydrogen cluster
ions~\cite{gobet}, the energy and the temperature are determined
from the size distribution of the fragments after collision of the
cluster with a Helium projectile. This is done using a method
introduced in Ref.~\cite{bonasera}. The reported caloric
curve~\cite{gobet} shows a plateau. Work along this line is in
progress and seems to show a negative specific heat
region~\cite{farizonassisi}, corresponding to a liquid-gas
transition.

In the third set of experiments on nuclear
fragmentation~\cite{dagostino}, the presence of negative specific
heat is inferred from the event by event study of energy
fluctuations in excited Au nuclei resulting from Au + Au collisions.
The data seem to indicate a negative specific heat at an excitation
energy around $4.5$ MeV/u. However, the signature corresponds to
indirect measurements, and the authors  cautiously use the word
``indication'' of negative specific heat.

\section{Equilibrium statistical mechanics: ensemble
inequivalence} \label{equilibriumsec}

Our purpose in this section is to propose a fully consistent
statistical mechanics treatment of systems with long-range
interactions. An overview of the different methods (saddle-point
techniques, large deviations, etc.) will be presented. Simple
mean-field models will be used both for illustrative purposes and as
concrete examples. Most of the features of long-range systems that
we will discuss are valid also for non mean-field Hamiltonians as
will be discussed in Sec.~\ref{perspectives}.

\subsection{Equivalence and inequivalence of statistical ensembles: physical
and mathematical aspects}\label{eqnoneq}

\subsubsection{Ensemble equivalence in short-range systems}
\label{equivshort}

In short-range systems, statistical ensembles are equivalent. A
thorough proof of this result can be found in the book by Ruelle
\cite{ruelle}. Let us first discuss the physical meaning of ensemble
equivalence in order to clarify why for long-range interacting
systems equivalence does not always hold.

The three main statistical ensembles are associated to the following
different physical situations:
\begin{itemize}
\item a completely isolated system at a given energy $E$: {\it microcanonical} ensemble;
\item a system that can exchange energy with a large thermal reservoir
characterized by the temperature $T$:
{\it canonical} ensemble;
\item a system that can exchange energy and particles with a reservoir
characterized by the
temperature $T$ and the chemical potential $\mu$: {\it grand
canonical} ensemble.
\end{itemize}

Equivalence of the ensembles relies upon two important physical
properties:
\begin{itemize}
\item[{\it i})] in the thermodynamic limit, excluding critical
points,  the relative fluctuations of the thermodynamic parameters
that are not held fixed (e.g. energy in the canonical ensemble)
vanish;

\item[{\it ii})] a macroscopic physical state that is realizable
in one ensemble can be realized also in another (equivalence at the
level of macrostates).
\end{itemize}

Let us concentrate our attention on the second item and let's refer
to the equivalence between microcanonical  and canonical ensemble:
an isolated system with a given energy has an average temperature,
i.e., an average kinetic energy. If instead we put the system in
contact with a thermal bath at that temperature, we have an average
energy equal to the energy of the isolated system. Therefore there
is a one-to-one correspondence between energy values and temperature
values. Actually, in the presence of phase transitions, this
statement has to be made more precise, as we will comment in
subsection \ref{maxwellinshort}.

The practical consequence of ensemble equivalence is that, for
computational purposes, one has the freedom to choose the ensemble
where calculations are easier, and typically this is not the
microcanonical ensemble (it is easier to integrate Boltzmann factors
than $\delta$-functions). Thus, in spite of its fundamental
importance in the construction of statistical mechanics, the
microcanonical ensemble is practically never used to perform
analytical calculations. On the contrary it is very much used in
numerical simulations, since it constitutes the fundamental
ingredient of molecular dynamics \cite{FrenkelSmith}.

Ensemble equivalence is mathematically based on certain properties
of the partition functions. To illustrate this point, we again
consider the microcanonical and canonical ensembles, referring the
reader to Ruelle \cite{ruelle} for a complete and rigorous
discussion. A more precise definition of the microcanonical
partition function (see formula (\ref{mic})) of a system in $d=3$
with $N$ particles confined in a volume $V$ is given by
\begin{equation}
\label{micro} \Omega (E,V,N) = \frac{1}{N!} \int_\Gamma \dd q^{3N}
\dd p^{3N} \, \delta \left( E - H(p,q)\right)\, ,
\end{equation}
where the domain of integration is the accessible phase space
$\Gamma$.

For lattice systems the definition of $\Omega$ is slightly
different. There is no explicit volume dependence (because volume is
fixed once the lattice constant and $N$ are given) and no $N!$ term
due to the distinguishability of the lattice sites (for more details
on the Gibbs paradox see ~\cite{Huang}). The microcanonical
partition function is in this case
\begin{equation}
\label{microlattice} \Omega_{lattice} (E,N) = \int \prod_i^N \dd
{\bf q}_i \prod_i^N \dd {\bf p}_i\, \delta \left( E - K(\{{\bf
p}_i\})-U(\{{\bf q}_i\}) \right)\,,
\end{equation}
where ${\bf q}_i$ and ${\bf p}_i$ are the conjugate variables
attached to site $i$ (see formula (\ref{potlat})) and $K$ the
kinetic energy.

The entropy is defined by
\begin{equation}
\label{entr} S(E,V,N) = \ln \Omega(E,V,N)\, .
\end{equation}
The thermodynamic limit corresponds to $N\rightarrow \infty$,
$E\rightarrow \infty$ and $V\rightarrow \infty$ such that $N/V
\rightarrow n$ and $E/N \rightarrow \varepsilon$, where the density
$n\ge 0$ and the energy per particle $\varepsilon$ are finite. The
limit
\begin{equation}\label{entrn}
s(\varepsilon,n)=\lim_{N\rightarrow \infty} \frac{1}{N}S(E,V,N)
\end{equation}
exists and gives the entropy per particle. The function
$s(\varepsilon,n)$ is continuous, increasing in $\varepsilon$ at
fixed $n$, so that the temperature
\begin{equation}
T= \left(\partial s/\partial \varepsilon\right)^{-1}
\label{deftemperaturemicro}
\end{equation}
is positive. Measuring the temperature by this formula seems hardly
feasible. However, for Hamiltonians with kinetic energy, it can be
shown that $T$ coincides with the average kinetic energy, which is
accessible experimentally~\cite{RUGH}. For short-range systems,
$s(\varepsilon,n)$ is a concave function of $\varepsilon$ at fixed
$n$, i.e.
\begin{equation}
\label{concpro} s\left(c\varepsilon_1 + (1-c)\varepsilon_2,n\right)
\ge c\,s(\varepsilon_1,n) +(1-c)\,s(\varepsilon_2,n)
\end{equation}
for any choice of $\varepsilon_1$ and $\varepsilon_2$, with $0\le c
\le 1$ (in lattice systems without kinetic energy the energy can be
bounded from above, and in turn this implies that $T$ can be
negative; however, concavity is still guaranteed if the interactions
are short-range). This property is important in connection with the
partition function of the canonical ensemble, given by
\begin{equation}
\label{canon} Z(\beta,V,N) = \frac{1}{N!} \int_\Gamma \dd q^{3N} \dd
p^{3N} \exp \left[ -\beta H(p,q)\right]\, ,
\end{equation}
with $\beta \geq 0$ the inverse temperature. A similar definition of
a canonical lattice partition function can be given, as done for the
microcanonical lattice partition function (\ref{microlattice}). In
the thermodynamic limit, the free energy per particle is
\begin{equation}\label{freen}
f(\beta,n) = -\frac{1}{\beta}\lim_{N\rightarrow \infty} \frac{1}{N}
\ln Z(\beta,V,N)~.
\end{equation}
Moreover, at fixed $n$, the function $\phi(\beta,n)\equiv \beta
f(\beta,n)$ is concave in $\beta$. In the following $\phi(\beta,n)$
will be called the {\em rescaled free energy}.

\begin{figure}[htbp!]
%%%\label{LegendreFenchel}
\begin{center}
\includegraphics[width=.7\textwidth]{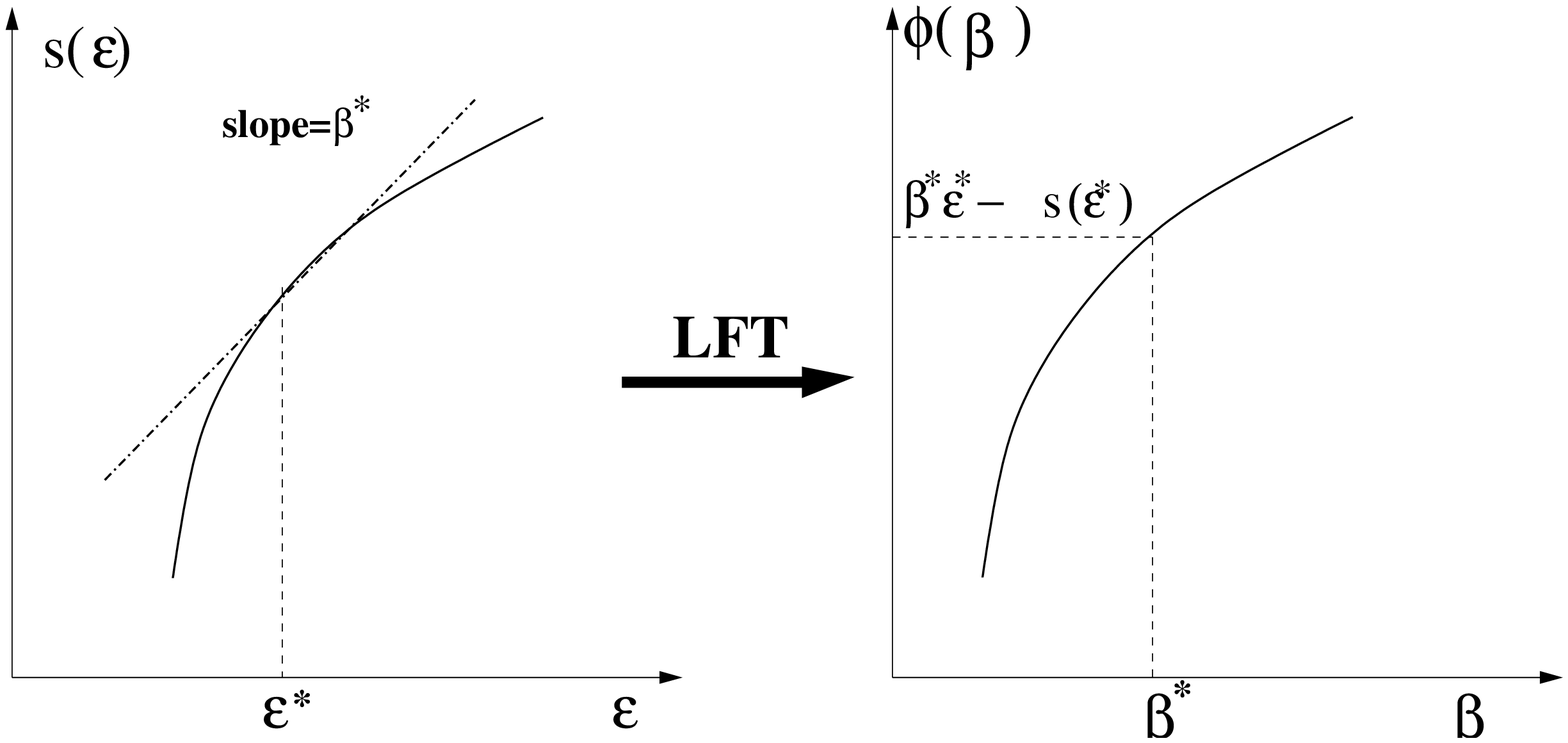}
\includegraphics[width=.7\textwidth]{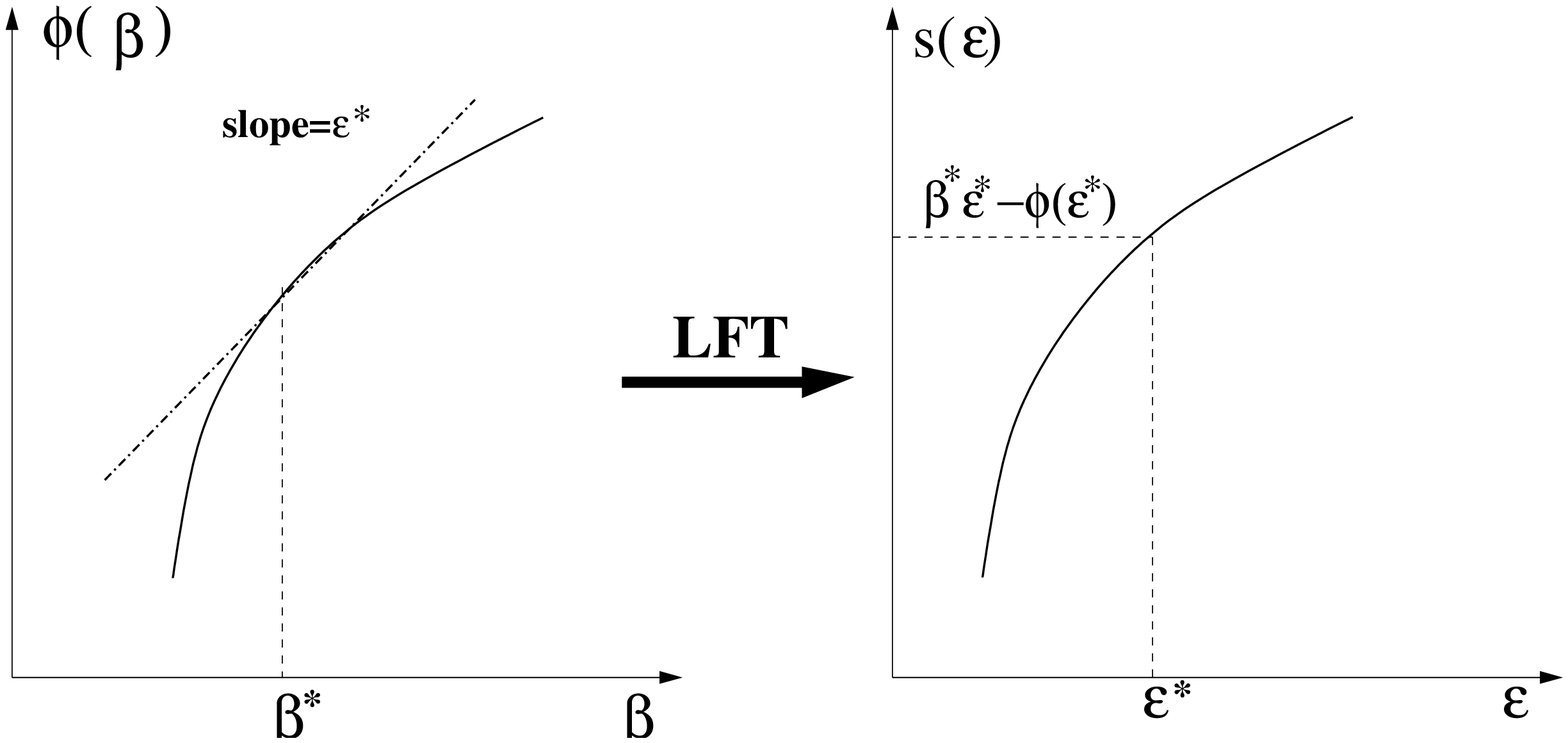}
\end{center}
\caption[] {Relation between the entropy per particle
$s(\varepsilon,n)$ and $\phi(\beta,n)=\beta f(\beta,n)$ (where
$f(\beta,n)$ is the free energy per particle) by the
Legendre-Fenchel Transform (LFT), $n$ is fixed.}
%%%\end{center}
\label{LegendreFenchel}
\end{figure}

The equivalence between microcanonical and canonical ensemble is a
consequence of the concavity of $\phi$ and $s$ and of the relation
between these two functions given by the Legendre-Fenchel Transform
(LFT). Indeed, one can prove that $\phi(\beta,n)$ is the LFT of
$s(\varepsilon,n)$
\begin{equation}
\label{legen1} \phi (\beta,n)=\beta f(\beta,n)= \inf_{\varepsilon}
\left[\beta \varepsilon - s(\varepsilon,n)\right]\,~.
\end{equation}
A brief sketch of the proof goes as follows
\begin{eqnarray}
\exp (-\beta N f(\beta,n))&=&Z(\beta,V,N) \\
 &=& \frac{1}{N!} \int \dd E \int \dd q^{3N} \dd p^{3N}\,
\delta (H(p,q)-E) \exp (-\beta E),\\
   &=& \int \dd E \, \Omega(E,V,N) \exp (- \beta E),\\
   &=& \int \dd E \, \exp \left(-N \left[\beta
\varepsilon -s (\varepsilon,n) \right] \right),
\label{addedforapp}
\end{eqnarray}
where the last equality is valid for large $N$. The saddle point of
the last integral gives formula (\ref{legen1}). Let us remark that
it would not have been necessary to hypothesize the concavity of
$\phi$ in $\beta$ from the beginning, because it follows from the
fact that the LFT of a generic function is a concave function. Also
the inverse LFT holds, since $s(\varepsilon,n)$ is concave in
$\varepsilon$
\begin{equation}
\label{legen2} s(\varepsilon,n)= \inf_{\beta} \left[\beta
\varepsilon - \phi(\beta,n)\right]\, .
\end{equation}
This indeed proves ensemble equivalence, because for each value of
$\beta$ there is a value of $\varepsilon$ that satisfies
Eq.~(\ref{legen1}), and, conversely, for each value of $\varepsilon$
there is a value of $\beta$ satisfying Eq.~(\ref{legen2}).
Fig.~\ref{LegendreFenchel} provides a visual explanation of the
relation between $s$ and $\phi$ and of the correspondence between
$\varepsilon$ and $\beta$.

Relations similar to those described in this subsection explain the
equivalence of other ensembles. For instance Van Hove \cite{vanhove}
proved that the equation of state for a short-range classical system
is the same in the canonical and grand-canonical ensemble. This
implies that isothermal compressibility is positive in both
ensembles.

\subsubsection{Phase separation and Maxwell construction in short-range systems}
\label{maxwellinshort}

It is important to discuss ensemble equivalence in the presence of
phase transitions. We will see that some interesting features arise.
Phase transitions are associated to singularities of thermodynamic
functions \cite{Huang}. Therefore, in the microcanonical and
canonical ensemble, they will be signaled by discontinuities in a
derivative of some order of the entropy $s$ or the rescaled free
energy $\phi$.

\begin{figure}[htbp!]
\begin{center}%
\includegraphics[width=.9\textwidth]{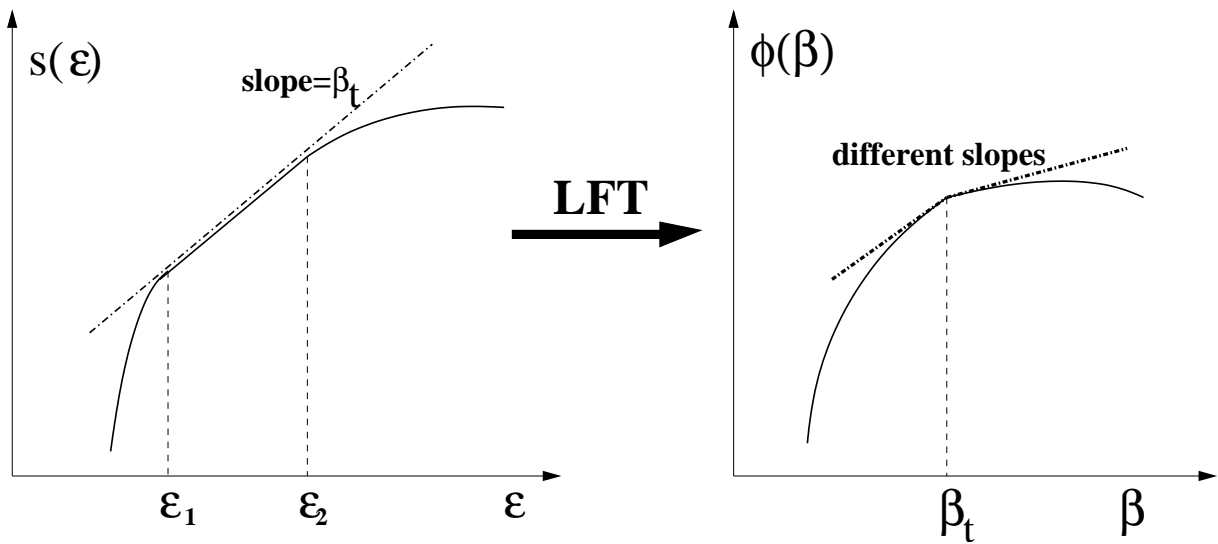}
\end{center}
\caption{Entropy $s(\varepsilon)$ and rescaled free energy
$\phi(\beta)$ in the case of a first order phase transition. The
inverse transition temperature is $\beta_t$ and
$[\varepsilon_1,\varepsilon_2]$ is the energy range of phase
coexistence.}

\label{phaseseparationshortrange}
\end{figure}

Let us concentrate on the dependence of $s$ and $\phi$ on
$\varepsilon$ and $\beta$, respectively. Consider for example a
model in which, for some choice of parameters, $s(\varepsilon)$ has
a zero curvature in some energy range
$[\varepsilon_1,\varepsilon_2]$. At both extremes of this interval
the second derivative of $s$ has a discontinuity (see
Fig.~\ref{phaseseparationshortrange}). Within the range
$[\varepsilon_1,\varepsilon_2]$, the function is not strictly
concave (i.e., Eq.~(\ref{concpro}) is satisfied with an equality).
All energy values in this range correspond to the same value of the
inverse temperature $\beta=\beta_t$, the slope of the straight
segment of $s(\varepsilon)$ in Fig.~\ref{phaseseparationshortrange}.
For each energy in this range, the system separates in two phases of
different energies $\varepsilon_1$ and $\varepsilon_2$. Therefore,
the energy will be given by
\begin{equation}
\label{enerphase} \varepsilon = c\, \varepsilon_1 + (1-c)\,
\varepsilon_2 \,
\end{equation}
where $c$ is the fraction of phase $1$ and, of course, $1-c$ the
fraction of phase $2$. We emphasize that this relation is a direct
consequence of additivity. It is easy to show that the
Legendre-Fenchel transform of $s(\varepsilon)$, the rescaled free
energy (shown in Fig.~\ref{phaseseparationshortrange}) has a
discontinuity in the first derivative with respect to $\beta$ at
$\beta_t$ ($(\dd \phi/\dd \beta)_\mp=\varepsilon_{1,2}$). Following
Ehrenfest's classification, this is a first order phase transition.

Let us remark that in this example there is no one-to-one
correspondence between $\varepsilon$ and $\beta$: several
microcanonical macroscopic states are represented by a single
canonical state. This shows that, in the presence of first-order
phase transitions the relation between the ensembles must be
considered with care. We would not say that the ensembles are
inequivalent in this case, which is a marginal one; therefore we do
not adopt the term used in mathematical physics in this case:
``partial equivalence'' \cite{TouchettePhysRep}.

Tuning the parameters of the model in such a way that the straight
segment in Fig.~\ref{phaseseparationshortrange} reduces to one
point, one recovers a strictly convex entropy function and a
one-to-one correspondence between $\varepsilon$ and $\beta$.
Consequently, the discontinuity in the first derivative of the
rescaled free energy is removed. This is the case of a second order
phase transition.

\begin{figure}[htb]
\begin{center}
\includegraphics[width=.7\textwidth]{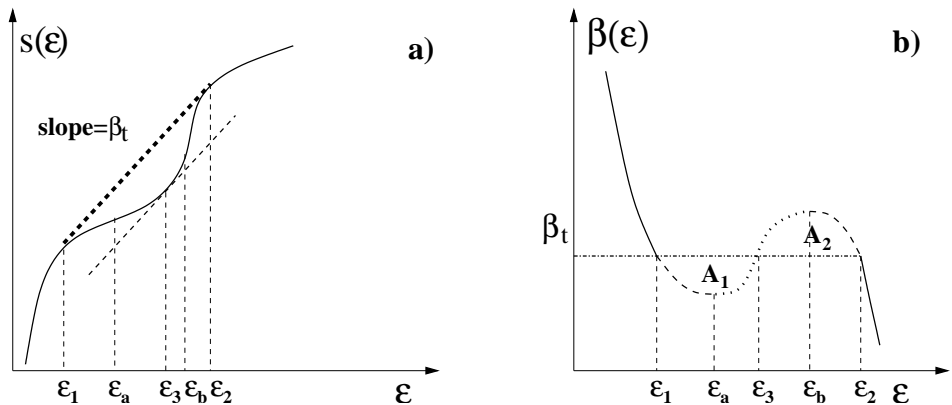}
\end{center}
\caption[]{a) Schematic shape of the entropy $s$ as a function of
the energy $\varepsilon$ (solid line) showing a ``globally" convex
in region in the range $[\varepsilon_1,\varepsilon_2]$, the thick
dashed line realizes the ``concave envelope". b) Inverse temperature
$\beta$ as a function of energy $\varepsilon$. According to the
Maxwell's constructions $A_1=A_2$. The curve $\beta(\varepsilon)$
represents states that are stable (solid line), unstable (dotted
line) and metastable (dashed lines).} \label{concavite}
\end{figure}

It could also happen that the entropy, instead of showing the
straight segment of Fig.~\ref{phaseseparationshortrange}, has a
convex region  (see the full line in Fig. \ref{concavite}a). For
short-range interactions this is what is observed for finite systems
near a phase transition. It has been shown numerically
\cite{Dieter,chomazdd,chomazassisi} and for simple models
\cite{MadameLyndenBell,MadameLyndenBellFirst} that by increasing
system size the entropy approaches the ``concave envelope" which is
constructed by replacing the full line in the energy range
$[\varepsilon_1,\varepsilon_2]$ by the straight thick dashed line in
Fig. \ref{concavite}a. In statistical mechanics, this procedure goes
under the name of Maxwell's construction and is mostly known in
connection with the Van der Waals theory of liquid-gas transition
\cite{Huang}. The relation between the construction of the ``concave
envelope" and the Maxwell construction can be easily established by
looking at the following relation
\begin{equation}\label{se1e2a}
s(\varepsilon_2) = s(\varepsilon_1) +
\int_{\varepsilon_1}^{\varepsilon_2} \dd \varepsilon \,
\beta(\varepsilon) \, ,
\end{equation}
which derives from the definition of the inverse temperature
$\beta=\dd s/\dd \varepsilon$, which is plotted in
Fig.~\ref{concavite}b as a function of energy. Another way of
obtaining $s(\varepsilon_2)$ is by integrating along the thick
dashed line in Fig. \ref{concavite}a.
\begin{equation}\label{se1e2b}
s(\varepsilon_2) = s(\varepsilon_1) + (\varepsilon_2 -
\varepsilon_1) \beta_t \,,
\end{equation}
where $\beta_t=\beta(\varepsilon_1)=\beta(\varepsilon_2)$. This
implies that
\begin{equation}\label{maxcos}
\int_{\varepsilon_1}^{\varepsilon_2} \dd \varepsilon \,
\beta(\varepsilon) = (\varepsilon_2 - \varepsilon_1) \beta_t \,.
\end{equation}
Splitting the integral in two intervals
$[\varepsilon_1,\varepsilon_3)$ and $[\varepsilon_3,\varepsilon_2]$
one gets
\begin{equation}
(\varepsilon_1-\varepsilon_3) \beta_t+
\int_{\varepsilon_1}^{\varepsilon_3} \dd \beta \,
\beta(\varepsilon)=(\varepsilon_2-\varepsilon_3) \beta_t+
\int_{\varepsilon_2}^{\varepsilon_3} \dd \beta \,
\beta(\varepsilon)~. \label{intermedMaxwell}
\end{equation}
The value $\varepsilon_3$ is the one obtained from the entropy by
looking where, in the convex region, the entropy has slope
$\beta_t$. Condition (\ref{intermedMaxwell}) is equivalent to the
equality of the areas $A_1$ and $A_2$ in Fig.~\ref{concavite}.
Introducing the {\it generalized free energy}, which is a function
of both energy and inverse temperature,
\begin{equation}
\widehat{f}(\beta,\varepsilon)=\varepsilon -\frac{1}{\beta}
s(\varepsilon), \label{generalizedf}
\end{equation}
one obtains from Eq.~(\ref{intermedMaxwell}) that
\begin{equation}
\widehat{f}(\beta_t,\varepsilon_1)=\widehat{f}(\beta_t,\varepsilon_2)~.
\label{gibbsconditionf}
\end{equation}

This shows that the requirement that the entropy is concave is
equivalent to Maxwell's equal areas construction and, in  turn,
equivalent to demand that the generalized free energies, computed at
the transition inverse temperature $\beta_t$ and at the two energies
$\varepsilon_1$ and $\varepsilon_2$ which delimit the coexistence
region, are equal (and looking at Fig.~\ref{concavite}a, also equal to $f(\beta_t)$.)

The Maxwell construction is related to the application of a maximum
entropy principle for additive systems. Indeed, for all energies in
the range $(\varepsilon_1, \varepsilon_2)$, the entropy
corresponding to the full line in Fig.~\ref{concavite}a is smaller
than the entropy corresponding to the dashed line at the same
energy. This latter entropy is related to a system which has
performed {\it phase separation} and is therefore obtained as a mixture
composed of a certain fraction of a state with energy
$\varepsilon_1$ and the remaining fraction with energy
$\varepsilon_2$, as in formula (\ref{enerphase}). Having this latter
system a larger entropy, the natural tendency will be to phase
separate. Hence the ``concave envelope" recovers maximum entropy
states.

It should be remarked that the truly ``locally" convex part of the
entropy is the one in the range $[\varepsilon_a,\varepsilon_b]$,
while the range $[\varepsilon_1,\varepsilon_2]$ is ``globally"
convex. One should therefore expect a difference in the properties
of the physical states in the various ranges. Indeed, states in the
range $[\varepsilon_a,\varepsilon_b]$ are {\it unstable} (dotted
line in Fig.~\ref{concavite}b), while states in the ranges
$[\varepsilon_1,\varepsilon_a]$ and $[\varepsilon_b,\varepsilon_2]$
are {\it metastable} (dashed lines in Fig.~\ref{concavite}b): at a
solid-liquid phase transition they would correspond to superheated
solids and supercooled liquids, respectively. While the unstable
states cannot be observed, the metastable states are observable but
are not true equilibrium states, because higher entropy phase
separated states are accessible.

%The important point to emphasize is the following. Since the
%concavity of the entropy
%for short-range systems is proved \cite{ruelle}, a nonconcave entropy
%like in Fig. \ref{concavite}
%can only rise as a consequence of an approximate calculation
%(very often the only possible one).
%The Maxwell construction helps us to recover the real physical
%situation; it would not be
%necessary if exact calculations were possible.

%The concavity properties that we have considered for the intensive quantities $s$ and $\phi$
%hold also for the corresponding extensive quantities. The only caveat might be that, in
%the case of a phase transition as in Figs.~\ref{concavite}, for finite $N$ there could be
%a portion of convexity, that becomes a straight line (in an exact calculation)
%for $N\rightarrow \infty$.

Considering again small systems, the nucleation of a bubble might
lead to creation of a convex entropy region. Indeed, once a bubble
of phase $1$ is nucleated in phase $2$, the energetic cost of the
interface is proportional to the surface, while the energetic gain
is proportional to the volume. If the system is small enough, these
two energies might be comparable, implying that the additivity
property is not satisfied \cite{Binder03,Binder04}.

\subsubsection{Ensemble inequivalence in long-range systems: negative specific heat}
\label{posneg}

As already anticipated in the Introduction, an important physical
property of systems with long-range interactions is that ensembles
can be inequivalent. This means that experiments realized in
isolated systems, described by the microcanonical ensemble, may give
different results from similar experiments performed with well
thermalized systems, for which the canonical ensemble is the
appropriate one. For instance, while the specific heat will turn out
to be always positive for a system in contact with a heat bath, it
might be negative for an isolated long-range system.

When the interactions are long-range, an entropy function with a
convex ``intruder" \cite{Dieter} like the one shown by the solid
line in Fig.~\ref{concavite}a can represent truly stable equilibrium
states. In Sec.~\ref{begmodel}, we will give a concrete example to
illustrate this important property. Here, we will develop some
general considerations which are not specific to a given model.

The construction which has led to the ``concave envelope" for
short-range systems cannot be realized for long-range systems. On
the one hand the same notion of phase is ill defined for long-range
systems (which are inherently inhomogeneous).  On the other hand,
even if a definition of phase would be possible, the lack of
additivity of long-range systems would not allow to obtain a mixed
state and, in particular, to derive relations like
(\ref{enerphase}).

The starting point for the construction of a consistent
thermodynamics of long-range systems is the calculation of
microcanonical entropy associated to a given {\it macrostate}. A
{\em microstate} is defined by the phase space variables of the
system, and thus it refers to a precise microscopic state, while a
{\em macrostate} is described in terms of a few macroscopic or
coarse-grained variables, and then it generally defines a large set
of microscopic states, all of them giving rise to the same values of
the macroscopic variables. The derivation of free energy from
microcanonical entropy using the Legendre-Fenchel transform
(\ref{legen1}) is still valid for long-range systems, ensuring that
the function $\phi$ is concave also for these systems. However, when
the entropy has a convex region, the inversion of the
Legendre-Fenchel transform, Eq.~(\ref{legen2}) does not give the
correct microcanonical entropy, but rather its ``concave envelope"
\cite{TouchettePhysRep}. Physically, this implies a lack of
equivalence of ensembles at the level of macrostates, i.e. all
microcanonical macrostates with energies between $\varepsilon_1$ and
$\varepsilon_2$ do not have a corresponding macrostate in the
canonical ensemble \cite{Touchette2003}.

The existence of a convex ``intruder" in the entropy-energy curve,
as in Fig.~\ref{concavite}a, is associated to the presence of
negative specific heat. Indeed,
\begin{equation}
\label{CV_entropy} \frac{\partial^2 S}{\partial E^2}=-\frac{1}{C_V
T^2}.
\end{equation}
where the heat capacity at fixed volume is $C_V=\partial E/\partial
T$.  Hence, in the energy range $[\varepsilon_a,\varepsilon_b]$, the
convexity of the entropy, $\partial^2 S/\partial E^2>0$, implies
that the heat capacity is negative $C_V<0$. This, in turn, implies
that the conveniently normalized specific heat $c_V=C_V/N$ is also
negative.

In the canonical ensemble, the specific heat is always positive,
even if the interactions are long-range. This is a straightforward
consequence of the concavity of the function $\phi$, which is given
by Eq.~(\ref{legen1}) also for long-range systems. Indeed,
\begin{equation}
\frac{\partial^2 \phi}{\partial \beta^2} = - \frac{c_V}{T^2} < 0~,
\end{equation}
implying that $c_V > 0$.  There is a subtlety related to the
calculation of $c_V$ at $\beta_t$. For a first order phase
transition, since there is a discontinuity of the first derivative
of $\phi$ at $\beta_t$, the specific heat is not well defined and
one rather speaks of latent heat, related to the jump
$(\varepsilon_2 - \varepsilon_1)$ of the energy as shown in
Fig.~\ref{concavite}b. At second order phase transitions, the second
derivative of $\phi$ is instead well defined and is discontinuous at
$\beta_t$.

As explained above, the presence of a convex ``intruder'' like in
Fig.\ref{concavite}a in the entropy of long-range systems does not
imply the appearance of singularities in the entropy, and therefore
it could be doubted that this behavior signals a true phase
transition in the microcanonical ensemble. Since this feature has
been found first for gravitational systems, it is sometimes called
``gravitational phase transition". We will better clarify this issue
analyzing what happens in the canonical ensemble. The rescaled free
energy $\phi(\beta)$ is again expressed by Eq.~(\ref{legen1}) and
the mean value of the energy is
\begin{equation}
\label{canen2} \varepsilon(\beta) = \frac{\partial \phi}{\partial
\beta}~.
\end{equation}
The plot of $\varepsilon(\beta)$ is obtained from the curve in
Fig.~\ref{concavite}b by considering the ordinate $\beta$ as the
control variable. The Maxwell construction is realized by the
horizontal dashed line at $\beta(\varepsilon_1)$. If, in the
canonical ensemble, we start from a value of $\beta$ such that the
energy of the system is less than $\varepsilon_1$ and we gradually
decrease $\beta$, the system will reach the energy $\varepsilon_1$
and then will jump to the energy $\varepsilon_2$, and after the jump
$\beta$ will decrease continuously. Therefore, while in the
microcanonical ensemble there is no singularity of the entropy, in
the canonical ensemble there is a discontinuity of the derivative of
the rescaled free energy $\phi$, corresponding to a jump in the
energy (associated to a latent heat). In the canonical ensemble the
system has therefore a first order phase transition. Equilibrium
macroscopic states with energies in the range
$[\varepsilon_1,\varepsilon_2]$ do not exist, since the lack of
additivity, as we noted, does not allow, contrary to short-range
systems, to have mixtures of states as in Eq.~(\ref{enerphase}). The
temperature of the phase transition in the canonical ensemble is
obtained by the Maxwell construction. Correspondingly,
microcanonical microstates exist in the energy range
$[\varepsilon_1,\varepsilon_2]$ and {\it phase separation} is not
thermodynamically favoured in this ensemble.

The fact that the presence of a canonical first order phase
transition is necessary to obtain ensemble inequivalence was
conjectured in Ref.~\cite{BMR}.  This statement has been put on a
more rigorous basis in Refs.~\cite{Touchette2003,julienfreddyjstat},
analyzing the convexity properties of the entropy $s(\varepsilon)$.
In fact, it has been shown \cite{Touchette2003} that if the rescaled
free energy $\phi(\beta)$ is differentiable, then the entropy
$s(\veps)$ can be obtained by its Legendre-Fenchel transform. This
applies also for second order phase transitions, when the second
derivative of $\phi(\beta)$ is discontinuous. Therefore, in the
presence of a second order phase transition in the canonical
ensemble, the microcanonical and canonical ensembles are equivalent.

In this subsection we have discussed in detail the case where no
singularity are present in the entropy. Although already showing all
the features of ensemble inequivalence, this case is not generic and
we'll discuss in the next subsection a model that has a second order
phase transition in the microcanonical ensemble and still a first
order transition in the canonical ensemble.

\begin{figure}[htb]
\begin{center}
\includegraphics[width=.5\textwidth]{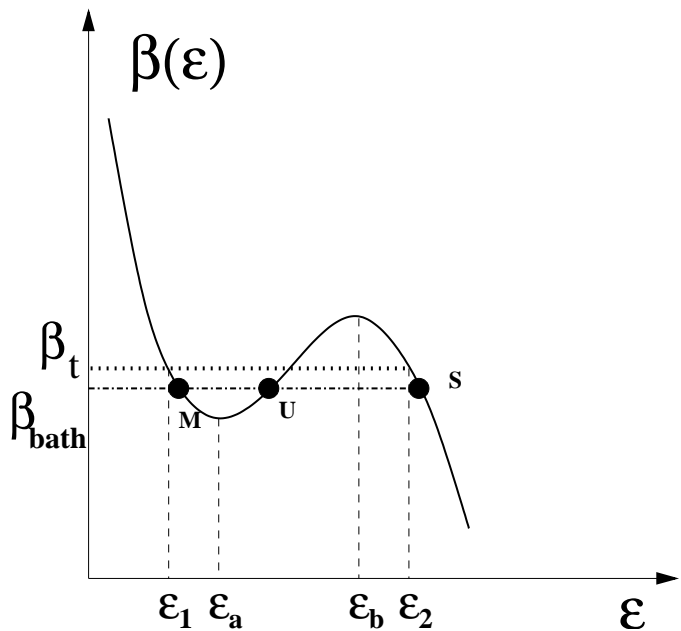}
\end{center}
\caption[]{Inverse temperature $\beta$ as a function of energy
$\varepsilon$. The Maxwell's construction is shown by the dotted
line, while the dash-dotted line indicates the inverse temperature
of the bath $\beta_{bath}$. $U$ denotes an unstable macroscopic
state with negative specific heat, while $M$ and $S$ are metastable
and stable, respectively.} \label{instabilite}
\end{figure}

Let's conclude this subsection with a remark. We have remarked that
energies between $\varepsilon_1$ and $\varepsilon_2$ correspond to
the same value of $\beta$ in the canonical ensemble. It is
interesting to figure out what happens if an initially isolated
system with negative specific heat and with an energy between
$\varepsilon_1$ and $\varepsilon_2$, is put in contact with a heat
bath that has its inverse temperature $\beta_{bath}$. Looking at
Fig.~\ref{instabilite} can be of help to understand the argument. We
consider the case where the energy of the system lies in the range
in which the specific heat is negative
$[\varepsilon_a,\varepsilon_b]$ when the system is put in contact
with the bath. Let us take for instance point $U$ in
Fig.~\ref{instabilite} as an initial point. We are interested to
study the behavior of the system subjected to small perturbations,
so that it can still be considered to be initially close to a
microcanonical system. We see immediately that the system becomes
unstable. In fact, if it gets a small amount of energy from the
bath, its temperature lowers (negative specific heat!), and
therefore further energy will flow from the bath to the system,
inducing a lowering of the system's temperature and then creating an
instability. If, on the contrary, the initial energy fluctuation
decreases the system's energy, its temperature rises, inducing a
further energy flow towards the bath, and, hence, an increase of
system's temperature. Thus, in contact with a heat bath, the system
does not maintain energies in which its microcanonical specific heat
is negative. The flow of energy started by the initial energy
fluctuations stops when the system reaches again the same
temperature of the bath, but at an energy for which its specific
heat is positive. Looking at Fig.~\ref{instabilite}, it is clear
that this could be either outside the range
$[\varepsilon_1,\varepsilon_2]$, i.e. point $S$, or inside this
range, point $M$. This feature is valid for all points $U$ inside
$]\varepsilon_a,\varepsilon_b]$. Once in $M$, the system will be in
a thermodynamically metastable state and a sufficiently large
fluctuation in the energy exchange with the bath will make it leave
this metastable state, ending up again in a state with energy
outside $[\varepsilon_1,\varepsilon_2]$, i.e. point $S$, which has
the same inverse temperature of the bath $\beta_{bath}$. If the
system instead jumps directly from $U$ to $S$, it will stay there
because this point lies on a thermodynamically stable branch.

\subsection{An analytical solvable example: the mean-field Blume-Emery-Griffiths model}
\label{begmodel}

We have presented above the main physical and mathematical aspects
related to ensemble equivalence or inequivalence in the study of
long-range systems. Other mathematical approaches and tools, that
exist, will be presented in connection with concrete examples.
Actually, this subsection is dedicated to a toy model that exhibits
all features that have been discussed so far, in particular ensemble
inequivalence and negative specific heat in the microcanonical
ensemble. Historically, the relation between first order phase
transition and negative specific heat for long-range systems in the
thermodynamic limit was first pointed out in Refs.~\cite{Antoni1,
Antoni2}. The phenomenology we are going to discuss in this section
has been heuristically described in Ref.~\cite{Antoni4}.

\subsubsection{Qualitative remarks}

The Blume-Emery-Griffiths (BEG) model is a lattice spin model with
infinite range, mean-field like interactions whose phase diagram can
be obtained analytically both within the canonical and the
microcanonical ensembles. This study enables one to compare the two
resulting phase diagrams and get a better understanding of the
effect of the non-additivity on the thermodynamic behavior of the
model.

The model we consider is a simplified version of the
Blume-Emery-Griffiths model~\cite{Blume}, known as the Blume-Capel
model, where the quadrupole-quadrupole interaction is absent. The
model is intended to reproduce the relevant features of
superfluidity in He$^3$-He$^4$ mixtures. Recently, it has also been
proposed as a realistic model for metallic
ferromagnetism~\cite{Ayuela}. It is a lattice system (\ref{potlat}),
and each lattice point $i$ is occupied by a spin-1 variable, i.e., a
variable $S_i$ assuming the values $S_i=0,\pm 1$. We will consider
the mean-field version of this model, for which all lattice points
are coupled with the same strength. The Hamiltonian is given by
\begin{equation}
H=\Delta\sum_{i=1}^N S_i^2 -\frac{J}{2N}\left(\sum_{i=1}^N \,
S_i\right)^2 \label{BEG}~,
\end{equation}
where $J>0$ is a ferromagnetic coupling constant and $\Delta>0$
controls the energy difference between the ferromagnetic
$S_i=1,\forall i$, or $S_i=-1,\forall i$, and the paramagnetic,
$S_i=0,\forall i$, states. In the following we will set $J=1$,
without loss of generality since we consider only ferromagnetic
couplings. The paramagnetic configuration has zero energy, while the
uniform ferromagnetic configurations have an energy $(\Delta-1/2)N$.
In the canonical ensemble, the minimization of the free energy
$F=E-TS$ at zero temperature is equivalent to the minimization of
the energy. One thus finds that the paramagnetic state is the more
favorable from the thermodynamic point of view if
$E(\{\pm1\})>E(\{0\})$, which corresponds to $\Delta>1/2$. At the
point $\Delta=1/2$, there is therefore a phase transition; it is a
{\em first} order phase transition since, it corresponds to a sudden
jump of magnetization from the ferromagnetic state to the
paramagnetic state.

{\setlength{\unitlength}{0.5pt}
\begin{picture}(100,320)(-100,0)
\put(60,90){\framebox(200,40){Ferromagnetic state}}
\put(360,90){\framebox(200,40){Paramagnetic state}}
\put(250,30){ $1^{st}$ order PT} %\put(235,10){Phase Transition}
\put(-60,245){$2^{nd}$ order} \put(-40,225){PT}
\put(600,150){$\Delta$} \put(305,150){$\frac{1}{2}$}
\put(40,130){\scriptsize 0}\put(20,280){
$T$}\put(60,235){\scriptsize $2/3$} \put(50,135){\vector(1,0){550}}
\put(50,135){\vector(0,1){150}} \put(310,135){\vector(0,-1){80}}
\put(50,240){\vector(-1,0){30}}
\end{picture}}
\begin{figure}[htb]
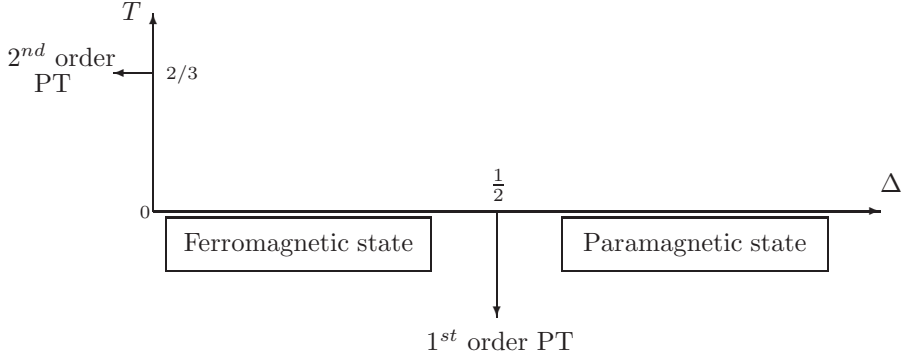

\caption{Elementary features of the phase diagram of the
Blume-Emery-Griffiths model, showing the phase transitions on the
temperature $T$ and local coupling $\Delta$ axis, respectively}
\label{simplebegdiag}
\end{figure}

For vanishingly small ratio $\Delta$, the first term of
Hamiltonian~(\ref{BEG}) can be safely neglected so that one recovers
the Curie-Weiss Hamiltonian~(\ref{hamiladditi}) with spin $1$,
usually introduced to solve the Ising model within the mean-field
approximation. It is well known that such a system has a {\em
second} order phase transition when $T=2/3$ (we remind that we are
adopting units for which $J=1,k_B=1$). Since one has phase
transitions of different orders on the $T$ and $\Delta$ axis
(see Fig.~\ref{simplebegdiag}), one
expects that the $(T, \Delta)$ phase diagram displays a {\em
transition line} separating the low temperature ferromagnetic phase
from the high temperature paramagnetic phase. The transition line is
indeed found to be first order at large $\Delta$ values, while it is
second order at small $\Delta$'s.

\subsubsection{The solution in the canonical ensemble}
The canonical phase diagram of this model in the $(T,\Delta)$ is
known since long time~\cite{Blume0,Capel,Blume}. The partition
function reads
\begin{equation}
Z(\beta,N) =\sum_{\{S_1,\ldots,S_N\}}\exp{\left(-\beta
\Delta\sum_{i=1}^N S_i^2 +\frac{\beta J}{2N}\left(\sum_{i=1}^N
S_i\right)^2\right)}. \label{partition}
\end{equation}
Using the Gaussian identity
\begin{equation}\label{hubb}
\exp(bm^2)=\sqrt{\frac{b}{\pi}}\int_{-\infty}^{+\infty} \dd x
\exp(-bx^2+2mbx),
\end{equation}
(often called the Hubbard-Stratonovich transformation) with
$m=\sum_iS_i/N$ and $b=N\beta J/2$, one obtains
\begin{equation}
Z(\beta,N)=\sum_{\{S_1,\ldots,S_N\}} \exp{\left(-\beta
\Delta\sum_{i=1}^N
S_i^2\right)}\sqrt{\frac{N\beta}{2\pi}}\int_{-\infty}^{+\infty} \dd
x \exp\left(-\frac{N\beta}{2}x^2+mN\beta x\right).
\label{partitionbis}
\end{equation}
One then easily gets
\begin{equation}
Z(\beta,N)=\sqrt{\frac{N\beta}{2 \pi}}\int_{-\infty}^{+\infty} \dd x
\exp(-N\beta \tilde{f}(\beta,x)) \label{part}
\end{equation}
where
\begin{equation}
 \tilde{f}(\beta,x)={1\over 2} x^2 -\frac{1}{\beta}\ln[1+e^{-\beta
\Delta}(e^{\beta x}+e^{-\beta x})]. \label{free}
\end{equation}
The integral in (\ref{part}) can be computed using the saddle point
method where $N$ is the large parameter. The free energy is thus
\begin{equation}
\label{minvx} f(\beta)=\inf_x \tilde{f}(\beta,x).
\end{equation}
It is not difficult to see that the spontaneous magnetization
$\langle m \rangle$ is equal to the value of $x$ at the extremum
which appears in Eq. (\ref{minvx}). We should also note that
$\tilde{f}(\beta,x)$ is even in $x$; therefore, if there is a value
of $x$ different from $0$ realizing the extremum, also the opposite
value realizes it. This means that if the minimum $\overline{x}$ is
equal to $0$ the system is in the paramagnetic phase, while if
$\overline{x} \ne 0$ the system is in the ferromagnetic phase, where
it can assume a positive or a negative magnetization. The phase
diagram, in the $(T,\Delta)$ plane, is then divided into a
paramagnetic region ($\overline{x}=0$) and a ferromagnetic one
($\overline{x}\ne 0$).

\begin{figure}[htb]
\begin{center}
\includegraphics[width=.45\textwidth]{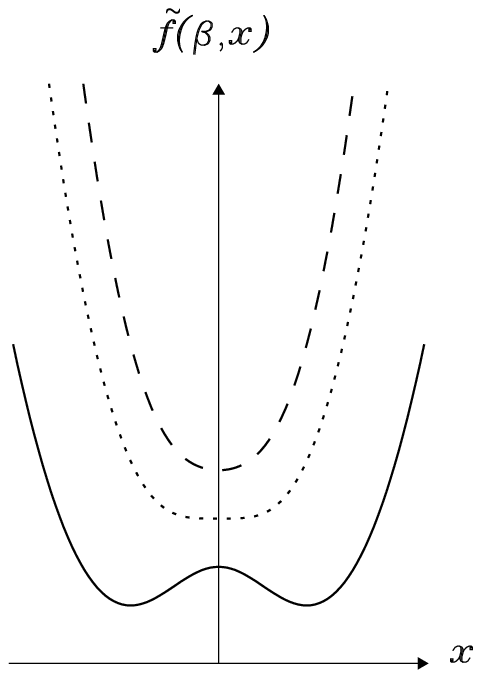}
\includegraphics[width=.45\textwidth]{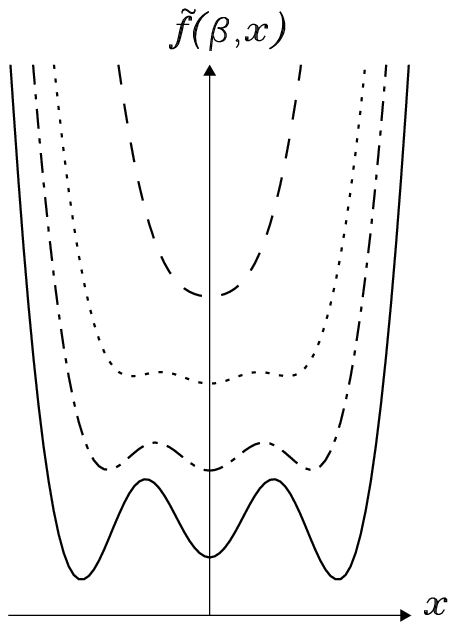}
\end{center}
\vskip -1truecm \caption{Free energy $\tilde{f}(\beta,x)$ {\em vs}
$x$ for different values of the inverse temperature $\beta=1/T$.
Left panel shows the case of a {\em second} order phase transition,
temperature values $T=$0.8 (dashed line), 0.63 (dotted), 0.4 (solid)
when $\Delta=0.1$ are displayed. Right panel shows the case of a
{\em first} order phase transition with $\Delta=0.485$ when $T=0.5$
(dashed), 0.24 (dotted), 0.21 (dash-dotted), 0.18 (solid).}
\label{ftilde}
\end{figure}

Let us now show that the two regions are divided by a second order
phase transition line and a first order phase transition line, which
meet at a tricritical point. As in the Landau theory of phase
transitions, we find a second order transition line by a power
series expansion in $x$ of the function $\tilde{f}(\beta,x)$ in Eq.
(\ref{free}). The second order line is obtained by equating to zero
the coefficient of $x^2$, i.e., by the relation
\begin{equation}\label{CriticaLine}
A_c \equiv \beta  - {1 \over 2} e^{\beta \Delta} -1 =0 \, ,
\end{equation}
provided that the coefficient of $x^4$ is positive, i.e., provided
that
\begin{equation}\label{coefx4}
B_c \equiv 4 - e^{\beta \Delta} >0 \, .
\end{equation}
The tricritical point is obtained when $A_c = B_c =0$. This gives
$\Delta=\ln(4)/3 \simeq 0.4621$ and $\beta=3$. The continuation of
the critical line after the tricritical point is the first order
phase transition line, which can be obtained by finding numerically
the local maximum value $\overline{x}\ne 0$ (magnetic phase) for
which $\tilde{f}(\beta,x)$ is equal to $\tilde{f}(\beta,0)$
(paramagnetic phase), i.e., by equating the free energies of the
ferromagnetic and the paramagnetic phases. The behavior of the
function $\tilde{f}(\beta,x)$ as $\beta$ varies is shown in
Fig.~\ref{ftilde}: panel a) represents the case of a second order
phase transition ($\Delta=0.1$) and panel b) the case of a first
order phase transition ($\Delta=0.485$).

\begin{figure}[htb]
\begin{center}
\includegraphics[width=.5\textwidth]{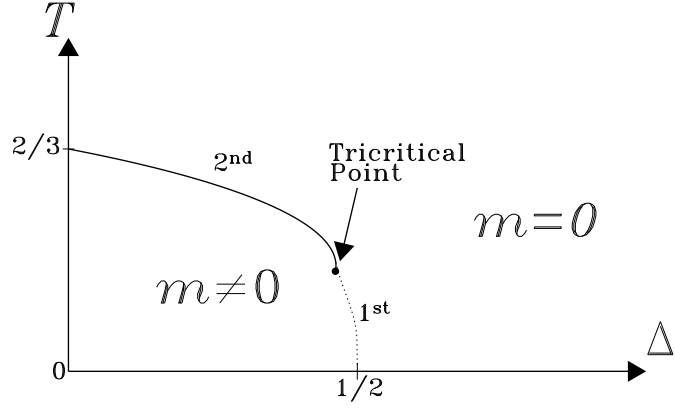}
\end{center}
%\vskip -1truecm
\caption{Phase diagram of the Blume-Emery-Griffiths model in the
canonical ensemble. The second order transition line (solid) ends at
the tricritical point ($\bullet$), where the transition becomes
first order (dotted).} \label{phasediagramBEG}
\end{figure}

A picture of the phase diagram is shown in
Fig.~\ref{phasediagramBEG}.

\subsubsection{The solution in the microcanonical ensemble}

The derivation of the phase diagram of the BEG model (\ref{BEG}) in
the {\it microcanonical ensemble} relies on a simple counting
problem \cite{BMR}, since all spins interact with equal strength,
independently of their mutual distance. A given macroscopic
configuration is characterized by the numbers $N_+, N_-,N_0$ of up,
down and zero spins, with $N_+ + N_- + N_0 = N$. The energy $E$ of
this configuration is only a function of $N_+, N_-$ and $N_0$ and is
given by
\begin{equation}
\label{Energy} E=\Delta Q - {1 \over {2N}}M^2~,
\end{equation}
where $Q=\sum_{i=1}^N S_i^2=N_+ + N_-$ (the quadrupole moment) and
$M=\sum_{i=1}^N S_i=N_+ - N_-$ (the magnetization) are the two order
parameters. The number of microscopic configurations $\Omega$
compatible with the macroscopic occupation numbers $N_+, N_-$ and
$N_0$ is
\begin{equation}
\label{Omega} \Omega = {{N!} \over {{N_+!}{N_-!}{N_0!}}}~.
\end{equation}
Using Stirling's approximation in the large $N$ limit, the entropy,
$S=\ln \Omega$, is given by
\begin{eqnarray}
\label{Entropy1}
S&=&- N \left[(1-q)\ln (1-q) + {1 \over 2}(q+m) \ln (q+m) %\nonumber \\ &+&
+{1 \over 2}(q-m) \ln (q-m) - q \ln 2\right]~,
\end{eqnarray}
where $q=Q/N$ and $m = M/N$ are the quadrupole moment and the
magnetization per site, respectively. Equation (\ref{Energy}) may be
written as
\begin{equation}
q = 2K\varepsilon +Km^2~, \label{qofm}
\end{equation}
where $K=1/{(2\Delta)}$. Using this relation, the entropy per site
$\tilde{s}=S/N$ can be expressed in terms of $m$ and $\varepsilon$,
as follows
\begin{eqnarray}\label{entpsite}
\tilde{s}(\varepsilon,m)&=&- (1-2K\varepsilon - Km^2)\ln
(1-2K\varepsilon - Km^2) -
{1 \over 2} (2K\varepsilon + Km^2 + m) \ln (2K\varepsilon + Km^2 + m) \nonumber \\
&&-{1 \over 2}(2K\varepsilon + Km^2 -m) \ln (2K\varepsilon + Km^2 -
m) + (2K\varepsilon + Km^2) \ln 2~.
\end{eqnarray}
At fixed $\varepsilon$, the value of $m$ which maximizes the entropy
corresponds to the equilibrium magnetization. The corresponding
equilibrium entropy
\begin{equation}
\label{entropyequilibriumBEG} s(\varepsilon)=\sup_m
\tilde{s}(\varepsilon,m)
\end{equation}
contains all the relevant information about the thermodynamics of
the system in the microcanonical ensemble. As usual in systems where
the energy per particle is bounded from above, the model has both a
positive and a negative temperature region: entropy is a one humped
function of the energy. In order to locate the continuous transition
line, one develops $\tilde{s}(\varepsilon,m)$ in powers of $m$, in
analogy with what has been done above for the canonical free energy
\begin{equation}
\label{EntropyExpansion} \tilde{s}=\tilde{s}_0 +A_{mc}\,m^2 +
B_{mc}\,m^4 + O(m^6)~,
\end{equation}
where
\begin{equation}
\label{s0}
\tilde{s}_0=\tilde{s}(\varepsilon,m=0)=-(1-2K\varepsilon)\ln(1-2K\varepsilon)
- 2K\varepsilon \ln (K\varepsilon)~,
\end{equation}
and
\begin{eqnarray}
\label{AB}
A_{mc}&=&-K \ln {K\varepsilon \over {(1-2K\varepsilon)}} -{1 \over {4K \varepsilon}}~,\\
B_{mc}&=&-{K \over {4 \varepsilon(1-2K\varepsilon)}}+{1 \over {8
K\varepsilon^2}}- {1\over {96 K^3 \varepsilon^3}}~.
\end{eqnarray}
In the paramagnetic phase both $A_{mc}$ and $B_{mc}$ are negative,
and the entropy is maximized by $m=0$. The continuous transition to
the ferromagnetic phase takes place at $A_{mc}=0$ for $B_{mc}<0$. In
order to obtain the critical line in the $(T,\Delta)$ plane, we
first observe that temperature is calculable on the critical line
($m=0$) using (\ref{deftemperaturemicro}) and (\ref{s0}). One gets
\begin{equation}
\label{Temperature} \frac{1}{T} = 2K \, \ln {{1- 2K\varepsilon}
\over K\varepsilon}~.
\end{equation}
Requiring now that $A_{mc}=0$, one gets the following expression for
the critical line
\begin{equation}
\label{MicroCritical} \beta =\frac{\exp[{{\beta}/{(2K)}}]}{2} +1~.
\end{equation}
Equivalently, this expression may be written as $\beta = 1
/(2K\varepsilon)$. The microcanonical critical line thus coincides
with the critical line (\ref{CriticaLine}) obtained for the
canonical ensemble. The tricritical point of the microcanonical
ensemble is obtained at $A_{mc}=B_{mc}=0$. Combining these equations
with Eq.~(\ref{Temperature}), one finds that, at the tricritical
point, $\beta$ satisfies the equation
\begin{equation}
\label{MicroTricritical} \frac{K^2}{2 \beta^2} \left[1+2 \exp
\left(-\frac{\beta}{2K} \right) \right] -\frac{K}{2 \beta} + {1
\over 12} =0~.
\end{equation}
Equations (\ref{MicroCritical}) and (\ref{MicroTricritical}) yield a
tricritical point at $K \simeq 1.0813$, $\beta= 3.0272$. This has to
be compared with the canonical tricritical point located at
$K=1/(2\Delta)=3/ \ln (16) \simeq 1.0820$, $\beta = 3$. The two
points, although very close to each other, do not coincide. The
microcanonical critical line extends beyond the canonical one. This
feature, which is a clear indication of ensemble inequivalence, was
first found analytically for the BEG model \cite{BMR} and later
confirmed for gravitational
models~\cite{ChavanisHouches,Chavanisreview2006}. The non coincidence
of microcanonical and canonical tricritical points is a generic feature,
as proven in Ref.~\cite{julienfreddyjstat}.

\subsubsection{Inequivalence of ensembles}
\label{ineqbeg}

\begin{figure}[htb]
\begin{center}
\includegraphics[width=.7\textwidth]{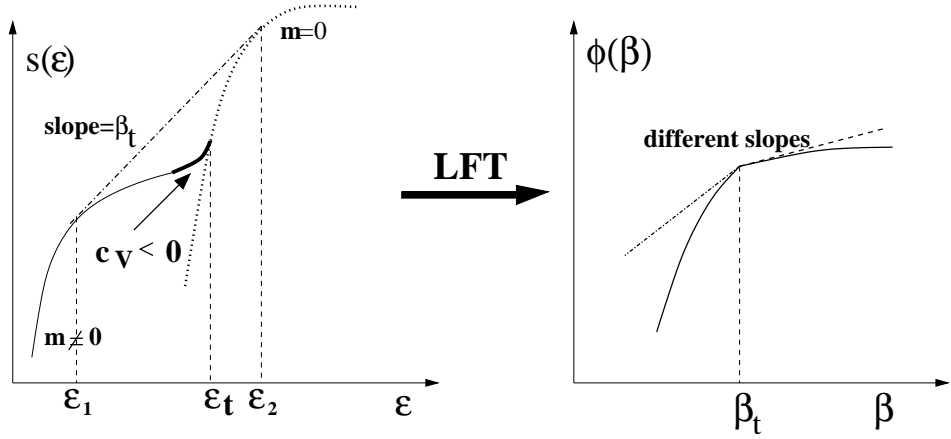}
\end{center}
\caption{Left graph: schematic plot of the entropy $s(\varepsilon)$
as a function of energy density $\varepsilon$ for the BEG model in a
case where negative specific heat coexists with a temperature jump.
The dash-dotted line is the concave envelope of $s(\varepsilon)$ and
the region with negative specific heat $c_V<0$ is explicitly
indicated by the thick line. Right graph: Rescaled free energy
$\phi(\beta)$: the first order phase transition point $\beta_t$ is
shown.} \label{entropyBEG}
\end{figure}

We have already discussed in general terms the question of ensemble
equivalence or inequivalence in Secs.~\ref{equivshort} and
\ref{posneg}. Inequivalence is associated to the existence of a
convex region of the entropy as a function of energy. This is
exactly what happens for the BEG model in the region of parameters
$1< K < 3 \ln (16)$. Since the interesting region is extremely
narrow for this model \cite{BMR}, it is more convenient to plot a
schematic representation of the entropy and of the free energy (see
Fig.~\ref{entropyBEG}). We show what happens in a region of $K$
where both a {\it negative specific heat} and a {\it temperature
jump} are present. The entropy curve consists of two branches: the
high energy branch is obtained for $m=0$ (dotted line), while the
low energy one is for $m \neq 0$ (full line). The $m=0$ branch has
been extended also in a region where it corresponds to metastable
states, just to emphasize that these correspond to a smaller entropy
and that it remains a concave function overall the energy range. We
have not extended the $m \neq 0$ branch in the high energy region
not to make the plot confusing: it would also correspond to a
metastable state. The two branches merge at an energy value
$\varepsilon_t$ where the left and right derivatives do not
coincide; hence microcanonical temperature is different on the two
sides, leading to a {\it temperature jump}. It has been proven in
Ref.~\cite{julienfreddyjstat}, that for all types of bifurcation the
temperature jump is always negative. In the low energy branch, there
is a region where entropy is locally convex (thick line in
Fig.~\ref{entropyBEG}), giving a {\it negative specific heat}
according to formula (\ref{CV_entropy}). The convex envelope, with
constant slope $\beta_t$ is also indicated by the dash-dotted line.
In the same figure, we plot the rescaled free energy $\phi(\beta)$,
which is a concave function, with a point $\beta_t$ where left and
right derivatives (given by $\varepsilon_1$ and $\varepsilon_2$
respectively) are different. This is the first order phase
transition point in the canonical ensemble.

\begin{figure}[htb]
\begin{center}
\includegraphics[width=.4\textwidth]{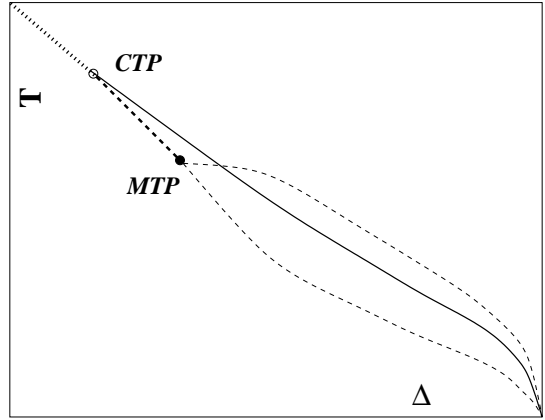}
\end{center}
\caption[]{A schematic representation of the phase diagram, where we
expand the region around the canonical (CTP) and the microcanonical
(MTP) tricritical points. The second order line, common to both
ensembles, is dotted, the first order canonical transition line is
solid and the microcanonical transition lines are dashed (with the
bold dashed line representing a continuous transition).}
\label{schematic}
\end{figure}

\begin{figure}[htb]
\begin{center}
\includegraphics[width=.9\textwidth]{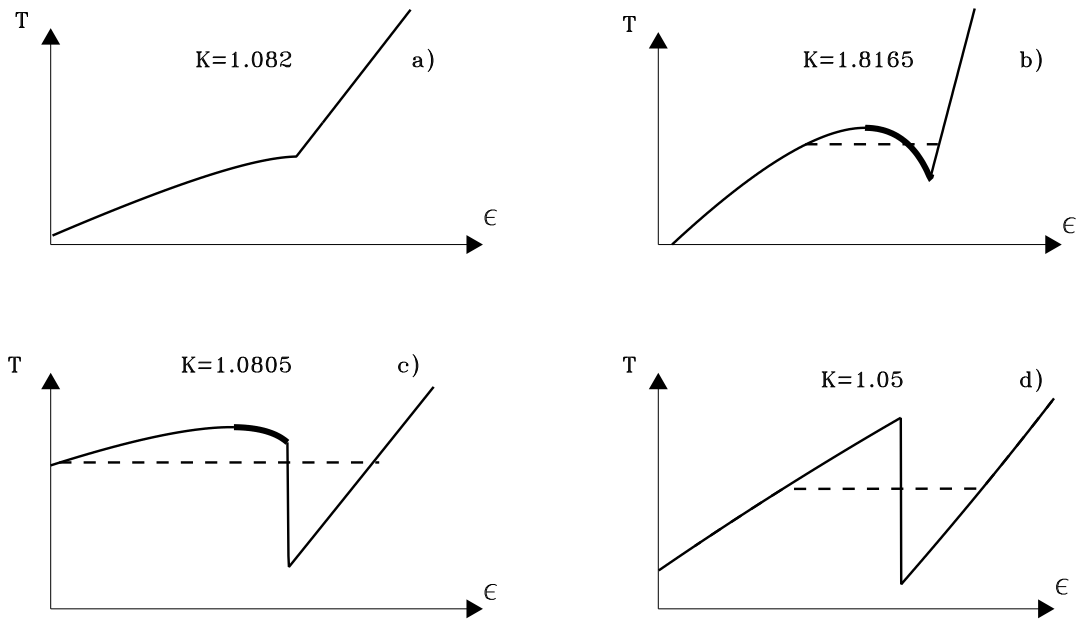}
\end{center}
\vskip -1truecm \caption[]{The temperature-energy relation in the
microcanonical ensemble for different values of $K$. The dashed
horizontal line is the Maxwell construction in the canonical
ensemble and identifies the canonical first order transition
temperature at the point where two minima of the free energy
coexist. Thick lines identify negative specific heat in the
microcanonical ensemble. We don't report the numerical values on the
axes for the readability of the figure.} \label{tvse}
\end{figure}

A schematic phase diagram near the canonical tricritical point (CTP)
and the microcanonical one (MTP) is given in Fig.~\ref{schematic}.
In the region between the two tricritical points, the canonical
ensemble yields a first order phase transition at a higher
temperature, while in the microcanonical ensemble the transition is
still continuous. It is in this region that negative specific heat
appears. Beyond the microcanonical tricritical point, temperature
has a jump at the transition energy in the microcanonical ensemble.
The two lines emerging on the right side from the MTP correspond to
the two limiting temperatures which are reached when approaching the
transition energy from below and from above (see Fig.~\ref{tvse}c
and \ref{tvse}d). The two microcanonical temperature lines and the
canonical first order phase transition line all merge on the $T=0$
line at $\Delta=1/2$.

To get a better understanding of the microcanonical phase diagram
and also in order to compare our results with those obtained for
self-gravitating systems~\cite{ChavanisHouches,Chavanisreview2006}
and for finite systems~\cite{chomazdd,chomazassisi,Dieter,grossdd},
we consider the temperature-energy relation $T(\varepsilon)$ (also
called in the literature ``caloric curve"). Also this curve has two
branches: a high energy branch (\ref{Temperature}) corresponding to
$m=0$, and a low energy branch obtained from
(\ref{deftemperaturemicro}) using the spontaneous magnetization
$m_s(\varepsilon)\neq 0$. At the intersection point of the two
branches, the two entropies become equal. However, their first
derivatives at the crossing point can be different, resulting in a
jump in the temperature, i.e. {\it a microcanonical first order
transition}. When the transition is continuous in the microcanonical
ensemble, i.e. the first derivative of the entropy branches at the
crossing point are equal, BEG model always displays, at variance
with what happens for gravitational systems, a discontinuity in the
second derivative of the entropy. This is due to the fact that here
we have a true symmetry breaking transition
\cite{julienfreddyjstat}. Fig.~\ref{tvse} displays the
$T(\varepsilon)$ curves for decreasing values of $K$. For
$K=3/\ln(16)$, corresponding to the canonical tricritical point, the
lower branch of the curve has a zero slope at the intersection point
(Fig.~\ref{tvse}a). Thus, the specific heat of the ordered phase
diverges at this point. This effect signals the canonical
tricritical point through a property of the microcanonical ensemble.
Decreasing $K$, down to the region between the two tricritical
points, a {\it negative specific heat} in the microcanonical
ensemble first arises ($\partial T/\partial \varepsilon <0$), see
Fig.~\ref{tvse}b. At the microcanonical tricritical point, the
derivative $\partial T/\partial \varepsilon$ of the lower branch
diverges at the transition point, yielding a vanishing specific
heat. For smaller values of $K$, a jump in the temperature appears
at the transition energy (Fig.~\ref{tvse}c). The lower temperature
corresponds to the $m=0$ solution (\ref{Temperature}) and the upper
one is given by $\exp(\beta/2K)= 2(1-q^*)/\sqrt{(q^*)^2-(m^*)^2}$,
where $m^*,q^*$ are the values of the order parameters of the
ferromagnetic state at the transition energy. The negative specific
heat branch disappears at even smaller values of $K$, leaving just a
temperature jump (see Fig.~\ref{tvse}d). In the $K \to 1$ limit the
low temperature branch, corresponding to $q=m=1$ in the limit,
shrinks to zero and the $m=0$ branch (\ref{Temperature}) occupies
the full energy range.

\subsection{Entropy and free energy dependence on the order parameter}

In this section, we will discuss in detail the dependence of both
the canonical free energy and the microcanonical entropy on the
order parameter. This will allow to understand more deeply the
relation between the two ensembles by revisiting Maxwell
constructions. Besides that, we will also discover an interesting
physical effect, {\it negative susceptibility}, of which we will
give an explicit example, see Sec.~\ref{generalizedHMF}.

\subsubsection{Basic definitions}
\label{orddepen}

Let us start from Eq.~(\ref{entpsite}), that gives the entropy per
site $\tilde{s}(\varepsilon,m)$, for the BEG model, as a function of
the energy per site $\varepsilon$ \emph{and} the magnetization $m$.
This entropy is proportional to the logarithm of the number of
configurations which have a given energy \emph{and} a given
magnetization. In the general expression Eq.~(\ref{microlattice}),
these configurations can be obtained by adding a further Dirac delta
function in the integrand, so that only the configurations with a
given $m$ would be counted. Thus one gets
\begin{equation}
\label{sofem} \tilde{s}(\veps,m) = \lim_{N\rightarrow \infty}
\frac{1}{N} \ln \int \prod_i^N \dd {\bf q}_i \, \delta
\left(E-U(\{{\bf q}_i\})\right) \delta \left(Nm-M(\{{\bf
q}_i\})\right)
\end{equation}
where $M$ is the total magnetization corresponding to configuration
$\{{\bf q}_i\}$.  For spin models, the local variable takes discrete
values ${\bf q}_i \equiv S_i$, hence $M=\sum_i S_i$ and the integral
in Eq.~(\ref{sofem}) is replaced by a discrete sum. Besides that,
there is no kinetic energy, hence only the potential energy appears
$H=U$. The calculation of entropy (\ref{sofem}) is often an
intermediate step in the calculation of $s(\varepsilon)$, with $M$
the order parameter. In order to get $s(\varepsilon)$, one computes
the global maximum of the constrained entropy (\ref{sofem}). In the
thermodynamic limit, this procedure is fully justified, since the
relative contribution of all configurations corresponding to values
of the order parameter that are different from the one realizing the
global maximum, vanishes. This is what has been done in our study of
the BEG model, e.g. in Eq.~(\ref{entropyequilibriumBEG}).

In the canonical ensemble, the computation of the partition function
for a given value of the order parameter, i.e. for the system at a
given temperature \emph{and} at a given magnetization $m$, can be
obtained by adding a Dirac delta function to the integrand in
Eq.~(\ref{canon})
\begin{equation}
\label{fofbetam} \tilde{f}(\beta,m)= -\frac{1}{\beta}
\lim_{N\rightarrow \infty} \frac{1}{N} \ln \sum_{\{S_1,\ldots,S_N\}}
\dd S_i \exp \left[-\beta H(\{ S_i\})\right] \delta \left( Nm-M(\{
S_i\})\right) \, .
\end{equation}
Thus, the free energy depends on both $\beta$ \emph{and} the
magnetization $m$. Finally, a relation analogous to
Eq.~(\ref{legen1}) holds between the entropy and free energy
\begin{equation}
\label{legen3} \tilde{f}(\beta,m)= \inf_{\varepsilon} \left[
\varepsilon - \frac{1}{\beta} \tilde{s} (\varepsilon,m)\right] \, ,
\end{equation}
valid, as before, for all systems, independently of the range of the
interactions.

We are therefore led to the introduction of the {\it generalized
free energy} (see Eq.(\ref{generalizedf}))
\begin{equation}
\label{freeext}
\widehat{f}(\beta,\varepsilon,m)=\varepsilon-\frac{1}{\beta}\tilde{s}(\varepsilon,m)
\, ,
\end{equation}
that will be used, as previously, to study the relation between
microcanonical and canonical equilibrium states. Needless to say, it
is not at all guaranteed that the function
$\tilde{s}(\varepsilon,m)$ can be easily derived, in general, as we
have done for the BEG model; nevertheless this general discussion is
useful to show how the properties of this function explain the
occurrence, or not, of ensemble equivalence.

In the microcanonical ensemble, the entropy $s(\varepsilon)$ of the
system at a given energy $\varepsilon$ is given by formula
(\ref{entropyequilibriumBEG}). In the canonical ensemble, the free
energy $f(\beta)$ of the system at a given inverse temperature
$\beta$ will be given by
\begin{equation}
\label{canoproblem} f(\beta)=\inf_{\varepsilon,m}
\widehat{f}(\beta,\varepsilon,m) = \inf_m \tilde{f}(\beta,m)=
\inf_{\varepsilon,m}\left[\varepsilon -\frac{1}{\beta}
\tilde{s}(\varepsilon,m)\right]~,
\end{equation}
as can be easily deduced by Eqs. (\ref{entropyequilibriumBEG}),
(\ref{legen1}), (\ref{legen3}) and (\ref{freeext}). The two extremal
problems (\ref{entropyequilibriumBEG}) and (\ref{canoproblem}), that
basically contain the single function $\tilde{s}(\varepsilon,m)$,
can be employed to study ensemble equivalence. Suppose we fix
$\beta$ and solve the extremal problem (\ref{canoproblem}), finding
the values of $\varepsilon$ and $m$ that realize an extremum. Then,
we will have ensemble equivalence if the same values will realize
the extremum in formula (\ref{entropyequilibriumBEG}) while at the
same time the derivative $\partial s /\partial \varepsilon$ will be
equal to the fixed value of $\beta$. In conclusion, we seek in both
extremal problems the solution of the following first order
conditions
\begin{eqnarray}
\frac{\partial \tilde{s}}{\partial m}&=&0 \label{variationala} \\
\frac{\partial \tilde{s}}{\partial \varepsilon}&=&\beta \, .
\label{variationalb}
\end{eqnarray}
We denote by $\varepsilon^*(\beta),m^*(\beta)$ the solution of the
variational problem (\ref{variationala}) and (\ref{variationalb}).
Using (\ref{variationalb}), it is straightforward to verify that
\begin{equation}
\frac{\dd(\beta f)}{\dd\beta}=\varepsilon^*(\beta),
\end{equation}
meaning that the value of $\varepsilon$ at the extremum is indeed
the canonical mean energy.

However, we have to consider also the stability of these extrema. We
denote derivatives by subscripts, e.g., $\tilde{s}_{m}$ is the first
derivative of $\tilde{s}$ with respect to $m$. The only condition
required by (\ref{entropyequilibriumBEG}) is that
$\tilde{s}_{mm}<0$. In order to discuss the stability of the
canonical solution, one has to determine the sign of the eigenvalues
of the Hessian of the function to be minimized in
(\ref{canoproblem}). The Hessian is
\begin{equation}
{\cal H}=-\frac{1}{\beta} \left( \begin{array}{cc}
\tilde{s}_{mm} & \tilde{s}_{m\varepsilon} \\
\tilde{s}_{\varepsilon m} & \tilde{s}_{\varepsilon \varepsilon}
\end{array} \right)~.
\end{equation}
The extremum is a minimum if and only if both the determinant and
the trace of the Hessian are positive
\begin{eqnarray}
-\tilde{s}_{\varepsilon \varepsilon}-\tilde{s}_{mm}&>&0 \label{condstab1}\\
\tilde{s}_{\varepsilon
\varepsilon}\tilde{s}_{mm}-\tilde{s}_{m\varepsilon}^2&>&0~.
\label{condstab2}
\end{eqnarray}
This implies that $\tilde{s}_{\varepsilon \varepsilon}$ and
$\tilde{s}_{mm}$ must be negative, and moreover
$\tilde{s}_{\varepsilon
\varepsilon}<-\tilde{s}_{m\varepsilon}^2/|\tilde{s}_{mm}|$. This has
strong implications on the canonical specific heat, which must be
positive, as it has been shown on general grounds in subsection
\ref{posneg}, see Eq. (\ref{cvpos}). Let us prove it by using the
variational approach, instead of the usual Thirring
argument~\cite{Thirringdd,Thirring}, that uses the expression of the
canonical partition sum. Indeed, taking the derivatives of
Eqs.~(\ref{variationala}) and (\ref{variationalb}) with respect to
$\beta$, after having substituted into them
$\varepsilon^*(\beta),m^*(\beta)$, one gets
\begin{eqnarray}
\tilde{s}_{\varepsilon \varepsilon}\frac{\dd \varepsilon^*}{\dd
\beta}+\tilde{s}_{\varepsilon m}\frac{\dd m^*}
{\dd \beta}&=&1\\
\tilde{s}_{m\varepsilon}\frac{\dd \varepsilon^*}{\dd
\beta}+\tilde{s}_{mm}\frac{\dd m^*}{\dd \beta}&=&0, \label{second}
\end{eqnarray}
where all second derivatives are computed at
$\varepsilon^*(\beta),m^*(\beta)$. Recalling now that the specific
heat per particle at constant volume is
\begin{equation}
c_V=\frac{\dd \varepsilon^*}{\dd T}=-\beta^2\frac{\dd
\varepsilon^*}{\dd \beta}~,
\end{equation}
one gets
\begin{equation}
c_V=\beta^2\frac{\tilde{s}_{mm}}{\tilde{s}_{\varepsilon
m}^2-\tilde{s}_{\varepsilon \varepsilon}\tilde{s}_{mm}}, \label{cv}
\end{equation}
which is always positive if the stability conditions
(\ref{condstab1}) and (\ref{condstab2}) are satisfied. Since the
stability condition in the microcanonical ensemble only requires
that $\tilde{s}_{mm}<0$, a canonically stable solution is also
microcanonically stable. The converse is not true: one may well have
an entropy maximum, $\tilde{s}_{mm}<0$, which is a free energy
saddle point, with $\tilde{s}_{\varepsilon \varepsilon}>0$. This
implies that the specific heat~(\ref{cv}) can be negative.

The above results are actually quite general, provided the canonical
and microcanonical solutions are expressed through variational
problems of the type (\ref{entropyequilibriumBEG}) and
(\ref{canoproblem}). The extrema, and thus the caloric curves
$T(\varepsilon)$, are the same in the two ensembles, but the
stability of the different branches is different. This aspect was
first discussed by  Katz~\cite{Katz} in connection with
self-gravitating systems (see also Ref.~\cite{ChavanisHouches}).

\subsubsection{Maxwell construction in the microcanonical ensemble}
\label{Maxwellisback}

We have already discussed Maxwell's construction in
subsection~\ref{maxwellinshort}. We have shown that, for short range
systems, where microcanonical and canonical ensembles are always
equivalent, Maxwell's construction derives from the concave envelope
construction for the microcanonical entropy. This in turn is a
consequence of {\it additivity} and of the presence of a first order
phase transition in the canonical ensemble. Here, we discuss
Maxwell's construction for long-range systems, where the
microcanonical entropy can have a stable convex intruder, leading to
ensemble inequivalence.

The study of the BEG model, in subsection \ref{begmodel}, has
emphasized the presence of an extremely rich phenomenology. In
particular, in a specific region of the control parameter $K$, the
canonical and microcanonical ensembles show a first order phase
transition, with a forbidden energy range in the former ensemble and
a temperature jump in the latter (see Fig.~\ref{tvse}c). Both the
$\varepsilon(\beta)$ curve and the $\beta(\varepsilon)$ one become
multiply valued if we include metastable and unstable states. Since
we know that the Maxwell construction leads to an equal area
condition for the $\beta(\varepsilon)$ curve, which defines the
phase transition inverse temperature $\beta_t$ in the canonical
ensemble, we wonder here whether a similar construction exists for
the $\varepsilon(\beta)$ relation which would lead to the
determination of the transition energy $\varepsilon_t$ in the
microcanonical ensemble.

\begin{figure}[htb]
\begin{center}
\includegraphics[scale=.9]{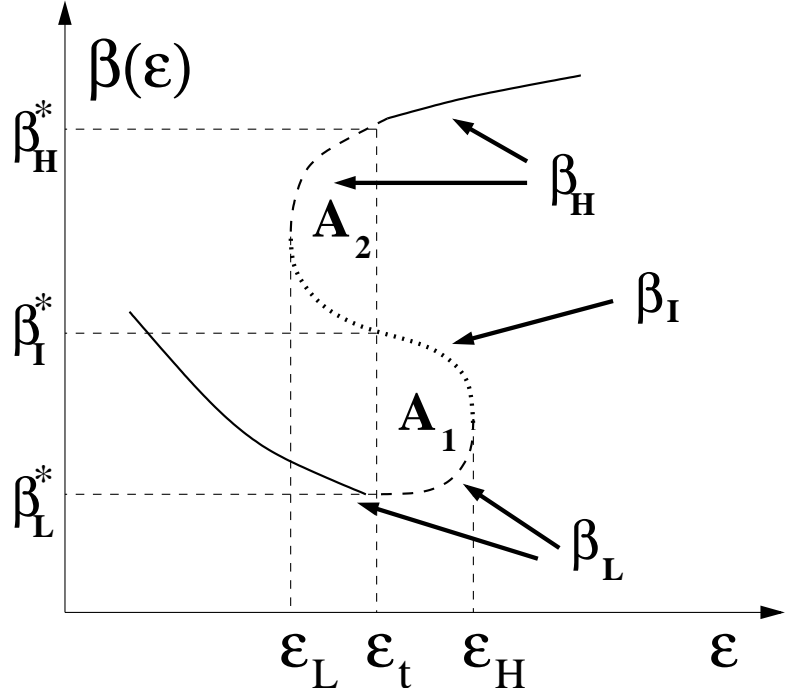}
\end{center}
\caption{Typical shape of the $\beta(\varepsilon)$ curve at a
microcanonical first order phase transition. The transition energy
$\epsilon_t$ is determined by an equal area $A_1=A_2$ Maxwell's
construction. All states are represented: stable (solid line),
metastable (dashed lines) and unstable (dotted line). The inverse
temperature jump is given by $\beta_H^*-\beta_L^*$.} \label{maxwell}
\end{figure}

In the following discussion of Maxwell's construction, it is crucial
to understand the mechanism that generates multiple branches of the
$\beta(\varepsilon)$ curve. Let us define
\begin{equation}
\tilde{\beta}(\varepsilon,m) = \frac{\partial \tilde{s}
(\varepsilon,m)}{\partial \varepsilon}~. \label{betatilde}
\end{equation}
We have explained that the equilibrium magnetization $m^*$, at any
energy $\varepsilon$, is the global maximum of the entropy per site
(\ref{entpsite}). Once $m^*$ has been computed, the equilibrium
inverse temperature is given by
$\beta(\varepsilon)=\tilde{\beta}(\varepsilon,m^*)$. However, also
local maxima, local minima and saddles of the entropy exist,
corresponding to different values of $m$. Following such critical
points as a function of $\varepsilon$, one determines the different
branches of $\beta(\varepsilon)$. In particular, we have the
continuation at energies lower than the microcanonical transition
energy $\varepsilon_t$ of the high energy $m=0$ branch (dashed part
of $\beta_H(\varepsilon)$ in Fig.~\ref{maxwell}) and the
continuation at higher energies of the magnetized branch
$\beta_L(\varepsilon)$. It's interesting to remark that the $m=0$
point remains an extremum for all values of $\varepsilon$ since
$\tilde{s}(\varepsilon,m)$ is even in $m$.

An example of inverse temperature $\beta(\varepsilon)$ relation is
plotted in Fig.~\ref{maxwell}. The lower branch $\beta_L$ starts at
low energy and ends at the energy $\varepsilon_H$, where its
derivative becomes infinite. The upper branch $\beta_H$ starts at
high energy and ends at the energy $\varepsilon_L$, where again its
derivative is infinite. These two branches are connected by the
vertical line at energy $\varepsilon_t$, and by the intermediate
branch $\beta_I$  that goes from $\varepsilon_L$ to $\varepsilon_H$.
The equilibrium state is given by the lower branch for $\varepsilon
< \varepsilon_t$ and by the upper branch for $\varepsilon >
\varepsilon_t$. The lower branch for $\varepsilon_t < \varepsilon <
\varepsilon_H$ (dashed) and the upper branch for $\varepsilon_L <
\varepsilon < \varepsilon_c$ (dashed) represent metastable states,
while the intermediate branch represents unstable states (dotted).
Therefore, increasing the energy, the equilibrium value of $\beta$
jumps from the lower to the upper branch (thus following the
vertical line) at the transition energy $\varepsilon_t$. It is easy
to show that the vertical line realizes a Maxwell construction,
i.e., that the two areas $A_1$ and $A_2$ are equal. The curve
$\beta(\varepsilon)$ has therefore three branches, that we denote by
$\beta_L(\varepsilon)$ (the low energy magnetized branch),
$\beta_I(\varepsilon)$ (the intermediate branch of unstable states)
and $\beta_H(\varepsilon)$ (the high energy paramagnetic branch).
Then, we have
\begin{eqnarray}
\label{maxwmicr} A_2 - A_1 &=& \int_{\varepsilon_t}^{\varepsilon_H}
\beta_L(\varepsilon) {\rm d}\varepsilon
+\int_{\varepsilon_H}^{\varepsilon_L} \beta_I(\varepsilon) {\rm
d}\varepsilon + \int_{\varepsilon_L}^{\varepsilon_t}
\beta_H(\varepsilon) {\rm d}\varepsilon \\
&=& \left( s_L(\varepsilon_H) - s_L(\varepsilon_t) \right) + \left(
s_I(\varepsilon_L) - s_I(\varepsilon_H) \right) + \left(
s_H(\varepsilon_t) - s_H(\varepsilon_L) \right)~,
\end{eqnarray}
where in the r.h.s. $s_i(\varepsilon)$ is the function whose
derivative gives the branch $\beta_i(\varepsilon)$, with $i=H,I,L$.
We use now the continuity property of the entropy, imposing that
$s_L(\varepsilon_H) =s_I(\varepsilon_H)$ and $s_I(\varepsilon_L) =
s_H(\varepsilon_L)$. Moreover, the transition occurs at the energy
where the entropies of the low energy branch and of the high energy
branch are equal, i.e., that $s_L(\varepsilon_t) =
s_H(\varepsilon_t)$. We then obtain that $A_1 = A_2$. It should
remarked that the values of the three branches of $\beta
(\varepsilon)$ at $\varepsilon_t$ determine the size of the
temperature jump. Indeed, $\beta_L(\varepsilon_t)=\beta_L^*$,
$\beta_I(\varepsilon_t)=\beta_I^*$ and
$\beta_H(\varepsilon_t)=\beta_H^*$ (see Fig.\ref{maxwell}).

The equal area condition implies that
\begin{equation}
\int_{\beta_L}^{\beta_H} \dd \beta \left[\varepsilon (\beta) -
\varepsilon_t \right ]=0~.
\end{equation}
Using $\varepsilon=\dd \phi/\dd \beta$, one gets
\begin{equation}
\phi(\beta_H) - \phi(\beta_L) -\varepsilon_t (\beta_H - \beta_L)=0~,
\end{equation}
which, after defining the {\it generalized entropy}
\begin{equation}
\widehat{s}(\beta,\varepsilon)=\beta \varepsilon - \phi(\beta)~,
\label{defshat}
\end{equation}
leads to
\begin{equation}
\widehat{s}(\beta_L,\varepsilon_t)=\widehat{s}(\beta_H,\varepsilon_t),
\label{gibbsconditions}
\end{equation}
which is the condition equivalent to (\ref{gibbsconditionf}).

The first time Maxwell's construction appears for self-gravitating
systems is in Ref.~\cite{Aronson}. It was later extended to
microcanonical phase transitions in
Refs.~\cite{chavanisaddit,MukamelHouches}. On the other hand canonical
and microcanonical caloric curves were studied in
Ref.~\cite{stahkkieslinschindler}. It is
also possible that the curve is made of several disconnected
branches. We then emphasize that for more complex
caloric curves than the one in Fig.~\ref{maxwell}, the evaluation of Maxwell's
areas must be done cautiously~\cite{ChavanisHouches,Chavanisreview2006}.

\subsubsection{Negative susceptibility}
\label{secnegsusc}

We have seen that ensemble inequivalence can give rise to negative
specific heat in the microcanonical ensemble. We show here that
another consequence of ensemble inequivalence is the existence of
equilibrium microcanonical states with a negative magnetic
susceptibility, a non negative quantity in the canonical ensemble.
We will follow a treatment close to that of subsection
\ref{orddepen}.

Fixing the energy and the magnetization, the entropy is given by Eq.
(\ref{sofem}). From the first principle of thermodynamics, that for
magnetic systems reads $T \dd S = \dd E - h \dd M$, with $E$ and $M$
the internal energy and the total magnetization of the system,
respectively, it is straightforward to prove the following formula
for the average effective magnetic field in the microcanonical
ensemble (see also \cite{campa2007})

\begin{equation}\label{hofem}
h(\veps,m) = - \frac{\displaystyle \frac{\partial
\tilde{s}}{\partial m}}{\displaystyle \frac{\partial \tilde{s}}
{\partial \veps}} = - \frac{1}{\beta(\veps,m)} \, \frac{\partial
\tilde{s}}{\partial m} \, .
\end{equation}
%We see that, in the case where we release the magnetization
%constraint and the entropy is given by $s(\veps) = \sup_m \tilde{s}(\veps,m)$,
%this magnetic field is zero.
Taking into account that, like $\beta$ is canonically conjugated to
$H$, $\beta h$ is canonically conjugated to $m$, it is natural to
define the following partition function
\begin{equation}\label{partbetam}
Z(\beta,h,N) =  \sum_{\{S_1,\ldots,S_N\}} \exp \left\{ -\beta \left[
H(\{ S_i \}) - h M(\{ S_i \})\right]\right\} \, ,
\end{equation}
where, as in Eq. (\ref{sofem}), $M(\{ S_i \})$ is the total
magnetization corresponding to configuration $\{ S_i \}$.
Analogously to Eq.~(\ref{canoproblem}), the free energy is
\begin{equation}
\label{fofbetah} f(\beta,h) = -\frac{1}{\beta}\lim_{N\rightarrow
\infty} \frac{1}{N} \ln Z(\beta,h,N) = \inf_{\veps,m}\left[\veps
-hm-\frac{1}{\beta}\tilde{s}(\veps,m)\right] \, .
\end{equation}
For $h=0$, we obviously recover Eq.~(\ref{canoproblem}). As in
subsection \ref{orddepen}, we see that the relation between the two
ensembles can be studied by analyzing the single function
$\tilde{s}(\veps,m)$. Taking into account Eq.~(\ref{hofem}), the
variational problem in Eq.~(\ref{fofbetah}), together with the
variational problem that defines $s(\varepsilon)$ can be solved by
imposing that
\begin{eqnarray}
\label{variational2a}
\frac{\partial \tilde{s}}{\partial m}&=& -h \beta \\
\frac{\partial \tilde{s}}{\partial \varepsilon}&=&+\beta \, ,
\label{variational2b}
\end{eqnarray}
that generalize the conditions (\ref{variationala}) and
(\ref{variationalb}) to $h\neq 0$, providing the functions
$\veps(\beta,h)$ and $m(\beta,h)$. Independently of the value of
$h$, the stability conditions in the canonical ensemble are the same
as those given in Eqs.~(\ref{condstab1}) and (\ref{condstab2}). In
the microcanonical ensemble, since we are not maximizing with
respect to $m$, we have no condition on the second derivative of
$\tilde{s}(\veps,m)$ with respect to $m$.

Magnetic susceptibility is defined as
\begin{equation}
\label{susc_can} \chi=\frac{\partial m}{\partial h}~.
\end{equation}
Deriving (\ref{variational2a}) and (\ref{variational2b}) with
respect to $h$, one obtains
\begin{eqnarray}
\tilde{s}_{m\veps}\frac{\partial \veps}{\partial
h}+\tilde{s}_{mm}\frac{\partial m}
{\partial h} &=& -\beta \, , \label{second1} \\
\tilde{s}_{\veps \veps}\frac{\partial \veps}{\partial
h}+\tilde{s}_{\veps m}\frac{\partial m} {\partial h} &=& 0~,
\label{second2}
\end{eqnarray}
from which we get
\begin{equation}
\label{suscan} \chi= -\beta \frac{\tilde{s}_{\veps \veps}}
{\tilde{s}_{\veps \veps}\tilde{s}_{mm} - \tilde{s}_{\veps m}^2}~.
\end{equation}
This formula is valid in both the canonical and microcanonical
ensemble. However, the results can differ in the two ensembles
because the quantities in this formula are computed at different
stationary points in the two ensembles. In the canonical ensemble,
$\chi$ is positive definite for all $h$ because of the stability
conditions ~(\ref{condstab1}) and (\ref{condstab2}).

As for the specific heat, the positivity of magnetic susceptibility
in the canonical ensemble can be derived on general grounds from
Eq.~(\ref{partbetam}), since it is easily shown that susceptibility
is proportional to the canonical expectation value $\langle
\left(M-\langle M\rangle \right)^2 \rangle$.

On the other hand, in the microcanonical ensemble, as already
remarked, no condition on the second derivatives of $\tilde{s}$ with
respect to $m$ is required, and therefore susceptibility can have
either sign. Indeed, in Ref.~\cite{campa2007} it is shown that a
simple $\phi^4$ model can display a negative microcanonical
susceptibility (see also subsection~\ref{modelphi4}).

With ensemble equivalence, as it always happens in short-range
systems, magnetic susceptibility is positive also in the
microcanonical ensemble. We note that this result is also a
byproduct of the convexity property discussed in
subsection~\ref{Convexity}, since the attainable region in the
$(\veps,m)$ plane is necessarily convex for short-range systems. In
fact, this implies that Eqs.~(\ref{condstab1}) and (\ref{condstab2})
are satisfied for all equilibrium values $(\veps,m)$.

\subsection{The Hamiltonian Mean Field model}
\label{exemple_HMF}

Very few examples are known in statistical mechanics where one can
explicitly compute microcanonical entropy in cases where the
variables are continuous. Everybody knows the perfect gas derivation
of microcanonical entropy \cite{Huang}. However, as soon as one
considers interactions, the task becomes unfeasible. On the other
hand, since for short-range systems microcanonical and canonical
ensemble are equivalent and it is much simpler to perform integrals
with Boltzmann weights rather than with Dirac delta functions, much
more attention has been devoted to compute free energies. An
exception is the study of gravitational systems for which, since it
was clarified that ensembles can be non equivalent
\cite{Thirringdd,Thirring}, some attention has been devoted to
calculations in different ensembles \cite{Horwitz77,Horwitz78,
Saslaw85,Padmanabhan90,paddy,ChavanisHouches,Chavanisreview2006}. In
particular mean-field models have been shown to be solvable using
saddle point methods \cite{Horwitz77,Horwitz78} and phase
transitions in the grand-canonical, canonical and microcanonical
ensembles have been studied \cite{Youngkins}. On another line of
research, simplified gravitational one-dimensional models have been
considered \cite{Henon64,Hohl67}, of which microcanonical solutions
have been found \cite{Rybicki71,Milanovic98}.

This section is devoted to the discussion of a mean-field model, the
Hamiltonian Mean Field (HMF) model, whose potential keeps only the
first mode of the Fourier expansion of the potential of
one-dimensional gravitational and charged sheet models. Indeed, it
turns out that the model is nothing but the mean-field version of
the XY model (see e.g. Chapter 6.1 of Ref.~\cite{Chaikin95}), which,
however, had never been studied in the microcanonical ensemble.

The HMF model has been extensively studied for more than a decade.
The simple mean-field interaction allows us to perform analytical
calculations, but maintains several complex features of long-range
interactions.

It has been shown that the behavior of certain wave-particle
Hamiltonians can be understood using the HMF model as a reference.
For instance, some equilibrium and non equilibrium properties of the
HMF model can be mapped onto those of the Colson-Bonifacio model of a
single-pass Free Electron Laser~\cite{FELPRE,bbdrjstatphys}.

In this section, we will derive both the canonical and the
microcanonical solutions, that we anticipate to be equivalent. The
microcanonical solution will be obtained by
three different methods: a straightforward one, inspired by the
solution of gravitational models (subsection~\ref{appendix}), and
two more involved ones, using a variational procedure
(subsection~\ref{minmaxsect}) or large deviations
(subsection~\ref{largedeviationsforHMF}). This latter, although more
complicated, will allow us to introduce a method of solution which
has a much wider range of applications.

Moreover, this model will serve as a paradigm for discussing
important dynamical behaviors typical of long-range interacting
systems, which will be presented in Sec.~\ref{kineticequation}.

\subsubsection{Introduction}

The Hamiltonian Mean Field model~\cite{Antoni95,Dauxois02,Judith} is
defined by the following Hamiltonian
\begin{equation}
H_N=\sum_{i=1}^{N}\frac{p_i^2}{2}+\frac{J}{2N} \sum_{i,j}\left[ 1
-\cos(\theta_i-\theta_j)\right]\quad, \label{Ham_HMF}
\end{equation}
where $\theta_i\in[0,2\pi[$ is the position (angle) of the $i$-th
unit mass particle on a circle and $p_i$ the corresponding
conjugated momentum. This system can be seen as representing
particles moving on a unit circle interacting via an infinite range
attractive ($J>0$) or repulsive ($J<0$) cosine potential or,
alternatively, as classical XY-rotors with infinite range
ferromagnetic ($J>0$) or antiferromagnetic ($J<0$) couplings. The
renormalization factor $N$ of the potential energy is kept not only
for historical reasons, but also because, as we have explained in
section~\ref{additivity}, in this way the energy per particle and
temperature are well defined in the $N\to\infty$ limit. In the
literature, some authors have treated the case in which the energy
is not extensive, i.e. they remove the factor $1/N$. This leads to
different thermodynamic limit behaviors~\cite{toral,celia}.

Historically, this model has been independently introduced in the
continuous time version in
Refs.~\cite{RuffoMarseille,Inagaki93a,Inagaki93b,diego98a,
diego98b}, and had been previously considered in its time discrete
version~\cite{konishi-kaneko}.

We will solve the HMF model in both the canonical and microcanonical
ensembles. Alternatively, the model can be solved using the maximum
entropy principle for the single particle distribution
function~\cite{Inagaki93a}, which is however suitable only for mean
field models.

It is very useful to rewrite the Hamiltonian (\ref{Ham_HMF}) in a
different form, using the definition of the $x$ and $y$ components
of the microscopic magnetization
\begin{equation}
\label{magnHMF} m_x = \frac{1}{N}\sum_{i=1}^N \cos \theta_i \quad
\quad \quad \quad \mbox{and} \quad \quad \quad \quad m_y =
\frac{1}{N}\sum_{i=1}^N \sin \theta_i \, .
\end{equation}
We then easily find that
\begin{equation}\label{Ham_HMFM}
H_N = \sum_{i=1}^{N}\frac{p_i^2}{2}
+\frac{NJ}{2}\left(1-m^2\right)~.
\end{equation}

In the following we will treat only the ferromagnetic case and we
will set $J=1$ without loss of generality. The antiferromagnetic
case is less interesting for what equilibrium properties are
concerned (the homogenous state is stable at all energies), but it
displays interesting dynamical features, like the formation of
collective modes under the form of ``biclusters"
\cite{DHR2000,BDR2001,BBDR2002,BBDR2002b,JeongChoi2006}.

\subsubsection{The canonical solution}

\begin{figure}[!htbp]
\centering \resizebox{0.8\textwidth}{!}{\includegraphics{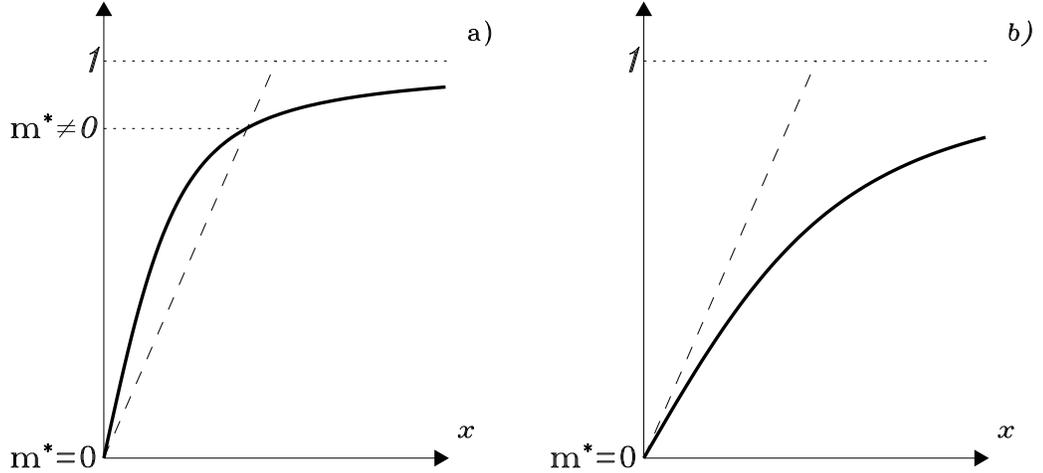}}
\vskip -1truecm
%\centering
%\includegraphics[bb=65 320 545 555,clip,scale=0.8]{HMFFIG.eps}
\caption{Graphical solution of Eq.~(\ref{solhmfcan}) for two values
of $\beta$, one above and one below the critical value $\beta_c=2$.
In both plots the straight dashed line is the bisectrix, while the
curved solid line is the ratio of modified Bessel functions on the
r.h.s. of Eq.~(\ref{solhmfcan}). a) $\beta=4$: the equilibrium
solution is $m^* \neq 0$; b) $\beta=1.5$: the only and equilibrium
solution is $m^*=0$.} \label{HMFFIG}
\end{figure}

The canonical solution of this model can be easily derived. After
the trivial Gaussian integration over the momenta, the canonical
partition function reads
\begin{eqnarray}\label{hmfcanon}
Z(\beta,N) &=& \exp
\left(-\frac{N\beta}{2}\right)\left(\frac{2\pi}{\beta}
\right)^{N/2} \times \nonumber \\
&&\int \dd \theta_1 \dots \dd \theta_N \exp \left\{
\frac{\beta}{2N}\left[\left(\sum_{i=1}^N \cos \theta_i \right)^2
+\left(\sum_{i=1}^N \sin \theta_i \right)^2\right]\right\} \, .
\end{eqnarray}
Using the Hubbard-Stratonovich transformation, see (\ref{hubb}),
this expression becomes
\begin{eqnarray}\label{hmfhubb1}
Z(\beta,N) &=&  \exp
\left(-\frac{N\beta}{2}\right)\left(\frac{2\pi}{\beta}
\right)^{N/2} \times \nonumber \\
&& \frac{N\beta}{2\pi} \int \dd x_1 \dd x_2 \exp
\left\{N\left[-\frac{\beta (x_1^2+x_2^2)}{2} + \ln
I_0(\beta(x_1^2+x_2^2)^{\frac{1}{2}})\right]\right\} \, ,
\end{eqnarray}
where $I_0(z)$ is the modified Bessel function of order $0$
\begin{equation}
\label{mbeszero} I_0(z)= \int_0^{2\pi} \dd \theta \exp \left(
z_1\cos \theta +z_2\sin \theta \right)= \int_0^{2\pi} \dd \theta
\exp \left( z\cos \theta \right) \, ,
\end{equation}
where $z \equiv \left(z_1^2 + z_2^2 \right)^{1/2}$. We can go to
polar coordinates in the $(x_1,x_2)$ plane, to obtain
\begin{equation}
\label{hmfhubb2} Z(\beta,N) = \exp
\left(-\frac{N\beta}{2}\right)\left(\frac{2\pi}{\beta} \right)^{N/2}
N\beta \int_0^{\infty} \dd x \exp \left\{N\left[-\frac{\beta x^2}{2}
+ \ln I_0(\beta x)\right]\right\} \, .
\end{equation}
In the thermodynamic limit $N\rightarrow \infty$, the integral in
(\ref{hmfhubb2}) can be computed using the saddle point method, so
that the rescaled free energy per particle is
\begin{equation}\label{freehmf}
\phi(\beta)=\beta f(\beta)= \frac{\beta}{2} -\frac{1}{2}\ln 2\pi
+\frac{1}{2}\ln \beta +\inf_{x\ge 0} \left[\frac{\beta x^2}{2}-\ln
I_0(\beta x)\right] \, .
\end{equation}
The extremal problem in $x$ in the last equation can be easily
solved graphically, by looking for the solution of the equation
\begin{equation}
\label{solhmfcan} x=\frac{I_1 (\beta x)}{I_0 (\beta x)} \, ,
\end{equation}
where $I_1(z)$ is the modified Bessel function of order $1$, which
is also the derivative of $I_0(z)$. The graphical solution is made
easier by the fact that, for real $z>0$, the function $I_1/I_0$ is
positive, monotonically increasing and with negative second
derivative (see, e.g., Ref. \cite{CGM}, where a generalization of
the HMF model is studied, and where these properties are proved for
an entire class of functions that include $I_1/I_0$).
Fig.~\ref{HMFFIG} shows the graphical solution of (\ref{solhmfcan})
for two different values of $\beta$. For $\beta \leq 2$ the solution
is given by $x=m^*=0$, while for $\beta \geq 2$ the solution
monotonically increases with $\beta$, approaching $m^*=1$ for $\beta
\rightarrow \infty$. The solution $m^*=0$ of (\ref{solhmfcan}),
present for all values of $\beta$, is not acceptable for $\beta >
2$, since it's not a minimum in formula (\ref{freehmf}).

It is not difficult to show that the value $m^*$ realizing the
extremum in Eq. (\ref{freehmf}) is equal to the spontaneous
magnetization. We emphasize that values of $x$ different from this
are not related to the magnetization of a state different from
equilibrium state, but that only the minimizer $m^*$, which is
always between $0$ and $1$, is the magnetization of the equilibrium
state. This should be suggested also by the fact that the
integration range in Eq.~(\ref{hmfhubb2}) goes from $0$ to $\infty$.

From the above procedure it is clear that the spontaneous
magnetization is defined only in its modulus $m^*$, while there is a
continuous degeneracy in its direction.

In conclusion, we have shown that the HMF model displays a second
order phase transition at $\beta_c=2$ ($T_c=0.5$). The derivative of
the rescaled free energy with respect to $\beta$ gives the energy
per particle
\begin{equation}
\label{enerhmfcan} \veps (\beta) = \frac{1}{2\beta} + \frac{1}{2}
-\frac{1}{2}(m^*(\beta))^2
\end{equation}
As already evident from the Hamiltonian, the lower bound of $\veps$
is $0$. At the critical temperature the energy is
$\veps_c=\frac{3}{4}$. A plot of the function $\phi(\beta)$ is shown
in the next subsection, after the microcanonical solution.

\subsubsection{The microcanonical solution}
\label{appendix}

Since we have shown that the HMF model has a second order phase
transition in the canonical ensemble, we could immediately conclude,
according to the remark in subsection~\ref{posneg}, that ensembles
are equivalent for this model. Deriving the entropy would be
straighforward using the inverse Legendre-Fenchel transform of the
free energy computed in the previous subsection (see formula
(\ref{freehmf})). However, since in the following we will need the
explicit expression of the microcanonical entropy of the HMF model
both as a function of the energy and of magnetization in order to
solve a generalized HMF model in the microcanonical ensemble, we
will devote this subsection, the following and
subsection~\ref{largedeviationsforHMF} to the derivation of the
entropy for the HMF model using several methods. The reason for such
a thorough derivation is mainly pedagogical: we want to show the
application of three different methods to a simple model. The first
method is mutuated from similar ones used in the context of
gravitational systems. The second one, more general, is a
variational method based on a strong hypothesis on the form the
partition function, and allows a straightforward derivation of the
entropy, also in cases of ensemble inequivalence. The third method
illustrates the application of large deviation theory, which is the
most general currently available tool to solve systems with
long-range interactions.

The microcanonical solution has been heuristically
obtained, under the hypothesis of concave entropy, in
Ref.~\cite{antoniHinrichsenruffo} and in a different form in
Ref.~\cite{velasquez}.

The simplicity of the Hamiltonian makes it possible to obtain
directly the thermodynamic limit of the entropy per particle, as we
now show. First of all, let us introduce the method usually applied
in self-gravitating systems \cite{binneytremaine}. Denoting by $K$
and $U$ the kinetic and the potential energy, respectively, the
number of microscopic configurations corresponding to the energy $E$
in a generic system is given by
\begin{eqnarray}
  \Omega(E,N)&=&\int\prod_i  \dd p_i \dd \theta_i\, \delta(E-H_N) \\
&=&\int\prod_i  \dd p_i \dd \theta_i\underbrace{\int \dd K  \,
\delta\left(K-\sum_i\frac{p_i^2}{2}\right)}_{=1}\,
  \delta\left(E-K-U(\{\theta_i\})\right) \\
&=&\int \dd K \underbrace{\int\prod_i  \dd p_i   \,
  \delta\left(K-\sum_i\frac{p_i^2}{2}\right)}_{ \Omega_{\rm kin}(K)}\,
 \underbrace{ \int\prod_i  \dd \theta_i \delta
\left(E-K-U(\{\theta_i\})\right)}_{ \Omega_{\rm conf}(E-K)}\quad.
\label{eq:tradwaysuite}
\end{eqnarray}

The factor $\Omega_{\rm kin}$, which is related to the surface of the
hypersphere with radius $R=\sqrt{2K}$ in $N$ dimensions, can be computed
straightforwardly using the properties of the Dirac $\delta$ function,
obtaining the expression: $\Omega_{\rm kin}={2\pi^{N/2}R^{N-2}}/{\Gamma(N/2)}$.
Using the asymptotic expression of the $\Gamma$-function, $\ln
\Gamma(N)\simeq (N-1/2)\ln N-N+(1/2)\ln (2\pi)$, and keeping only
the terms that do not give a vanishing contribution to the entropy
per particle in the thermodynamic limit, one obtains
\begin{eqnarray}
\Omega_{\rm kin}\left(K\right)&\stackrel{N\to+\infty}{\sim}&
\exp\left(\frac{N}{2}\left[1+\ln \pi-\ln \frac{N}{2}+\ln
(2K)\right]  \right) \\
&=&  \exp\left(\frac{N}{2}\left[1+\ln (2\pi)+\ln u\right]
\right)\quad, \label{omegazero}
\end{eqnarray}
where $u=2K/N$. Defining the configurational entropy per particle
$s_{\rm conf}(\tilde{u})=(\ln \Omega_{\rm conf}(N\tilde{u}))/N$,
where $\tilde{u}=U/N=(E-K)/N=\veps -u/2$,
Eq.~(\ref{eq:tradwaysuite}) can be rewritten as
\begin{eqnarray}
\Omega(N\veps,N)&\stackrel{N\to+\infty}{\sim}& \frac{N}{2} \int \dd
u \exp\left[{N}\left(\frac{1}{2}+\frac{\ln (2
\pi)}{2}+\frac{1}{2}\ln u +s_{\rm conf}(\tilde{u})\right)
\right]\quad. \label{eq:tradwaysuitetekkr}
\end{eqnarray}
Hence, solving the integral in the saddle point approximation, we
obtain the entropy
\begin{eqnarray}
s(\veps)&=&\lim_{N\to +\infty}\frac{1}{N}\ln \Omega_N(\veps N)  \\
&=&\frac{1}{2}+\frac{1}{2}\ln (2\pi)+\sup_u\left[\frac{1}{2}\ln u +
s_{\rm conf}(\tilde{u}) \right] \, . \label{final}
\end{eqnarray}
We note that this expression is quite general, in the sense that it
is valid for any system in which the kinetic energy assumes the
usual quadratic form. To proceed further, we need an explicit
expression for the configurational entropy $s_{\rm conf}$, something
which is generally not easily feasible.

\begin{figure}[!htbp]
\centering
\resizebox{0.9\textwidth}{!}{\includegraphics{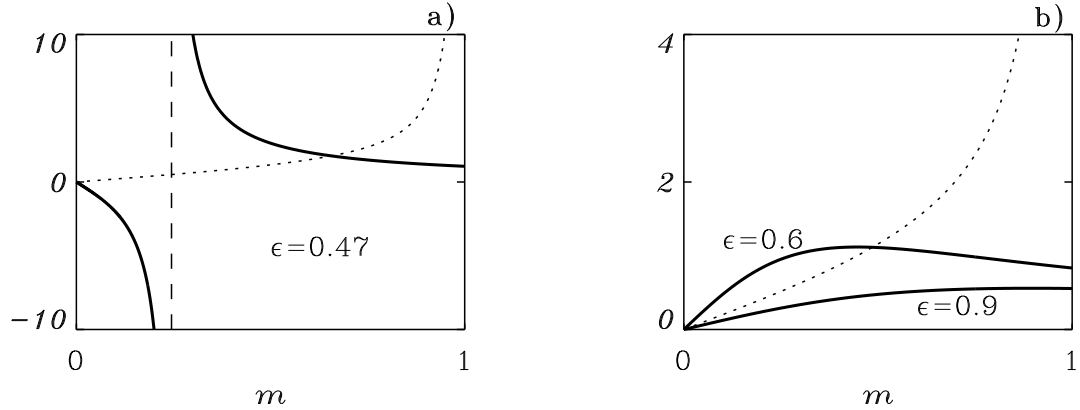}}
\caption{Graphical solution of Eq.~(\ref{solbinv}) for three values
of the energy $\veps$, one above and two below the critical value
$\veps_c=3/4$. In both plots the dotted curve is the function
$B_{inv}(m)$, while the solid curve is the function of the first
term in Eq.~(\ref{solbinv}). This last function diverges for
$m^2=1-2\veps$, which is between $0$ and $1$ for $0\le \veps \le
1/2$. The corresponding asymptote is the vertical dotted line in the
left panel. a) $\veps=0.47$: the relevant solution is the one with
$m>0$; b) $\veps=0.6$ and $\veps=0.9$: in the first case the
relevant solution is the one with $m>0$, while in the second case
the only solution is $m=0$.} \label{HMFFIGMICRO}
\end{figure}

Now we use the fact that the potential energy of the HMF model, as
evident in (\ref{Ham_HMFM}), is a very simple function of the
microscopic magnetization $\mathbf{m}=(m_x,m_y)$, with a one to one
correspondence between the value of the potential energy $U$ and the
modulus of the microscopic magnetization $m^2=m_x^2+m_y^2$. In fact,
if we define
\begin{equation}\label{deltamagn}
\Omega_m(m) = \int \prod_i \dd \theta_i \, \delta \left(\sum_i \cos
\theta_i - Nm \right) \delta \left( \sum_i \sin \theta_i \right) \,
,
\end{equation}
we have that this function will be proportional to $\Omega_{{\rm
conf}}(U)$ for $\tilde{u}=U/N=(1/2 - m^2/2)$. The coefficient of
proportionality will give a vanishing contribution to $s_{\rm
conf}(\tilde{u})$ in the thermodynamic limit. We note that, as in the
canonical case, there is a continuous degeneracy on the direction of
the spontaneous magnetization; therefore, we do not lose generality
by choosing the spontaneous magnetization in the direction of the
$x$ axis. The integral in (\ref{deltamagn}) can be computed using
the Fourier representation of the $\delta$ function. We therefore
have
\begin{eqnarray}\label{deltamagn2}
\Omega_m(m) &=& \left(\frac{1}{2\pi}\right)^2
\int_{-\infty}^{\infty} \dd q_1 \int_{-\infty}^{\infty} \dd q_2 \int
\prod_i \dd \theta_i \, \exp \left[iq_1 \left(\sum_i \cos \theta_i -
Nm \right)\right] \exp \left[iq_2 \left( \sum_i \sin \theta_i
\right)\right] \\
&=& \left(\frac{1}{2\pi}\right)^2 \int_{-\infty}^{\infty} \dd q_1
\int_{-\infty}^{\infty} \dd q_2 \, \exp \left\{N \left[-iq_1m + \ln
J_0((q_1^2+q_2^2)^{\frac{1}{2}})\right]\right\} \, ,
\end{eqnarray}
where $J_0(z)$ is the Bessel function of order $0$
\begin{equation}
\label{beszero} J_0(z)= \int_0^{2\pi} \dd \theta \exp \left( iz\cos
\theta \right) \, .
\end{equation}
To solve the integral in (\ref{deltamagn2}) with the saddle point
method, we have to consider $q_1$ and $q_2$ as complex variables.
Using that the derivative of $J_0$ is $-J_1$, the opposite of the
Bessel function of order $1$, the saddle point has to satisfy the
following equations
\begin{eqnarray}
\label{sadhmfmicro} -im
-\frac{J_1}{J_0}((q_1^2+q_2^2)^{\frac{1}{2}})
\frac{q_1}{(q_1^2+q_2^2)^{\frac{1}{2}}} &=& 0 \nonumber \\
-\frac{J_1}{J_0}((q_1^2+q_2^2)^{\frac{1}{2}})
\frac{q_2}{(q_1^2+q_2^2)^{\frac{1}{2}}} &=& 0 \, .
\end{eqnarray}
The solution of these equations is $q_2=0$ and $q_1=-i \gamma$,
where $\gamma$ is the solution of the equation
\begin{equation}
\label{solhmfmicro} \frac{I_1(\gamma)}{I_0(\gamma)}=m \,.
\end{equation}
Here, we have used the properties $J_0(iz)=I_0(z)$ and
$J_1(iz)=iI_1(z)$. Denoting by $B_{inv}$ the inverse function of
$I_1/I_0$, we can write in the thermodynamic limit
\begin{equation}
\label{solomegm} s_{\rm
conf}\left(\frac{1}{2}-\frac{1}{2}m^2\right)=\lim_{N\to
+\infty}\frac{1}{N}\ln \Omega_m(m) = -mB_{inv}(m) + \ln
I_0(B_{inv}(m)) \, .
\end{equation}
We can now substitute (\ref{solomegm}) in (\ref{final}), using that
$u=2(\veps -1/2 +m^2/2)$, and performing equivalently a maximization
over $m$ instead of that over $u$, to obtain
\begin{eqnarray}\label{final2}
s(\veps)&=& \frac{1}{2}+\frac{1}{2}\ln (2\pi) + \frac{1}{2}\ln 2
\nonumber \\&+& \sup_{m\ge m_0}\left[\frac{1}{2}\ln \left( \veps
-\frac{1}{2} +\frac{1}{2}m^2 \right) -mB_{inv}(m) + \ln
I_0(B_{inv}(m)) \right] \, ,
\end{eqnarray}
with $m_0^2 = \sup [0,1 - 2\veps ]$. The maximization problem over
$m$ is solved graphically, looking for the solutions of the equation
\begin{equation}
\label{solbinv} \frac{m}{2\veps -1 + m^2} - B_{inv}(m)=0 \, .
\end{equation}
The graphical solution $m=m(\veps)$ is shown in
Fig.~\ref{HMFFIGMICRO}, and it gives the following results. For $0
\le \veps \le 3/4$, the magnetization $m(\veps)$ monotonically
decreases from $1$ to $0$, while for $\veps > 3/4$ the solution is
always $m=0$. At $\veps =3/4$, there is a second order phase
transition, a first signature that the two ensembles give equivalent
predictions.

Indeed, it's easy to prove that $s(\veps)$ is concave. First, we
easily derive from~(\ref{final2}) that the inverse temperature
$\beta(\veps)$ is
\begin{equation}
\label{betamicrohmf} \beta(\veps)= \frac{\dd s}{\dd \veps} =
\frac{1}{2\veps -1 + m^2(\veps)}~.
\end{equation}
If $\beta(\veps)$ has a negative derivative, then $s(\veps)$ is
concave. The negativity of this derivative is trivial for $\veps >
3/4$, when $m(\veps)= 0$. For $\veps \le 3/4$ we can proceed as
follows. We note that, using the last equation, we can write
(\ref{solbinv}) also as
\begin{equation}\label{solbinv2}
m = \frac{I_1(m\beta(\veps))}{I_0(m\beta(\veps))} \, .
\end{equation}
We have just proved graphically that (\ref{solbinv}), and thus
(\ref{solbinv2}), have a unique solution for each $\veps$, that
decreases if $\veps$ increases. Studying (\ref{solhmfcan}) in the
canonical ensemble, we have found that this unique solution has the
property that $m$ decreases when $\beta$ decreases. Therefore, in
the present case an increase in $\veps$  results in a decrease in
$\beta(\veps)$, also when $m(\veps)>0$. This proves the concavity of
the entropy $s(\veps)$ and therefore ensemble equivalence. The
concavity property of the microcanonical entropy (\ref{final2})
ensures that it could also be computed by the Legendre-Fenchel
transform of $\phi$, the rescaled free energy computed in the
canonical ensemble and reported in Eq.~(\ref{freehmf}).

\begin{figure}[htb]
\resizebox{0.8\textwidth}{!}{\includegraphics{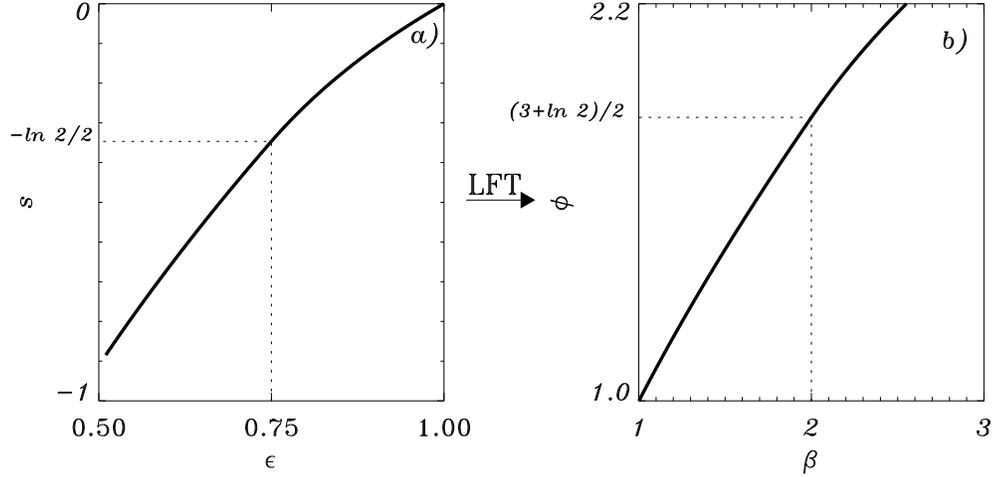}}
\caption{Entropy versus energy (a) and rescaled free energy versus
inverse temperature (b) for the HMF model~(\ref{Ham_HMF}) with
$J=1$. The dotted lines are traced at the phase transition point.}
\label{hmfentropie}
\end{figure}

\begin{figure}[htb]
\resizebox{0.8\textwidth}{!}{\includegraphics{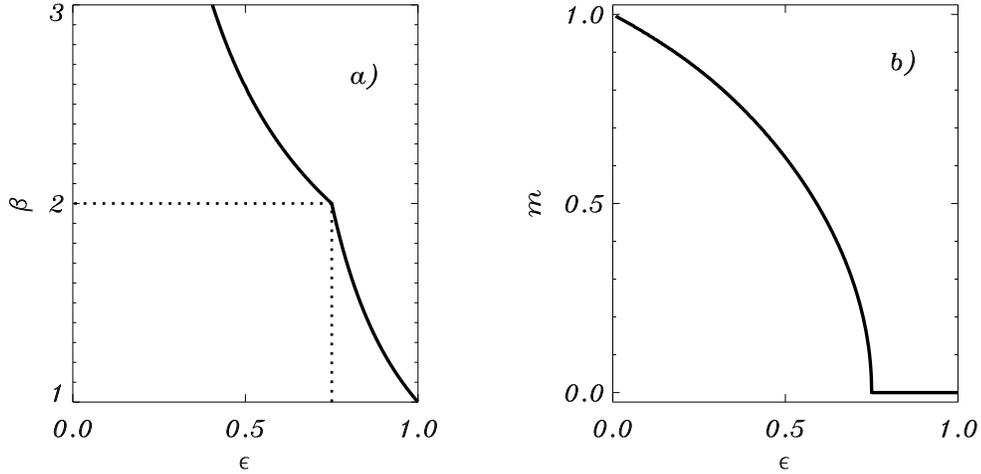}}
\caption{Inverse temperature versus energy (a) and magnetization
versus energy (b) for the HMF model~(\ref{Ham_HMF}) with $J=1$. The
dotted lines are traced at the phase transition point.}
\label{beteeth}
\end{figure}

We conclude this section devoted to the explicit calculation of the
canonical and microcanonical solution of the HMF model by showing
the plots of the relevant thermodynamic variables. In
Fig.~\ref{hmfentropie} we show the full dependence of the entropy on
the energy (see Eq.~(\ref{final2})) and the rescaled free energy
versus the inverse temperature (see Eq.~(\ref{freehmf})). In
Fig.~\ref{beteeth} we plot the caloric curve (see
Eq.~(\ref{betamicrohmf})) and the dependence of the order parameter
on the energy, obtainable from Eq.~(\ref{solbinv}).

\subsubsection{Min-max procedure}
\label{minmaxsect}

We have already emphasized that the microcanonical partition
function is much more difficult to obtain than the canonical one.
Although we have been able to obtain it by a direct counting for the
BEG model, this procedure is applicable only for discrete variables.
In this subsection we will show how the microcanonical entropy can
be obtained when the canonical free energy has been derived through
an optimization procedure of the type shown in (\ref{freehmf}). At
the same time, this discussion will allow us to understand how
ensemble inequivalence can arise. However, it should be remarked
that the derivation presented here is not rigorous and relies on
strong assumptions. We will follow the argument presented in
Ref.~\cite{leyvruf}, and later generalized in
Ref.~\cite{campaorder}, although we will give here a somewhat
different proof of the result.

It is crucial to assume that the canonical partition function,
similarly to what has been done in Eq.~(\ref{hmfhubb2}) for the HMF
model, is given by the following integral
\begin{equation}
\label{zminmax} Z(\beta,N) = \int_{-\infty}^{+\infty} \dd x \, \exp
\left[ -N \tilde{\phi}(\beta,x)\right] \, ,
\end{equation}
with $\tilde{\phi}(\beta,x)$ a differentiable function of $\beta$
for ($\beta\ge 0$) and a dummy variable $x$. This is a crucial
assumption, which is at the core of the applicability of the method.
We do not explicitly give any clue to what is the nature and
physical meaning of the variable $x$, but of course we have in mind
the integration variable which appears in the Hubbard- Stratonovich
transformation. Moreover, it should be observed that the variable
$x$ has nothing to do with the order parameter $m$. Even the
variation range of these two variables is different: in principle
the range of $x$ is the full real axis and the integral in
Eq.~(\ref{zminmax}) is required to converge (for example, in the
Hubbard-Stratonovich trasformation there is a dependence of the
integrand of the kind $\exp ( -x^2 )$).

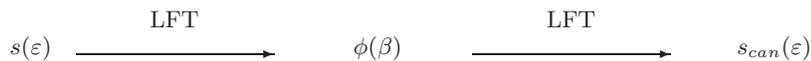
\begin{figure}[htb]
{\setlength{\unitlength}{0.5pt}
\begin{picture}(100,125)(-100,-50)
\put(-400,0){$s(\veps)$}\put(-300,20){ LFT}
\put(-350,0){\vector(1,0){150}} \put(-140,0){$\phi(\beta)$}
\put(00,20){ LFT} \put(-50,0){\vector(1,0){150}} \put(150,0){$\Large
s_{can}(\veps)$}
\end{picture}}
\caption{Relation between microcanonical entropy $s(\veps)$,
rescaled free energy $\phi(\beta)$ and canonical entropy
$s_{can}(\veps)$. LFT indicates the Legendre-Fenchel
transform.}\label{doublelegendre}
\end{figure}

In the thermodynamic limit, we therefore
have
\begin{equation}
\label{freeminmax} \phi(\beta) = \inf_x \tilde{\phi} (\beta,x)~.
\end{equation}
The {\it canonical entropy} is defined by the following variational
principle,
\begin{equation}
\label{canentrgen} s_{can}(\veps) = \inf_{\beta \ge
0}\left[\widehat{s} (\beta,\veps) \right] \,
\end{equation}
where $\widehat{s}$ is defined in formula (\ref{defshat}). This is
nothing but the Legendre-Fenchel transform of $\phi(\beta)$. We
observe that this expression is valid also in the case in which
$\veps$ is such that there is no $\beta$ value for which the
derivative of $\phi(\beta)$ is equal to $\veps$. In our case, we can
insert Eq. (\ref{freeminmax}) to obtain
\begin{equation}
\label{canentrminmax} s_{can}(\veps) = \inf_{\beta \ge 0}\left\{
\sup_x \left[ \beta \veps -\tilde{\phi}(\beta,x)\right]\right\} \, .
\end{equation}
One should remark that this relation is valid only if the assumption
made for the partition function in Eq.~(\ref{zminmax}) is fulfilled.
In particular the $x$-dependence of $\tilde{\phi}$ cannot be
neglected in order for the integral in Eq.~(\ref{zminmax}) to
converge. The relation between $s(\veps)$, $\phi(\beta)$ and
$s_{can}(\veps)$ is summarized in Fig.~\ref{doublelegendre}.

When ensembles are equivalent $s_{can}(\veps)=s(\veps)$. We will
show here that a necessary and sufficient condition for the
ensembles to be inequivalent is that $s(\veps)$ is strictly smaller
than $s_{can}(\veps)$. This is for instance what happens for the BEG
model, for which $s_{can}(\veps)$ is the concave envelope of
$s(\veps)$, as shown in Fig.~\ref{entropyBEG}.

We will prove here that, once the function $\tilde{\phi}(\beta,x)$
is known, one can introduce its Legendre-Fenchel transform
\begin{equation}
\label{inversionvarphi} \tilde{s}(\veps,x)=\inf_\beta \left[ \beta
\veps - \tilde{\phi}(\beta,x)\right]
\end{equation}
and obtain microcanonical entropy by the following formula
\begin{equation}
s(\veps)= \sup_x [\tilde{s}(\veps,x)] = \sup_x \left\{\inf_{\beta
\geq 0}\left[\beta\varepsilon-\tilde{\phi}(\beta,x)\right]\right\}~.
\label{entrminmax1}
\end{equation}
The only difference in expressions ~(\ref{canentrminmax}) and
(\ref{entrminmax1}) is just the order in which the minimum with
respect to $\beta$ and the maximum with respect to $x$ are taken.
Although this might seem a detail, it determines a different result
(see Appendix A, where this is proven in full generality). It turns
out that
\begin{equation}
s(\veps)\le s_{can}(\veps)~. \label{ineq_entr}
\end{equation}
Let us remark that the entropy defined in formula
(\ref{inversionvarphi}) does not coincide with $\tilde{s}(\veps,m)$.
In particular, one can easily verify that $\tilde{s}(\veps,x)$ is
always concave in $\veps$ for all $x$ values, since it is obtained
from a Legendre-Fenchel transform. On the contrary,
$\tilde{s}(\veps,m)$ can be non concave in $\veps$.

Let us briefly sketch the proof of formula (\ref{entrminmax1}),
which is based on the analysis of microcanonical partition function.
Using the Laplace representation of the Dirac delta function, this
latter can be expressed as
\begin{eqnarray}
\Omega (E,N)&=&  \sum_{\{S_1,\ldots,S_N\}} \ \delta (E-H(\{ S_i \})) \label{laplacea}\\
            &=& \frac{1}{2 \pi i} \int_{\beta -i \infty}^{\beta +i \infty} \dd \lambda
        \sum_{\{S_1,\ldots,S_N\}}\exp [\lambda (E - H(\{ S_i \})] \label{laplaceb}\\
        &=& \frac{1}{2 \pi i} \int_{\beta -i \infty}^{\beta +i \infty} \dd \lambda \,
         Z(\lambda,N) \exp (\lambda E) \label{laplacec}
        \end{eqnarray}
where $\beta=\mbox{Re}(\lambda)>0$ is the inverse temperature.
We use $\lambda$, instead of $\beta$, as an integration variable,
because we are considering the analytical continuation of $Z(\lambda,N)$
to the complex plane.
The last integral cannot be solved, in the thermodynamic
limit, using the saddle point method because, after expressing
$Z(\beta,N) \sim \exp [-N\beta f(\beta)]$, the function $\beta
f(\beta)$ is not in general differentiable for all $\beta$. In spite
of this, we can heuristically argue that the integral will be
dominated by the value of the integrand at a real value of $\lambda$
($\lambda=\beta+i\lambda_I$), otherwise we would obtain an
oscillatory behavior of $\Omega(E,N)$, giving rise to
negative values, which are impossible for the density of states.
To have a proof of this we can proceed as follows.

We are assuming to be in cases where $Z(\beta,N)$, for real $\beta$, can be
expressed as in (\ref{zminmax}), with $\tilde{\phi}$ analytic. Then, this integral
representation will be valid, in the complex $\lambda$ plane, for at
least a strip that includes the real axis, let's say for
$|\lambda_I|< \Delta$, with $\Delta >0$.

We now divide the integral in (\ref{laplacec}) in three intervals,
defined by $\lambda_I < -\delta$, $-\delta < \lambda_I < \delta$ and
$\lambda_I > \delta$, respectively, with $0 < \delta < \Delta$. In
Appendix B, we show that the contribution to the integral in
$\lambda$ coming from values of $\lambda_I$ outside the strip,
i.e. for values of $\lambda_I$ with
$|\lambda_I| > \Delta$, is exponentially small in $N$. Therefore the
calculation of the microcanonical partition function reduces to
performing the following integral
\begin{equation}
\label{stripintegral}
\Omega (E,N) =\frac{1}{2\pi i} \int_{\beta -i\delta}^{\beta
+i\delta} \dd \lambda \, e^{N\lambda\veps}Z(\lambda,N)
=\frac{1}{2 \pi i}
\int_{-\infty}^{+\infty} \dd x
        \int_{\beta -i \delta}^{\beta +i \delta} \dd \lambda \,
        \exp \left( N[\lambda \veps -\tilde{\phi} (\lambda,x) ] \right)~,
\end{equation}
where $0<\delta<\Delta$, and where, in the second equality, we have used the fact
that inside the strip $|\lambda_I|< \Delta$ we can represent $Z(\beta,N)$
as in (\ref{zminmax}).
Since $\tilde{\phi}$ is analytic in all the
domain of integration, we can perform the integral in $\lambda$ in
the large $N$ limit using the saddle point method. It can be shown
that for each value of $x$ the real part of the argument of the
exponential in (\ref{stripintegral}) is larger if computed on the
real axis $\lambda_I=0$. Indeed, we have
\begin{eqnarray}
\int_{-\infty}^{+\infty} \dd x \, \exp \{ N[\lambda \veps
-\tilde{\phi} (\lambda,x) ] \}&=&
\exp (N \lambda \veps) Z(\lambda,N) \\
&=& \exp (N \lambda \veps) \sum_{\{S_1,\ldots,S_N\}} \exp \{ -\beta
H(\{ S_i \})
 -i \lambda_I H(\{S_i \})\}\\
&=& \exp (N \beta \veps)  \sum_{\{S_1,\ldots,S_N\}} \exp \{ -\beta
H(\{ S_i \})
-i \lambda_I H(\{S_i \})+i\lambda_I N \veps \} \\
&=& \exp (N \beta \veps) \langle \exp \left(i \lambda_I [N \veps -
H(\{ S_i \})] \right) \rangle Z(\beta,N)~,
\end{eqnarray}
where in the last expression the average $\langle \cdot \rangle$ is
performed with Boltzmann weight $\exp [\beta H(\{ S_i \})]$. Using
now the definition $Z (\beta, N)= \int \dd x \exp [-N
\tilde{\phi}(\beta,x)]$, one can rewrite $\Omega(E,N)$ as
\begin{equation}
\Omega (E,N)=\frac{1}{2 \pi i} \int_{-\infty}^{+\infty} \dd x
        \int_{\beta -i \delta}^{\beta +i \delta} \dd \lambda
        \exp \left( -N \tilde{\phi} (\beta,x) +N \beta \veps + \ln \langle
        \exp \left( i \lambda_I [N \veps -H(\{ S_i \})] \right) \rangle \right)~.
\end{equation}
The real part of the exponent in the integral is obtained by
replacing $\ln (\langle \cdot \rangle)$ with $\ln (|\langle \cdot
\rangle|)$. The maximum of this logarithm is obtained for
$\lambda_I=0$, since for all other nonzero values of $\lambda_I$,
when performing the average in $\langle \cdot \rangle$, one would
sum unit vectors $\exp (i \lambda_I \alpha)$ with different values
of the phase $\alpha$, depending on the specific configuration $\{
S_i \}$, obtaining a result which will have certainly a smaller
modulus than summing all the vectors in phase with $\lambda_I=0$.
Therefore, if there is a saddle point on the real axis, it will
certainly give a larger real part for the argument of the
exponential than other saddle points eventually present outside the
real axis. Moreover, we have just proven that the saddle is
necessarily a maximum along the imaginary $\lambda$ direction.
Therefore, from the general properties of holomorphic functions, it
will be a minimum along the $\beta=\mbox{Re}(\lambda)$ axis. Once
the integral over $\lambda$ is performed, the remaining integral
over $x$ can be computed using again the saddle point method, now
for a real function, which gives a maximum over $x$. Combining all
this,  one gets the formula for the microcanonical entropy
(\ref{entrminmax1}).

Another way of arguing~\cite{leyvruf} is to remark that, since
$\tilde{\phi}(\lambda,x)$ is obtained by analytically continuing a
real function, if saddle points are present in the complex $\lambda$
plane, they necessarily appear in complex conjugate pairs. This
would induce oscillations in the values of $\Omega(E,N)$ as a
function of $N$, which would imply that $\Omega(E,N)$  could even
take negative values. This is absurd and leads to exclude the
presence of saddle points out of the real $\lambda$ axis.

We have already remarked that ensemble inequivalence is a
consequence of inequality (\ref{ineq_entr}), which in turn derives
from the different order of the minimum in $\beta$ with respect to
the maximum in $x$ in expressions~(\ref{canentrminmax}) and
(\ref{entrminmax1}). Let us now study in more detail the extrema
defined by the two different variational problems. The first order
stationarity conditions are
\begin{eqnarray}
\label{systminmaxa}
\frac{\partial \tilde{\phi}}{\partial \beta} &=& \veps \\
\frac{\partial \tilde{\phi}}{\partial x} &=& 0 \,~,
\label{systminmaxb}
\end{eqnarray}
which are of course the same for the two ensembles. However, the
stability conditions deriving from the two problems are different.
For what concerns the canonical entropy $s_{can}(\veps)$, we have
the conditions
\begin{eqnarray}
\label{systcanminmaxa}
\frac{\partial^2 \tilde{\phi}}{\partial x^2} &>& 0 \\
\frac{\partial^2 \tilde{\phi}}{\partial \beta^2}\frac{\partial^2
\tilde{\phi}}{\partial x^2} -\left(\frac{\partial^2
\tilde{\phi}}{\partial \beta \partial x}\right)^2 &<& 0 \, ,
\label{systcanminmaxb}
\end{eqnarray}
while for the microcanonical entropy $s(\veps)$ we have
\begin{eqnarray}
\label{systmicrminmaxa}
\frac{\partial^2 \tilde{\phi}}{\partial \beta^2} &<& 0 \\
\frac{\partial^2 \tilde{\phi}}{\partial \beta^2}\frac{\partial^2
\tilde{\phi}}{\partial x^2} -\left(\frac{\partial^2
\tilde{\phi}}{\partial \beta \partial x}\right)^2 &<& 0 \, .
\label{systmicrminmaxb}
\end{eqnarray}
A necessary and sufficient condition to satisfy
(\ref{systcanminmaxa}), (\ref{systcanminmaxb}),
(\ref{systmicrminmaxa}) and (\ref{systmicrminmaxb}) is that
$\partial^2 \tilde{\phi}/\partial \beta^2 < 0$ and $\partial^2
\tilde{\phi} / \partial m^2 > 0$. However, since the conditions are
different in the two ensembles, one can find values of $\beta$ and
$x$ that correspond to stable states in one ensemble but are
unstable in the other. It should be noted that one could find more
than one stationary point in a given ensemble, satisfying the
corresponding stability condition. Obviously, in this case one has
to choose the global extremum. If the global stable extrema are
different in the two ensembles, then we have ensemble inequivalence.
Tightly linked to stability is the sign of specific heat. Indeed,
using the expression
\begin{equation}
c_V=- \beta^2 \frac{\partial \tilde{\phi}(\beta,x(\beta))}{\partial
\beta^2}~,
\end{equation}
where $x(\beta)$ is obtained by solving (\ref{systminmaxa}) and
(\ref{systminmaxb}), one can obtain an expression for the specific
heat which is valid in both the ensembles
\begin{equation}\label{cvminmax}
c_V=-\beta^2\frac{\displaystyle \frac{\displaystyle \partial^2
\tilde{\phi}}{\partial \beta^2}\frac{\displaystyle \partial^2
\tilde{\phi}}{\partial x^2} -\left(\frac{\displaystyle \partial^2
\tilde{\phi}}{\partial \beta \partial x}\right)^2} {\displaystyle
\frac{\displaystyle \partial^2 \tilde{\phi}}{\partial x^2}} \,.
\end{equation}
We see from (\ref{systcanminmaxa}) and (\ref{systcanminmaxb}) that
this expression is positive in the canonical ensemble. However, in
the microcanonical ensemble the conditions (\ref{systmicrminmaxa})
and (\ref{systmicrminmaxb}) do not determine the sign of $\partial^2
\tilde{\phi} /\partial x^2$, and thus the specific heat can have
either sign in the microcanonical ensemble.

Let us check that, using the method described in this subsection, we
can rederive the microcanonical entropy of the HMF model, Eq.
(\ref{final2}). For the HMF model, the function
$\tilde{\phi}(\beta,x)$ can be obtained from (\ref{hmfhubb2})
\begin{equation}
\label{phihmf} \tilde{\phi}(\beta,x) =
\frac{\beta}{2}-\frac{1}{2}\ln 2\pi +\frac{1}{2} \ln \beta +
\frac{\beta x^2}{2} -\ln I_0(\beta x) \, .
\end{equation}
The stationarity conditions (\ref{systminmaxa}) and
(\ref{systminmaxb}) read in this case
\begin{eqnarray}
\label{systminmaxhmfa} \frac{\partial \tilde{\phi}}{\partial \beta}
&=& \frac{1}{2}+\frac{1}{2\beta}+\frac{1}{2}x^2-x
\frac{I_1(\beta x)}{I_0(\beta x)} = \veps \\
\frac{\partial \tilde{\phi}}{\partial x} &=& \beta x -\beta
\frac{I_1(\beta x)}{I_0(\beta x)}= 0 \, . \label{systminmaxhmfb}
\end{eqnarray}
Inserting the second equation in the first, we find that $\beta^{-1}
= 2\veps -1 + x^2$. Substituting back in the second equation, and
using the function $B_{inv}$, the inverse function of $I_1/I_0$
previously defined, we have
\begin{equation}
\label{solbinvhmf} B_{inv}(x) - \frac{x}{2\veps -1 + x^2} =0 \, ,
\end{equation}
which is identical to Eq.~(\ref{solbinv}). The computation of the
second derivatives gives
\begin{eqnarray}
\label{derminmaxhmfa} \frac{\partial^2 \tilde{\phi}}{\partial
\beta^2}
=-\frac{1}{2\beta^2} - x \frac{\partial}{\partial x}\left(\frac{I_1(\beta x)}{I_0(\beta x)}\right)&<& 0 \\
\frac{\partial^2 \tilde{\phi}}{\partial x^2} =\beta -\beta
\frac{\partial}{\partial x}\left(\frac{I_1(\beta x)}{I_0(\beta
x)}\right) &>& 0 \, . \label{derminmaxhmfb}
\end{eqnarray}
These inequalities are both satisfied at the stationary points
determined by (\ref{systminmaxhmfa}) and (\ref{systminmaxhmfb}).
Actually, Eq.~(\ref{derminmaxhmfa}) is identically satisfied, since
the derivative of $I_1/I_0$ is positive definite. This confirms
ensemble equivalence for the HMF model.

Finally, Eqs. (\ref{canentrminmax}) and (\ref{entrminmax1}) give
\begin{eqnarray}
\label{entrminmaxhmf} s(\veps)=s_{can}(\veps) &=& \frac{1}{2}
+\frac{1}{2}\ln (2\pi) +\frac{1}{2}\ln 2 +\frac{1}{2}\ln \left(
\veps -\frac{1}{2} +\frac{1}{2}x^2\right)
\nonumber \\
&& - \frac{x^2}{2\veps -1 +x^2} +\ln I_0\left(\frac{x^2}{2\veps -1
+x^2}\right) \, ,
\end{eqnarray}
with $x$ satisfying Eq. (\ref{solbinvhmf}). We thus obtain an
expression identical to (\ref{final2}), taking into account
Eq.~(\ref{solbinv}).

\subsection{The large deviation method and its applications}
\label{Methode_generale}

The mathematical theory of large deviations is a field in its own,
and obviously it is outside the scope of this review to give details
of this theory and to treat at the level of mathematical rigor what
will be presented about it. The methods based on large deviation
theory and their applications to problems in statistical mechanics
have been popularized among theoretical physicists in several books
and review papers \cite{Ellis85,Dembo,TouchettePhysRep}. We would
like to cite here a few interesting works in which physical systems
with long-range interactions have been studied using this method.
Michel and Robert~\cite{MichelRobert94} successfully used large
deviations techniques to rigorously prove the applicability of
statistical mechanics to two-dimensional fluid mechanics, proposed
earlier~\cite{Miller90,Robert91,RobertSommeria91}. Ellis {\em et al.}~\cite{Ellis99}
pursued the approach of Robert and Sommeria~\cite{RobertSommeria91} to solve
two-dimensional geophysical systems.

Many particle systems with long-range interactions often offer a
relatively simple field of application of the theory of large
deviations. This is very interesting, especially if one considers
that the outcome of the calculation is the entropy function.

The structure of this section is the following. In
subsection~\ref{generlarge}, where we introduce the method, we will
show how to obtain the entropy for a class of systems in which the
Hamiltonian can be expressed in a way that is often realized in
long-range systems. In subsection~\ref{Pottsmodel} we will show the
simple application to the 3-states Potts model, where the result can
be compared with the direct computation of the entropy. This simple
discrete system already shows ensemble inequivalence. In
subsection~\ref{largedeviationsforHMF}, we will treat again the HMF
model to introduce the application of the method to systems with
continuous variables. Interesting applications will be treated in
subsections~\ref{generalizedHMF} and \ref{modelphi4}, both devoted
to systems presenting ensemble inequivalence: a generalized HMF
model and the so called $\phi^4$ model. Finally, in
subsection~\ref{Colson-Bonifaciomodel}, we will present the
equilibrium solution of the Colson-Bonifacio model of a linear Free
Electron Laser.

\subsubsection{The computation of the entropy for long-range systems}
\label{generlarge}

Denoting again collectively with $x \equiv (\{ p_i \},\{ q_i \})$
the phase space variables of a Hamiltonian system, let us suppose
that the energy per particle $\veps(x)=H(x)/N$ can be expressed as a
function of (few) global ``mean-fields'' $\mu_1(x),\dots,\mu_n(x)$;
i.e., we suppose that it is possible to write
\begin{equation}
\label{findmu} \veps(x)= \bar{\veps}(\mu_1(x),\dots,\mu_n(x)) \,.
\end{equation}
This is a situation often realized in long-range systems. Actually,
one could require that formula (\ref{findmu}) be valid only
asymptotically for $N \to \infty$, while for large but finite $N$
the right hand side could contain a remainder $R(x)$ that can be
neglected in the thermodynamic limit. However, in mean-field systems
representation (\ref{findmu}) is exact for all $N$. With the
specification of the microscopic configuration $x$, we define what
is generally indicated as a microstate of the system. On the
contrary, specifying that the system is in a state in which the
global variables have given values $\mu_1,\dots,\mu_n$, we are
defining what is called a macrostate of the system. Once the
macrostate is chosen, the microscopic configuration is not
determined, since all $x$ that satisfy $\mu_k(x)=\mu_k$ for
$k=1,\dots,n$ belong to the same macrostate.

The identification, in a concrete system, of the global variables is
the first step in the application of the large deviation method.

The second step will be the computation of the entropy of the
different macrostates, i.e., the calculation of the function
\begin{equation}
\label{entrmustate} \bar{s}(\mu_1,\dots,\mu_n)= \lim_{N\to \infty}
\frac{1}{N} \ln \int \dd x \,
\delta(\mu_1(x)-\mu_1)\dots\delta(\mu_n(x)-\mu_n)~.
\end{equation}
Leaving aside for the moment the problem of computing
$\bar{s}(\mu_1,\dots,\mu_n)$, that at first sight does not seem to
be any simpler than computing the entropy function $s(\veps)$, it is
easy to see how this last function can be obtainable from
$\bar{s}(\mu_1,\dots,\mu_n)$. In fact, we have
\begin{eqnarray}\label{entrepsmu}
\int \dd x \delta \left[N(\veps(x)-\veps)\right] &=& \int \dd x \,
\delta \left[N(\bar{\veps}(\mu_1(x),\dots,\mu_n(x))-\veps)\right] \\
&=&\int \dd x \dd \mu_1\dots \dd \mu_n \, \delta(\mu_1(x)-\mu_1)
\dots \delta(\mu_n(x)-\mu_n) \delta \left[N(\bar{\veps}
(\mu_1,\dots,\mu_n)-\veps)\right]\\
&\stackrel{N\to+\infty}{\sim}& \int \dd \mu_1\dots \dd \mu_n \, \exp
\left[N\bar{s}(\mu_1, \dots,\mu_n)\right] \delta \left[N(\bar{\veps}
(\mu_1,\dots,\mu_n)-\veps)\right] \, .
\end{eqnarray}
We therefore have
\begin{equation}\label{entrepsmu1}
s(\veps)=
\sup_{[\mu_1,\dots,\mu_n|\tilde{\veps}(\mu_1,\dots,\mu_n)=\veps]}
\bar{s}(\mu_1,\dots,\mu_n)~.
\end{equation}
The solution of this extremal problem constitutes the third and
final step of the computation. We are left with the problem of
computing $\bar{s}(\mu_1,\dots,\mu_n)$, i.e., with the actual
implementation of the second step. Let us introduce the following
canonical partition function
\begin{equation}\label{canpartmu}
\bar{Z}(\lambda_1,\dots,\lambda_n) = \int \dd x \, \exp
\left[-N\left( \lambda_1 \mu_1(x) + \dots + \lambda_n
\mu_n(x)\right) \right] \, .
\end{equation}
Few steps completely analogous to those relating $\phi(\beta)=\beta
f(\beta)$ to $s(\veps)$ show that the free energy associated with
$\bar{Z}(\lambda_1,\dots,\lambda_n)$ is given by the
(multi-dimensional) Legendre-Fenchel transform of
$\bar{s}(\mu_1,\dots,\mu_n)$:
\begin {equation}
\label{canpartmumin} \bar{\phi}(\lambda_1,\dots,\lambda_n) \equiv
-\lim_{N \to \infty} \frac{1}{N} \ln
\bar{Z}(\lambda_1,\dots,\lambda_n) = \inf_{\mu_1,\dots,\mu_n} \left[
\lambda_1 \mu_1 + \dots + \lambda_n \mu_n -
\bar{s}(\mu_1,\dots,\mu_n)\right] \, .
\end{equation}
We know that in general $s(\veps)$ is not concave and it cannot be
obtained by the Legendre-Fenchel transform of $\phi(\beta)$. We
would expect the same difficulty in the inversion of
(\ref{canpartmumin}). However, if it happens that
$\bar{\phi}(\lambda_1,\dots,\lambda_n)$ is differentiable for real $\lambda$ (see Appendix
C), then we are guaranteed that $\bar{s}(\mu_1,\dots,\mu_n)$ can be obtained
by the inversion of (\ref{canpartmumin}) and that it is therefore
concave \cite{Touchette2003}
\begin{equation}
\label{entrmuinv} \bar{s}(\mu_1,\dots,\mu_n) =
\inf_{\lambda_1,\dots,\lambda_n} \left[ \lambda_1 \mu_1 + \dots +
\lambda_n \mu_n - \bar{\phi}(\lambda_1,\dots,\lambda_n)\right] \, .
\end{equation}

Obviously the practical usefulness of this method resides in the
fact that generally, even in the presence of phase transitions, the
function $\bar{\phi}(\lambda_1,\dots,\lambda_n)$ happens to be
differentiable. In Appendix C, we present a brief proof of the differentiability
of $\bar{\phi}(\lambda_1,\dots,\lambda_n)$ and of the validity of
(\ref{entrmuinv}) for the cases in which the global variables
$\mu_k(x)$ are sums of one-particle functions.

Finally, it is not difficult to see how $\phi(\beta)$ is related to
$\bar{s}(\mu_1,\dots,\mu_n)$. Again, with steps analogous to those
linking $\phi(\beta)$ to $s(\veps)$, we find that $\phi(\beta)$ is
given by the following extremal problem
\begin{equation}
\label{freeminsmu} \phi(\beta)=\beta f(\beta)=
\inf_{\mu_1,\dots,\mu_n}\left[\beta \bar{\veps}(\mu_1,\dots,\mu_n) -
\bar{s}(\mu_1,\dots,\mu_n)\right]~.
\end{equation}
The two variational problems (\ref{entrepsmu1}) and
(\ref{freeminsmu}) express the microcanonical entropy and the
canonical free energy as a function of $\bar{s}(\mu_1,\dots,\mu_n)$
and of the energy function $\bar{\veps}(\mu_1,\dots,\mu_n)$, and
offer a tool to study ensemble equivalence in concrete systems.

It is important to emphasize that the large deviation method allows
to reduce the statistical mechanics study of a model to an
optimization problem. The method of global variables reduces exactly
(or sometimes approximately) the search of the equilibrium solution
to a variational problem. This approach often drastically simplifies
the derivation of the statistical mechanics properties and is not
very well known among physicists.

However, such a procedure does not apply to all long-range
interacting systems. In particular, those for which statistical
mechanics cannot be reduced (even approximately) to a mean-field
variational problem are excluded. As shown, the method is strongly
dependent on the possibility to introduce global or coarse-grained
variables (examples are the averaged magnetization, the total
kinetic energy, {\em etc.}) such that the Hamiltonian can be
expressed as a function of these variables, as in
Eq.~(\ref{findmu}), modulo a remaining term $R(x)$ whose relative
contribution vanishes in the thermodynamic limit. If also short
range interactions are present the procedure will not be possible.
One might still be able to express the Hamiltonian as a function of
coarse-grained variables plus a rest, but the rest will not vanish
in the thermodynamic limit.

The global variables $\{ \mu_i \}$ could be given by fields. For
instance, they could correspond to a local mass density in a
gravitational system, or a coarse-grained vorticity density in 2D
turbulence. In these cases they would be mathematically infinite
dimensional variables. However, with natural extensions, the steps
that we have showed can be repeated
\cite{MichelRobert94,ellisdd,Ellis02}.

Finally, we note that we could be interested in entropy functions
depending not only on the energy density $\veps$, but also on other
quantities, see e.g. Eq.~(\ref{sofem}) for the dependence on
magnetization. In this case, also these additional quantities should
be expressed as functions of the global variables
$\mu_1,\dots,\mu_n$  (e.g. $m(x)=\bar{m}(\mu_1,\dots,\mu_n)$), while
the variational problem (\ref{entrepsmu1}) will be constrained to
fixed values of $\bar{\veps}(\mu_1,\dots,\mu_n)=\veps$ and
$\bar{m}(\mu_1,\dots,\mu_n)=m$. Examples of this sort will be
discussed in the following.

\subsubsection{The three-states Potts model: an illustration of the method}
\label{Pottsmodel}

Here, we apply the large deviation method to the three-state Potts
model with infinite range interactions \cite{bbdrjstatphys}. This
simple example has been used as a toy model to illustrate peculiar
thermodynamic properties of long-range systems~\cite{Ispolatov01a}.
The diluted three-state Potts model with short-range interactions
has also been studied in connection with ``Small'' systems
thermodynamics by Gross~\cite{Gross00}.

The Hamiltonian of the three-state Potts model is
\begin{equation}
 H_N = -\frac{J}{2N}\sum_{i,j=1}^{N} \delta_{S_i,S_j}~.
\label{Hpotts3}
\end{equation}
Each lattice site $i$ is occupied by a spin variable $S_i$, which
assumes three different states $a$, $b$, or $c$. A pair of spins
gives a ferromagnetic contribution $-J$ ($J>0$) to the total energy
if they are in the same state, and no contribution otherwise. It is
important to stress that the energy sum is extended over {\em all}
pairs $(i,j)$: the interaction is infinite range.

Let us apply the method, following the three steps described in the
following. The first step of the method consists in associating, to
every microscopic configuration $x$, global (coarse-grained)
variables, such that the Hamiltonian can be expressed as a function
of them. For Hamiltonian (\ref{Hpotts3}) the appropriate global
variables are
\begin{equation}
\label{muvarPotts} \mu \equiv (\mu_a,\mu_b)=(n_a,n_b) \quad ,
\end{equation}
where $(n_a,n_b)$ are the fractions of spins in the two different
states $a,b$. As a function of the microscopic configuration, we
have
\begin{equation}
\label{globalpotts} \mu_k = \frac{1}{N}\sum_{i=1}^N \delta_{S_i,k}
\, ,
\end{equation}
with $k=a,b$. The Hamiltonian is expressed in terms of the global
variables as
\begin{equation}
\label{Hpottpoly} \bar{\veps}(n_a,n_b) =
-\frac{J}{2}\left[n_a^2+n_b^2+(1-n_a-n_b)^2\right]\quad,
\end{equation}
which is exact for any $N$.

The second step is the computation of $\bar{s}(\mu_1,\mu_2)$
According to what we have shown in the previous subsection, we have
to compute first the partition function
\begin{equation}
\label{canpartmupotts} \bar{Z}(\lambda_a,\lambda_b) = \sum_{S_i}
\exp \left( -\lambda_a \sum_{i=1}^N \delta_{S_i,a} -\lambda_b
\sum_{i=1}^N \delta_{S_i,b} \right) \, ,
\end{equation}
where the integral over the configurations is a discrete sum in this
case. The last expression is easily solved, to find
\begin{equation}
\label{canpartmupotts1} \bar{Z}(\lambda_a,\lambda_b) = \left(
e^{-\lambda_a} + e^{-\lambda_b} +1 \right)^N \, .
\end{equation}
We therefore obtain
\begin{equation}\label{phipotts}
\bar{\phi}(\lambda_a,\lambda_b)= - \ln \left( e^{-\lambda_a} +
e^{-\lambda_b} +1 \right) \, .
\end{equation}
As this function is evidently analytic, we can compute the function
$\bar{s}$ by a Legendre-Fenchel transform. The entropy function
$\bar{s}(n_a,n_b)$ will then be
\begin{equation}
\label{entrpotts1} \bar{s}(n_a,n_b)=\inf_{\lambda_a,\lambda_b}
\left[ \lambda_a n_a +\lambda_b n_b + \ln \left( e^{-\lambda_a} +
e^{-\lambda_b} +1 \right)\right] \, ,
\end{equation}
which is easily solved to get
\begin{equation}
\label{entrpotts2} \bar{s}(n_a,n_b)= -n_a \ln n_a -n_b \ln n_b
-\left( 1-n_a-n_b\right) \ln \left( 1-n_a-n_b \right) \, .
\end{equation}

\begin{figure}[htb]
\resizebox{0.4\textwidth}{!}{\includegraphics{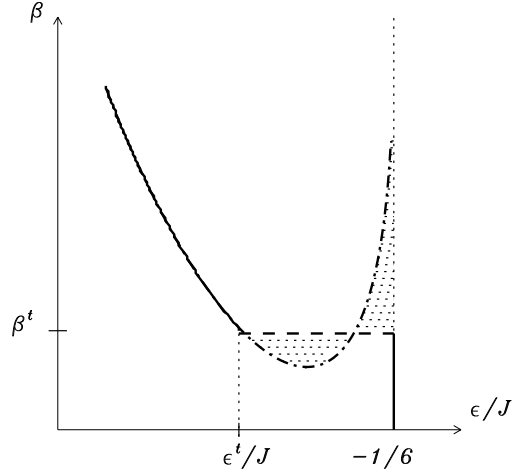}}
\caption{Caloric curve (inverse temperature vs. energy density) for
the three-states infinite range Potts model. The canonical solution
is represented by a solid line. The microcanonical solution
coincides with the canonical one for $\varepsilon\leq \varepsilon^t$
and is instead indicated by the dash-dotted line for
$\varepsilon^t\leq \varepsilon<-J/6$. The increasing part of the
microcanonical dash-dotted line corresponds to a negative specific
heat region. In the canonical ensemble, the model displays a first
order phase transition at $\beta^t$. The two dotted regions bounded
by the dashed line and by the microcanonical dash-dotted line have
the same area (Maxwell's construction).} \label{potts_fig}
\end{figure}

We now proceed to the third and final step of the calculation of the
entropy $s(\veps)$. The variational problem (\ref{entrepsmu1})
becomes
\begin{equation}
\label{entropie_potts} s(\veps) =\sup_{n_a,n_b} \Bigl( -n_a\ln
n_a-n_b\ln n_b-(1-n_a-n_b) \ln(1-n_a-n_b)\ \Bigl|
       -\frac{J}{2}\left(n_a^2+n_b^2+(1-n_a-n_b)^2\right)=\varepsilon \Bigr)\quad.
\end{equation}
As we have anticipated, this expression could be obtained by direct
counting. The variational problem (\ref{entropie_potts}) can be
solved numerically. The microcanonical inverse temperature can then
be derived and it is shown in Fig.~\ref{potts_fig} in the allowed
energy range $[-J/2,-J/6]$. Ispolatov and Cohen~\cite{Ispolatov01a}
have obtained the same result by determining the density of states.
A negative specific heat region appears in the energy range
$[-0.215\,J,-J/6]$.

Let us now consider the canonical ensemble. Applying
Eq.~(\ref{freeminsmu}) to Hamiltonian (\ref{Hpotts3}), we have
\begin{equation}
\label{elibre_potts} \phi(\beta)=\inf_{n_a,n_b} \Bigl( n_a\ln
n_a+n_b\ln n_b +(1-n_a-n_b) \ln (1-n_a-n_b) -\frac{\beta J}{2}
\left(n_a^2+n_b^2+(1-n_a-n_b)^2\right) \Bigr)\quad.
\end{equation}
To obtain the caloric curve, one has to compute $\veps=\dd \phi/\dd
\beta$. Figure~\ref{potts_fig} shows that at the canonical
transition inverse temperature $\beta^t \simeq 2.75$, corresponding
to the energy $\veps^t/J \simeq -0.255$, a first order phase
transition appears, with an associated latent heat. The low energy
``magnetized'' phase becomes unstable, while the high energy
``homogeneous'' phase, which has the constant energy density,
$\veps/J=-1/6$, is stabilized. In Fig.~\ref{potts_fig}, the two
dotted regions have the same area, respecting Maxwell's
construction.  At the inverse transition temperature, there is also
a jump in the global variables $(n_a,n_b,1-n_a-n_b)$, which are the
order parameters of the model.

This extremely simple example shows already ensemble inequivalence.
In the microcanonical ensemble, there is no phase transition and the
specific heat becomes negative. On the other hand, in the canonical
ensemble, there is a first order phase transition with a latent
heat. The caloric curves do not coincide. We observe that in the
energy range of ensemble inequivalence, microcanonical temperatures
do not coincide with any canonical one.

\subsubsection{The HMF model: dealing with continuous variables}
\label{largedeviationsforHMF}

In this subsection, we consider again the HMF model to show the
implementation of the large deviation method to a system with
continuous variables. Let us write again here Hamiltonian
(\ref{Ham_HMFM}) using the components of the magnetization, $m_x$
and $m_y$, defined in (\ref{magnHMF}) explicitly evidenced:
\begin{equation}\label{Ham_HMFMb}
H_N = \sum_{i=1}^{N}\frac{p_i^2}{2} +\frac{N}{2}
\left(1-m_x^2-m_y^2\right) \, .
\end{equation}
By a direct inspection of Hamiltonian~(\ref{Ham_HMFMb}), one can
identify the global variables $u=\frac{1}{N}\sum_i p_i^2$, $m_x$ and
$m_y$. Since $v=\frac{1}{N}\sum_i p_i$, the average momentum, is a
conserved quantity with respect to the dynamics defined by the
Hamiltonian, it is convenient to included it among the global
variables. Therefore, we will compute the entropy function
$s(\veps,v)$.

The first step of the procedure therefore consists in the
identification of the following global variables
\begin{equation}
\label{globalHMF} \mu =(u,v,m_x,m_y)\quad.
\end{equation}
We note that also in this case the expression of the Hamiltonian as
a function of the three global variables $u$, $m_x$ and $m_y$ is
exact for each $N$, since we have
\begin{equation}
\label{enermuhmf} \bar{\veps}(u,m_x,m_y)=
\frac{1}{2}\left(u+1-m_x^2-m_y^2\right) \, ,
\end{equation}
without a reminder. Let us now go to the second step, i.e. the
computation of $\bar{s}(u,v,m_x,m_y)$. We know, by now, that first
we have to compute the partition function
\begin{equation}
\label{canpartmuhmf}
\bar{Z}(\lambda_u,\lambda_v,\lambda_x,\lambda_y)= \int \left(
\prod_i \dd \theta_i \dd p_i\right) \, \exp \left(
-\lambda_u\sum_{i=1}^N p_i^2 -\lambda_v\sum_{i=1}^N p_i
-\lambda_x\sum_{i=1}^N \cos \theta_i -\lambda_y\sum_{i=1}^N \sin
\theta_i \right)\, ,
\end{equation}
which is solved to get
\begin{equation}\label{canpartmuhmf1}
\bar{Z}(\lambda_u,\lambda_v,\lambda_x,\lambda_y)= \left[
e^{\lambda_v^2/4\lambda_u}\sqrt{\frac{\pi}{\lambda_u}}\,
I_0\left(\sqrt{\lambda_x^2+\lambda_y^2}\right)\right]^N \, ,
\end{equation}
where $I_0$ is the modified Bessel function of order~0. We note that
the existence of the integral in (\ref{canpartmuhmf}) requires that
$\lambda_u >0$. Then, we have
\begin{equation}
\label{phihmflarge}
\bar{\phi}(\lambda_u,\lambda_v,\lambda_x,\lambda_y)=
-\frac{\lambda_v^2}{4\lambda_u}-\frac{1}{2}\ln \pi +\frac{1}{2}\ln
\lambda_u -\ln I_0\left(\sqrt{\lambda_x^2+\lambda_y^2}\right) \, .
\end{equation}
The analyticity of this function allows us to write that
\begin{eqnarray}
\label{entrhmflarge}
\bar{s}(u,v,m_x,m_y)=\inf_{\lambda_u,\lambda_v,\lambda_x,\lambda_y}
\left[ \lambda_u u +\lambda_v v +\lambda_x m_x +\lambda_y m_y
+\frac{\lambda_v^2}{4\lambda_u}+\frac{1}{2}\ln \pi \right.\nonumber \\
\left. -\frac{1}{2}\ln \lambda_u +\ln
I_0\left(\sqrt{\lambda_x^2+\lambda_y^2}\right)\right] \, .
\end{eqnarray}
This variational problem can be solved explicitly, and this can be
done for the ``kinetic'' subspace $(\lambda_u,\lambda_v)$ separately
from the ``configurational'' one $(\lambda_x,\lambda_y)$, giving
\begin{equation}
\label{entrhmflarge1}
\bar{s}(u,v,m_x,m_y)=\bar{s}_{kin}(u,v)+\bar{s}_{conf}(m_x,m_y)~.
\end{equation}
We note in addition that $\bar{s}_{conf}(m_x,m_y)$ does not depend
on $m_x$ and $m_y$ separately, but on the modulus
$m=\sqrt{m_x^2+m_y^2}$. This was expected, since, as already
emphasized, we have degeneracy with respect to the direction of the
spontaneous magnetization. Using the function $B_{inv}$ previously
introduced, i.e., the inverse function of $I_1/I_0$, we obtain
\begin{eqnarray}
\label{entrhmflarge2}
\bar{s}_{kin}(u,v)&=&\frac{1}{2}+\frac{1}{2}\ln \pi
+\frac{1}{2}\ln 2(u-v^2) \\
\bar{s}_{conf}(m) &=& -m B_{inv}(m) + \ln I_0(B_{inv}(m)) \, .
\label{entrhmflarge3}
\end{eqnarray}
Let us remark that Cauchy-Schwarz inequality implies that $u\geq
v^2$.

The third step of the procedure gives the entropy function
\begin{eqnarray}\label{casmicrocanonique}
s(\veps,v)&=&\sup_{u,m} \left[\bar{s}(u,v,m)\Biggr|
\frac{u}{2}+\frac{1}{2}-\frac{m^2}{2}=\veps \right] \\
&=&\sup_{u,m} \left[\bar{s}_{kin}(u,v)+\bar{s}_{conf}(m)\Biggr|
\frac{u}{2}+\frac{1}{2}-\frac{m^2}{2}=\veps \right] \\
&=&\frac{1}{2}+\frac{1}{2}\ln (4\pi) + \frac{1}{2}\ln
\left(\veps-\frac{1}{2}
+\frac{1}{2}m^2-\frac{1}{2}v^2\right)%\nonumber \\&&
 -m B_{inv}(m) + \ln I_0(B_{inv}(m)) \, ,
\end{eqnarray}
where in the last equality $m$ satisfies the equation
\begin{equation}
\label{constraintmlarge} \frac{m}{2\veps -1 +m^2 -v^2} -
B_{inv}(m)=0 \, .
\end{equation}
This function depends on energy and on momentum. Maximizing with
respect to $v$, we obtain $s(\veps)$. It is easy to find that this
maximum is obtained when $v=0$ for each $\veps$, and that the
entropy $s(\veps)$ is given by
\begin{eqnarray}
\label{entrlargehmf} s(\veps) &=& \frac{1}{2} +\frac{1}{2}\ln (4\pi)
+\frac{1}{2}\ln \left( \veps -\frac{1}{2} +\frac{1}{2}m^2\right)
\nonumber \\
&& - \frac{m^2}{2\veps -1 +m^2} +\ln I_0\left(\frac{m^2}{2\veps -1
+m^2}\right) \, ,
\end{eqnarray}
with $m$ satisfying Eq.~(\ref{constraintmlarge}) taken at $v=0$. We
have therefore recovered the previous expressions, i.e.,
Eq.~(\ref{final2}) with Eq.~(\ref{solbinv}) or
Eq.~(\ref{entrminmaxhmf}) with Eq.~(\ref{solbinvhmf}).

For completeness, we also derive the rescaled free energy
$\phi(\beta)=\beta f(\beta)$. Applying Eq.~(\ref{freeminsmu}),
taking into account (\ref{enermuhmf}), we get
\begin{equation}
\label{freelargehmf} \phi(\beta)=\beta f(\beta) = \inf_{u,v,m}
\left[\beta \left( \frac{u}{2} +\frac{1}{2} -\frac{1}{2}m^2\right)
-\tilde{s}(u,v,m) \right] \, .
\end{equation}
This variational problem can be easily solved to get
(\ref{freehmf}).

\subsubsection{A generalized HMF model: ergodicity breaking}
\label{generalizedHMF}

A generalized HMF model has been recently introduced with
Hamiltonian~\cite{debuyl,bdmrPRE}
\begin{equation}
H_N=\sum_i^N \frac{p_i^2}{2}+U~, \label{hamiltonianmdeuxmquatre}
\end{equation}
where
\begin{equation}
U=N \, W(m)=N \left( -\frac{J}{2}m^2-K\frac{m^{4}}{4} \right)~,
\label{potmdeuxmquatre}
\end{equation}
with $m=\sqrt{m_x^2+m_y^2}$, being $m_x,m_y$ defined in formula
(\ref{magnHMF}). Models of this kind have been used to describe the
physics of nematic liquid crystals~\cite{Lebwohl}

\begin{figure}[htbp]
\begin{center}
\resizebox{0.45\textwidth}{!}{\includegraphics{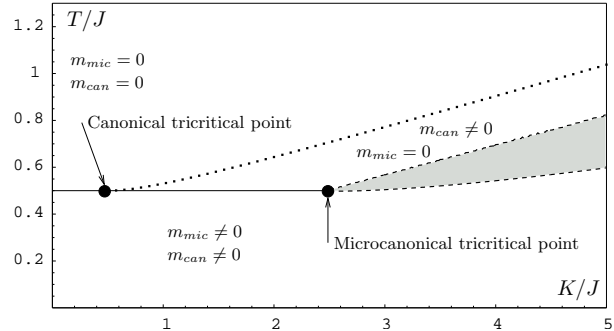}}
\end{center}
\caption{Phase diagram of model (\ref{hamiltonianmdeuxmquatre}). The
canonical second order transition line (solid line at $T/J=1/2$)
becomes first order (dotted line, determined numerically) at the
canonical tricritical point. The microcanonical second order
transition line coincides with the canonical one below $K/J=1/2$ but
extends further right to the microcanonical tricritical point at
$K/J=5/2$. At this latter point, the transition line bifurcates in
two first order microcanonical lines, corresponding to a temperature
jump. The behavior of the order parameter in the two ensembles is
also shown in the figure, to highlight the striking difference in
the predictions of the two ensembles.} \label{phasediagrammquatre}
\end{figure}

\paragraph{Statistical mechanics}
Let us now study the statistical mechanics of model
(\ref{hamiltonianmdeuxmquatre}) with $J$, $K >0$ using large
deviation techniques, in a similar way as it was done for the HMF
model in subsection~\ref{largedeviationsforHMF}. Like the HMF model
this model has four global variables
\begin{equation}
\mu=(m_x,m_y,u,v)
\end{equation}
and its microcanonical entropy is given by the same formula
(\ref{casmicrocanonique}) as for the HMF model. However, the
consistency equation (\ref{constraintmlarge}) is replaced by the
following
\begin{equation}
B_{inv}(m)=\frac{Jm+Km^3}{2 \veps + Jm^2 +Km^4/2-v^2}~.
\label{cons_mquatre}
\end{equation}
One solves analytically this relation for second order phase
transition and numerically for first order phase transition. Taking
then into account the standard relation between inverse temperature
and energy, one gets the phase diagram and the caloric curves
discussed below. Similarly to the HMF model, one can also get the
rescaled canonical free energy
\begin{equation}
\phi(\beta)=\inf_x \left[\frac{\beta Jm^2}{2}+\frac{3 \beta K
m^4}{4} -\ln (2 \pi I_0 [\beta (Jm+Km^3)]) \right]
\label{freeenergymquatre}
\end{equation}
In Fig.~\ref{phasediagrammquatre}, we report the phase diagram of
model (\ref{hamiltonianmdeuxmquatre}) in the parameter space
$(T/J,K/J)$ in both the canonical and the microcanonical ensemble.
As expected~\cite{BMR}, the two phase diagrams differ in the region
where the canonical transition line is first order. The second order
phase transition lines at $T/J=1/2$ coincide up to the canonical
tricritical point, located at $K/J=1/2$, but then the predictions of
the two ensembles differ: while the canonical ensemble gives a first
order phase transition, in the microcanonical ensemble the critical
line at $T/J=1/2$ persists up to the microcanonical tricritical
point at $K/J=5/2$. Even more striking is the difference between the
two ensembles for $K/J>5/2$. Here, two different temperatures
coexist at the transition energy in the microcanonical ensemble,
showing what we have called a temperature jump. This corresponds to
a microcanonical first order phase transition, whose ``phase
coexistence'' region is shown by the shaded area in
Fig.~\ref{phasediagrammquatre}. As for the order parameter $m$, the most
impressive difference between the two ensembles is in the
intermediate region between the canonical first order phase
transition line and the upper microcanonical first order line, where
the canonical ensemble predicts a non vanishing order parameter
$m_{can}\neq 0$, while the microcanonical ensemble gives
$m_{mic}=0$.

\begin{figure}[htbp]
\centering
\resizebox{0.45\textwidth}{!}{\includegraphics{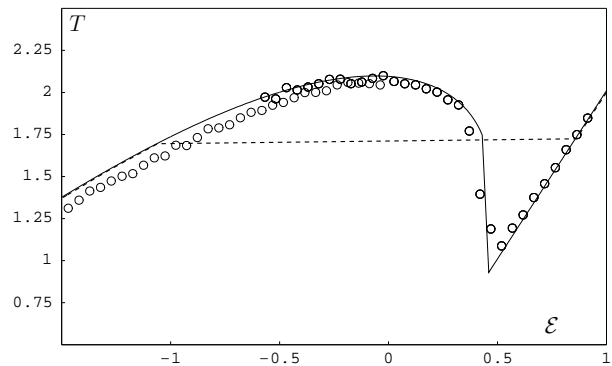}}
\caption{Caloric curve $T(\veps)$ at $K/J=10$. The microcanonical
ensemble (solid line) predicts a region of negative specific heat,
where temperature $T$ decreases as the energy is increased.
Moreover, a temperature jump is present at the transition energy. In
the canonical ensemble, we have a first order phase transition
(dashed line). The points are the result of a molecular dynamics
simulation performed by solving numerically the equations of motion
given by Hamiltonian (\ref{hamiltonianmdeuxmquatre}) with
$N=100$.}\label{caloriccurve}
\end{figure}

Figure~\ref{caloriccurve} displays temperature vs. energy, so-called
caloric curve, in the region where both the canonical and the
microcanonical ensemble predict a first order phase transition.
However, the microcanonical ensemble gives two temperatures at the
transition energy. Both a region of negative specific heat and a
temperature jump are present in the microcanonical ensemble. The
specific heat is always positive in the canonical ensemble and
latent heat is released. In a microcanonical simulation performed
with $N=100$ particles (shown by the points in
Fig.~\ref{caloriccurve}), the temperature jump is smoothed by the
finite size effects and negative specific heat is well reproduced.
The agreement with microcanonical thermodynamics predictions is very
good.

We have shown how a simple generalization of the HMF model can give
rise to all the features of ensemble inequivalence displayed by the
BEG model, {\it i.e.} negative specific heat and temperature jumps
in the microcanonical ensemble. A novel characteristic of this model
with respect to the BEG model is the fact that the variables take
continuous values. Moreover, the model possess a true Hamiltonian
dynamics which allows us to confirm the theoretical predictions
obtained in the thermodynamic limit also for finite $N$ and to study
non equilibrium features.

\paragraph{Parameter space convexity}
\label{phasediagramconvex}

We here discuss a concrete example where we can show that the space
of thermodynamic parameters is not convex, as already discussed in
Sec.~\ref{Convexity}. Historically, this phenomenon has been first
observed for a spin chain with asymmetric
coupling~\cite{borgonovi1,borgonovi2,celardo}, and for an Ising
model with both nearest neighbour and mean-field
interactions~\cite{schreiber}, see Sec.~\ref{Isinglongplusshort}.
However, it has been later realized that this occurs more generally,
and even in the simple model studied here
(\ref{hamiltonianmdeuxmquatre}). We will consider the case $J=-1$
and $K>0$~\cite{bdmrPRE} for which the $m^2$ term in the Hamiltonian
(\ref{hamiltonianmdeuxmquatre}) is antiferromagnetic. We will study
the structure of the set of accessible states in the space of
thermodynamic parameters in the microcanonical ensemble.

Intuitively, we expect that, for large values of $K$, the system is
ferromagnetic while, for small values of $K$, the antiferromagnetic
coupling will dominate and makes the system paramagnetic. As we
shall demonstrate below, there exists a range of values of the
parameter $K$ for which the model exhibits a first order
microcanonical phase transitions between a paramagnetic phase at
high energies and a ferromagnetic phase at low energies. In both
phases there are regions in the $(\veps,K)$ plane in which the
accessible magnetization interval exhibits a gap, resulting in
breaking of ergodicity.

\begin{figure}[htbp]
\centering
\resizebox{0.45\textwidth}{!}{\includegraphics{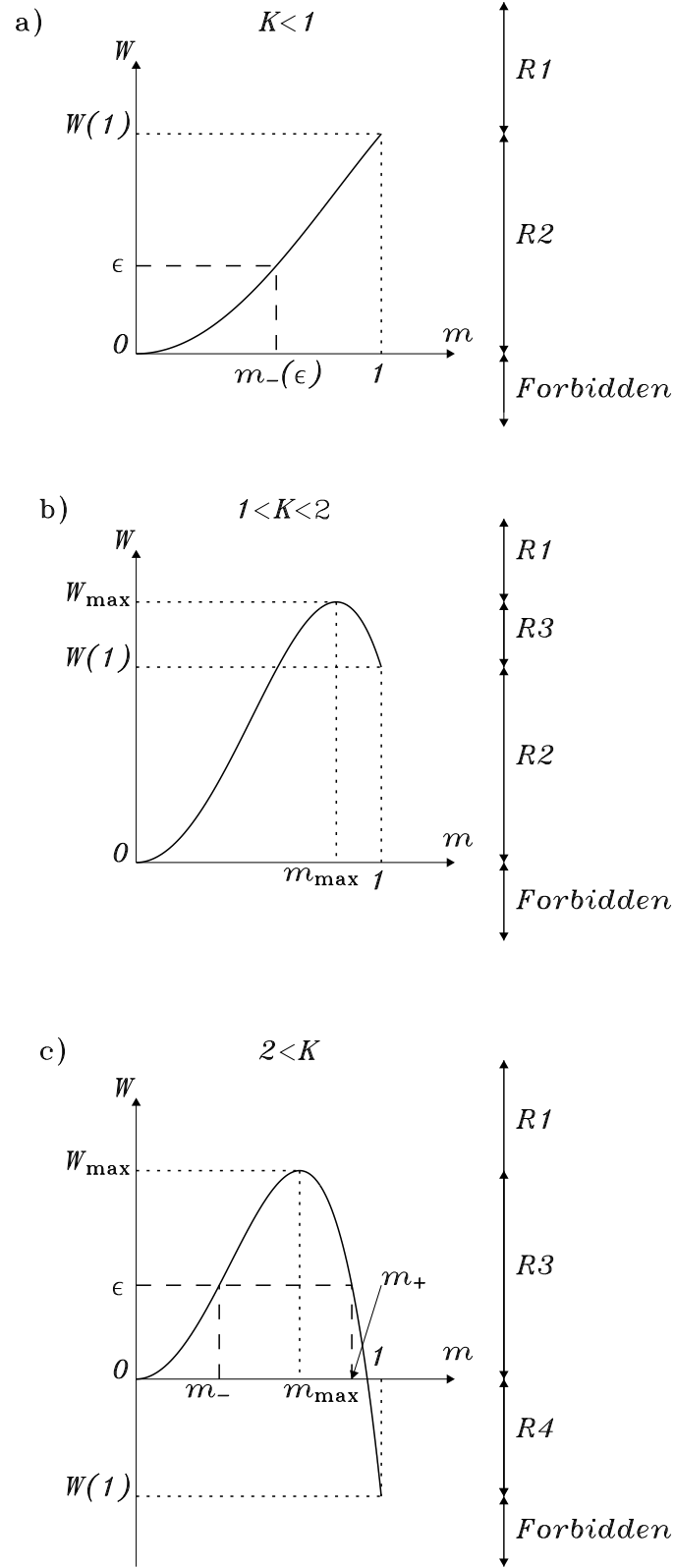}}
\bigskip
\caption{Specific potential energy $W$ vs. magnetization $m$ for
three different cases:  $K<1$ (panel a), $1<K<2$ (panel b) and $2<K$
(panel c). The location of the maximal magnetization $m_{\max}$ and
the corresponding potential energy $W_{\max}$ are shown (see text).
In panels (a) and (c), two examples of the location of the critical
magnetization $m_\pm(\veps,K)$ is indicated for energy density
values $\veps$ in the intermediate regions. }
\label{fig:diffmagnetization}
\end{figure}

The specific kinetic energy $u/2=\veps-W(m)$ is by definition a non
negative quantity, which implies that
\begin{equation}
\label{inequality} \veps \geq W(m)=m^{2}/2-Km^{4}/4.
\end{equation}
We will show that as a result of this condition not all the values
of the magnetization $m$ are attainable in a certain region in the
$(\veps,K)$ plane; a disconnected magnetization domain is indeed a
typical case. As explained below, this situation is the one of
interest. Let us characterize the accessible domains in the
$(\veps,K)$ plane more precisely by analyzing the different values
of $K$ (all this will become clear after looking at
Fig.~\ref{fig:diffmagnetization}).

\begin{figure}[htbp]
\resizebox{0.45\textwidth}{!}{\includegraphics{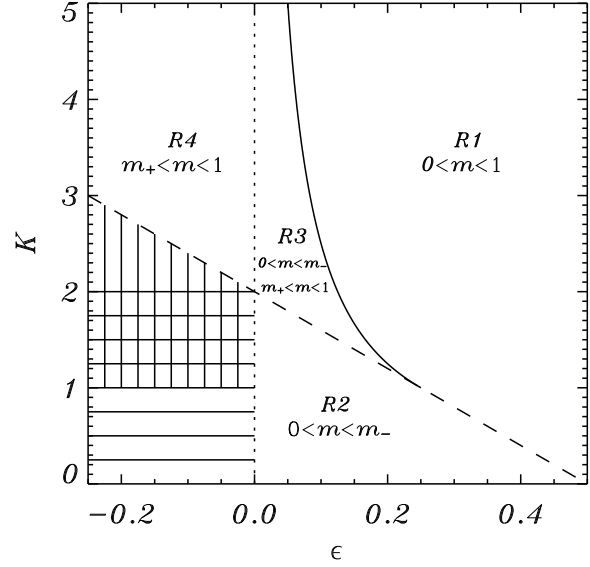}}
\caption{The $(\veps,K)$ plane is divided in several regions. The
solid curve corresponds to $K=1/(4\veps)$, the oblique dashed line
to $K=2-4\veps$, while the dotted one to $K=1$. The vertically
shaded, quadrilled, and horizontally shaded regions are forbidden.
The accessible magnetization interval in each of the four regions is
indicated (see text for details).} \label{fig:energy_range}
\end{figure}

For $K<1$, the local maximum of the potential energy $W$ is not
located inside the magnetization interval $[0,1]$ (see
Fig.~\ref{fig:diffmagnetization}a). The potential being a strictly
increasing function of the magnetization, the maximum is reached at
the extremum $m=1$. The complete interval $[0,1]$ is thus
accessible, provided the energy $\veps$ is larger than $W(1)$: the
corresponding domain is in {\em R1} defined and illustrated in
Figs.~\ref{fig:diffmagnetization} and~\ref{fig:energy_range}. The
horizontally shaded region is forbidden, since the energy is lower
than the minimum of the potential energy $W(0)=0$. Finally, the
intermediate region $0<\veps<W(1)$ defines region~{\em R2}: it is
important to emphasize that only the interval $\left[0,m_{-}\left(
\veps,K\right)\right]$, where $m_{\pm }\left( \veps,K\right)
=[({1\pm \sqrt{1-4\veps K}})/{K}]^{1/2}$, is accessible. Larger
magnetization values correspond to a potential energy~$W(m)$ larger
than the energy density~$\veps$, which is impossible.
Figure~\ref{fig:diffmagnetization}a also displays the value $m_-$
corresponding to an energy in the intermediate region {\em R2}.

For $K \geq 1$, the specific potential energy $W$ has a maximum
$W_{\max }=1/4K$ which is reached at $0<m_{\max }=1/\sqrt{K}\leq1$.
Figures~\ref{fig:diffmagnetization}b and
~\ref{fig:diffmagnetization}c, where the potential energy per
particle defined in Eq~(\ref{potmdeuxmquatre}) is plotted {\em vs}
magnetization $m$, display such cases. For an energy $\veps$ larger
than the critical value $W_{\max }$, condition ~(\ref{inequality})
is satisfied for any value of the magnetization $m$. The complete
interval $\left[ 0,1\right]$ is thus accessible for the
magnetization $m$. This  region is {\em R1} represented in
Fig.~\ref{fig:energy_range}.

Let us now consider the cases for which $\veps \leq W_{\max}$. As
discussed above, the minimum $W_{\min}$ of the potential energy is
also important to distinguish between the different regions. For
$1\leq K\leq 2$ (see Fig.~\ref{fig:diffmagnetization}b), the minimum
of $W(m)$ corresponds to the non-magnetic phase $m=0$ where
$W(0)=0$. The quadrilled region shown in
Fig.~\ref{fig:energy_range}, which corresponds to negative energy
values, is thus not accessible. On the contrary, positive energy
values are possible and correspond to very interesting cases, since
only sub-intervals of the complete magnetization interval $[0,1]$
are accessible. There are however two different cases
\begin{itemize}

\item for $0<W(1)<\veps<W_{\max }$, the domain of possible
magnetizations is $\left[0,m_{-}\left( \veps,K\right)\right]
\cup\left[ m_{+}\left(\veps,K\right) ,1 \right]$. The above
conditions are satisfied in {\em R3} of Fig.~\ref{fig:energy_range}.

\item for $0<\veps<W(1)$, only the interval $\left[0,
m_{-}\left(\veps,K\right)\right]$ satisfies
condition~(\ref{inequality}). This takes place within {\em R2} of
Fig.~\ref{fig:energy_range}.

\end{itemize}

For  the domain $2\leq K$, the minimum of the potential energy is
attained at the extremum, $m=1$, implying $\veps>W(1)=1/2-K/4$. The
vertically shaded region is thus forbidden. In the accessible
region, two cases can be identified

\begin{itemize}
\item for $W(1)<\veps<0$, only the interval $\left[ m_{+}\left(
\veps,K\right) ,1\right] $ satisfies condition~(\ref{inequality}).
It is important to note that $m_{+}\left(\veps,K\right) \leq 1 $
provided $\veps\geq 1/2-K/4 $. These cases correspond to region {\em
R4}.

\item for $0\leq \veps\leq W_{\max}$, the two intervals
$\left[0, m_{-}\left(\veps,K\right)\right] $ and $ \left[
m_{+}\left( \veps,K\right) ,1 \right]$ satisfy
condition~(\ref{inequality}), corresponding to {\em R3} of
Fig.~\ref{fig:energy_range}.

\end{itemize}

\begin{figure}[htbp]
\resizebox{0.45\textwidth}{!}{\includegraphics{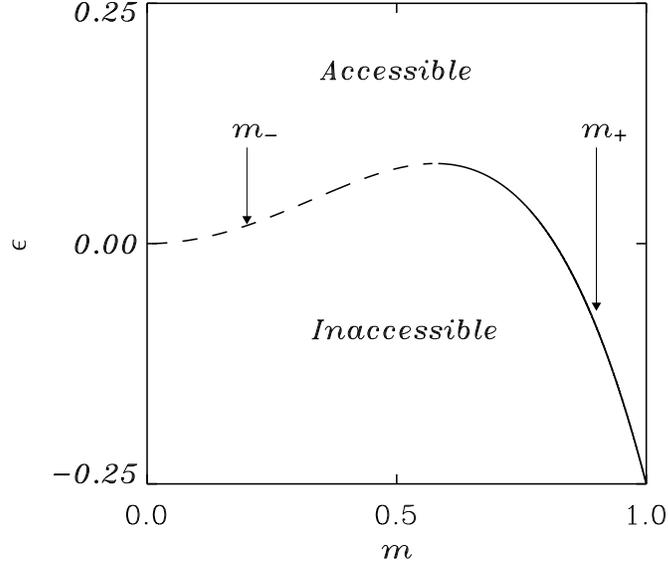}}
\vskip 0.5truecm
%\centering\includegraphics[width=7cm]{limitemagnetization.ps}
\caption{Accessible region in the $(m,\veps)$ plane for $K=3$. For
energies in a certain range, a gap in the accessible magnetization
values is present and defined by the two boundaries
$m_\pm(\veps,K)$.} \label{fig:limitelagnetization}
\end{figure}

In summary, the complete magnetization interval  $[0,1]$ is
accessible only in the region {\em R1}. In {\em R2}, only
$\left[0,m_{-}\right]$ is accessible, while only
$\left[m_{+},1\right]$ is accessible in {\em R4}. Finally, we note
that the phase space of the system is not connected in the region
{\em R3}. Indeed, the magnetization cannot vary continuously from
the first interval $\left[0,m_{-}\right]$ to the second one
$\left[m_{+},1\right]$, although both are accessible. These
restrictions yield the accessible magnetization domain shown in
Fig.~\ref{fig:limitelagnetization}. The fact that for a given energy
the space of the thermodynamic parameter $m$ is disconnected implies
ergodicity breaking for the Hamiltonian dynamics. It is important to
emphasize that the discussion above is independent of the number of
particles and ergodicity is expected to be broken even for a finite
$N$.

\paragraph{Phase diagram in the microcanonical ensemble}
\label{phasediagram}

We have thus found that in certain regions in the $(\veps,K)$ plane,
the magnetization cannot assume any value in the interval $[0,1]$.
For a given energy there exists a gap in this interval to which no
microscopic configuration can be associated. Let us now study the
phase diagram in the microcanonical ensemble, i.e. in the parameter
space $(\veps,K)$.

\begin{figure}[htbp]
\resizebox{0.45\textwidth}{!}{\includegraphics{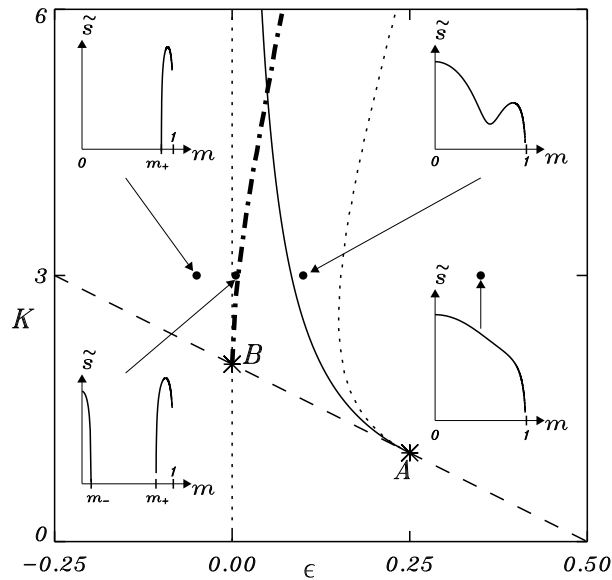}} \vskip
0.5truecm
%\centering\includegraphics[width=8cm]{mecastat.ps}
\caption{Phase diagram of the mean field model
(\ref{hamiltonianmdeuxmquatre}) with $J=-1$. The dash-dotted curve
corresponds to the first order phase transition line, issued from
the point $B(0,2)$. As in Fig.~\ref{fig:energy_range}, the solid
curve indicates  the right border of the region {\em R3}, where the
space of thermodynamic parameters is disconnected. The dashed line
corresponds to the $K=2-4\veps$. The dotted line issued from the
point $A(1/4,1)$ represents the metastability line for the
magnetized state, while the $\veps=0$ vertical dotted line is also
the metastability line for the paramagnetic state. The four insets
represent the entropy $\tilde{s}$ versus the magnetization $m$ for
the four energies: $\veps=$-0.05, 0.005, 0.1 and 0.35, when $K=3$.}
\label{fig:phase_diagram}
\end{figure}

First, by comparing the low and high energy regimes, it is possible
to show that a phase transition is present between the two regimes.
In the domain {\em R4} of Fig.~\ref{fig:energy_range}, for very low
energy $\veps$ (close to the limiting value $1/2-K/4$), the
accessible range for $m$ is a small interval located close to $m=1$
(see Fig.~\ref{fig:diffmagnetization}c). The maximum in $m$ of the
entropy~$\tilde{s}(\veps,m)$ corresponds therefore to a magnetized
state located very close to $m=1$ (see the top-left inset in
Fig.~\ref{fig:phase_diagram}). On the contrary, in the very large
energy domain ($\veps \gg m \simeq 1)$, the variations of the
entropy with respect to $m$ are dominated by the variations of the
configurational entropy $\tilde{s}_{conf}$, since the kinetic
entropy
\begin{equation}
\tilde{s}_{kin}(\veps,m)=\frac{1}{2} \ln \left( \veps -
\frac{m^2}{2} + K \frac{m^4}{4} \right) \label{eqentrokin}
\end{equation}
is roughly a constant $\tilde{s}\simeq (\ln \veps)/2$ when $m$ is of
order one. As expected, the configurational entropy is  a decreasing
function of the magnetization: the number of microstates
corresponding to a paramagnetic macrostate being much larger than
the same number for a ferromagnetic state. The configurational
entropy has therefore a single maximum located at $m^*=0$. A phase
transition takes place between the paramagnetic state at large
energy and a magnetic state at small energy. Moreover, as the
paramagnetic state is possible only for positive energies $\veps$
(see Figs.~\ref{fig:diffmagnetization}), the transition line is
located in the domain $\veps\geq 0$. In this region, the quantity
$\partial_{m}^{2}\tilde{s}(\veps,0)=-(1+1/(2\veps))$ is negative,
which ensures that, for any value of $\veps$ and $K$, the
paramagnetic state $m^*=0$ is a local entropy maximum. The latter
argument allows us to exclude a second order phase transition at a
positive critical energy, since the second derivative
$\partial_{m}^{2}\tilde{s}(\veps,0)$ would have to vanish, which is
impossible. The above argument leads to the conclusion that the
phase transition must be {\em first order}.

Let us now focus on the behavior of the entropy in the region {\em
R3}, where the accessible range for $m$ is the union of two
disconnected intervals $\left[ 0,m_{-} \right] \cup \left[
m_{+},1\right]$. As discussed above, the entropy
$\tilde{s}(\veps,m)$ has a local maximum in the first interval
$\left[ 0,m_{-} \right] $ located at $m^*=0$ and associated with the
entropy $\tilde{s}_{\max}^{1}=\tilde{s}(\veps,0) =\log (\veps)/2$.
In the second interval $\left[ m_{+},1\right]$, a maximum is also
present with $\tilde{s}_{\max}^{2}=\tilde{s}(\veps,m^*)$ where
$m^*\geq m_+>0$. As $\tilde{s}^{1}_{\max }(\veps)$ diverges to
$-\infty$ when $\veps$ tends to 0, a magnetized state is expected on
the line $\veps=0$, as long as $\tilde{s}_{\max }^{2}$ remains
finite. Since $K=2$ is the only value for which $\tilde{s}_{\max
}^{2}(0,K)$ diverges, the first order transition line originates at
the point $B(0,2)$ in Fig.~\ref{fig:phase_diagram}. Although it is
possible to study analytically the asymptotic behavior of the
transition line near this point, we can rather easily compute
numerically the location of the first order transition line,
represented by the dash-dotted line in Fig.~\ref{fig:phase_diagram}.

Figure~\ref{fig:phase_diagram} also shows the metastability line
(the dotted line starting at point $A(1/4,1)$), for the magnetized
state $m^* \neq 0$. To the right of this metastability line, there
is no metastable state (local entropy maximum for any $m>0$, see
bottom-right inset in Fig.~\ref{fig:phase_diagram}) while a
metastable state (local maximum) exists at some non vanishing
magnetization on the other side (see top-right inset in
Fig.~\ref{fig:phase_diagram}). Finally, the vertical dotted line
$\veps=0$ corresponds to the metastability line of the paramagnetic
state $m^*=0$.

%\begin{figure}[htbp]
%\resizebox{0.45\textwidth}{!}{\includegraphics{superposentropie.ps}}
%%\centering{\includegraphics[width=8cm]{superposentropie.ps}}
%\caption{Entropy $\tilde{s}$ versus magnetization $m$ for $K=3$ and
%several energy values. The different curves correspond, from top to
%bottom, to $\veps=0.35$ (dotted), 0.155 (dash-triple dotted,
%metastability limit for the magnetized state), 0.1 (dashed),
%$\veps=1/(4K)=1/12$ (solid, appearance of the gap), 0.0089
%(dash-dotted, first order phase transition), -0.05 (long dashed).
%This picture demonstrates that gaps in the accessible states develop
%as the energy is lowered.} \label{fig:superposentropie}
%\end{figure}

One of the key issues we would like to address is the possible links
between the breakdown of connectivity, and thus ergodicity breaking,
on the one hand, and the phase transition, on the other hand.
Obvious general properties do exist: a region of parameters where
the space is disconnected corresponds to a region where metastable
states do exist. Let us justify this statement.  At the boundary of
any connected domain,  when the order parameter is close to its
boundary value $m_b$, there is a single accessible state. In a model
with continuous variables, like the one we discuss here, this leads
to a divergence of the entropy. In this case the singularity of the
entropy is proportional to $\ln(m-m_b)$ (see for instance equation
(\ref{eqentrokin})). For a model with discrete variables, like an
Ising model, the entropy would no more reach $-\infty$ as $m$ tends
to $m_b$, but would rather take a finite value. However, the
singularity would still exist and would then be proportional to
$(m-m_b)\ln(m-m_b)$. In both cases, of discrete and continuous
variables, at the boundary of any connected domain, the derivative
of the entropy as a function of the order parameter tends generally
to $\pm\infty$. As a consequence, entropy extrema cannot be located
at the boundary. Thus, a local entropy maximum (metastable or
stable) does exist in a region of parameters where the space is
disconnected.

Hence, there is an entropy maximum (either local or global) in any
connected domain of the space. For instance, considering the present
model, in Fig.~\ref{fig:phase_diagram}, the area {\em R3} is
included in the area where metastable states exist (bounded by the
two dotted lines and the dashed line). In such areas where
metastable states exist, one generically expects first order phase
transitions. Thus the breakdown of phase space connectivity is
generically associated to first order phase transition, as
exemplified by the present study. However, this is not necessary,
one may observe metastable states without first order phase
transitions, or first order phase transitions without connectivity
breaking.

A very interesting question is related to the critical points $A$
and $B$ shown in Fig.~\ref{fig:phase_diagram}. As observed in the
phase diagram, the end point for the line of first order phase
transition (point $B$) corresponds also to a point where the
boundary of the region where the space is disconnected is not
smooth. Similarly, the end point for the line of appearance of
metastable states (point $A$) is also a singular point for the
boundary of the area where the space is disconnected. It is thus
possible to propose the conjecture that such a relation is generic,
and that it should be observed in other systems where both first
order phase transitions and phase space ergodicity breaking do
occur.

\paragraph{Equilibrium dynamics}
\label{equilibriumdynamics}

The feature of disconnected accessible magnetization intervals,
which is typical of systems with long-range interactions, has
profound implications on the dynamics. In particular, starting from
an initial condition which lies within one of these intervals, local
dynamics is unable to move the system to a different accessible
interval. A macroscopic change would be necessary to carry the
system from one interval to the other. Thus the ergodicity is broken
in the microcanonical dynamics even at finite $N$.

\begin{figure}[htbp]
\resizebox{0.45\textwidth}{!}{\includegraphics{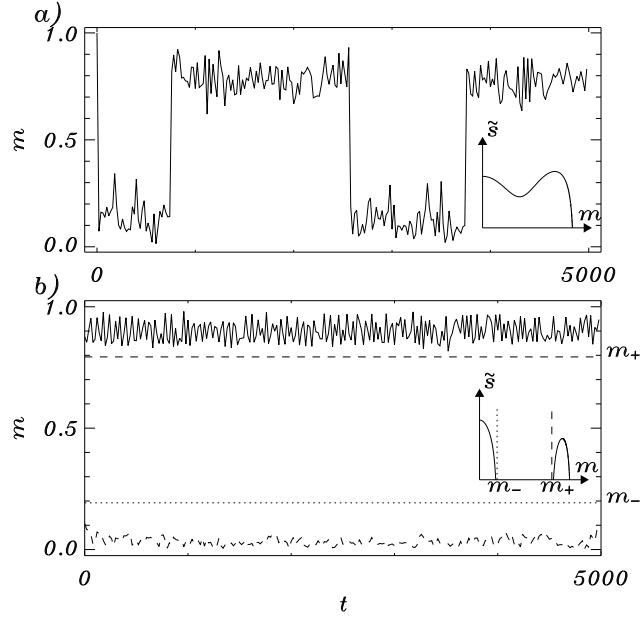}}
%\centering\includegraphics[width=8cm]{dynamic.ps}
\caption{Time evolution of the magnetization $m$ (the entropy of the
corresponding cases is plotted as an inset). Panel a) corresponds to
the case $\veps=0.1$ and $K=8$, while panel b) to $\veps=0.0177$ and
$K=3$. In panel b), two different initial conditions are plotted
simultaneously: the solid line corresponds to $m(t=0)=0.1$ while the
dashed line to $m(t=0)=0.98$. The dashed (resp. dotted) line in
panel b) corresponds to the line $m=m_+\simeq0.794$ (resp.
$m=m_-\simeq0.192$).} \label{fig:tracem}
\end{figure}

In Ref.~\cite{schreiber}, this point has been demonstrated using the
microcanonical Monte-Carlo dynamics suggested by
Creutz~\cite{creutz}, see Sec.~\ref{Isinglongplusshort}. Here, we
use the Hamiltonian dynamics given by the equation of motions
\begin{eqnarray}
\dot{\theta_n}&=&+\frac{\partial H_N}{\partial p_n}=p_n \\
\dot p_n&=&-\frac{\partial H}{\partial\theta_n}=
N\left(1-Km^2\right)\left(\sin\theta_nm_x-\cos\theta_nm_y\right).
\end{eqnarray}
We display in Fig.~\ref{fig:tracem} the evolution of the
magnetization for two cases, since we have shown above that the gap
opens up when $\veps$ decreases. The first case corresponds to the
domain {\em R1}, in which the accessible magnetization domain is the
full interval $[0,1]$. Fig.~\ref{fig:tracem}a presents the time
evolution of $m$. The magnetization switches between the
paramagnetic metastable state $m^*=0$ and the ferromagnetic stable
one $m^*>0$. This is possible because the number of particles is
small ($N=20$) and, as a consequence, the entropy barrier (see the
inset) can be overcome. Considering a system with a small number of
particles allows to observe flips between local maxima, while such
flips would be less frequent for larger $N$ values.

In the other case, we consider a stable $m^*=0$ state which is
disconnected from the metastable one. This makes the system unable
to switch from one state to the other. Note that this feature is
characteristic of the microcanonical dynamics, since an algorithm
reproducing canonical dynamics would allow the crossing of the
forbidden region (by moving to higher energy states, which is
impossible in the microcanonical ensemble). The result of two
different numerical simulations is reported in
Fig.~\ref{fig:tracem}b. One is initialized with a magnetization
within $[0,m_-]$, while the other corresponds to an initial
magnetization close to $m(0)=1$ ({\em i.e.} within $[m_+,1]$). One
clearly sees that the dynamics is blocked in one of the two possible
regions, and not a single jump is visible over a long time span.
This is a clear evidence of ergodicity breaking.

\subsubsection{The mean-field $\phi^4$ spin model: negative susceptibility}
\label{modelphi4}

We here show how to solve the so called mean-field $\phi^4$ spin
model using large deviations. The Hamiltonian of the model is the
following
\begin{equation}
\label{ham_phi4} H_N = \sum_{i=1}^N \left( \frac{p_i^2}{2}
-\frac{1}{4}q_i^2 +\frac{1}{4}q_i^4\right)
-\frac{1}{4N}\sum_{i,j=1}^N q_i q_j \, .
\end{equation}
It is a system of unit mass particles moving on a line. These
particles are subjected to a local double-well potential, and they
interact with each other through a mean-field (infinite range)
interaction given by the all-to-all coupling in the double sum.
%Here again, we shall use $x$ to denote a point of
%the phase space of the system, i.e., $x=(\{p_i\},\{q_i\})$.

This model was introduced by Desai and Zwanzig \cite{desai1978}.
More recently, the canonical solution was obtained
\cite{dauxois2003}, showing that the system exhibits a second-order
ferromagnetic phase transition at a critical temperature $T_c \simeq
0.264$, corresponding to a critical energy $\veps_c = T_c/2 \simeq
0.132$. Later, the entropy of the $\phi^4$ model in presence of an
extra magnetic field was computed \cite{campa2006}. Finally, a
calculation of the entropy, as a function of energy and
magnetization
\begin{equation}
\label{magnphi4} m = \frac{1}{N} \sum_{i=1}^N q_i \,~,
\end{equation}
has been performed in Ref.~\cite{hahn2005,hahn2006,campa2007}. We
note that in this model the modulus of this quantity is not bounded
inside the interval $[-1,1]$.

We will perform the three steps of the large deviation method, in
order to compute the microcanonical entropy as a function of the
energy $\veps$ and of the magnetization $m$; i.e., we will compute
the function $\tilde{s}(\veps,m)$ defined in Eq.~(\ref{sofem}). More
details can be found in Ref.~\cite{campa2007}.

%We should note that, contrary to the HMF model, the additional
%variables, other than the energy $\veps$, on which the entropy depends, is not
%an integral of the motion, since the magnetization $m(x)$ is not conserved by
%the temporal evolution determined by the Hamiltonian of the $\phi^4$ model.

The first step consists in the identification of the global
variables in terms of which we can express the energy $\veps$. In
addition to magnetization other global variables are $u$, twice the
average kinetic energy, and
\begin{equation}
\label{onsitephi4} z=\frac{1}{4N}\sum_{i=1}^N \left( q_i^4-q_i^2
\right) \, ,
\end{equation}
related to the local potential. We easily see that
\begin{equation}\label{enermuphi4}
\bar{\veps}(u,z,m) = \frac{1}{2}u + z -\frac{1}{4}m^2~.
\end{equation}

The second step begins with the computation of the partition
function
\begin{equation}\label{canpartmuphi4}
\bar{Z}(\lambda_u,\lambda_z,\lambda_m)= \int \left( \prod_i \dd q_i
\dd p_i\right) \, \exp \left[ -\lambda_u\sum_{i=1}^N p_i^2
-\lambda_m\sum_{i=1}^N q_i -\lambda_z\sum_{i=1}^N \left( q_i^4 -
q_i^2 \right) \right]\, .
\end{equation}
Also in this case the calculation splits into the one of the kinetic
part and of the potential part, and we have
\begin{equation}\label{canpartmuphi41}
\bar{Z}(\lambda_u,\lambda_z,\lambda_m)= \bar{Z}_u(\lambda_u)
\bar{Z}_{z,m}(\lambda_z,\lambda_m)= \left[
\sqrt{\frac{\pi}{\lambda_u}}\right]^{\frac{N}{2}}
\left[\int_{-\infty}^{+\infty} \dd q \, e^{-\lambda_m q-\lambda_z
\left( q^4 - q^2\right)} \right]^N \, ,
\end{equation}
where the two terms on the right hand side define
$\bar{Z}_u(\lambda_u)$ and $\bar{Z}_{z,m}(\lambda_z,\lambda_m)$,
respectively. The existence of the integral requires $\lambda_u >0$
and $\lambda_z >0$. From the previous expression, we find
\begin{equation}\label{phiphi4large}
\bar{\phi}(\lambda_u,\lambda_z,\lambda_m)= \bar{\phi}_u(\lambda_u)+
\bar{\phi}_{z,m}(\lambda_z,\lambda_m) \, ,
\end{equation}
with
\begin{eqnarray}\label{phiphi4large2}
\bar{\phi}_u(\lambda_u) &=& -\frac{1}{2}\ln \pi +\frac{1}{2}
\ln \lambda_u  \\
\bar{\phi}_{z,m}(\lambda_z,\lambda_m) &=& -\ln
\left[\int_{-\infty}^{+\infty} \dd q \, e^{-\lambda_m q-\lambda_z
\left( q^4 - q^2\right)} \right]\, .
\end{eqnarray}
Both these functions are analytic, and therefore we can write
\begin{equation}\label{entrphi4large}
\bar{s}(u,z,m)=\bar{s}_{kin}(u) + \bar{s}_{conf}(z,m) \, ,
\end{equation}
where
\begin{eqnarray}\label{entrphi4large1}
\bar{s}_{kin}(u)&=&\inf_{\lambda_u} \left[ \lambda_u u
-\bar{\phi}_u(\lambda_u)\right]
=\frac{1}{2}+\frac{1}{2}\ln (2\pi)+\frac{1}{2}\ln u  \\
\bar{s}_{conf}(z,m)&=&\inf_{\lambda_z,\lambda_m} \left[ \lambda_z z
+\lambda_m m -\bar{\phi}_{z,m}(\lambda_z,\lambda_m) \right] \, .
\end{eqnarray}

\begin{figure}[htbp]
\centering
\resizebox{0.98\textwidth}{!}{\includegraphics{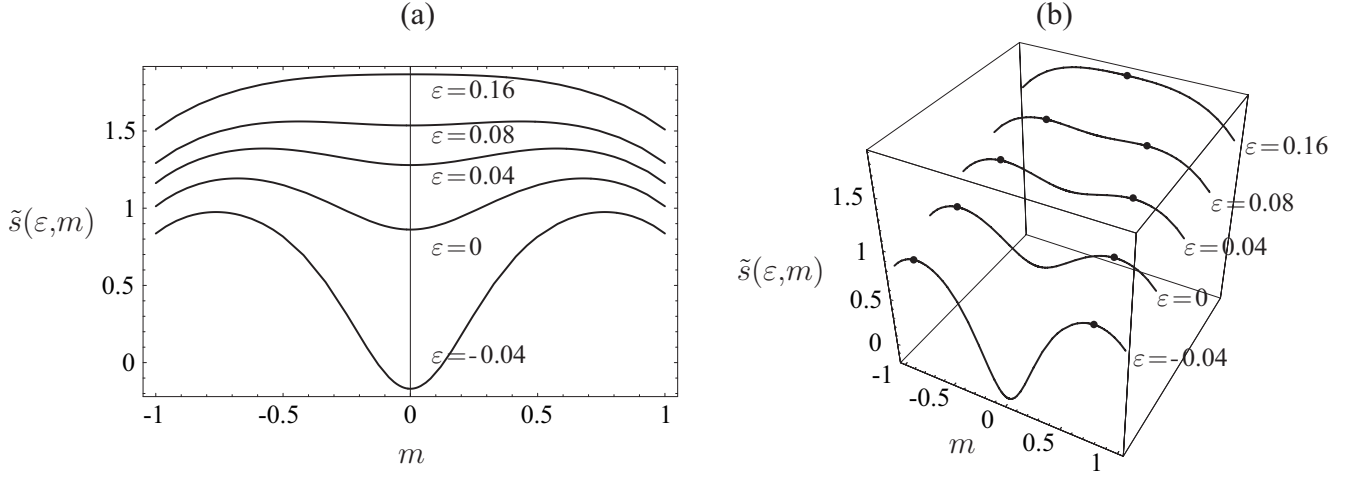}}
\caption{(a) Entropy $\tilde{s}$ as a function of magnetization $m$
for different values of the energy $\veps$. (b) 3D view of
$\tilde{s}(\veps,m)$. The black dots show the location of the
equilibrium values $m^*$ in the microcanonical ensemble in which the
magnetization constraint is released, and therefore the equilibrium
at a given $\veps$ is realized for the $m^*$ value maximizing
$\tilde{s}(\veps,m)$.} \label{sem3fig}
\end{figure}

The third step of the procedure gives the entropy function
\begin{equation}\label{entropyphi4large}
\tilde{s}(\veps,m)=\sup_{u,z} \left[\bar{s}_{kin}(u) +
\bar{s}_{conf}(z,m) \Biggr| \frac{u}{2}+z-\frac{m^2}{4}=\veps
\right] \, ,
\end{equation}
that can be rewritten as
\begin{equation}
\label{entropyphi4large1} \tilde{s}(\veps,m)
=\frac{1}{2}+\frac{1}{2}\ln (4\pi) + \sup_z
\left[\frac{1}{2}\ln \left(\veps -z +\frac{m^2}{4}\right) +
\bar{s}_{conf}(z,m) \right] \, .
\end{equation}
This variational problem can be solved numerically and the result is
plot in Fig.~\ref{sem3fig} for different values of~$\varepsilon$.

The purpose of this subsection, besides the possibility to offer a
further implementation of the large deviation method, is to show an
example of ensemble inequivalence which is not related to the
presence of a negative specific heat in the microcanonical ensemble,
but rather to the negativity of another quantity that in the
canonical ensemble is positive definite, namely the magnetic
susceptibility at constant temperature (see Sec.\ref{secnegsusc}).
Although striking, this feature appears for other long-range systems
whose entropy depends on two macroscopic variables. For instance, it
has been recently discussed in connection with models of two-dimensional geophysical
flows \cite{venaillebouchet}, for which entropy depends on total circulation
and energy.

The inequivalence has to be considered with respect to the canonical
ensemble, whose partition function is
\begin{equation}\label{canpartmuphi4h}
Z(\beta,h)= \int \left( \prod_i \dd q_i \dd p_i\right) \, \exp
\left( -\beta\left[\sum_{i=1}^N \left(
\frac{p_i^2}{2}-\frac{q_i^2}{4} +\frac{q_i^4}{4}\right)
-\frac{1}{4N}\sum_{i,j=1}^N q_i q_j -h \sum_{i=1}^N q_i \right]
\right) \, .
\end{equation}
From this we derive the rescaled free energy
\begin{equation}
\label{fofbetahphi4} \phi(\beta,h) =\beta f(\beta,h) =
-\lim_{N\rightarrow \infty} \frac{1}{N} \ln Z(\beta,h) =
\inf_{\veps,m}\left[\beta \veps -\beta hm-\tilde{s}(\veps,m)\right]
\, .
\end{equation}
It is straightforward to apply to this case expression
(\ref{freeminsmu}), obtaining the free energy in the framework of
the large deviation method. In the concrete case of the $\phi^4$
model we have
\begin{equation}
\label{freeminsmuphi4} \beta f(\beta,h)= \inf_{u,z,m}\left[\beta
\bar{\veps}(u,z,m) -\beta h m - \bar{s}(u,z,m)\right]
\end{equation}

Our presentation in subsection~\ref{secnegsusc} already contains all
we need to study magnetic susceptibility. We repeat here the basic
formula in Eq.~(\ref{suscan}) and apply it to the $\phi^4$ model.
\begin{equation}
\label{suscanbis} \chi=\left(\frac{\partial m}{\partial h}\right)=
-\beta \frac{\tilde{s}_{\veps \veps}}{\tilde{s}_{\veps
\veps}\tilde{s}_{mm} - \tilde{s}_{\veps m}^2}~,
\end{equation}
which is valid for both the canonical and the microcanonical
ensemble. It turns out that $\tilde{s}_{\veps\veps}$ is always
negative, and therefore the sign of the susceptibility is related to
the sign of $\tilde{s}_{mm}$. We see that, whenever this quantity is
positive (something that cannot happen in the canonical ensemble),
the susceptibility is negative, and, by continuity, it will be
negative also in regions where $s_{mm}$ is negative.

In the left picture of Fig.~\ref{sem3fig} we plot
$\tilde{s}(\veps,m)$ as a function of $m$ for different values of
$\veps$, clearly showing the regions of convexity of $\tilde{s}$.
The right picture offers a 3D view of $\tilde{s}(\veps,m)$. The plot
of $\tilde{s}(\veps,m)$ versus $m$ has a bimodal shape, and a region
of convexity, for $\veps < \veps_c$.

Just for the sake of clarity and readability, we can repeat an
argument analogous to the one presented in subsection~\ref{posneg},
but now with the magnetization $m$ playing the role there played by
the energy $\veps$. In that case a region of convexity in the
microcanonical $s(\veps)$ was associated to the presence of a first
order phase transition in the canonical ensemble, with an energy
jump at a given temperature, without any phase transition in the
microcanonical ensemble. Now, a region of convexity in
$\tilde{s}(\veps,m)$, seen as a function of $m$, is associated again
with a first order transition in the canonical ensemble. The
transition is present at temperatures lower than the critical
temperature $T_c$, and it is expressed by the jump from a positive
equilibrium magnetization to a negative equilibrium magnetization,
when the magnetic field $h$ is varied from $0^+$ to $0^-$. Since the
Hamiltonian of this model has a symmetry for $q_i \to -q_i$, the
critical magnetic field, in the canonical ensemble, is $h_c=0$.
Again, no phase transition is present in the microcanonical
ensemble. The analogy between the two cases is not surprising, since
they are both related to the non concavity of the entropy function,
as seen with respect to two different variables.

\subsubsection{The Colson-Bonifacio model for the Free Electron Laser}
\label{Colson-Bonifaciomodel}

In this subsection we discuss an extremely simplified model of the
Free Electron Laser (FEL)~\cite{Colson1976,Bonifacio84}, which can
be explicitly solved using large deviations
\cite{FELPRE,bbdrjstatphys}. It should be remarked that a model
of this kind was historically introduced by Zaslavsky and coworkers
\cite{Zaslavsky_77} to describe structural phase transitions in crystals.

In the linear FEL, a relativistic electron beam propagates through a
spatially  periodic magnetic field, interacting with the
co-propagating electromagnetic wave. Lasing occurs when the
electrons bunch in a subluminar beat wave~\cite{Bonifacio90}.
Scaling away the time dependence of the phenomenon and introducing
appropriate variables, it is possible to catch the essence of the
asymptotic state by studying the classical Hamiltonian
\begin{equation}
H_N=\sum_{j=1}^N\frac{p_j^2}{2} -N \delta A^2 +2 A \sum_{j=1}^N
\sin(\theta_j-\varphi) \label{eq:Hamiltonien}.
\end{equation}
The $p_i$'s are related to the energies relative to the center of
mass of the $N$ electrons and the conjugated variables~$\theta_i$
characterize their positions with respect to the co-propagating
wave.  The complex electromagnetic field variable, $\mathbf{A}=A\,
e^{i\varphi}$, defines the amplitude and the phase of the dominating
mode ($\mathbf{A}$ and $\mathbf{A}^\star$ are conjugate variables).
The parameter $\delta$  measures the average deviation from the
resonance condition. In addition to the energy $H_N$, the total
momentum $P=\sum_j p_j + NA^2$ is also a conserved quantity. Most of
the studies of this model have concentrated on the numerical
solution of Hamiltonian (\ref{eq:Hamiltonien}), starting from
initial states with a small field amplitude $A$ and the electrons
uniformly distributed with a small kinetic energy.  Then, the growth
of the field has been observed and its asymptotic value determined
from the numerics. Our study below allows to find the asymptotic
value of the field analytically.

In the first step, similarly to the HMF case,
Hamiltonian~(\ref{eq:Hamiltonien}) can be rewritten as
\begin{eqnarray}
H_N\simeq  NH(\mu) &=& N\left(\frac{u}{2}-\delta A^2 +2A
\left(-m_x\sin{\varphi} +m_y\cos{\varphi} \right) \right)
\end{eqnarray}
where $m_x$, $m_y$, are defined in Eq.~(\ref{magnHMF}) and $u$ just
below Eq.~(\ref{Ham_HMFMb}), together with $v$. Defining the phase
of the mean field~$\varphi^{\prime}$ as $m_x+im_y=m
\exp{(i\varphi^{\prime})}$, the global variables are $\mu
=(u,v,m,\varphi',A, \varphi)$.

Going to step two we should first remark that the two field
variables $A$, $\varphi$, give a contribution of order $1/N$ to the
entropy, which can be neglected. Hence, the $\bar{\phi}$ function
reduces to the one of the HMF model, see
formula~(\ref{phihmflarge}). Analogously, one obtains the same
contributions to the kinetic and configurational entropies of the
HMF model, shown in formulas~(\ref{entrhmflarge2})
and~(\ref{entrhmflarge3}).

Performing finally the third step, after defining the total momentum
density as $\sigma=P/N$, the microcanonical variational problem to
be solved is
\begin{equation}
\label{eq:entropieLEL} s(\varepsilon,\sigma,\delta )=
\sup_{\mu}\Biggl[ \frac{1}{2}\ln{(u-v^2)} +\tilde{s}_{conf}(m) \,
\Biggr| \, \varepsilon=
\frac{u}{2}+2Am\sin{\left(\varphi^{\prime}-\varphi\right)} -\delta
A^2, \, \sigma=v+A^2 \Biggr].
\end{equation}

Using the constraints of the variational problem, one can express
$u$ and $v$ as functions of the other variables, obtaining the
following form of the entropy
\begin{equation}
\label{eq:entropieLELbb} s(\varepsilon,\sigma,\delta )=  \sup_{A,
\varphi,m,  \varphi'  }\Biggl[
\frac{1}{2}\ln{\left[2\left(\varepsilon-\frac{\sigma^2}{2}\right)-
4Am\sin\left(\varphi^{\prime}-\varphi\right) +2(\delta-\sigma)
    A^2-A^4\right]} +s_{conf}(m)
 \Biggr].
\end{equation}
The extremalization over the variables $\varphi$ and
$\varphi^{\prime}$ is straightforward, since by direct inspection of
formula (\ref{eq:entropieLELbb}), it is clear that the entropy is
maximized when $\varphi^{\prime}-\varphi=-\pi/2$. Then
\begin{equation}
\label{eq:entropieLEL2} s(\varepsilon,\sigma,\delta)=\sup_{A,  m
}\Biggl[
\frac{1}{2}\ln{\left[2\left(\varepsilon-\frac{\sigma^2}{2}\right)+4Am
+2(\delta-\sigma) A^2-A^4\right]} +\tilde{s}_{conf}(m) \Biggr]\equiv
\sup_{A, m} \tilde{s}(A,m).
\end{equation}
The non zero $\sigma$ case can be reduced to the vanishing $\sigma$
problem using the identity
$s(\varepsilon,\sigma,\delta)=s(\varepsilon-\sigma^2/2,0,\delta-\sigma)$.
From now on, we will discuss only the zero momentum case since we
can absorb non zero momenta into the definition of~$\varepsilon$ and
$\delta$. This has also a practical interest, because it is the
experimentally relevant initial condition.

The conditions for having a local stationary point are
\begin{eqnarray}
\frac{\partial \tilde{s}}{\partial A} &=& \frac{2\left(\delta A-A^3
+m\right)}
{2\varepsilon+2\delta A^2+4Am-A^4}=0, \label{eq:derivee1a} \\
\frac{\partial \tilde{s}}{\partial m} &=& \frac{2A}
{2\varepsilon+2\delta A^2+4Am-A^4} -B_{inv}(m)=0,
\label{eq:derivee1b}
\end{eqnarray}
where $B_{inv}$ is defined in formula (\ref{constraintmlarge}).  It
is clear that $m=A=0$ is a solution of  conditions
(\ref{eq:derivee1a}) and (\ref{eq:derivee1b}): it exists only for
positive $\varepsilon$. We will limit ourselves to study its
stability. It must be remarked that this is the typical initial
condition studied experimentally in the FEL: it corresponds to
having a beat wave with zero amplitude and the electrons uniformly
distributed. The lasing phenomenon is revealed by an exponential
growth of both $A$ and the electron bunching parameter $m$.

The second order derivatives of the entropy $\tilde{s}(A,m)$,
computed on this solution, are
\begin{eqnarray}
\frac{\partial^2 \tilde{s}}{\partial A^2}(0,0) &=& \frac{\delta}{e},
\quad \frac{\partial^2 \tilde{s}}{\partial m^2}(0,0) = -2, \quad
\frac{\partial^2 \tilde{s}}{\partial A \partial m}(0,0) =
\frac{1}{\varepsilon}.
\end{eqnarray}
The two eigenvalues of the Hessian are the solutions of the equation
\begin{equation}
x^2-x\left(-2+\frac{\delta}{\varepsilon}\right)-\frac{2\delta}{\varepsilon}-\frac{1}{\varepsilon^2}=0.
\end{equation}
The stationary point is a maximum if the roots of this equation are
both negative. This implies that their sum
$(-2+{\delta}/{\varepsilon})$ is negative and their product
$(-{2\delta}/{\varepsilon}-{1}/{\varepsilon^2})$ is positive.
Recalling that we restrict to positive $\varepsilon$ values, the
condition for the sum to be negative is $\varepsilon>\delta/2$ and
the one for the product to be positive is $\varepsilon>-1/(2\delta)$
with $\delta<0$. The second condition is more restrictive, hence the
only region where the solution $m=A=0$ exists and is stable is
$\varepsilon>-1/(2\delta)$ with $\delta<0$. When crossing the line
$\varepsilon=-1/(2\delta)$ ($\delta<0$), a non zero bunching
solution ($m \neq 0$) originates continuously from the zero bunching
one, producing a second order phase transition. This analysis fully
coincides with the one performed in the canonical ensemble in
Ref.\cite{Firpo00,FirpoThese}. The maximum entropy solution in the
region complementary to the one where the zero bunching solution is
stable can be obtained by solving numerically
Eqs.~(\ref{eq:derivee1a}) and~(\ref{eq:derivee1b}). This corresponds
to having a non zero field intensity and bunching.

The equilibrium solution of the Colson-Bonifacio model has been
shown to describe well the system only in certain energy
ranges~\cite{barrethesis}. Dynamical effects must be carefully taken
into account. The dynamics can be studied in the $N \to \infty$
limit by a Vlasov equation, see Sec.\ref{outofequilibrium}. The
homogeneous zero field state is linearly Vlasov stable for $\delta >
\delta_c=(27/4)^{1/3}$ \cite{Farina}, hence, in this region of
parameters one cannot expect that the system will relax to the
equilibrium state. Moreover, as we will discuss, Vlasov equation can
display an infinity of quasi-stationary states, and the system can
be trapped in such states for a time that diverges with $N$. The
value of bunching parameter and field intensity can differ from the
equilibrium ones significantly \cite{FELPRE}.

\subsection{The origin of singularities of thermodynamic functions}
\label{topological}

We have seen that phase transitions have a paramount importance in
the study of ensemble equivalence and inequivalence. Indeed, we have
emphasized that inequivalence occurs when phase transitions have
different properties in the microcanonical and in the canonical
ensemble. The BEG model has shown many aspects of this (see
Sec.~\ref{begmodel}): the microcanonical transition lines do not
coincide with the canonical ones; there is a region in the space of
thermodynamic parameters where there is a first order phase
transition in the canonical ensemble while the transition is second
order in the microcanonical ensemble, which moreover presents a
negative specific heat. There is another region where both ensembles
have a first order phase transition, but with different properties,
i.e. an energy jump in the canonical ensemble and a temperature jump
(with sometimes also a negative specific heat) in the microcanonical
ensemble. The $\phi^4$ model has shown another kind of
inequivalence: instead of being associated to the thermodynamically
conjugated pair $(\veps,T)$, we have found that it is related to the
pair $(m,h)$.

Obviously, phase transitions are present also in short-range
systems, but in that case ensembles are equivalent. We remind that,
as emphasized in subsection~\ref{maxwellinshort}, we adopt here the
physical point of view about ensemble equivalence at first-order
phase transitions for short-range systems. At variance with the
mathematical physics literature \cite{TouchettePhysRep}, we do not
use the term partial equivalence in cases where there is not a
one-to-one correspondence between energy in the microcanonical
ensemble and temperature in the canonical ensemble.

Then, the questions may arise about a possible qualitative
difference between phase transitions in short-range and in
long-range systems. This section is devoted to a brief discussion of
some results that concern this problem. We will give only few
details, for two reasons: first, this subject lies outside the
purpose of this review; second, two recent reviews and a book have
appeared dealing with this problem where the interested reader can
find more details~\cite{Casetti2000,kastnerrevmodphys,Pettinibook}.
We will discuss the origin of the analytical properties of entropy
and of free energy that, in turn, are associated to the presence of
phase transitions.

The $\phi^4$ model can be studied without the magnetization
constraint in the microcanonical ensemble, and without the presence
of an external magnetic field in the canonical ensemble. This is
what has been done in Ref.~\cite{dauxois2003}, where, as we have
mentioned in subsection~\ref{modelphi4}, a second order phase
transition has been found. According to what we have emphasized at
the end of subsection~\ref{posneg}, a second order phase transition
implies that the two ensembles are equivalent. Nevertheless, the
presence of the transition means that the second derivatives of
$s(\veps)$ and of $\phi(\beta)$ are discontinuous, the first at the
transition energy $\veps_c$ and the second at the corresponding
transition inverse temperature $\beta_c$. The non analyticity of the
free energy is the usual signature of phase transitions. What we are
interested in here is the origin of this non analyticities.

The entropy function $\tilde{s}(\veps,m)$ of the $\phi^4$ model,
previously studied, is an analytic function in both arguments. We
have also seen that the inequivalence with the canonical ensemble,
where a first order phase transition is present, is due to the
concavity properties of $\tilde{s}(\veps,m)$. If we study the system
without the magnetization constraint, we can use, as before, the
large deviation method, with the only difference that at the third
step of the procedure, we maximize in Eq.~(\ref{entropyphi4large1})
also with respect to $m$. This is simply because the entropy
$s(\veps)$ is, for each $\veps$, the maximum over $m$ of
$\tilde{s}(\veps,m)$. Since there is a point of non analyticity in
$s(\veps)$, at the energy $\veps_c$ of the second order phase
transition, this means that the process of maximization of the
analytic function $\tilde{s}(\veps,m)$ leads to a non analytic
function $s(\veps)$. This is due to the bimodal shape of
$\tilde{s}(\veps,m)$, see Fig.~\ref{sem3fig}. This bimodal shape
would not be possible in a short-range system, where
$\tilde{s}(\veps,m)$ has to be concave with respect to $m$ at each
value $\veps$. This argument has been recently used by Kastner and
collaborators~\cite{kastnerrevmodphys}, with particular emphasis on
the $\phi^4$ model~\cite{hahn2005,hahn2006} and on the mean-field
spherical
model~\cite{KastnerSchnetz2006,Kastner2006,Casetti2006,kastnerprl2007}.

In subsection \ref{equivshort}, we have emphasized only the
concavity with respect to $\veps$. However, in short-range systems
the entropy is concave also with respect to other macroscopic
variables, e.g. magnetization \cite{gallavotti}. Of course, there
might be ranges where the concavity is not strict, associated to
first order phase transitions. This means that, for short-range
systems, a possible non analyticity in $s(\veps)$ could not derive
from the maximization of an analytic $\tilde{s}(\veps,m)$, but that
already the function $\tilde{s}(\veps,m)$, over which the
maximization is performed, is non analytic. The mechanism that can
give rise to non analyticities of the entropy in short-range systems
has been emphasized by Pettini and coworkers (see
~\cite{Casetti2000,Pettinibook} and references therein). These
authors prove \cite{franzosipettini1,franzosipettini2,
franzosipettini3} that a {\it necessary} condition for having a
phase transition in short-range system is the presence of
topological changes in the configurational space subset ${\cal
M}(v)= \left( \{ q_i \} | U(\{ q_i\} )/N \le v \right)$, which is
nothing but the level set $v$ of the potential energy $U$. This
topological change, according to Morse theory, is signalled by the
presence  of ``critical points" in ${\cal M}(v)$, i.e. points where
the differential of $U$ vanishes. This idea that phase transitions
have an origin in some topological and analytical properties of the
phase space could of course be extended to systems with long-range
interactions. Indeed, it has been extended, and detailed
calculations of the critical points of the HMF model
\cite{Casetti1999,Casetti2003} and of the $k$-trigonometric model
\cite{Angelani2003,Angelani2005,Angelani2007} have been performed,
allowing to show analytically that a topological invariant (the Euler
characteristic) has a jump at the phase transition point. However,
results in this field raise apparent paradoxes. Indeed, by
considering the mean-field $\phi^4$ model with varying strength of
the mean-field term, Kastner~\cite{hahn2005,hahn2006} prove that
nothing special appears in the analytical and topological properties
of the potential energy at the second-order phase transition point.
Hence, the mechanism here at work to create a phase transition must
be different and it is natural to think that it is indeed related to
the extremalization of a thermodynamic analytic function (entropy or
free energy), independently of the presence of critical points in
the potential.

However, although phase transitions in long-range systems, as just
explained, can derive also from a mechanism that is not present in
short-range systems, still they could also derive from topological
changes. The problem to solve is to select which of the ``critical
points", whose number grows with $N$, will turn out to determine the
few non analytic points of the thermodynamic functions. Recently, it
has been shown that the non analyticities due to ``critical points"
can become infinitely weak in the thermodynamic limit
\cite{kastnerprl2007} (solving also a contradiction arisen in a
study of the spherical model \cite{Ana2004,Ana2005}), i.e., the
order of the derivative of the entropy which is discontinuous
becomes larger and larger when $N$ grows. In order to explain how a
discontinuity in a first or second order derivative of the entropy
can survive in the thermodynamic limit, Kastner
\cite{kastnerprl2008} introduces a criterion of ``sufficient
flatness" of the critical point. An application to the HMF and to
the $k$-trigonometric model confirms this criterion, selecting the
correct phase transition point in the thermodynamic limit
\cite{kastnerprl2008}.

All these results concern {\it necessary} conditions and the much
harder problem of finding {\it sufficient conditions} has not yet
been solved. For instance, for a generalization of the
$k$-trigonometric model, it has been found \cite{Angelani2007} that
a phase transition occurs at an energy where no topological change
of the phase space manifold is present.

\section{Out-of-equilibrium dynamics: Quasi-stationary states}
\label{outofequilibrium}

In this section, we will give a brief introduction to the kinetic
equations used for long-range systems, i.e. Klimontovich, Vlasov and
Lenard-Balescu equations. We will present the numerical evidence of
the existence of {\it quasi-stationary} states in the $N$-particle
dynamics of the Hamiltonian Mean Field (HMF) model. These states,
that are out-of-equilibrium and have a lifetime that increases with
system size $N$, turn out to be related to stable stationary states
of the Vlasov equation. Selecting the dynamics of a single particle
and treating the rest as a bath, we will find that the one-particle
velocity distribution function obeys a Fokker-Planck equation with
nonlinear diffusion coefficient. Solving this equation, we find
cases where diffusion can be normal or otherwise anomalous. A
theory, introduced by Lynden-Bell for gravitational systems, which
relies on a ``fermionic" entropy, allows us finally to predict the
one-particle distribution function in the quasi-stationary states
for a specific class of initial conditions.

%We will introduce a simple model of the Free-Electron-Laser (FEL),
%which shares many features of its phenomenology with the HMF model,
%including the existence of quasi-stationary states. Finally,
%long-time relaxation to thermal equilibrium can exhibit different
%scaling laws with $N$: this will be discussed at the end of this
%section.

\subsection{Kinetic equations}
\label{kineticequation}

\subsubsection{Klimontovich, Vlasov and Lenard-Balescu equations}
\label{KlimoVlasovLenard}

Here, we will discuss the equations that describe the evolution of
the one-particle distribution function for systems with long-range
interactions.

The first kinetic equation for a Hamiltonian $N$-body system was
derived by Boltzmann~\cite{Boltzmann}. Boltzmann's theory, in which
particles interact only through binary collisions, describes diluted
gases with short-range interactions. In order to take into account
collisions determined by long-range Coulomb and gravitational
forces, Landau~\cite{Landau} and Chandrasekhar~\cite{chandrasekhar}
modified Boltzmann's collisional term. Mean-field collective effects
were first considered by Vlasov \cite{originalVlasovpaper} and led
to the Vlasov equation, later refined by Landau himself
\cite{Landaucontour}. A treatment of collision terms in the context
of the Vlasov-Landau approach was developed by Lenard and Balescu
~\cite{Lenard,Balescu} (see Ref.~\cite{FeixBertrand} for a review):
collective effects were effectively taken into account by a
dielectric function. These theoretical approaches have also led to
the development of kinetic theories for point vortices in
two-dimensional hydrodynamics \cite{DubinOneil99,ChavanisHouches} and for non
neutral plasmas confined by a magnetic field
\cite{DubinJin,Dubin03}. A detailed recent discussion of the
derivation of Landau, Vlasov and Lenard-Balescu equations and of the
approximations involved can be found in Ref.~\cite{ChavanisAssisi}.

The relation between the $N$-particle dynamics and the solution of
the Vlasov equation has been studied for mean-field systems by
Neunzert~\cite{Neunzert,Neunzertbis} and Braun and
Hepp~\cite{BraunHepp,Spohn_Livre}. The latter proved that, for
sufficiently smooth potentials, the distance between two initially
close solutions of the Vlasov equation increases at most
exponentially in time. If we apply this result to a large $N$
particle approximation of a continuous distribution the error at
$t=0$ is typically of order $1/\sqrt{N}$, thus for any $\epsilon$
and any particle number $N$, there is a time $t$ up to which the
dynamics of the original Hamiltonian and its Vlasov description
coincide within an error bounded by $\epsilon$. The theorem implies
that this time $t$ increases at least as $\ln N$.

For what concerns the methods to derive the Vlasov equation and higher
order kinetic equations,
two main perturbative approaches have been followed. The first one
begins with the Liouville equation and, either by the usual BBGKY
expansion~\cite{Huang,Balescubook1} or by some projector operator
formalism \cite{WillisPicard, Kandrup} leads to the kinetic equation
for the one-particle distribution function making some particle
decorrelation hypothesis. The second approach is based on the
Klimontovich equation~\cite{Klimontovich,landau,nicholson}. In both
approaches, the small parameter of the perturbative expansion turns
out to be $1/N$, where $N$ is the number of particles. These two
methods give completely equivalent results. In the following, we
will discuss the method based on the Klimontovich equation. The
Vlasov equation will come out at leading order, while the
Lenard-Balescu correction term will appear at order $1/N$.

\subsubsection{Derivation of Klimontovich equation}
\label{Klimontovichderivation}

The general Hamiltonian we will discuss is of the form
\begin{equation}
  \label{eq:hamiltonian}
  H_N = \sum_{j=1}^{N} \frac{P_{j}^{2}}{2}
  +U(\{\Theta_j\}),
  %- \frac{1}{2N} \sum_{i,j=1}^{N} \cos(\theta_{i}-\theta_{j})
\end{equation}
where we have used the coordinate $\Theta_j$ for particle $j$
because, in the following, we will mainly treat models where
positions are specified by an angle and
\begin{equation}
U(\Theta_1,\dots,\Theta_N)= \sum_{i<j}^N V(\Theta_i-\Theta_j).
\end{equation}
We do not use here Kac's scaling of the potential (which amounts
here to put a $1/N$ prefactor in the above formula), because we want
to maintain the derivation at the general level, which includes both
short and long-range interactions. When relevant, we will make
comments on where Kac's scaling would produce effects for long-range
interactions.

The state of the $N$-particle system can be described by the {\em
discrete} one-particle time-dependent density function
\begin{equation}
 f_d\left(\theta,p,t\right)= \displaystyle \frac{1}{N} \sum_{j=1}^N\delta
\left(\theta -\Theta _{j}\left( t\right)\right)\delta \left(
  p-P_{j}\left( t\right) \right),
  \label{discretedensityfunction}
\end{equation}
where $\delta$ is the Dirac function, $(\theta,p)$ the Eulerian
coordinates of the phase space and  $(\Theta_i,P_i)$ the Lagrangian
coordinates of the $N$-particles whose dynamics is given by the $2N$
equations of motions
\begin{eqnarray}
\label{eqmotionun}
\dot \Theta_j&=&P_j,\\
  \dot P_j&=&- \frac{\partial
  U}{\partial\Theta_j}
\label{eqmotionbis}.
\end{eqnarray}
Differentiating with respect to time the one-particle
density~(\ref{discretedensityfunction}) and using
Eqs.~(\ref{eqmotionun}) and~(\ref{eqmotionbis}), one gets
\begin{equation}\label{derivedensitybis}
  \frac{\partial f_d(\theta,p,t)}{\partial t}=-\frac{1}{N}\sum_j P_j \frac{\partial}{\partial\theta} \delta
\left(\theta -\Theta _{j}\left( t\right)\right)\delta \left(
  p-P_{j}\left( t\right) \right)+\frac{1}{N}\sum_j  \frac{\partial U}{\partial\Theta_j}
  \frac{\partial}{\partial p}\delta
\left(\theta -\Theta _{j}\left( t\right)\right)\delta \left(
  p-P_{j}\left( t\right) \right)~.
\end{equation}
Taking advantage of the property of the Dirac $\delta$-function,
$a\delta(a-b)=b\delta(a-b)$, it is possible to rewrite this equation
as
\begin{eqnarray}\label{derivedensityter}
  \frac{\partial f_d(\theta,p,t)}{\partial t}&=&-\frac{1}{N}\sum_j p \frac{\partial}{\partial \theta} \delta
\left(\theta -\Theta _{j}\left( t\right)\right)\delta \left(
  p-P_{j}\left( t\right) \right)+\frac{1}{N}\sum_j \frac{\partial v}{\partial\theta} \frac{\partial}{\partial p}\delta
\left(\theta -\Theta _{j}\left( t\right)\right)\delta \left(
  p-P_{j}\left( t\right) \right),
\end{eqnarray}
where
\begin{equation}
v(\theta,t)=N \int \dd \theta' \dd p' V(\theta-\theta')
f_d(\theta',p',t)~, \label{averagepotential}
\end{equation}
which leads to the Klimontovich equation
\begin{equation}
\frac{\partial f_d}{\partial t}+p\frac{\partial f_d}{\partial
\theta} -\frac{\partial v}{\partial \theta} \frac{\partial
f_d}{\partial p}=0.\label{equationfdiscrete}
\end{equation}
One notes that the above derivation is {\em exact} even for a finite
number of particles $N$. However, this equation contains the
information about the orbit of every single particle (since $f_d$
depends on the $2N$ Lagrangian coordinates of each particle,
$(\Theta_i,P_i)$) which is far more than we need. Hence,
Klimontovich equation is especially useful as a starting point for
the derivation of approximate equations that describe the average
properties of the system. This is what we discuss in the following
subsections.

The more widely used approach to kinetic equations begins with the
Liouville equation. This equation governs the time evolution of the
probability density in the full $2N$ dimensional phase space
$(\theta_1,p_1,\ldots,\theta_N,p_N)$. To arrive at the Liouville
equation, let us first define the Klimontovich distribution function
\begin{equation}
 F_d\left(\theta_1,p_1,\theta_2,p_2,\dots,\theta_N,p_N,t\right)= \displaystyle  \prod_{j=1}^N\delta
\left(\theta_i -\Theta _{j}\left( t\right)\right)\delta \left(
  p_j-P_{j}\left( t\right) \right),
  \label{Liouvilledensityfunction}
\end{equation}
where $(\Theta_i(t),P_i(t))$ are again the Lagrangian coordinates.
Deriving (\ref{Liouvilledensityfunction}) with respect to time, and
using the equations of motion, one obtains
\begin{equation}
\frac{\partial F_d}{\partial t}+\sum_{i=1}^N p_i \frac{\partial
F_d}{\partial \theta_i} - \sum_{i=1}^N \frac{\partial
U(\theta_1,\ldots,\theta_N)}{\partial \theta_i} \frac{\partial
F_d}{\partial p_i}=0. \label{equationLiouville}
\end{equation}
This is the Liouville equation for $F_d$. It has been derived for a
particular distribution function, which describes a single point in
the $2N$ dimensional space. The Liouville equation for a generic
distribution function is obtained by an averaging procedure which
leads to the introduction of a smooth density
$\rho(\theta_1,p_1,\theta_2,p_2,\dots,\theta_N,p_N,t)$, which can be
shown to obey the same equation (\ref{equationLiouville}). This
smoothing procedure is similar to the one that leads to the Vlasov
equation, that we will describe just below. However, while the
Liouville equation for $\rho$ is exact (as the Klimontovich equation
for $f_d$), the Vlasov equation is not exact and we will indeed
discuss finite $N$ corrections. It is well known that the fluid
described by the density $\rho$ is incompressible.

\subsubsection{Vlasov equation: collisionless approximation of the
Klimontovich equation} \label{Vlasovderivation}

Determining $f_d(\theta,p,t)$, which characterizes whether a point
particle is to be found at a given point $(\theta,p)$ in Eulerian
phase space, would imply to solve the equations of motion
(\ref{eqmotionun},\ref{eqmotionbis}) with initial conditions
$(\{\Theta_i(0),P_i(0)\})$. In general, this is a difficult task and
typically not feasible for nonlinear systems. Alternatively, one can
define an averaged one-particle density function using an infinite
number of realizations prepared according to some prescription. One
could for instance consider a large number of initial conditions,
close to the same macroscopic state. Let's define the density of
such initial macroscopic state as $f_{in}(\{\Theta_i(0),P_i(0)\})$.
The average one-particle density function $f_0$ is obtained as
\begin{equation}\label{f0definition}
f_0(\theta,p,t)\equiv\langle f_d(\theta,p,t)\rangle=\int \prod_i \,
\dd \Theta_i(0) \dd P_i(0) \, f_{in}(\{\Theta_i(0),P_i(0)\})
f_d(\theta,p,t)~,
\end{equation}
where the dependence of $f_d$ on $(\{\Theta_i(0),P_i(0)\})$ comes
from the solution of the equations of motion that enter the
definition of $f_d$, see Eq.~(\ref{discretedensityfunction}).

The equation for the time evolution of the smoothed distribution
$f_0$ is obtained by again averaging over $f_{in}$. Before doing it,
let us introduce the definition of the fluctuations $\delta f$
around the smooth distribution
\begin{equation}
f_d(\theta,p,t)=f_0(\theta,p,t)+\frac{1}{\sqrt{N}}\delta
f(\theta,p,t). \label{deffetdeltaf}
\end{equation}
Evidently $f_0$ is independent of the detailed microscopic
properties of the initial state and depends on it only through
$f_{in}$, while $\delta f$, like $f_d$ depends on all Lagrangian
variables of the initial state. Obviously, the average over $f_{in}$
of $\delta f$ is zero. The introduction of the prefactor
$1/\sqrt{N}$ takes into account the typical size of relative
fluctuations. Indeed, $f_d - f_0$ is the difference between a
singular distribution, containing Dirac deltas, and a smooth
function; therefore the statement that this difference is of the
order $1/\sqrt{N}$ has to be interpreted physically. Its meaning is
the following: if we integrate $f_d - f_0$ in $\theta$ and $p$ in a
volume which is small compared to the total available volume, but
large enough to contain many particles, then the value of this
integral, both at equilibrium and out-of-equilibrium, is of the
order $1/\sqrt{N}$.

Inserting Eq.~(\ref{deffetdeltaf}) into Eq.~(\ref{averagepotential})
leads to
\begin{equation}
v(\theta,t)=\langle v\rangle (\theta,t)+\frac{1}{\sqrt{N}}\delta
v(\theta,t), \label{defVetdeltaV}
\end{equation}
where the first term comes from the average over $f_{in}$, and
coincides with
\begin{equation}
\langle v \rangle (\theta,t)=N \int \dd \theta' \dd p'
V(\theta-\theta') f_0(\theta',p',t)~, \label{potaverage}
\end{equation}
while the second term defines $\delta v$, which depends on all the
details of the initial state. If we had used Kac's scaling the $N$
prefactor in formula (\ref{potaverage}) would be absent, making the
average potential intensive. This latter scaling is commonly
used~\cite{Spohn_Livre} and would be more appropriate to derive the
Boltzmann equation.

Inserting both expressions~(\ref{deffetdeltaf})
and~(\ref{defVetdeltaV}) in the Klimontovich
equation~(\ref{equationfdiscrete}), one obtains
\begin{eqnarray}
\frac{\partial f_0}{\partial t}+p\frac{\partial f_0}{\partial
\theta} -\frac{\partial \langle v\rangle}{\partial \theta}
\frac{\partial f_0}{\partial
p}=-\frac{1}{\sqrt{N}}\left(\frac{\partial \delta f}{\partial
t}+p\frac{\partial \delta f}{\partial \theta} -\frac{\partial \delta
v}{\partial \theta} \frac{\partial f_0}{\partial p} -\frac{\partial
\langle v\rangle}{\partial \theta} \frac{\partial \delta f}{\partial
p}\right) +\frac{1}{{N}}\frac{\partial \delta v }{\partial
\theta}\frac{\partial \delta f}{\partial p} .
\label{equationpourfremplace}
\end{eqnarray}
Taking the average over $f_{in}$ of this equation leads to
\begin{eqnarray}
\frac{\partial f_0}{\partial t}+p\frac{\partial f_0}{\partial
\theta} -\frac{\partial \langle v\rangle}{\partial \theta}
\frac{\partial f_0}{\partial p}
&=&\frac{1}{N}\left\langle\frac{\partial \delta v }{\partial
\theta}\frac{\partial \delta f}{\partial p}\right\rangle.
\label{equationpourfzero}
\end{eqnarray}
It is important to stress that above equation is still exact. Up to
now we have not even made any hypothesis about the long or
short-range properties of the potential $V$. In standard kinetic
theory, Eq.~(\ref{equationpourfzero}) would correspond to the first
equation of the BBGKY hierarchy (see e.g. Chapter 4 of
Ref.~\cite{nicholson}). For short-range interactions, the r.h.s. of
this equation would originate the leading contribution to the
collision term of the Boltzmann equation, while the third term of
the l.h.s. would be negligible close to equilibrium. On the
contrary, for long-range interactions, it will turn out that the
r.h.s is of order $1/N$ \cite{Landau,Lenard,Balescu}. Indeed, the
scaling with $1/\sqrt{N}$ in Eqs. (\ref{deffetdeltaf}) and
(\ref{defVetdeltaV}) is appropriate for long-range interactions.
Therefore, for long-range interactions, the third term on l.h.s.
will be the leading term and, in the limit $N \to \infty$, one ends
up with the {\it Vlasov equation}
\begin{eqnarray}
\frac{\partial f_0}{\partial t}+p\frac{\partial f_0}{\partial
\theta} -\frac{\partial \langle v \rangle}{\partial \theta}
\frac{\partial f_0}{\partial p}=0. \label{equationpourfzeroici}
\end{eqnarray}
To distinguish the effects due to discreteness (finite value of $N$)
from the {\em collective} effects grouped on the l.h.s. of
Eq.~(\ref{equationpourfzero}), the r.h.s. is usually referred to as
{\em collisional} term. This is the origin of the name {\it
collisionless Boltzmann equation} sometimes used for the Vlasov
equation. It is however important to underline that there are no
true collisions for long-range systems: granular effects,
discreteness effects or finite $N$ corrections would be more
appropriate names.

The Vlasov equation has wide applications in gravitational systems
\cite{binneytremaine} and in plasma physics \cite{nicholson}. A
typical question that is posed is the one of stability of the
stationary solutions. In particular, the stability of stationary
homogeneous solutions has important applications in plasma physics
for the description of Landau damping of for beam-plasma
instabilities. In the following subsection, we will discuss in some
detail the stability of stationary homogenous solution for the
Hamiltonian Mean Field model.

\subsubsection{Stationary stable solutions of the Vlasov equation: application to the HMF model}
\label{sectiondiffdistrib}

Let us concentrate on homogenous states, for which the one-particle
distribution function does not depend on $\theta$, so that $f_0 =
f_0(p,t)$.  For these distributions, using Eq.~(\ref{potaverage}),
one immediately gets that $\langle v \rangle= const.$, and
consequently the Vlasov equation~(\ref{equationpourfzeroici})
reduces to
\begin{equation}
  \frac{\partial f_0}{\partial t}(p,t)=0.
  \label{Vlasovhomogeneous}
\end{equation}
Therefore, homogeneous distributions are stationary. However,
stationarity does not imply {\it stability}. Indeed, the stability
of stationary spatially homogeneous solutions can be studied by
subtracting Eq.~(\ref{equationpourfzero}) from
Eq.~(\ref{equationpourfremplace}). One gets
\begin{equation}
\frac{\partial \delta f}{\partial t}+p\frac{\partial \delta
f}{\partial \theta} - \frac{\partial \delta v}{\partial \theta}
\frac{\partial f_0}{\partial p}  -\frac{\partial \langle v
\rangle}{\partial \theta} \frac{\partial \delta f}{\partial p} =
\frac{1}{\sqrt{N}}\biggl[\frac{\partial \delta v}{\partial
\theta}\frac{\partial \delta f}{\partial p}-\left\langle
\frac{\partial \delta v}{\partial \theta} \frac{\partial \delta
f}{\partial p} \right\rangle \biggr]. \label{equationavantmoy}
\end{equation}
For times much shorter than $\sqrt{N}$ (or equivalently for $N \to
\infty$), we may drop the r.h.s. of Eq.~(\ref{equationavantmoy}),
which contains quadratic terms in the fluctuations. As the last term
of the l.h.s. vanishes, since $\langle v \rangle=const.$, the
fluctuating part of $f_d$, $\delta f$, obeys the {\it linearized
Vlasov
equation}~\cite{originalVlasovpaper,Landaucontour,BraunHepp,Spohn_Livre,spohn}
\begin{equation}
\frac{\partial \delta f}{\partial t}+p\frac{\partial \delta
f}{\partial \theta}- \frac{\partial \delta v}{\partial \theta}
\frac{\partial f_0}{\partial p}=0 .\label{Vlasovlinear}
\end{equation}
This equation could also be easily obtained by linearizing directly
the Vlasov equation (\ref{equationpourfzeroici}). Looking for plane
wave solutions
\begin{eqnarray}
\delta f(\theta,p,t)&=&\hat{f}(p)\ e^{\displaystyle i(k\theta-\omega t)}\label{stabpot1}\\
\delta v(\theta,t)&=&\hat{A}\  e^{\displaystyle i(k\theta-\omega
t)}, \label{stabpot2}
\end{eqnarray}
we obtain
\begin{eqnarray}
-i\omega\,\hat{ f}(p)+p\,ik\,\hat{f}(p)-i k \,\hat{A} f_0'(p)=0
\label{Vlasovlinearstab}
\end{eqnarray}
which leads to
\begin{eqnarray}
\hat{ f}(p)=\frac{k\,\hat{A}}{pk-\omega}\,f_0'(p).
\label{Vlasovlinearstabbis}
\end{eqnarray}
This analysis is completely general and applies to all
one-dimensional models.

To be more specific, let us consider the Hamiltonian Mean Field
model (\ref{Ham_HMF}). In this particular case, the potential
appearing in the Vlasov equation Eq.~(\ref{equationpourfzeroici}) is
\begin{equation}
\label{defpotential}
  \langle v \rangle(\theta,t)=\int_0^{2\pi}\!\!\dd
  \alpha\int_{-\infty}^{+\infty}\!\!\dd p \
  [1-\cos(\theta-\alpha)]\, f_0(\alpha,p,t).
\end{equation}
We now use
Eqs.~(\ref{stabpot1},\ref{stabpot2},\ref{Vlasovlinearstabbis}),
where for the HMF model $k$ is an integer because the potential is
$2 \pi$-periodic, in the definition of the fluctuations of the
potential
\begin{equation}
\label{fluctpotential} \delta v(\theta,t)=\int_0^{2\pi}\!\!\dd
\alpha\int_{-\infty}^{+\infty}\!\!\dd p\ [1-\cos(\theta-\alpha)]\,
\delta f(\alpha,p,t)~.
\end{equation}
We obtain
\begin{eqnarray}
\label{fouriertransformofdeltaVsuiteintermed} {\delta
v}(\theta,t)=\hat{A}\  e^{\displaystyle i(k\theta-\omega t)}
&=&-{\pi k}{}\left(\delta_{k,1}+\delta_{k,-1}\right)
\int_{-\infty}^{+\infty}\!\!\dd p\
\frac{f_0'(p)}{pk-\omega}\,\hat{A} \  e^{\displaystyle
i(k\theta-\omega t)}.
\end{eqnarray}
Introducing the so-called ``plasma response dielectric function"
(see Chapter 6 of Ref.~\cite{nicholson})
\begin{equation}
\tilde D(\omega,k)\equiv1+{\pi k
}{}\left(\delta_{k,1}+\delta_{k,-1}\right)
\int_{-\infty}^{+\infty}\!\!\dd p\ \frac{f_0'(p)}{pk-\omega},
\label{dielectricfunction}
\end{equation}
Eq.~(\ref{fouriertransformofdeltaVsuiteintermed}) can be rewritten
as $\tilde D(\omega,k) \hat{A}\, \exp (i(k\theta-\omega t))=0$. In
order to get non vanishing solutions for $\hat{A}$, the relation
\begin{equation}
\tilde D(\omega,k)=0 \label{dielectric}
\end{equation}
must be satisfied, which leads to the dispersion relation
$\omega=\omega(k)$, linking the frequency with the wavevector. From
the definition~(\ref{dielectricfunction}), it is evident that the
only possible collective modes are $k =\pm 1$. The denominator in
the integral on the r.h.s of Eq.~(\ref{dielectricfunction}) must be
treated with care. Indeed, a difficulty arises if one attempts to
obtain solutions $\omega(k)$ of Eq.~(\ref{dielectric}) corresponding
to purely oscillatory solutions of the linearized Vlasov equation
(\ref{Vlasovlinear}), i.e. with Im($\omega$)=0. These solutions are
important because they lie at the boundary between stable and
unstable modes, which correspond to negative and positive values of
Im($\omega$), respectively. If one makes the substitution of real
values of $\omega$ into the integral mentioned above, one notes that
the integrand has a pole at $p=\omega/k$ and therefore the integral
is not well defined. The prescription for performing this singular
integral is to deform integration contour of $p$ in such a way to
circulate around the pole $p=\omega/k$. This method was introduced
in plasma physics by Landau~\cite{Landaucontour} and the deformed
contour is called {\it Landau contour}. The deformation of the
contour is equivalent to the displacement of the pole
$\omega\rightarrow\omega\pm i0$. More rigorously this is expressed
by the Plemelj formula
\begin{equation}
\lim_{\gamma\rightarrow0}\frac{1}{x-a\pm i|\gamma|}={\cal
P}\frac{1}{x-a}\mp i\pi\delta(x-a) \label{Plemeljformula}
\end{equation}
where ${\cal P}$ denotes the principal value, which applies when
integrating over $x$. The second term on the r.h.s comes from
integrating on a semi-circle around the pole. The direction to be
chosen for the displacement of the pole is determined if one uses
the Laplace transform in time, suited for the solution of the
initial value problem associated to our kinetic equation. We will
use this technique in the next subsection, treating the
Lenard-Balescu equation. The Laplace transform is defined for
Im$(\omega)>0$. From this, we derive that $\tilde D(\omega,k)$ is
defined by Eq. (\ref{dielectricfunction}) for Im$(\omega)>0$, and by
its analytic continuation for Im$(\omega)\le 0$. In particular, this
means that for Im$(\omega)=0$ the rule for encircling the
singularity at $p=\omega/k$ is obtained by the Plemelj formula
(\ref{Plemeljformula}) by posing $\omega \rightarrow \omega + i0$.
Applying this rule, we find that for $k=\pm 1$ the dielectric
function for real values of $\omega$ is given by
\begin{equation}
\label{realplusimagparts} \tilde D(\omega,\pm 1)= 1 + \pi {\cal P}
\int_{-\infty}^{+\infty}\!\!\dd p \, \frac{f_0'(p)}{p \mp \omega}
\pm {i\pi^2} f_0'(\pm \omega) \, .
\end{equation}
We see in particular that $\tilde D(-\omega,-1)=\tilde
D^*(\omega,1)$. As we mentioned, the real solutions $\omega(k)$
obtained from equating to zero the dielectric function correspond to
the boundary between stable and unstable modes. From the last
equation, posing equal to zero both the real and the imaginary
parts, we see that in order to have a real $\omega$ as a solution of
$\tilde D(\omega,\pm 1)=0$, it is necessary that $f_0'(\pm
\omega)=0$. If one considers initial distributions with a single
maximum at the origin in momentum space, one obtains, for both
$\tilde D(\omega,1)$ and $\tilde D(\omega,-1)$, $\omega=0$. Hence,
the real part of Eq.~(\ref{realplusimagparts}) will be zero if
\begin{equation}
I \equiv 1+ \pi \int_{-\infty}^{+\infty}\frac{f_0'(p)}{p}\ \dd p
=0~. \label{therholdcon}
\end{equation}
This gives the stability boundary of the homogeneous distribution
$f_0(p)$. To deduce where the stability region is located in the
functional space of $f_0(p)$, one can perform an analysis based on
the Nyquist criterion, from which one finds that the stability
condition is $I > 0$ \cite{Inagaki93b}.

The linear stability analysis discussed above appeared in a series
of different
papers~\cite{Inagaki93b,Pichon,ChoiChoi1,ChoiChoi2,Judith,fredthierPRE},
The stability boundary can be also derived using the so-called
energy-Casimir method~\cite{yoshi}.

Condition~(\ref{therholdcon}) is a functional equation in $f_0(p)$.
By a convenient parametrization of  $f_0(p)$, one can obtain a
stability boundary in a finite dimensional control parameter space.
A relevant parametrization for the HMF model is the one in terms of
the energy per particle $\varepsilon$. Therefore, we will briefly
discuss below some examples where a critical energy density
$\varepsilon^c$ appears, above which the homogeneous solution is
stable.

\begin{figure}[htb]
\begin{center}
\includegraphics[height=6truecm]{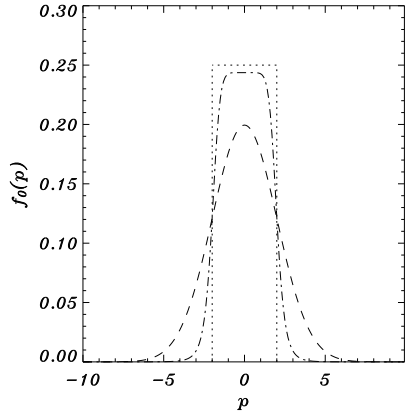}
\end{center}
\caption{Three examples of stationary homogenous solutions of the
Vlasov equation. The Gaussian (dashed), the water-bag (dotted) and
the power-law (Eq.~\ref{eq:powertail}) in the case $\nu=8$
(dash-dotted).} \label{figdiffzero}
\end{figure}

\begin{itemize}
\item The first one is the Gaussian distribution
\begin{equation}f_\text{g}(p)=\frac{1}{2\pi}\sqrt{\frac{\beta}{2\pi}} \exp({-\beta p^2/2})
\label{distgaussienne}
\end{equation}
(see Fig.~\ref{figdiffzero}). One gets $I=1-\beta/2$. The condition
$I=0$ coincides with the equilibrium statistical mechanics result
that the critical inverse temperature is $\beta_\text{g}^c=2$, and
its associated critical energy is $\veps_\text{g}^c=3/4$
\cite{Inagaki93a,Inagaki93b,Antoni95}.

\item The second example is the water-bag distribution
\begin{eqnarray}
f_{\text{wb}}(p)=\frac{1}{2\pi}\frac{1}{2p_0}\left[\Theta(p+p_0)-
\Theta(p-p_0)\right]\label{distwaterbag}
%=\left\{\begin{array}{cc}
%   1/2p_0 & \mbox{if}\ |p|<p_0\\
%   0 & \mbox{if}\ |p|<p_0
%  \end{array}\right.
  \end{eqnarray}
where $\Theta$ is the Heavyside function. This distribution, also
shown in Fig.~\ref{figdiffzero},  has been often used in the past to
test numerically the out-of-equilibrium properties of the HMF model.
In this case, one obtains $I=1-1/(2p_0^2)$ which leads to a smaller
critical energy $\veps_\text{wb}^c=7/12$.

\item Another example studied in the literature is the $q$-Gaussian
\begin{equation}
f_\text{T}(p)\sim [1-\alpha(1-q)p^2]^{\frac{1}{1-q}}\, ,
\label{distTsallis}
\end{equation}
with $\alpha$ positive. This function has a compact support for
$q<1$, while it decays with a power law tail for $q>1$; we recover
the Gaussian for $q=1$. The water-bag distribution is recovered for
$q \to -\infty$. For $q\ge 5/3$, the second moment of this
distribution, and thus the average kinetic energy, is not finite,
and therefore a cut-off to keep the energy finite is used in actual
implementations. For the cases with finite kinetic energy, i.e., for
$q<5/3$, one gets~\cite{Vallejos,Judith,CampaChavanis}
$\veps_\text{T}^c=\frac{3}{4}+\frac{q-1}{{2(5-3q)}}$. One easily
verifies that the previous values of the critical energy for the
Gaussian and the water-bag distribution are recovered for the proper
values of $q$.

\item The last example~\cite{yoshiJSM} is a distribution with power-law tails
\begin{equation}
\label{eq:powertail} f_{\text{pl}}(p) =
\frac{A_\nu}{1+|p/p_{0}|^{\nu}},
\end{equation}
characterized by the exponent~$\nu$. The unity, added in the
denominator to avoid the divergence at the origin $p=0$, does not
affect neither the asymptotic form, nor the theoretical predictions.
The parameter  $p_{0} = \sqrt{\frac{\sin(3\pi/\nu)}{\sin(\pi/\nu)}
\frac{K}{N}}$ controls the kinetic energy density $K/N$ and the
normalization factor is $A_\nu=\nu\sin(\pi/\nu)/(2\pi p_{0})$ . The
exponent $\nu$ must be greater than $3$ to get a finite kinetic
energy: we have used $\nu=8$ (see Fig.~\ref{figdiffzero}). Note that
the power-law distribution cannot be included in the
$q$-exponentials family \cite{Tsallisjsp}, although it has similar
power law tails at large $|p|$. Distribution~(\ref{eq:powertail}) is
stable above the critical energy density
$\veps_\text{pl}^{c}=\frac{1}{2}+\frac{\sin(\pi/\nu)}{4\sin(3\pi/\nu)}$,
which corresponds to $\veps_\text{pl}^{c}=0.75,~0.625$ and
$0.60355\dots$ for $\nu=4,~6$ and $8$ respectively.

\item Finally let us consider a more general distribution, namely
a mixed distribution between $ f_{\text {wb}} $ and $ f_{\text {g}}
$, defined as
\begin{equation}\label{eq:fa}
f_{a}(p) = (1-a) f_{\text {wb}}(p) + a f_{\text {g}}(p).
\end{equation}
Thanks to the linearity of the quantity~(\ref{therholdcon}) with
respect to the distribution, the critical energy density for this
mixed distribution $ f_{a} $ is obtained~\cite{yoshi} as a linear
combination of the previous results.

\end{itemize}

The condition~(\ref{therholdcon}) defines therefore a critical
energy $\veps^c$ which is, in general, different from the critical
energy $3/4$ where the second-order phase transition of equilibrium
statistical mechanics is located. However, as expected, the two
values coincide for a Gaussian distribution.

All the above distributions are thus stationary solutions of the
Vlasov equation~(\ref{equationpourfzeroici}).  However, it is
important to realize that they are Vlasov stable stationary
solutions among {\em infinitely} many others and there is no reason
to emphasize one more than the others.

\subsubsection{The Lenard-Balescu equation}
\label{seclenardbalescu}

We have so far concentrated on the collisionless dynamics as
described by the Vlasov equation. When using this equation, one
implicitly assumes that the particles of the system  move under the
influence of the average potential generated by all the other
particles. This means that the acceleration of all the particles is
given by a single function, i.e., by $-\partial \langle v \rangle
/\partial \theta$. However, this assumption is not valid for
arbitrarily long times. The presence of individual ``collisions''
invalidate the assumption that the acceleration of all the particles
derives from a single mean field function: ``collisions'' perturb
particles away from the trajectories they would have taken if the
distribution of particles in the system were perfectly smooth. In
long-range systems, however, we have seen that for times shorter
than $N$ (thus for times that can be very long), the collisionless
approximation is valid. For larger times, the Vlasov equation is not
valid and we have to consider the effect of ``collisions". Under
this effect, particles will deviate from the orbits determined by
the Vlasov equation on a characteristic time that is called {\it
relaxation time}. This can be justified from the fact that a stable
stationary solution of the Vlasov equation, as determined from the
analysis described in the previous subsection, will eventually be
perturbed by the effect of ``collisions".

Let us then go back to the exact Eq.~(\ref{equationpourfzero}) and
consider the right hand side. At the next level of approximation,
i.e., the level $1/N$, we can determine the right hand side using
the solutions for $\delta v$ and $\delta f$ as determined by the
collisionless dynamics, given by Eq. (\ref{Vlasovlinear}). If again
we restrict to the case in which $f_0$, a stable stationary solution
of the Vlasov equation, does not depend on $\theta$, at this level
of approximation our kinetic equation will therefore be
\begin{equation}
\frac{\partial f_0}{\partial t} =
\frac{1}{N}\left\langle\frac{\partial \delta v}{\partial \theta}
\frac{\partial \delta f}{\partial p}\right\rangle \, .
\label{equationpourfzerocolli}
\end{equation}
This equation is valid up to a time where the right hand side makes
$f_0$ leave its stability basin as determined by the Vlasov
equation. As already anticipated, one can expect this time to be of
order $N$. This equation is called the Lenard-Balescu equation, and
its solution requires the determination of the correlation function
on the right-hand-side of Eq.~(\ref{equationpourfzerocolli}). To
achieve this goal, it is necessary to use the spatio-temporal
Fourier-Laplace transform. Using this technique, one is able to
determine the correlation expressing the collisional term as a
function of its value at the initial time. We thus have to solve an
initial value problem.

The spatio-temporal Fourier-Laplace transform of the fluctuation of
the density $\delta f$ is defined by
\begin{equation}
\widetilde{\delta f}(k,p,\omega)  =\int_0^{2\pi} \frac{{\dd
\theta}}{2\pi} \int_{0}^{+\infty} \dd t\ e^{-i(k\theta-\omega
  t)}\ \delta f(\theta,p,t) \, ,
\label{FourierLaplacetransform}
\end{equation}
associated with a similar expression for the fluctuation of the
potential $\delta v$. As usual with Laplace transforms,
$\widetilde{\delta f}(k,p,\omega)$ is defined by the last equation
only for Im$(\omega)$ sufficiently large. For the remaining part of
the complex $\omega$ plane, it is defined by an analytic
continuation. The inverse transform is
\begin{equation}
\delta f(\theta,p,t) =\sum_{k= -\infty}^{+\infty} \int_{\cal C}
\frac{\dd \omega}{2\pi}\ e^{i(k\theta-\omega t)}\ \widetilde{\delta
f}(k,p,\omega) \, , \label{invFourierLaplacetransform}
\end{equation}
where the Laplace contour ${\cal C}$ in the complex $\omega$ plane
must pass above all poles of the integrand. The inverse transform
has a sum over the integer values of $k$ since the coordinates
$\theta$ take values from $0$ to $2\pi$. In fact, we will determine
the right hand side of Eq. (\ref{equationpourfzerocolli}) taking as
an example the HMF model, as we have done in the previous
subsection. More generally, there will be an integral over $k$, but
the analysis will be perfectly equivalent.

If we multiply Eq.~(\ref{Vlasovlinear}) by $e^{-i(k\theta-\omega
t)}$ and integrate over $\theta$ from $0$ to $2\pi$ and over $t$
from $0$ to $\infty$, we obtain
\begin{eqnarray}
-\widehat{\delta f}(k,p,0)-i\omega\,\widetilde{\delta
f}(k,p,\omega)+ikp\, \widetilde{\delta f}(k,p,\omega)-i k
\,\widetilde{\delta v}(k,\omega) f_0'(p)=0,
\label{Vlasovlineartransfs}
\end{eqnarray}
where the first term is the spatial Fourier transform of the initial
value
\begin{equation}
\widehat{\delta f}(k,p,0)  =\int_0^{2\pi} \frac{{\dd \theta}}{2\pi}
\ e^{-ik\theta}\ \delta f(\theta,p,0) \, ,
\label{spaceFouriertransform}
\end{equation}
and it arises from the integration by parts in obtaining the
Laplace-Fourier transform of ${\partial \delta f}/{\partial t}$. The above equation
can be rewritten as
\begin{eqnarray}
\widetilde{\delta f}(k,p,\omega)= \frac{k\,f_0'(p)}{pk-\omega} \,
\widetilde{\delta v}(k,\omega) +\frac{\widehat{\delta
f}(k,p,0)}{i(pk-\omega)}, \label{Vlasovlineartransf}
\end{eqnarray}
where one identifies a first ``collective'' term depending on the
perturbation of the potential, and a second one which depends on the
initial condition. Combining Eq.~(\ref{Vlasovlineartransf}) with the
Laplace-Fourier transform of the fluctuations of the
potential~(\ref{fluctpotential}), i.e.
\begin{eqnarray}
\label{fouriertransformofdeltaVsuiteinter} \widetilde{\delta
v}(k,\omega) &=&-\pi
\left(\delta_{k,1}+\delta_{k,-1}-2\delta_{k,0}\right)\int_{-\infty}^{+\infty}\!\!
\dd p\ \widetilde{\delta f}(k,p,\omega),
\end{eqnarray}
and integrating~(\ref{Vlasovlineartransf}) over the $p$ variable,
gives
\begin{eqnarray}
\int_{-\infty}^{+\infty}\!\!\dd p\ \widetilde{\delta
f}(k,p,\omega)\left[1+ \pi k \left(\delta_{k,1}+\delta_{k,-1}\right)
\int_{-\infty}^{+\infty}\!\! \dd p'\
\frac{f_0'(p')}{(p'k-\omega)}\right]=\int_{-\infty}^{+\infty}\!\!\dd
p\ \frac{\widehat{\delta f}(k,p,0)}{i(pk-\omega)}~,
\label{Vlasovlineartransfsui}
\end{eqnarray}
where one recognizes  the plasma response dielectric
function~(\ref{dielectricfunction}) in the parenthesis.
Equation~(\ref{fouriertransformofdeltaVsuiteinter}) can thus be
rewritten as
\begin{eqnarray}
\label{fouriertransformofdeltaVsuiteert} \widetilde{\delta
v}(k,\omega)&=&-\frac{\pi
\left(\delta_{k,1}+\delta_{k,-1}-2\delta_{k,0}\right)} {\tilde
D(\omega,k)}\int_{-\infty}^{+\infty}\!\!\dd p\ \frac{\widehat{\delta
f}(k,p,0)}{i(pk-\omega)}.
\end{eqnarray}
We see that the Laplace contour ${\cal C}$ for the inversion formula
must pass above all zeroes of $\tilde D(k,\omega)$. We can consider
that these zeros will all be located in the half-plane
Im$(\omega)<0$, since otherwise the problem of the $1/N$
perturbations to a stable stationary solution of the Vlasov equation
would not make sense~\cite{landau}.

One can then use these expressions to compute the collisional term
appearing on the right-hand-side of
Eq.~(\ref{equationpourfzerocolli}). Forgetting temporarily the
factor $1/N$ and the derivative with respect to the variable $p$,
one has
\begin{eqnarray}
\left\langle\frac{\partial \delta v }{\partial \theta}  \delta
f\right\rangle&=& \left\langle\sum_{k=-\infty}^{+\infty} \int_{{\cal
C}}\frac{\dd \omega}{2\pi}\ ik\,e^{i(k\theta-\omega t)}\
\widetilde{\delta v}(k,\omega)\sum_{k'=-\infty}^{+\infty}
\int_{{\cal C'}} \frac{\dd \omega'}{2\pi}\ e^{i(k'\theta-\omega'
t)}\ \widetilde{\delta f}(k',p,\omega')\right\rangle  \\
&=&\frac{1}{(2\pi)^2} \sum_{k=-\infty}^{+\infty}\,
\sum_{k'=-\infty}^{+\infty}\, \int_{{\cal C}}\dd \omega\,\int_{{\cal
C'}}\dd \omega' ik\,e^{i[(k+k')\theta-(\omega+\omega') t]}\
\left\langle\widetilde{\delta v}(k,\omega)\, \widetilde{\delta
f}(k',p,\omega')\right\rangle \label{equaciter}~,
\end{eqnarray}
which relies on evaluating the correlation $\langle\widetilde{\delta
v}(k,\omega)\, \widetilde{\delta f}(k',p,\omega')\rangle$. The
presence of the factor $k$ in the last equation allows us to forget
the $k=0$ contribution to $\widetilde{\delta v}(k,\omega)$ in Eq.
(\ref{fouriertransformofdeltaVsuiteert}); this was expected, since
the constant component of $\delta v$ cannot contribute to the force.
Using Eq.~(\ref{Vlasovlineartransf}), one finds
\begin{eqnarray}
\left\langle\widetilde{\delta v}(k,\omega)\, \widetilde{\delta
f}(k',p,\omega')\right\rangle &=& \frac{k'f_0'(p)}{pk'-\omega'}
\left\langle\widetilde{\delta v}(k,\omega)\,\widetilde{\delta
v}(k',\omega')\right\rangle +\frac{\left\langle\widetilde{\delta
v}(k,\omega)\,\widehat{\delta
f}(k',p,0)\right\rangle}{i(pk'-\omega')}\, . \label{distinguishterm}
\end{eqnarray}
The first term on the r.h.s. corresponds to the self-correlation of
the potential, while the second one to the correlation between the
fluctuations of the potential and of the distribution at time $t=0$.
Let us consider separately the two terms of the last equation.

From Eq.~(\ref{fouriertransformofdeltaVsuiteert}), neglecting the
$k=0$ contribution and taking the statistical average, we have
\begin{eqnarray}
\langle \widetilde{\delta v}(k,\omega)\widetilde{\delta
v}(k',\omega')\rangle&=&\frac{\pi^2
\left(\delta_{k,1}+\delta_{k,-1}\right)\left(\delta_{k',1}+\delta_{k',-1}\right)}
{\tilde D(\omega,k)\tilde D(\omega',k')}\int_{-\infty}^{+\infty}\,
\dd p\int_{-\infty}^{+\infty}\, \dd p'\ \frac{\langle
\widehat{\delta f}(k,p,0)\widehat{\delta
f}(k',p',0)\rangle}{i(pk-\omega)i(p'k'-\omega')}  \\
&=&\frac{\pi}{2}\frac{\delta_{k,-k'}
\left(\delta_{k,1}+\delta_{k,-1}\right)} {\tilde D(\omega,k)\tilde
D(\omega',-k)}\int_{-\infty}^{+\infty}\, \dd p \,
\int_{-\infty}^{+\infty}\, \dd p' \, \frac{\left[ f_0(p)\delta
(p-p')+\mu(k,p,p')\right] }{(pk-\omega)(p'k+\omega')}~,
\label{Potautocorrelation}
\end{eqnarray}
where we have replaced the autocorrelation of the fluctuation of the
distribution at $t=0$ (see Appendix D) by
\begin{equation}
\langle \widehat{\delta f}(k,p,0)\widehat{\delta f}(k',p',0)\rangle
= \frac{\delta_{k,-k'}}{2\pi}\left[ f_0(p)\delta
(p-p')+\mu(k,p,p')\right]~, \label{decouplage}
\end{equation}
with the first term expressing the single particle contribution to
the correlation, while in the second term the function $\mu(k,p,p')$
comes from the contribution of different particles. This last
function is smooth, but otherwise arbitrary, since it is related to
initial conditions of our initial value problem. However, it can be
shown \cite{landau} that, going back in the time domain with the
inverse Laplace-Fourier transform, the contribution of this function
to the correlation decays in time. Therefore we can consider only
the first term in Eq. (\ref{decouplage}), obtaining
\begin{equation}
\langle \widetilde{\delta v}(k,\omega)\widetilde{\delta v}
(k',\omega')\rangle =\frac{\pi}{2}\frac{\delta_{k,-k'}
\left(\delta_{k,1}+\delta_{k,-1}\right)} {\tilde D(\omega,k)\tilde
D(\omega',-k)}\int_{-\infty}^{+\infty}\, \dd p \,
\frac{f_0(p)}{(pk-\omega)(pk+\omega')}. \label{Potautocorrelation2}
\end{equation}
Considering again only the contributions that, after integration in
$\omega$ and $\omega'$, do not decay in time, it can be shown
\cite{landau}, through the use of the Plemelj
formula~(\ref{Plemeljformula}), that
$[(pk-\omega)(pk+\omega')]^{-1}$ can be substituted by
$(2\pi)^2\delta(\omega+\omega')\delta(\omega-pk)$. In addition,
using the property $\tilde D(\omega,k)=\tilde D^*(-\omega,-k)$, one
finally ends up with the result
\begin{equation}
\left\langle  {\delta v}(k,\omega){\delta
v}(k',\omega')\right\rangle =2\pi^3
\delta_{k,-k'}\left(\delta_{k,1}+\delta_{k,-1}\right)\,
\frac{\delta(\omega+\omega')} {\left|\tilde
D(\omega,k)\right|^2}\,\int\dd p\, f_0(p)\delta(\omega-pk),
\label{correldeltaVfinal}
\end{equation}
which vanishes except for $|k|=1$: only the cases $k=-k'=\pm1$
therefore contribute.

We now consider the second term of Eq.~(\ref{distinguishterm}).
Using again Eq.~(\ref{fouriertransformofdeltaVsuiteert}) without the
$k=0$ term we have
\begin{eqnarray}
\frac{\left\langle\widetilde{\delta v}(k,\omega)\,\widehat{\delta
f}(k',p,0)\right\rangle}{i(pk'-\omega')}=
-\frac{\pi\left(\delta_{k,1}+\delta_{k,-1}\right)}{\tilde
D(\omega,k)}\int_{-\infty}^{+\infty} \, \dd p'\,
\frac{\left\langle\widehat{\delta f}(k,p',0)\,\widehat{\delta
f}(k',p,0)\right\rangle}{i(p'k-\omega)i(pk'-\omega')} \, .
\label{PotForcecorrelation}
\end{eqnarray}
As for the analysis of the first term of
Eq.~(\ref{distinguishterm}), we substitute the initial time
correlation in the last integral with the first term in Eq.
(\ref{decouplage}), and afterwards we substitute
$[(pk-\omega)(pk+\omega')]^{-1}$ with
$(2\pi)^2\delta(\omega+\omega')\delta(\omega-pk)$. We therefore
obtain
\begin{eqnarray}
\frac{\left\langle\widetilde{\delta v}(k,\omega,k)\,\widehat{\delta
f}(k',p,0)\right\rangle}{i(k'p-\omega')}
&=&-\frac{2\pi^2\delta_{k,-k'}\left(\delta_{k,1}+\delta_{k,-1}\right)}
{\tilde D(\omega,k)} f_0(p)\delta(\omega+\omega')\delta(\omega-pk)
\, . \label{PotForcecorrelation2}
\end{eqnarray}

From Eq. (\ref{correldeltaVfinal}), one gets the contribution to
(\ref{equaciter}) of the first term of Eq. (\ref{distinguishterm}).
Exploiting the presence of the factors $\delta_{k,-k'}$ and
$\delta(\omega+\omega')$, this contribution is
\begin{eqnarray}
\frac{i \pi}{2} \sum_{k=-\infty}^{+\infty}&&\!\!\!\!\! \!\!\!\!\!
\int_{{\cal C}}\dd \omega\left(\delta_{k,1}+\delta_{k,-1}\right)\,
\frac{k^2 f_0'(p)}{pk-\omega} \frac{1}{\left|\tilde
D(\omega,k)\right|^2}\,\int\dd p'\, f_0(p')\delta(\omega-p'k)
\nonumber\\ &=&
 \frac{i \pi}{2}\int_{{\cal C}}\dd \omega\, f_0'(p)
\left[\frac{1}{p-\omega}\frac{f_0(\omega)}{\left|\tilde
D(\omega,1)\right|^2}
-\frac{1}{p+\omega}\frac{f_0(-\omega)}{\left|\tilde
D(\omega,-1)\right|^2}
\right]  \\
&=&\frac{i \pi}{2}\int_{-\infty}^{+\infty} \dd \omega\, f_0'(p)
\left[\left({\cal P}\frac{1}{p-\omega}-i\pi\delta(p-\omega)\right)
\frac{f_0(\omega)}{\left|\tilde D(\omega,1)\right|^2} -\left({\cal
P}\frac{1}{p+\omega}+i\pi\delta(p+\omega)\right)
\frac{f_0(-\omega)}{\left|\tilde D(\omega,-1)\right|^2}\right]  \\
&=&\pi^2\int_{-\infty}^{+\infty} \dd \omega\, \frac{1}{\left|\tilde
D(\omega,1)\right|^2} f_0'(p) f_0(\omega) \delta(p-\omega) \, .
\label{equaciter1}
\end{eqnarray}
Passing from the second to the third line we have used again the
Plemelj formula, that allows us to integrate on the real $\omega$
axis. Note that in this case the rule of singularity encircling is
the opposite of the usual one, i.e., it is $\omega \to \omega -i0$,
since the $\omega$ of the $(pk-\omega)$ term in the denominator in
the first line comes from the integration in $\omega'$, that gives
$\omega=-\omega'$ \cite{landau}. Passing from the third to the
fourth line, the variable of integration has been changed from
$\omega$ to $-\omega$ in the last two terms of the third line, and
we have also used the property $\tilde D(\omega,k)=\tilde
D^*(-\omega,-k)$. The contribution of the second term of
(\ref{distinguishterm}) to Eq. (\ref{equaciter}) is obtained from
(\ref{PotForcecorrelation2}). Exploiting again the factors
$\delta_{k,-k'}$ and $\delta(\omega+\omega')$, this contribution is
\begin{eqnarray}
\frac{-i}{2} \sum_{k=-\infty}^{+\infty}\, \int_{{\cal C}}\dd
\omega\, f_0(p) \left(\delta_{k,1}+\delta_{k,-1}\right)\,
\frac{\delta(\omega-pk)}{\tilde D(\omega,k)}  &=&\frac{-i}{2}
\int_{-\infty}^{+\infty}\dd \omega\, f_0(p)
\frac{\delta(\omega-p)}{\left|\tilde D(\omega,1)\right|^2}
\left[\tilde D^*(\omega,1) -\tilde D^*(-\omega,-1)\right] \\
&=&-\pi^2\int_{-\infty}^{+\infty} \dd \omega\, \frac{1}{\left|\tilde
D(\omega,1)\right|^2} f_0(p) f_0'(\omega) \delta(p-\omega) \, .
\label{equaciter2}
\end{eqnarray}

We now have all elements for the determination of the right hand
side of the Lenard-Balescu equation (\ref{equationpourfzerocolli}).
Inserting again the $1/N$ factor we obtain the following result
which is valid only in one dimension
\begin{eqnarray}
\frac{\partial f_0}{\partial t} &=&\frac{\pi^2}{N}\frac{\partial
}{\partial p}\int_{-\infty}^{+\infty} \frac{\dd \omega}{|\tilde
D(\omega,1)|^2}\delta\left(p-\omega\right)
\left(f_0(\omega)\frac{\partial f_0(p)}{\partial p}
-f_0(p)\frac{\partial f_0(\omega)}{\partial \omega} \right) =0 \, .
\label{LenardBalescu1D}
\end{eqnarray}
We thus see that, in one dimension, the Lenard-Balescu operator
vanishes: the diffusion term (first term in the r.h.s.) is exactly
balanced by the friction term (second term in the r.h.s.).
Consequently the collisional evolution is due to terms of higher
order in $1/N$, and the Vlasov equation is valid for a longer time
than previously expected. This remark was raised long ago in plasma
physics~\cite{KadomtsebPogutse}.

The previous computation can be performed for a general (long-range)
two-body potential in $d$ dimensions. The result is the
following general Lenard-Balescu equation~\cite{Balescubook,chavanisEPJB2006}
\begin{equation}
\frac{\partial f(\mathbf{v},t)}{\partial t} =\frac{\pi(2\pi)^d}{m^2}
\frac{\partial}{\partial \mathbf{v}} \cdot \int \dd \mathbf{v}_1 \dd
\mathbf{k} \
\mathbf{k}\frac{\hat{u}(|\mathbf{k}|)^2}{|\tilde
D(\mathbf{k},\mathbf{k}\cdot\mathbf{v}_1)|^2}
\delta\left(\mathbf{k}\cdot(\mathbf{v}-\mathbf{v}_1)\right) \left[
\mathbf{k}\cdot \left(f(\mathbf{v}_1,t)\frac{\partial
f(\mathbf{v},t)}{\partial \mathbf{v}} -f(\mathbf{v},t)\frac{\partial
f(\mathbf{v}_1,t)}{\partial \mathbf{v}_1} \right)\right] \, ,
\label{LenardBalescuGeneral}
\end{equation}
where the boldface variables are $d$-dimensional vectors, and where
$\hat{u}(|\mathbf{k}|)$ is the real $d$-dimensional Fourier
transform of the two-body potential; $m$ is the mass of the
particles and $f$ is normalized to $1$. The dielectric function
$\tilde D(\omega,k)$ in this general case is given by
\begin{equation}
\tilde D(\omega,\mathbf{k})=1+\frac{(2\pi)^d}{m}
\hat{u}(|\mathbf{k}|) \int\dd \mathbf{v} \,
\frac{\mathbf{k}\cdot\frac{\partial f(\mathbf{v})}{\partial
\mathbf{v}}} {\omega-\mathbf{k}\cdot\mathbf{v}} \, .
\label{generdielect}
\end{equation}
The $1/N$ ``smallness'' of the right hand side of Eq.
(\ref{LenardBalescuGeneral}) is incorporated in the two-body
potential. From this general expression we see that, while in $d=1$
the collisional term vanishes, it will instead be present for $d>1$.
Thus, in a two-dimensional Coulombian plasma or for
three-dimensional Newtonian interactions and in plasma physics, the
Lenard-Balescu collisional term gives a contribution already at
order $1/N$. Nicholson has derived its expression for
general potentials for homogeneous cases \cite{nicholson}; Chavanis has
continued this line of research ~\cite{chavanisPhysicaI2006,chavanisPhysicaII2006}
and obtained preliminary results for the inhomogeneous
case~\cite{chavanisPhysicaIII2008,chavanisPhysicaIV2008}. Note however that
collective effects are not taken into account in the latter works.

The Landau approximation, often used in plasma physics, corresponds
to neglect collective effects in Eq.~(\ref{LenardBalescuGeneral}),
which amounts to take $\tilde D(\omega,\mathbf{k})=1$, so that the
structure of the Landau equation does not depend on the potential.

One notes that in a one-dimensional framework, the Lenard-Balescu
and the Landau equations coincide: both of them reduce to the Vlasov
equation. The collisional evolution is due to terms of higher order
in $1/N$, implying that the collisional relaxation time scales as
$N^\delta$ with $\delta>1$.
Therefore, the system can remain frozen in a stationary solution of
the Vlasov equation for a very long time, larger than $N$. Only
non-trivial three-body correlations can induce further evolution of
the system, as also remarked in
Ref.~\cite{Chavanisnewaddi} for 2D hydrodynamics. This is the very
reason of the so long relaxation time
emphasized by numerical simulations~\cite{yoshi}.

\subsection{Quasi-stationary states, diffusion and entropies}
\label{QSSdiff}

\subsubsection{Numerical evidence of quasi-stationary states}
\label{Numericalevidence}

For the HMF model, but this is also true for any one-dimensional
long-range system, we have thus proven that Vlasov stable
homogeneous distribution functions do not evolve on time scales of
order smaller or equal to $N$.

\begin{figure}[htb]
\begin{center}
\includegraphics[height=6truecm]{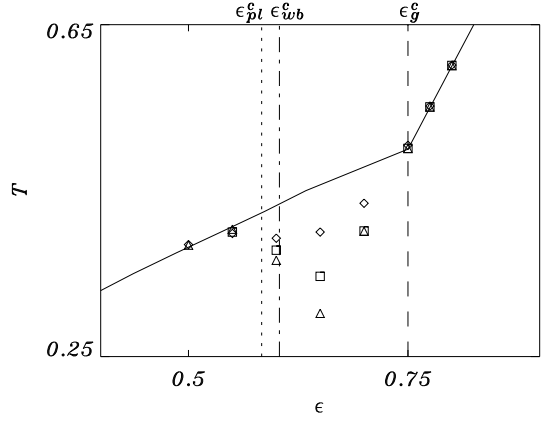}
\end{center}
\vskip -0.5truecm \caption{Caloric curve of the HMF Hamiltonian
(\ref{Ham_HMF}). The solid line is
  the equilibrium result in both the canonical and the microcanonical
  ensemble. The second order phase transition is revealed by the kink
  at $\varepsilon_g^c=3/4$. The three values of the energy indicated by the
  vertical lines are the stability thresholds for the homogeneous
  Gaussian (dashed), power-law of Eq.~(\ref{eq:powertail}) with
  $\nu=8$ (dash-dotted) and water-bag (dotted) initial momentum
  distribution. The Gaussian stability threshold coincides with the
  phase transition energy. The points are the results of constant
  energy (microcanonical) simulations for the Gaussian (losanges), the
  power-law (squares) and the water-bag (triangles). Simulations were
  performed with $N=5000$.   }
\label{figcaloric}
\end{figure}

\begin{figure}[htb]
\begin{center}
\includegraphics[height=6truecm]{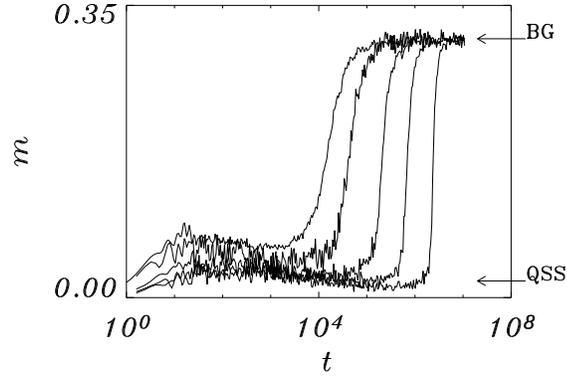}
\end{center}
\vskip -0.5truecm \caption{Time evolution of the modulus of the
magnetization $m(t)$ for different particle numbers: $N=10^3$,
$2.10^3$, $5.10^3$, $10^4$ and $2.10^4$ from left to right
($\varepsilon=0.69$). In all cases, an average over several samples
has been taken. Two values of the magnetization, indicated by
horizontal arrows, can be identified in this figure: the upper one
(labeled BG) corresponds to the expected equilibrium result for the
magnetization, while the lower one, labelled QSS, represents the
value of $M$ in the quasi-stationary state.} \label{figdiffN}
\end{figure}

The above result is an illuminating explanation of the numerical
strong disagreement which was reported in
Refs.~\cite{Antoni95,lrt2001} between constant energy molecular
dynamics simulations and canonical statistical mechanics
calculations. For energies slightly below the second order phase
transition energy (see Fig.~\ref{figcaloric}), numerical simulations
in the microcanonical ensemble show that the system is trapped for a
long time, whose duration increases with the number $N$ of
particles, in a state far from that predicted by equilibrium
statistical mechanics. Since the latter were initially derived using
the canonical ensemble, these results had been believed, at first,
to be the fingerprint of inequivalence between microcanonical and
canonical ensembles. However, although the interaction is
long-range, both ensembles lead to the same results for the HMF
model, where only a second order phase transition occurs as shown in
Sec.~\ref{exemple_HMF}. More careful numerical
experiments~\cite{yoshi} have revealed the tendency of the
simulation points, i.e., of the out-of-equilibrium state in which
the system is trapped, to lie on the continuation to lower energies
of the supercritical branch with zero magnetization of the caloric
curve. These states have been called quasi-stationary states (QSS).
More systematic simulations~\cite{yoshi} have determined a $N^{1.7}$
scaling law for the duration of the QSS, at the end of which the
system eventually evolves towards the Boltzmann-Gibbs equilibrium
state. In Fig.~\ref{figdiffN}, we display the time evolution of the
magnetization, $m(t)$, with increasing particle number, showing the
increase of the duration of the QSS: the power-law increase is
evidenciated by the choice of the logarithmic scale in the abscissa.
Since this scaling law has been found when the system is initially
prepared in a state that is a stable stationary solution of the
Vlasov equation (in particular, for initial states homogeneous in
$\theta$), this numerical evidence is in agreement with the result
derived above, that Vlasov stable homogeneous distribution functions
do not evolve on time scales of order smaller or equal to $N$.
Obviously, even if the initial state is Vlasov stable, finite $N$
effects will eventually drive the system away from it and towards
the Boltzmann-Gibbs equilibrium state.

If the initial condition is Vlasov unstable, a rapid evolution will
take place. Simulations have been performed also in this case
\cite{lrt2001}, showing that, after an initial transient, the
systems remains trapped in other types of QSS, with different
scaling laws in $N$ for their duration. The existence of an infinite
number of Vlasov stable distributions is actually the key point to
explain the out-of-equilibrium QSS observed in the HMF dynamics.
Let us show that the system evolves through other stable stationary states.
In order to check the stationarity and the stability of an initial
distribution $f_0(\theta,p)$, it is possible to study the temporal
evolution of the magnetization, which is constant if the system is
stable and stationary. Other
possible macrovariables are the moments of the distribution
function. It can be easily shown that any distribution of the form
$f(\theta,p,0)=f_0(\theta,p)=F(p^2/2 - m_x \cos \theta - m_y \sin \theta) \equiv f_0(e)$,
with generic $F$, is a stationary solution of the Vlasov equation, provided
$m_x$ and $m_y$ are determined self-consistently from Eq.~(\ref{potaverage}).
Obviously this does not yet mean that this
solution is also stable. This latter observation has suggested to study
numerically the stability of $f_0$ by checking the stationarity of
the first few moments $\mu_{n}=\average{e^{n}}_{N}$. Since the
stationarity of the moments is a necessary condition for stability,
the vanishing, for a long time lapse, of the time derivatives
$\dot{\mu}_{n}=\dd \mu_{n}/\dd t$, for $n=1,2$ and $3$, has been
used as a numerical suggestion that the system is in a QSS and that
the distribution $f_0(\theta,p)$ is a stable stationary solution of
the Vlasov equation. On the contrary, large derivatives clearly
indicate a non-stationary state.

\begin{figure}[htbp]
\centering \subfigure[~Distribution with power-law tails: Stable]
{\includegraphics[width=7.5cm]{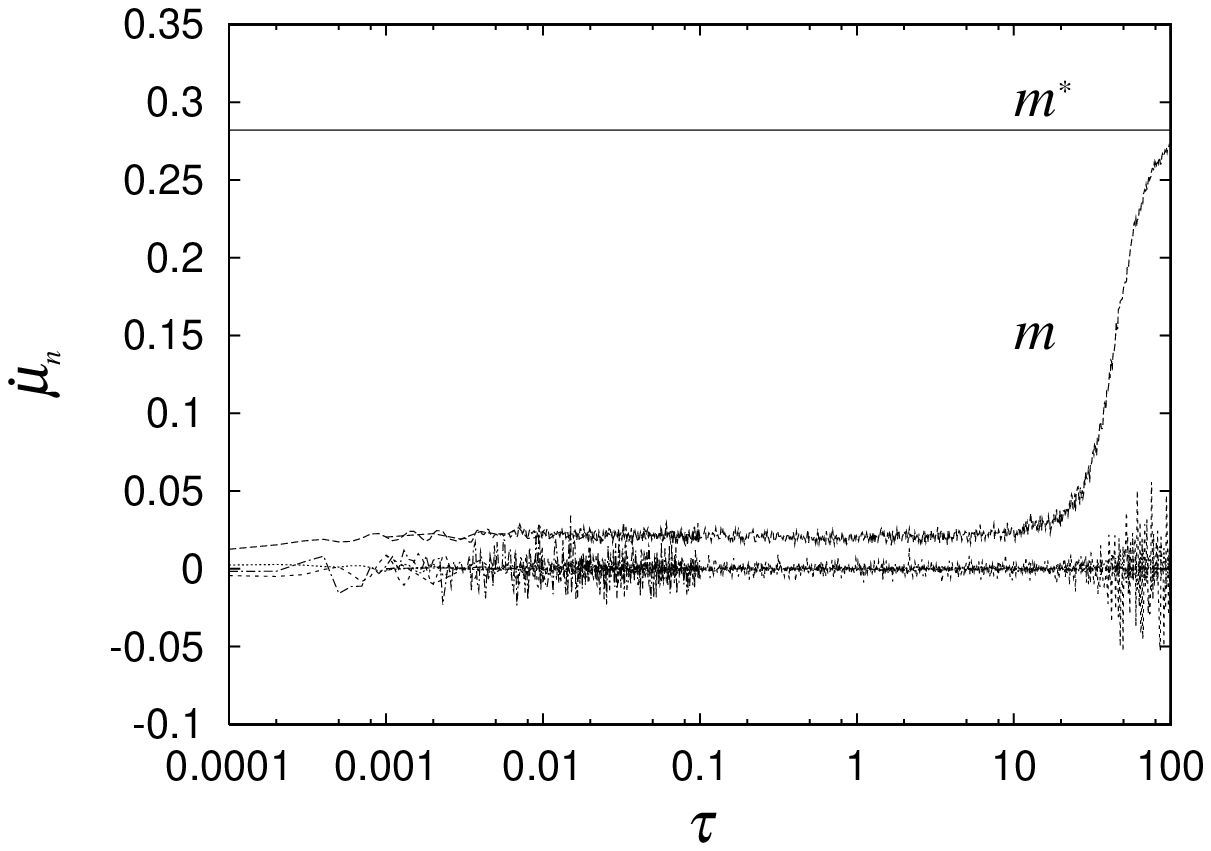}}
\centering \subfigure[~Gaussian distribution: Unstable]{
\includegraphics[width=7.5cm]{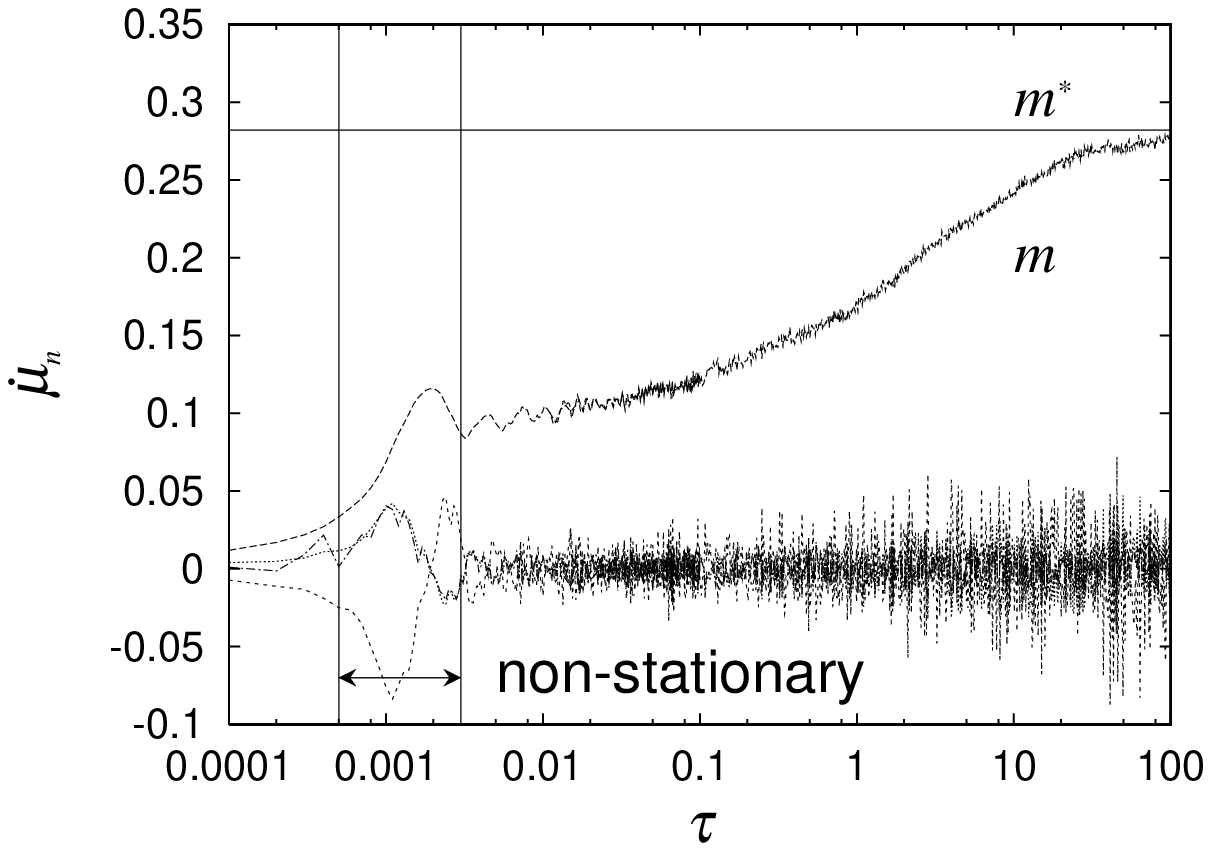}}
\caption{Time evolution of the magnetization $m$. The time is
rescaled as $\tau=t/N$. The quantities $\dot{\mu}_{n}~(n=1,2,3)$,
which detect the stationarity, are also plotted (multiplied by a
factor $100$ for graphical purposes). The equilibrium value of the
magnetization is $m^*$. Panel (a) corresponds to a stable
homogeneous initial distributions with power-law tails, while panel
(b) shows an unstable initial condition with gaussian tails. In both
panels, energy is $\varepsilon=0.7$. All numerical curves are
obtained by averaging over $20$ initial conditions with $N=10^{4}$.}
\label{fig:stationary}
\end{figure}

Figure~\ref{fig:stationary} presents the temporal evolution of these
quantities, together with the temporal evolution of the modulus~$m$
of the magnetization, for power-law ($\nu=8$) and gaussian initial
distributions in the case of an energy $\varepsilon$ in the interval
$[\varepsilon_\text{pl}^c,\varepsilon_\text{g}^c]$. In the stable
case (a), the stationarity holds throughout the computed time since
one notices that the three quantities $\dot{\mu}_{n}$ have vanishing
small fluctuations around zero. On the contrary, in the unstable
case (b), the system is first in an unstable stationary state
($\tau<0.0005$), before becoming non-stationary
($0.0005<\tau<0.003$) and finally reaches stable stationary states
($\tau>0.003$). Consequently the system evolves among different
Vlasov stationary states. In the stable case (a), the magnetization
$m$ stays around zero before taking off  around $\tau=20$ to reach
the equilibrium value $m^*$. In the unstable case,
Fig.~\ref{fig:stationary}(b) shows that after experiencing unstable
stationary and non-stationary states, the system presents a slow
quasi-stationary evolution across the infinite number of stationary
and stable Vlasov states. In Ref.~\cite{yoshi}, a careful numerical
study has shown that this slow characteristic timescale associated
to this final relaxation toward the Boltzmann-Gibbs equilibrium is
proportional to $N^{1.7}$. However, this law might be dependent on
the energy $\varepsilon$ and on the initial distributions. The
previous subsection allows however to claim that this relaxation
timescale is larger than $N$ at least. This is thus a very slow
process in comparison to the relaxation from an initially unstable
state. Recently, this slow evolution of the HMF system through
different stable stationary states of the Vlasov equation has been
more systematically studied, with the aim to determine how the
Vlasov stable solutions characterizing the system during the
out-of-equilibrium dynamics can be parametrized
\cite{CGM2007,CampaChavanis}. It has been found that, starting from
a homogeneous distribution given by a $q$-Gaussian with compact
support, at an energy slightly below the second order phase
transition energy, the evolution of the system during the QSS is
well approximated by distribution functions of the same type, but
with varying $q$, until the system heads towards Boltzmann-Gibbs
equilibrium.

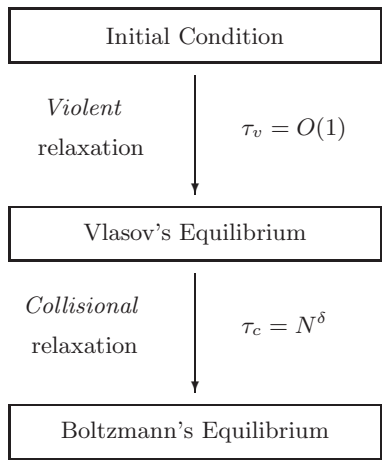
\begin{figure}[htb]
\begin{center}
{\setlength{\unitlength}{0.5pt}
\begin{picture}(200,350)(10,20)
\put(0,320){\framebox(280,40){Initial Condition}}
\put(0,170){\framebox(280,40){Vlasov's Equilibrium}}
\put(0,20){\framebox(280,40){Boltzmann's Equilibrium}}
\put(175,265){$\tau_v=O(1)$} \put(175,115){$\tau_c=N^\delta$}
\put(25,280){{\em Violent} } \put(15,250){ relaxation}
\put(10,130){{\em Collisional}} \put(10,100){ relaxation}
\put(140,160){\vector(0,-1){90}} \put(140,310){\vector(0,-1){90}}
\end{picture}}
\end{center}
\caption{Schematic description of the typical dynamical evolution of
systems with long-range interactions. $\tau_v$ and $\tau_c$ are the
{\it violent relaxation} and the {\it collisional relaxation}
timescales, respectively.} \label{schematicdescription}
\end{figure}

Let us stress that the above scenario is consistent with what
happens generically for systems with long-range
interactions~\cite{Lyndenwood68,ChavanisHouches,BBDRY2006}. In a
first stage, called {\it violent relaxation}, the system goes from a
generic initial condition, which is not necessarily Vlasov stable,
towards a Vlasov stable state. This is a fast process happening
usually on a fast timescale, {\em independent} of the number of
particles. In a second stage, named {\it collisional relaxation},
finite $N$ effects come into play and the Vlasov description is no
more valid for the discrete systems. The timescale of this second
process is strongly dependent on $N$. One generally considers that
it is a power law $N^\delta$. A typical example is the Chandrasekhar
relaxation time scale for stellar systems, which is proportional to
$N/\ln N$. This scenario of the typical evolution of long-range
systems is summarized in Fig.~\ref{schematicdescription}.

It is important to remark that, recently, Caglioti and
Rousset~\cite{CagliottiRousset} rigorously proved that for a wide
class of potentials, particles starting close to a Vlasov stable
distribution remain close to it for times that scale at least like
$N^{1/8}$: this result is consistent with the power law conjectured
for collisional relaxation. Unfortunately, apart from a recent
progress ~\cite{haurayjabin}, very few rigorous results exist in the
case of singular potentials, which would be of paramount importance
for Coulomb and gravitational interactions.

Stronger divergences with system size in long-range systems are
observed in connection with metastable
states~\cite{Antoni5,chavanisexpN,schreiber} where the relaxation
time increases exponentially with~$N$.

In summary, quasi-stationary states observed in the $N$-particle
dynamics of the HMF model are nothing but Vlasov stable
stationary states, which evolve because of {\it collisional}, finite
$N$, effects. There is an {\it infinity} of Vlasov stable
homogeneous (zero magnetization) states corresponding to different
initial velocity distributions $f_0(t=0,p)$, whose stability domain
in energy are different. The $q$-Gaussians in momentum homogeneous
distributions are Vlasov stable stationary states in a certain
energy region where QSS are observed in the HMF model. However, they
are not special in any respect, among an {\it infinity} of others.
In the HMF model at finite $N$, all of them converge sooner or later
to the Boltzmann-Gibbs equilibrium. However, the relaxation time is
shown numerically to diverge with a power-law $N^\delta$, with
$\delta\simeq 1.7$ for the homogeneous water-bag state.

The time scale $\tau = t/N$  is thus the appropriate one to study
momentum autocorrelation functions and diffusion in angle. We will
consider such issues  in the next subsection, where {\it weak} or
{\it strong anomalous diffusion} for angles is predicted, both at
equilibrium and for QSS.

\subsubsection{Fokker-Planck equation for the stochastic process of
a single particle} \label{fokkerplancksubsection}

Let us now consider the relaxation properties of a test particle,
initially with a momentum $p_1$ and an angle $\theta_1$, immersed in
a homogeneous background of $N$ particles; the
latter is consequently a thermal bath, or a reservoir, for the test
particle. The description of the motion for a so-called test
particle in a system with identical particles is a classical problem
in kinetic theory. Initially the test particle is in a given
microscopic state, while the other particles are distributed
according to the distribution $f_0$ and the test particle is assumed
not to affect the reservoir. The interaction with the fluid induces
a complicated stochastic process.

The test particle distribution is initially not in equilibrium, and
not necessarily close to it. However, it is natural to expect that
the distribution of the test particle will eventually correspond to
the distribution of the bath generated by all the other particles.
How it evolves from the initial Dirac distribution $f_1(\theta,p,0)=
\delta(\theta-\theta_1)\delta(p-p_1)$ toward the equilibrium
distribution is thus of high interest. We will show that the
distribution is a solution of a Fokker-Planck equation that can be
derived analytically. This equation describes the dynamical coupling
with the fluctuations of the density of particles, which induce
fluctuations in the potential: this is the origin of the underlying
stochastic process.

We analyze therefore the relaxation properties of a test particle,
indexed by 1, surrounded by a background system of $(N-1)$ particles
with a homogeneous in angle distribution $f_0(p)$. The averaged
potential $\langle v \rangle$ still vanishes for a homogeneous
distribution, so that the particle only feels the fluctuations of
the potential, that, according to Eq.~(\ref{defVetdeltaV}), is given
by $\delta v(\theta)/\sqrt{N}$. The potential felt by the test
particle at the position $\theta_1(t)$ is therefore $\delta
v(\theta)/\sqrt{N}$ computed at $\theta=\theta_1(t)$. We thus expect
that the instantaneous force on the test particle will be of order
$1/\sqrt{N}$. The equations of motion of the test particle are
therefore
\begin{equation}
\label{eqmotionofdyedparticle}
  \frac{\dd \theta_1}{\dd t}=p_1\qquad{\rm and}
\qquad \frac{\dd p_1}{\dd t}=
-\frac{1}{\sqrt{N}}\left.\frac{\partial \delta v(\theta,t)}{\partial
\theta} \right|_{\theta=\theta_1(t)} \, ,
\end{equation}
%\label{fluctpotential}
%\label{defVetdeltaV}
the integration of which leads to (omitting the index $1$ for the
sake of simplicity)
\begin{eqnarray}
\theta(t)&=&\theta(0)+p(0)\,t-\frac{1}{\sqrt{N}}\int_0^{t}\!\!\dd
u_1\int_0^{u_1}\!\!\dd u_2 \frac{\partial \delta v}{\partial
\theta}(\theta(u_2),u_2)
\label{devlexppdet1} \\
 p(t)&=&p(0)-\frac{1}{\sqrt{N}}\int_0^{t}\!\!\dd u
\frac{\partial \delta v}{\partial \theta}(\theta(u),u) \, ,
\label{devlexppdet}
\end{eqnarray}
where, again for simplicity, we have indicated directly inside the
dependence of $\delta v$ the test particle variable. The key point
of this approach is that we do not limit the study to the usual
ballistic approximation, in order to have an expansion exact at
order $1/N$. Therefore it is of paramount importance here to treat
accurately the essential {\em collective effects}.

By introducing iteratively the expression for the variable $\theta$
in the right-hand-side of Eq. (\ref{devlexppdet}) and by expanding
the derivatives of the potential, one gets the result at order $1/N$
of the momentum dynamics
\begin{eqnarray}
\label{equation_p} p(t)&=&p(0)-\frac{1}{\sqrt{N}}\int_0^{t}\dd u
\frac{\partial  \delta v}{\partial
\theta}\left(\theta(0)+p(0)u,u\right)
\nonumber \\
&&+\frac{1}{{N}}\int_0^{t}\dd u \frac{\partial^2 \delta v}{\partial
\theta^2} \left(\theta(0)+p(0)u,u\right)\int_0^{u}\dd u_1
\int_0^{u_1}\dd u_2 \frac{\partial \delta v}{\partial \theta}
(\theta(0)+p(0)u_2,u_2) \, .
\end{eqnarray}
As the changes in the momentum are small (of order
$1/\sqrt{N}$), the description of the momentum dynamics is well
represented by a stochastic process governed by a Fokker-Planck
equation~\cite{VanKampen}. If we denote by $f_1(p,t)$ the
distribution function at time $t$ of the test particle momentum,
then the general form of this equation is
\begin{eqnarray}
\frac{\partial f_1(p,t)}{\partial t} =-\frac{\partial}{\partial
p}\left[A(p,t) f_1(p,t)\right]
+\frac{1}{2}\frac{\partial^2}{\partial p^2} \left[B(p,t)
f_1(p,t)\right] \label{generalFokkerPlanckequation} \, ,
\end{eqnarray}
with
\begin{eqnarray}
A(p,t)&=&\lim_{\tau \to 0}\frac{1}{\tau}\langle \left(
p(t+\tau)-p(t)\right)\rangle_{p(t)=p}
\label{momentsFokkerPlanck1}\\
B(p,t)&=&\lim_{\tau \to 0}\frac{1}{\tau}\langle \left(
p(t+\tau)-p(t)\right)^2\rangle_{p(t)=p} \, ,
\label{momentsFokkerPlanck}
 \end{eqnarray}
where the expectation values are conditioned by $p(t)=p$. This
equation is therefore characterized by the time behavior of the
first two moments, called Fokker-Planck coefficients. We approximate
these two coefficients by an expression which is valid in the range
of time $t$ defined by $1 \ll t \ll N$. In this time range, using a
generalization of formula~(\ref{correldeltaVfinal}) that takes into
account that the initial coordinates of the test particles are
given, it is possible to obtain (see Appendix~E)
\begin{eqnarray}
\label{differentmoments1ici} A(p,t) &\sim&\!\!
\frac{1}{N}\left(\frac{\dd D}{\dd p}(p)
+\frac{1}{f_0}\frac{\partial f_0}{\partial p}D(p)\right) \\
B(p,t) &\sim&\!\! \frac{2}{N} D(p) \, , \label{differentmoments2}
\end{eqnarray}
where the diffusion coefficient is given by
\begin{eqnarray}
D(p)&=&2 \,\mbox{Re}\int_{0}^{+\infty}\dd t\ e^{ipt}\,\left\langle
\widehat{\delta v}(1,t)\widehat{\delta v}(-1,0)\right\rangle = \pi^2
\ \frac{f_0(p)}{ \left|\tilde D(p,1)\right|^2} \, . \label{difftyu}
\end{eqnarray}
Substituting~(\ref{differentmoments1ici})
and~(\ref{differentmoments2}) in the general form of the
Fokker-Planck equation (\ref{generalFokkerPlanckequation}), one ends
up with
\begin{eqnarray}
\frac{\partial f_1(p,t)}{\partial t}
=\frac{1}{N}\frac{\partial}{\partial p}\left[  D(p)\left(
\frac{\partial f_1(p,t)}{\partial p}- \frac{1}{f_0}\frac{\partial
f_0}{\partial p} f_1(p,t)\right)\right]
\label{FokkerPlanckequation}.
 \end{eqnarray}
We thus recover what has been established in plasma physics (see
e.g. Ref.~\cite{ichimaru}): the evolution of the velocity
distribution $f_1(p, t)$ of the test particle is governed by a
Fokker-Planck equation that takes a form similar to the
Lenard-Balescu equation~(\ref{LenardBalescu1D}) provided that we
replace the distribution $f_0(p, t)$ of the bath by $f_1(p,t)$. The
integro-differential equation is thus transformed in the
Fokker-Planck differential equation. Similar results in higher
dimension have been obtained later in Refs.~\cite{chavanisPhysicaII2006,
chavanisEPJB2006}.

For one-dimensional systems, it has been shown in
Sec.~\ref{seclenardbalescu} that the Lenard-Balescu collision term
cancels out so that the distribution function does not evolve on a
time scale of order $N$. Since, on the other hand, the Fokker-Planck
equation~(\ref{FokkerPlanckequation}) shows that the relaxation time
of a test particle towards the distribution of the bath is of order
$N$, this implies that we can assume that the distribution of the
particles $f_0$ is stationary when one studies the relaxation of a
test particle. This is true for any distribution function $f_0(p)$
that is a stable stationary solution of the Vlasov equation. This is
not true in higher dimensions, except for the Maxwellian
distribution.% Then, in our one-dimensional case, we can assume that
%Eq.~(\ref{FokkerPlanckequation}) is valid at least for times of
%order $N$ with a constant distribution~$f_0$.

%With the initial condition $f_1(0,p)=\delta(p-p_1)$,
%Eq.~(\ref{FokkerPlanckequation}) describes the evolution of a test
%particle in a potential $-\ln f_0(p)$ created by the other
%particles. We note that the above framework is valid for any
%distribution of the bath~$f_0$, provided that it is stable with
%respect to the Vlasov equation.
Equation~(\ref{FokkerPlanckequation}) emphasizes that the momentum
distribution of particle $1$ evolve on timescales of order $N$. We
will thus introduce the timescale $\tau=t/N$, so that the
Fokker-Planck equation can be rewritten as
\begin{equation}
\label{fokkerplanckequation2} \frac{\partial f_1}{\partial
\tau}=\frac{\partial }{\partial p}\left[D(p)\left(\frac{\partial
f_1}{\partial p}-\frac{1}{f_0}\frac{\partial f_0}{\partial
p}f_1\right)\right],
\end{equation}
valid for times $\tau$ at least of order $1$. We see that the time
derivative of $f_1$ vanishes if the distribution function $f_1$ of
the test particle is equal to the quasi-stationary distribution
$f_0$ of the surrounding bath. We expect that $f_1$, governed by
this Fokker-Planck equation, will converge to $f_0$ in a time $\tau$
of order $1$: this means that the distribution function $f_1$ of the
test particle converges towards the quasi-stationary distribution
$f_0$ of the surrounding bath. Thus, it does not converge towards
the equilibrium Gaussian distribution, in complete agreement with
the result that $f_0$ is stationary for times scales of order~$N$.
This result is not valid in higher
dimensions~\cite{chavanisEPJB2006}. It is important to stress here
that collective effects are taken into account through the presence
of the dielectric response function $\tilde D(p,1)$ in the
denominator of the diffusion coefficient (see Eq.~(\ref{difftyu})).

\begin{figure}[htb]
\begin{center}
\resizebox{6truecm}{!}{\includegraphics{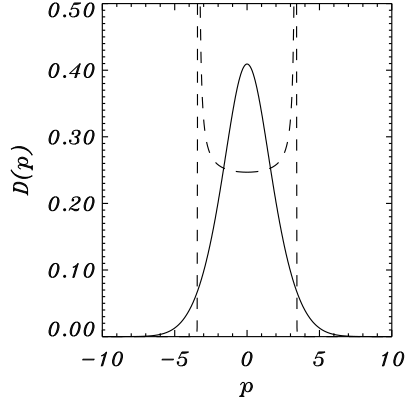}}
\end{center}
\caption{Diffusion coefficient $D(p)$ in the case $\varepsilon=2$
for a Boltzmann thermal bath (solid line) and a waterbag
distribution (dashed line).} \label{figplotDdep}
\end{figure}

Analyzing the stochastic process of equilibrium fluctuations in the
particular case of homogeneous Gaussian distribution,
Bouchet~\cite{BouchetPRE} derived the diffusion coefficient of a
test particle in a equilibrium bath. His result is recovered when
considering a homogeneous Gaussian
distribution~(\ref{distgaussienne}) in expression~(\ref{difftyu})
since one gets
\begin{eqnarray}
D(p) &=& \pi^2   \ \frac{\frac{1}{2\pi}\sqrt{\frac{\beta}{2\pi}} \,
e^{-\beta p^2/2}}{ \displaystyle \left[ 1-\frac{\beta}{2}
+\frac{1}{2}\beta^{3/2}\,p\, e^{-\beta p^2/2}
\int_0^{p\sqrt{\beta}}e^{t^2}/2 \dd t
\right]^2+\frac{1}{8}\pi\beta^3 p^2 \ e^{-\beta p^2}} \, ,
\label{diffusioncoefficientcitersui}
\end{eqnarray}
plotted in Fig.~\ref{figplotDdep}. It is important to stress that
such an expression leads to a diffusion coefficient with
Gaussian-like tails $D(p)\sim \sqrt{{\pi\beta}/{8}}\ e^{-{\beta
p^2}/{2}}.$

The above general derivation~\cite{fredthierPRE} allows however to
study {\em any} arbitrary distribution.  Let us carry on here the
calculation of the diffusion coefficient for the Vlasov-stable water
bag distribution~(\ref{distwaterbag}). It has an interesting
behavior, since the dielectric response $\tilde D(\omega,1)$ has
zeroes on the real axis, contrary to any even distributions {\em
strictly} decreasing for positive values of the frequency $\omega$.
One gets
\begin{eqnarray}
D(p)  &=& \pi^2   \ \frac{\frac{1}{4\pi
p_0}\left[\Theta(p+p_0)-\Theta(p-p_0)\right]}{ \displaystyle \left[
1-\frac{1}{2}\frac{1}{p_0^2-p^2} \right]^2+\left[\frac{\pi}{4
p_0}\left(\delta(p+p_0)-\delta(p-p_0)\right)\right]^2} \, ,
\label{diffusioncoefficientcitersuity}
\end{eqnarray}
which is also plotted in Fig.~\ref{figplotDdep}. One can obtain
similar results for $q$-exponential
distributions~(\ref{distTsallis})
(see Ref.~\cite{Judith}) or power-law tails
distributions~(\ref{eq:powertail}).

\subsubsection{Long-range temporal correlations and diffusion}
\label{temporalcorrelations}

Since the Fokker-Planck equation for the single particle
distribution function $f_1(p)$ has a variable diffusion coefficient,
the relaxation towards the Boltzmann distribution can be slowed
down, especially if the diffusion coefficient decreases rapidly with
momentum. It is thus important to study this Fokker-Planck
equation for different distribution functions of the bath. A similar
study was earlier performed for 2D vortices in Ref.~\cite{Chavanisnewaddi}.
One consequence is that velocity correlation functions can decrease
algebraically with time instead of exponentially, a behavior which
might lead to anomalous diffusion as we will show below.

By introducing the appropriate change of variable $x=x(p)$, defined
by $\dd x/\dd p=1/\sqrt{D(p)}$, and the associated distribution
function $\widehat{f}_1$, defined by $\widehat{f}_1(\tau ,x)\dd x =
f_1(\tau,p)\dd p$, one can
map~\cite{marksteiner,farago,micciche,fredthierPRE,chavanisEPJB2007}
the Fokker-Planck equation (\ref{fokkerplanckequation2}) to the
constant diffusion coefficient Fokker-Planck equation
\begin{equation}\label{fokkerplanckequationrescaledss}
\frac{\partial \widehat{f}_1}{\partial \tau}=\frac{\partial
}{\partial x}\left(\frac{\partial \widehat{f}_1}{\partial
x}+\frac{\partial \psi}{\partial x} \widehat{f}_1\right),
\end{equation}
where the potential $\psi(x)$ is given by
\begin{eqnarray}
\psi(x) &=& - \ln\left(\sqrt{D(p(x))}f_0(p(x))\right).
\end{eqnarray}
Using the property $\tilde D(p,1) \stackrel{|p|\to\infty}{\sim} 1$,
that implies, by Eq.~(\ref{difftyu}), that $D(p)
\stackrel{|p|\to\infty}{\sim} \pi^2 f_0(p)$, one gets $\psi(x)
\stackrel{x\to\pm\infty}{\sim} - \frac{3}{2}\ln f_0(p(x))$. From
this, one derives that for many classes of distribution functions
$f_0$, the potential $\psi(x)$ is asymptotically equivalent to a
logarithm. In fact, we have
\begin{equation}
\label{aproxpsilargex}
\psi(x)\stackrel{x\to\pm\infty}{\sim}\alpha\ln |x|,
 \end{equation}
with $\alpha=3$ if $f_0(p)$ decreases to zero more rapidly than
algebraically for large $p$, and $\alpha <3$ if $f_0(p)$ decreases
to zero algebraically; more precisely, $\alpha= 3 \nu/(2+\nu)$ if
$f_0(p)$ decays at large $p$ as $p^{-\nu}$. For weakly confining
potentials $\psi(x)$, i.e., when $\nu < 1$ and thus $\alpha < 1$,
Eq.~(\ref{fokkerplanckequationrescaledss}) has a non-normalizable
ground state. The example of the heat equation, which corresponds to
$\psi(x)= 0$, describes a diffusive process leading to an asymptotic
self-similar evolution. In such a case, the spectrum of the
Fokker-Planck equation is purely continuous. By contrast, a strongly
confining  potential $\psi(x) $ (for instance, the
Ornstein-Ulhenbeck process with a quadratic potential) would lead to
exponentially decreasing distributions and autocorrelation
functions, linked to the existence in the spectrum of a gap above
the ground state. The logarithmic potential~(\ref{aproxpsilargex})
is a limiting case between the two behaviors. The normalizable
ground state is unique and coincides with the bottom of the
continuous spectrum. The absence of a gap forbids {\em a priori} any
exponential relaxation. To illustrate this result, we evaluate the
asymptotic behavior explicitly in two cases.
\begin{itemize}
  \item
Let us first consider distribution functions $f_0(p)$ with fast
(more than algebraically) decreasing tails so that
\begin{equation}\label{stretch_exponential}
f_0(p) \stackrel{|p|\to\infty}{\sim} C\exp(-\beta p^\delta),
\end{equation}
which includes not only the Gaussian  $(\delta = 2)$ and exponential
tails $(\delta = 1)$, but also stretched-exponential ones with any
arbitrary positive exponent~$\delta$. From the change of variable
$\dd x/\dd p=1/\sqrt{D(p)}$, asymptotic analysis leads to $p(x)
\stackrel{|x|\to\infty}{\sim} \left({2}
\ln|x|/\beta\right)^{1/\delta}$ and to $\psi(x)
\stackrel{x\to\pm\infty}{\sim} 3\ln |x|$. These estimates are
sufficient (see Ref.~\cite{fredthierPRE} for details) to evaluate
the long time behavior of the momentum autocorrelation function
\begin{eqnarray}
\langle p(\tau)p(0)\rangle \stackrel{\tau\to+\infty}{\propto}
\frac{(\ln \tau)^{2/\delta}}{\tau},\label{correlinpasymp2stretched}
\end{eqnarray}
which proves the existence of long-range temporal momentum
autocorrelation for all values of $\delta$, and therefore also in
the case of Boltzmann equilibrium, $\delta=2$.

  \item Let us now consider a distribution function $f_0(p)$ with algebraic
tails
\begin{equation}
f_0(p) \stackrel{|p|\to\infty}{\sim} C|p|^{-\nu}\, .
\label{powerlawtails}
\end{equation}
In this case, one has $p(x) \stackrel{|x|\to\infty}{\sim}
C'x^{2/(2+\nu)}$ and the asymptotic behavior~(\ref{aproxpsilargex})
with, as we said, $\alpha=3\nu/(2+\nu)$. We consider only cases
where $\nu > 3$, to ensure that the second moment of the
distribution $f_0$ (i.e., the average kinetic energy) does exist.
For $\nu >3$ one has that $9/5 < \alpha < 3$. The result for the
momentum autocorrelation function is
\begin{eqnarray}
\langle p(\tau)p(0)\rangle \stackrel{\tau\to+\infty}{\propto}
\tau^{(3-\nu)/(2+\nu)} \, , \label{correlinpasymp2}
\end{eqnarray}
which characterizes an algebraic asymptotic behavior.

\end{itemize}

From the momenta autocorrelation, one usually derives the angle
diffusion $ \langle (\theta(\tau)-\theta(0))2\rangle=2D_\theta
\,\tau$ where $D_\theta$ is defined via the Kubo formula
\begin{equation}
\label{kubo} D_\theta=\int_0^{+\infty}\dd \tau\ \langle
p(\tau)p(0)\rangle \, .
\end{equation}
However, since the exponent $({3-\nu})/({2+\nu}) = -1+ 5/(2+\nu)$ is
larger than $-1$, the asymptotic result~(\ref{correlinpasymp2})
shows that the integral~(\ref{kubo}) diverges. The asymptotic
result~(\ref{correlinpasymp2stretched}) leads also to a divergent
integral, although less singular. It is thus natural to expect
anomalous diffusion for the angles. This extremely small anomaly
(logarithmic) for distribution functions with Gaussian or stretched
exponential tails induced difficulties to detect numerically
anomalous diffusion~\cite{lrr,yamaPRE}.

Note that, generalizing the theory of Potapenko et
al.~\cite{Potapenko1997}, Chavanis and
Lemou~\cite{ChavanisLemou2005} studied how the structure and the
progression of the distribution function tails, also called fronts,
depends on the behavior of the diffusion coefficient for large
velocities. They showed that the progression of the front is
extremely slow (logarithmic) in that case so that the convergence
towards the equilibrium state is peculiar.

\begin{figure}[htbp]
  \centering
  \subfigure[~Correlation]
  {\includegraphics[width=6.5cm]{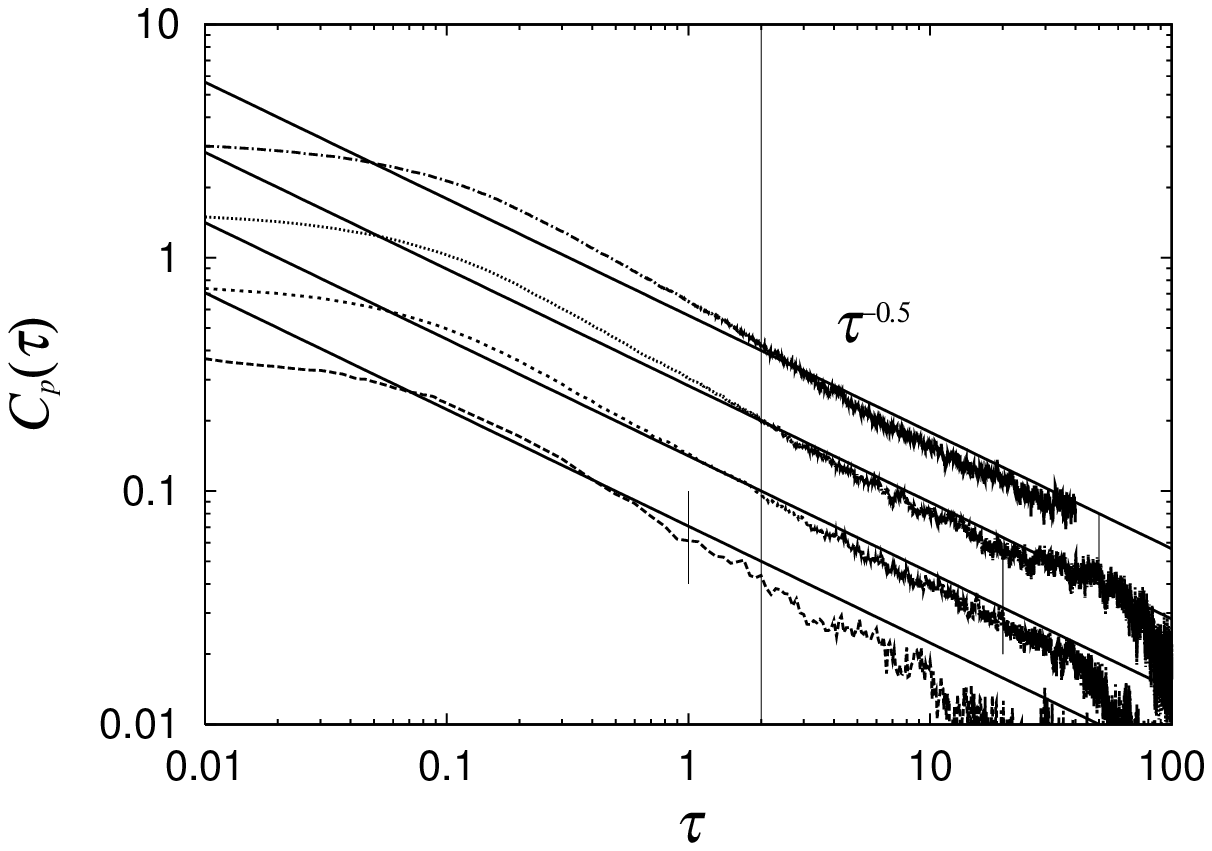}}
  \subfigure[~Anomalous Diffusion]
  {\includegraphics[width=6.5cm]{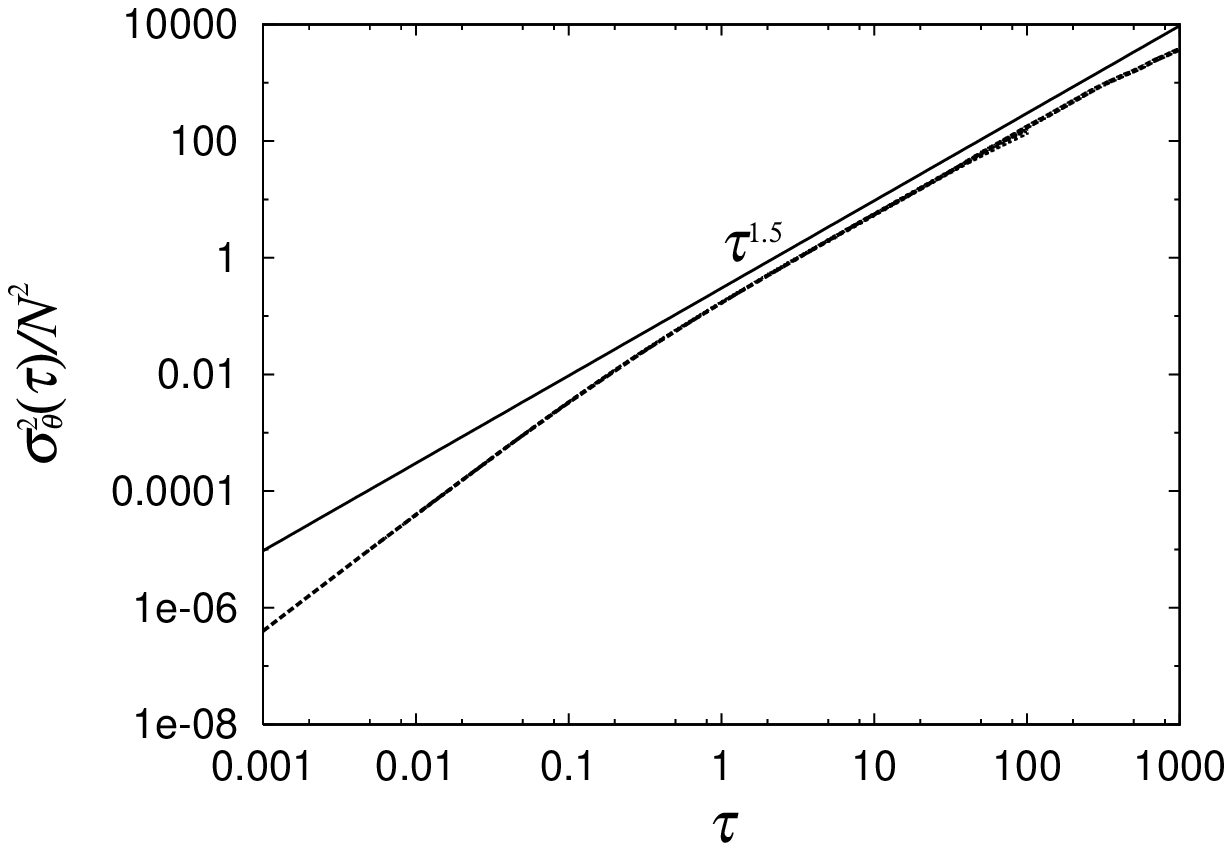}}
  \caption{Check of the theoretical prediction for stable initial
    distributions with power-law tails, in the case $\varepsilon=0.7$.
    Points are numerically obtained by averaging $20,20,10$ and $5$
    realizations for $N=10^{3},10^{4},2.10^{4}$ and $5.10^{4}$ respectively.
    In panel (a), four curves represent the correlation functions of momenta,
    while the straight lines with the slope $-1/2$
    represent the theoretical prediction.
    The curves and the lines are multiplied from the original
    vertical values by $2,4$ and $8$ for $N=10^{4},2.10^{4}$
    and $5.10^{4}$ for graphical purposes.
   % The vertical line indicates the valid time region of the theory starts.
    Similarly, panel (b) presents the diffusion of angles,
    while the straight line with the slope $3/2$
    is theoretically predicted.
    The four curves for the four different values of $N$
      are reported and almost collapse.
  }
  \label{fig:Cp-power}
\end{figure}

Using the time rescaling $\tau=t/N$, which introduces a factor
$1/N^{2}$, the angles diffusion
$\sigma_{\theta}^{2}(\tau)=\average{[\theta(\tau)-\theta(0)]^{2}}$
can be rewritten as
\begin{eqnarray}
\dfrac{\sigma_{\theta}^{2}(\tau)}{N^{2}}
  &=& \int_{0}^{\tau} \mbox{d}\tau_{1} \int_{0}^{\tau} \mbox{d}\tau_{2}
  \average{p(\tau_{1})p(\tau_{2})}  \\
  &=& 2 \int_{0}^{\tau} \mbox{d}s \int_{0}^{\tau-s} \mbox{d}\tau_{2}
  \average{p(s+\tau_{2})p(\tau_{2})} \, ,
\end{eqnarray}
in which the new variable $s=\tau_{1}-\tau_{2}$ has been introduced to
take advantage of the division of the square domain into two
isoscale triangles corresponding to $s>0$ and $s<0$. In the
quasi-stationary states, the integrand
$\average{p(s+\tau_{2})p(\tau_{2})}$ does not depend on $\tau_{2}$
(the QSS evolves on a time scale much larger than $N$) and hence
diffusion can be simplified~\cite{yamaPRE} as
\begin{equation}
\label{eq:sigma_Cp} \dfrac{\sigma^{2}_{\theta}(\tau)}{N^{2}} = 2
\int_{0}^{\tau} \dd s \, (\tau-s) \average{p(s)p(0)} \, .
\end{equation}
A distribution with power law tails~(\ref{powerlawtails}) will
therefore correspond to
\begin{equation}
\label{diffusion_angles_algebrique} \langle
(\theta(\tau)-\theta(0))^2\rangle \stackrel{\tau\to+\infty}{\propto}
\tau^{1+\frac{5}{2+\nu}}.
\end{equation}
A comparison of this predicted~\cite{fredthierPRE} anomalous
diffusion for angles with direct numerical computation of the HMF
dynamics is a tough task because of the scaling with $N$ of the time
dependence of the autocorrelation function. However, it has been
recently confirmed by numerical simulations in Ref.~\cite{yoshiJSM}.
For the stable case ($\varepsilon=0.7$) with power-law tails
($\nu=8$), the theory predicts that the correlation function decays
algebraically with the exponent $-1/2$ (see
Eq.~(\ref{correlinpasymp2})). Figure~\ref{fig:Cp-power}(a) shows
that the theoretical prediction agrees well with numerical
computations. The expression~(\ref{diffusion_angles_algebrique}) can
be rewritten in that case as
$\sigma_{\theta}^{2}(\tau)\sim\tau^{3/2}$:
Figure~\ref{fig:Cp-power}(b), in which the four curves for the four
different values of $N$ almost collapse, attests also the validity
of this prediction. This is a clear example of anomalous diffusion.

For the stretched exponential distribution
function~(\ref{stretch_exponential}), one ends up with
\begin{equation}
\label{diffusion_angles_algebriquebis} \langle
(\theta(\tau)-\theta(0))^2\rangle \stackrel{\tau\to+\infty}{\propto}
 \tau (\ln\tau)^{2/\delta+1}.
\end{equation}
As for power law tails, the diffusion is again anomalous, although
with a logarithmically small anomaly. Consequently, the anomalous
diffusion for angles also occurs for the Gaussian distribution which
corresponds to the special case $\delta=2$. This weak anomalous
diffusion, i.e. normal diffusion with logarithmic corrections, has
also been confirmed~\cite{yoshiJSM}. Since Gaussian distributions
correspond to equilibrium distributions in the microcanonical and
the canonical ensembles, it is important to realize that anomalous
diffusion might thus be encountered for both equilibrium and
out-of-equilibrium initial conditions. An analogous behavior has been
observed for point vortices~\cite{ChavanisHouches,chavanisEPJB2007}.

\subsubsection{Lynden-Bell's entropy}
\label{Lyndenbellentropy}

In subsection \ref{Numericalevidence} we have discussed in detail,
focusing on the HMF model, the two-stage relaxation process that is
often observed in systems with long-range interactions. It has been
indeed realized, beginning with a seminal paper by H\'enon
\cite{Henon64} on globular clusters, that the dynamical evolution is
divided in two well separated phases. A first phase, called {\it
dynamical mixing}, where an initial fast (violent) evolution leads
to a {\it quasi-stationary state}, and a second phase, called {\it
relaxation phase}, where ``collisions" have the cumulative effect to
drive the systems towards statistical equilibrium. If the number of
particles is large, the two phases are well separated in time. It
was also observed that the quasi-stationary state was strongly
dependent on the initial condition \cite{Henon64}. In subsection
\ref{sectiondiffdistrib} we have discussed the existence of
stationary stable and unstable one-particle distributions for the
Vlasov equation and we have presented (subsection
\ref{Numericalevidence}) the interpretation of quasi-stationary
states in terms of ``attractive" Vlasov equilibria. A statistical
approach that explains the existence of Vlasov equilibria has been
proposed long ago by Lynden-Bell~\cite{Lyndenbell67}. He begins by
remarking that Vlasov equation, which represents the evolution of an
incompressible fluid, obeys Liouville theorem in six dimensions. We
will here restrict to a two-dimensional phase space for simplicity.
This implies that the ``mass" of phase elements between $f$ and $f +
\dd f$ is conserved (remind that $f$ is the one-particle
distribution function). If we discretize the one-particle
distribution function into a set of $k$ quantized levels $\eta_i$,
$i=1, \ldots, k$, this means that the area corresponding to the
$i$-th level $m(\eta_i)=\int \dd \theta \dd p \, \delta
(f(\theta,p,t) -\eta_i)$ is a constant of the motion. In the limit
of a continuous distribution of levels, one obtains an infinity of
conservation laws. It is easy to prove that this implies that any
functional of the form $\int \dd \theta \dd p \, C(f)$ is conserved:
these functionals are called Casimirs. For instance, Gibbs entropy
is a particular Casimir: $S_{Gibbs} = -\int \dd \theta \dd p \, f
\ln f$. This specifically implies that, in terms of the {\it fine
grained} one-particle distribution, Gibbs entropy cannot increase.
Let us then define a {\it coarse grained} distribution
\begin{equation}
\bar{f}(\theta,p)= \sum_{i=1}^k \rho(\theta,p,\eta_i) \, \eta_i~,
\end{equation}
where $\rho(\theta,p,\eta_i) \, \dd \theta \dd p$ is the probability
of finding level $\eta_i$ in the phase-space macrocell
$D_{macro}=[\theta, \theta + \dd \theta] \times [p, p + \dd p]$.
Obtaining this probability directly from the microscopic dynamics is
an extremely difficult problem, and its solution would constitute a
significant step forward in the understanding of the nature of
quasi-stationary states. Lynden-Bell's proposal is to evaluate
$\rho(\theta,p,\eta_i)$ by using a sort of ``Boltzmann principle"
\cite{Robert90,Robert91,RobertSommeria91,MichelRobert94,Ellis99}.
The evaluation of $\rho(\theta,p,\eta_i)$ can be done for any number
of levels $k$, but to make the calculation simpler, let us
approximate the one-particle distribution with only two levels
$\eta_1=0$ and $\eta_2=f_0$ (the case with many levels has been treated in
Refs.~\cite{Lyndenbell67,Chavanisreview2006}). Alternatively, one can
consider a fine-grained distribution function which, at $t=0$, has only two
levels: these are called {\it water-bags} in astrophysics and plasma
physics. The exact time evolution of the water-bag is such that the
shape of the area occupied by the Vlasov ``fluid" with level $f_0$
is deformed, stretched and folded due to the hyperbolicity
originated by the nonlinear dynamics of the Vlasov equation (see
e.g. Fig.~2 in Ref.~\cite{Lyndenbell67}). However, due to
Liouville's theorem, this area is conserved. Let us divide each
macrocell $D_{macro}$ into $\nu$ microcells of volume $\omega$ and
consider a microscopic configuration in which the $i$-th macrocell
is occupied by $n_i$ microcells with level $f_0$ and $\nu-n_i$ with
level zero. The total number of occupied microcells is ${\cal N}$,
such that the total mass is $m={\cal N} \omega f_0$. This latter is
also equal to the normalization of the fine grained distribution
$m=\int \dd \theta \dd p f(\theta,p)$. The ${\cal N}$ occupied
microcells are first placed into macrocells. There are ${\cal N}! /
\prod_i n_i!$ ways to do this. Within the $i$-th cell, one can
distribute the first of the $n_i$ occupied microcells in $\nu$ ways,
the second in $\nu-1$ and so on. The number of ways of assigning the
$n_i$ occupied microcells is thus $\nu! / (\nu -n_i)!$. Then, the
total number of microstates compatible with the macrostate where
$n_i$ microcells are occupied in macrocell $i$ is given by the
product of these two factors
\begin{equation}
W(\lbrace n_{i} \rbrace)=\frac{{\cal N}!}{\prod_i n_i !} \times
\prod_{i} \frac{\nu!}{(\nu - n_{i})!}~.
\end{equation}
The first factor is calculated exactly as for Boltzmann gas
\cite{Huang}, because the occupied microcells are {\it
distinguishable}, while the second factor reminds Fermi-Dirac
statistics and derives from an {\it exclusion principle}, which is a
consequence of fluid incompressibility: one microcell cannot be
occupied more than once by a fluid element of level $f_0$. Apart
from this latter constraint, fluid microcells are let to distribute
freely among the different macrocells: this corresponds to making an
assumption of ergodicity. This doesn't happen for the true Vlasov
dynamics and is sometimes referred to as the hypothesis of {\it
efficient mixing}. Indeed, it has been found that dynamical effects
can hinder mixing
\cite{Antoniazzi07_1,Antoniazzi07_2,Bachelard08,Levin1,Levin2,brands,ChavanisAssisi}.
Using Stirling's approximation and expressing $n_i$ in terms of the
average probability to find level $f_0$ in cell $i$,
$\rho_i(f_0)=n_i/\nu$, one obtains, neglecting an additive constant,
\begin{equation}
\ln W = \nu  \, \sum_i \rho_i \ln \rho_i + (1-\rho_i) \ln (1 -
\rho_i),
\end{equation}
which can also be rewritten in terms of the coarse grained
distribution function, $\rho_i=\bar{f}_i/f_0$. Taking the continuum
limit $\sum_i \to \int \dd \theta \dd p /(\omega \nu)$, one finally
gets
\begin{equation}
s_{LB}[\bar{f}]= - \int \frac{1}{\omega} \, \dd \theta \dd p \left[
\frac{\bar{f}}{f_0} \ln \frac{\bar{f}}{f_0} + \left( 1 -
\frac{\bar{f}}{f_0} \right) \ln \left( 1 - \frac{\bar{f}}{f_0}
\right) \right] \label{LyndenBellentropy}
\end{equation}
Following the standard procedure, inspired by large deviation theory
\cite{TouchettePhysRep}, one then maximizes $s_{LB}$ subject to the
constraint of conserving energy $E$, mass $m$ and other global
invariants like momentum (or angular momentum for higher
dimensions).

Historically, the first applications of Lynden-Bell's ideas
encountered both confirmations and failures
\cite{Hohl,Lecar,Henon68,SakagamiGouda}, although the crucial point
of necessarily performing simulations with a large number of
particles was never clearly addressed, due to computer time
limitations. Very recently, a careful analysis of the
one-dimensional self-gravitating sheet model has shown that
Lynden-Bell statistics applies for initial data with virial ratio
close to unity~\cite{yama}. A statistical theory similar to
Lynden-Bell's was independently developed for the Euler equation
\cite{Miller90,Robert91,RobertSommeria92} and, later on, the deep
analogy between the Vlasov-Poisson system and the Euler equation was
for the first time clearly stressed in
Ref.~\cite{ChavanisSommeriaRobert}. This approach was then used in
the context of mean-field models in
Refs.~\cite{FELPRE,Antoniazzi06}, with an application to the
Colson-Bonifacio model of the Free Electron Laser (see subsection
\ref{Colson-Bonifaciomodel}). As already discussed in section
\ref{Numericalevidence}, the presence of quasi-stationary states for
the HMF model was recognized and characterized numerically in
Ref.~\cite{yoshi}. Lynden-Bell's theory was then shown to predict
the main features of the one-particle distribution function of
quasi-stationary states of the HMF model in Ref.~\cite{Antoniazi}.
Let's finish this subsection by discussing in some detail this
latter result, which clearly shows the power of Lynden-Bell's
approach.

\begin{figure}[th]
\vskip 1.25truecm
\resizebox{0.65\textwidth}{!}{\includegraphics{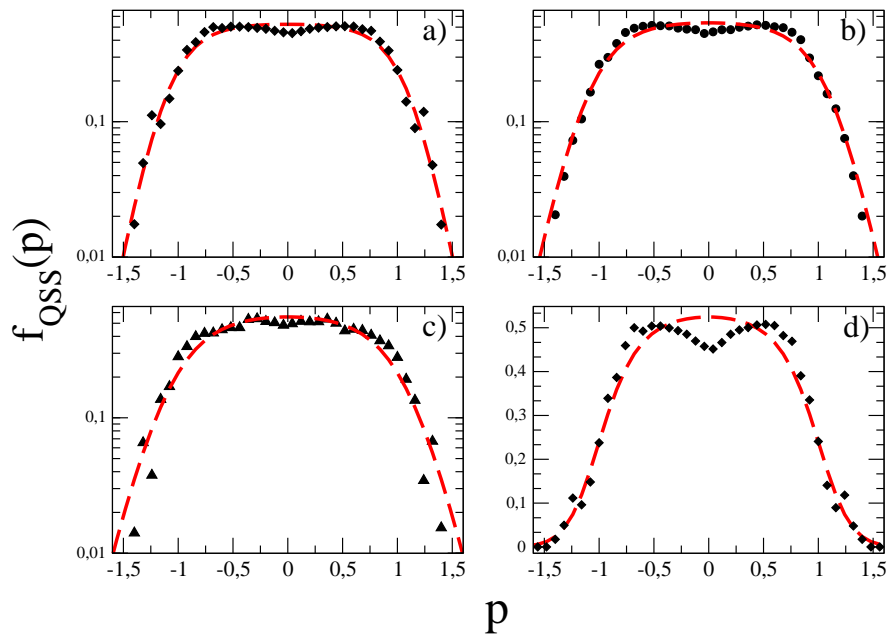}}
\caption{\label{velocityprofiles} Velocity distribution functions in
the quasi-stationary state for the HMF model with $e=0.69$ and
different values of $M_0$. Symbols refer to numerical simulations,
while dashed solid lines stand for the theoretical profile
(\ref{eq:veloc_distr}). Panels a), b) and c) present the three cases
${M_0}=0.3$, ${M_0}=0.5$ and ${M_0}=0.7$ in lin-log scale, while
panel d) shows the case ${M_0}=0.3$ in lin-lin scale.  The numerical
curves are computed from one single realization with $N=10^7$ at
time $t=100$.}
\end{figure}

Let us first recall the Vlasov equation for the HMF model
\begin{equation}
\frac{\partial f}{\partial t} + p\frac{\partial f}{\partial \theta}
- \frac{d V}{d \theta} \frac{\partial f}{\partial p}=0\quad ,
\label{eq:VlasovHMF}
\end{equation}
where $f(\theta,p,t)$ is the {\it fine grained} one-particle
distribution function and
\begin{eqnarray}
V(\theta)[f] &=& 1 - M_x[f] \cos(\theta) - M_y[f] \sin(\theta) ~, \\
M_x[f] &=& \int_{-\pi}^{+\pi} \int_{-\infty}^{+\infty}  f(\theta,p,t)
\, \cos{\theta}  {\mathrm d}\theta
{\mathrm d}p\quad , \\
M_y[f] &=& \int_{-\pi}^{+\pi} \int_{-\infty}^{+\infty}  f(\theta,p,t)
\, \sin{\theta}{\mathrm d}\theta {\mathrm d}p\quad .
\label{eq:pot_magn}
\end{eqnarray}
The globally conserved quantities are energy
\begin{equation}
h[f]=\int\!\!\!\! \int \frac{p^2}{2} f(\theta,p,t) \, \dd \theta \dd p -
\frac{M_x^2+M_y^2 - 1}{2}~,
\end{equation}
and momentum
\begin{equation}
P[f]=\int\!\!\!\! \int p f(\theta,p,t) \dd \theta \dd p ~.
\end{equation}
As for the initial distribution, we consider a {\it water bag} with
rectangular shape in the $(\theta,p)$ plane. The distribution $f$
takes only two distinct values, namely $f_0=1/(4 \Delta \theta
\Delta p)$, if the angles (velocities) lie within an interval
centered around zero and of half-width $\Delta {\theta}$ ($\Delta
{p}$), and zero otherwise (``mass" $m$ is normalized to one and
momentum $P[f]$ is zero). There is a one to one relation between the
parameters $\Delta \theta$ and $\Delta p$ and the initial values of
magnetization and energy
\begin{equation}
M_0 = \frac{\sin (\Delta \theta)}{\Delta \theta} \, \, \, , \, \, \,
e = \frac{(\Delta p)^2}{6} + \frac{1 -(M_0)^2}{2}.
\end{equation}
While $h[f]=e$ and $P[f]=0$ are constants of the motion,
magnetization $M=\sqrt{M_x^2+M_y^2}$ evolves with time.
Lynden-Bell's maximum entropy principle is then defined by the
following constrained variational problem
\begin{eqnarray}
s(e) = \max_{\bar{f}} \biggl(  s(\bar{f}) \biggr|
 h(\bar{f})=e;\;\!\! P(\bar{f})=0;\; \!\!\!%\nonumber\\
\int \!\! \dd \theta \dd p \, \bar{f}=1\biggr)~.
\label{eq:problemevar}
\end{eqnarray}
The problem is solved by introducing three Lagrange multipliers
$\beta/f_0$, $\lambda/f_0$ and $\mu/f_0$ for energy, momentum and
mass normalization. This leads to the following analytical form of
the distribution
\begin{equation}
\label{eq:barf} \bar{f}(\theta,p)= \frac{f_0} {1+\exp\left[{\beta (p^2/2  -
M_y[\bar{f}]\sin\theta
 - M_x[\bar{f}]\cos\theta)+\lambda p+\mu}\right]}.
\end{equation}
This distribution differs from the Boltzmann-Gibbs one because of
the ``fermionic'' denominator. Inserting expression (\ref{eq:barf})
into the energy, momentum and normalization constraints and using
the definition of the magnetization, it can be straightforwardly
shown that the momentum multiplier vanishes, $\lambda=0$. Moreover,
defining $x=e^{-\mu}$ and ${\mathbf m}=(\cos \theta, \sin \theta)$,
yields the following system of implicit equations in the unknowns
$\beta$, $x$, $M_x$ and $M_y$
\begin{eqnarray}
\label{eq:cond0} f_0 \frac{x}{\sqrt{\beta}} \int {\mathrm d} \theta\
e^{\beta {\mathbf M} \cdot {\mathbf m}}\ F_0\left(x e^{\beta
{\mathbf M} \cdot {\mathbf
m}}\right) &=& 1 \\
\label{eq:cond3} f_0 \frac{x}{2 \beta^{3/2}} \int {\mathrm d}
\theta\ e^{\beta {\mathbf M} \cdot {\mathbf m}} F_2\left(x e^{\beta
{\mathbf M} \cdot {\mathbf m}}\right)
&=& e+\frac{M^2 - 1}{2} \\
\label{eq:cond1} f_0 \frac{x}{\sqrt{\beta}} \int {\mathrm d} \theta\
\cos \theta\,\ e^{\beta {\mathbf M} \cdot {\mathbf m}}\ F_0\left(x
e^{\beta {\mathbf M} \cdot {\mathbf m}}\right)
&=& M_x  \\
\label{eq:cond2} f_0 \frac{x}{\sqrt{\beta}} \int {\mathrm d} \theta\
\sin \theta\, e^{\beta {\mathbf M} \cdot {\mathbf m}}\ F_0\left(x
e^{\beta {\mathbf M} \cdot {\mathbf m}}\right) &=& M_y
\end{eqnarray}
with $F_0(y) = \int \exp (-v^2/2)/(1+y \exp (-v^2/2)){\mathrm d}v$
and $F_2(y) = \int v^2 \exp (-v^2/2)/(1+y \exp (-v^2/2)){\mathrm
d}v$. This system of equations can be solved numerically and, given
the parameters that fix the initial conditions, i.e. energy,
momentum and $M_0$, univocally determines the values of the Lagrange
multipliers $\beta$ and $\mu$, the values of $M_x$ and $M_y$ where
Lynden-Bell's entropy is extremal and, finally, the distribution
$\bar{f}$ in formula (\ref{eq:barf}) itself, {\it with no adjustable
parameter}. For $e=0.69$, the maximum entropy state has zero
magnetization for  $M_0<M_{crit}=0.897$, a value at which
Lynden-Bell's theory predicts a second order phase transitions
\cite{Antoniazi,Antoniazzi07_2,chavanisEPJB2006}. Interpreting $\bar{f}$ as the
distribution in the quasi-stationary state (QSS), we obtain in this
case
\begin{equation}
\label{eq:veloc_distr} \bar{f}=f_{QSS}(p)= \frac{f_0} {1+\exp\left[{\beta
p^2/2+\mu}\right]},
\end{equation}
with $\beta$ and $\mu$ to be determined from the knowledge of $M_0$.
Velocity profiles predicted by (\ref{eq:veloc_distr}) are displayed
in Fig.~\ref{velocityprofiles} for different values of the initial
magnetization.

\begin{figure}[htbp]
\centering
\includegraphics[width=12cm]{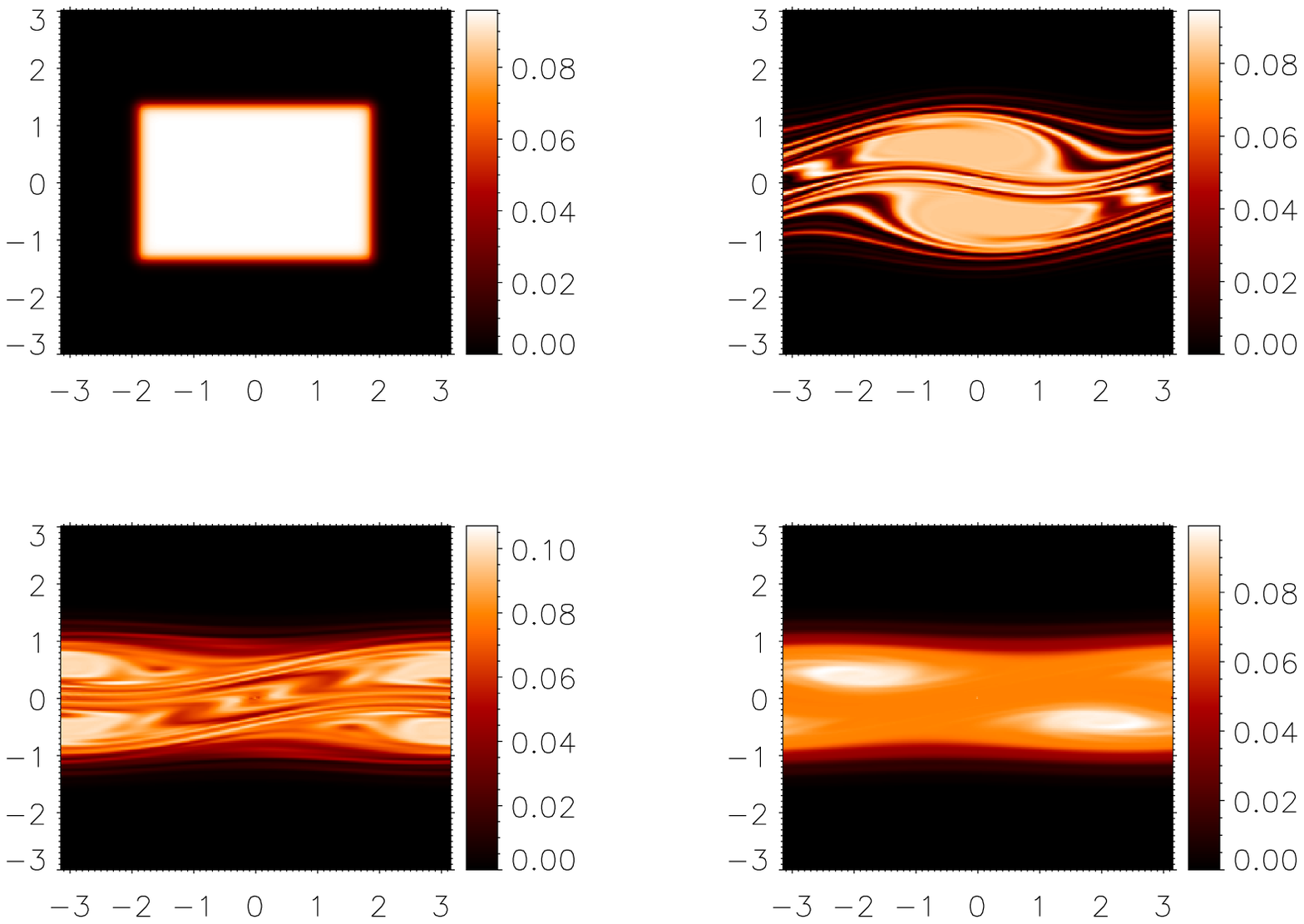}
%  \vspace*{2.5em}
\caption{Simulation of the Vlasov equation (\ref{eq:VlasovHMF}) that
starts from a {\it water-bag} initial condition with $e=0.69$ and
$M_0=0.5$. The final snapshot (lower right panel) is the
quasi-stationary state} \label{vlasovsimulation}
\end{figure}

Although not a single free parameter is used, one finds an excellent
qualitative agreement. The presence of two symmetric bumps in the
velocity distributions is not predicted by Lynden-Bell's theory and
is a consequence of a collective phenomenon which leads to the
formation of two {\it clusters} in the $(\theta,p)$ plane. This is
shown by the direct simulation of the Vlasov equation
(\ref{eq:VlasovHMF}) presented in Fig.~\ref{vlasovsimulation}. The
bumps represent an intrinsic peculiarity of QSS and have been
characterized dynamically in Ref.~\cite{Bachelard08}.

More recently, other authors have obtained similar encouraging
results for Lynden-Bell's theory in a model of non-neutral plasma
\cite{Levin1} and for radially symmetric solutions of a
self-gravitating system \cite{Levin2}. They also find that, when
dynamical effects lead to collective oscillations of the mean-field,
Lynden-Bell's theory shows a disagreement with numerical data and
propose a modification of the theory. Work along this line is in
progress.

The theoretical approach proposed by Lynden-Bell \cite{Lyndenbell67}
allows to predict the presence of a phase transition line in the
$(M_0,e)$ and in the $(f_0,e)$ control parameter
planes~\cite{Antoniazzi07_2,Chavanis_phasetrans}. At equilibrium the
At equilibrium the phase
transition does not depend on the choice of $M_0$ and is located at
$e=3/4$. On the contrary, in the out-of-equilibrium QSS the phase
transition line is located at smaller energies joining the
transition points $(0,7/12)$ \cite{yoshi} and $(1,3/4)$ in the
$(M_0,e)$ plane. Moreover the phase transition changes from first to
second order. These latter results are resumed in
Ref.~\cite{Chavanis_pharo}.

The Lynden-Bell approach turns out to be a good way to attack the
problem of quasi-stationary states. It gives predictions for both
averages and distributions functions which compare quite well with
numerical simulations. When the approach fails to describe detailed
features, they are viable ways of modifying it taking into account
dynamical properties.

\section{Generalization to non mean-field models}
\label{perspectives}

Although substantial progress has been done recently in the
understanding of long-range interactions, the hardest questions
about physical systems where the interactions weakly decay with the
distance, and also shows singularities at short distances, remain
open (see the introductory section~\ref{physicalexamples}). In
section~\ref{equilibriumsec}, we have shown how simple mean-field
models, that are explicitly solvable in both the canonical and the
microcanonical ensemble, already  show several features previously
encountered in gravitational systems, e.g. negative specific heat.
These simple models also display peculiar out-of-equilibrium
dynamical effects, like the presence of quasi-stationary states,
which resemble those found for realistic systems, see
section~\ref{outofequilibrium}.

In this section, we make a little step in the direction of the study
of Hamiltonians that are not fully mean-field. In
subsection~\ref{splusl} we introduce, for both a mean-field Ising
and an XY model, a nearest-neighbour interaction term. We show that
the phase diagram in canonical and microcanonical ensembles
preserves the features of non equivalence found for pure mean-field
Hamiltonians. In particular, we mention the non coincidence of the
microcanonical and canonical tricritical points and the presence of
negative specific heat. In subsection~\ref{decayinginteraction}, we
consider an Ising model in one dimension with weakly decaying
coupling and a modification of the HMF model which includes a
coupling with the same properties. For both models, we show that the
mean-field properties extend to the weakly decaying case and that
one can obtain analytically the free energy and the entropy. Since
the corresponding mean-field limit of both models shows ensemble
equivalence, this property persists also for such models. In
subsection~\ref{taka}, we show the analytical solution in both the
canonical and microcanical ensembles of a self-gravitating system of
particles moving in one dimension. The phase diagram is different in
the two ensembles and show all features of ensemble inequivalence.
In an appropriate limit, the model reduces to the HMF model and
therefore ensemble inequivalence can be turned off by varying a
parameter. Finally, in subsection~\ref{ramaz}, we conclude with a
summary of a very recent result concerning a spin system with
dipolar interactions whose Hamiltonian can be reduced in appropriate
limits to that of the XY model with nearest neighbour and mean-field
interactions presented in subsection~\ref{splusl}. This result opens
the possibility to verify experimentally some of the striking and
counterintuitive features of long-range interactions.

\subsection{Systems with short- and long-range interactions: the transfer integral method}
\label{splusl}

In this subsection, we will present both the canonical and the
microcanonical solutions of an Ising model and of an XY model with
nearest neighbour and mean-field interactions. We will obtain the
phase diagram in both ensembles. All the features of mean-field
models discussed in Sec.~\ref{equilibriumsec} will appear again in
this context: ensemble inequivalence, negative specific heat,
temperature jumps, ergodicity breaking. This shows that the addition
of a short-range term to a long-range Hamiltonian does not remove
the interesting behaviors described before. For what concerns
methodological aspects, the use of transfer integral combined with
the Hubbard-Stratonovich transformations and the min-max method
allows us to treat both models with discrete and continuous
variable. The models we will consider are defined on a
one-dimensional lattice (indeed a ring because of the periodic
boundary conditions), but there is no reason of principle that
forbid to extend the calculation to strip and bar geometries.

\subsubsection{Ising model}
\label{Isinglongplusshort}

\paragraph{Introduction}

Nagle introduced an interesting Ising model combining long- with
short-range interactions~\cite{Nagle}. Further elaborations of the
model were proposed by Khardar~\cite{Khardar,Khardar83}. The
Hamiltonian is
\begin{equation}
H_N=-\frac{1}{2N}\left(\sum_{i=1}^NS_i\right)^2
-\frac{K}{2}\sum_{i=1}^N\left(S_iS_{i+1}-1\right),
\label{hamilkardarnagel}
\end{equation}
where $S_i=\pm 1$. In this one-dimensional spin chains, the first
term has an infinite range and is the typical one of the Curie-Weiss
Hamiltonian~(\ref{hamiladditi}). This term is responsible for the
non additive properties of the model, see Sec.~\ref{additivity}.
Note that the prefactor $J$ present in
Hamiltonian~(\ref{hamiladditi}) has been set to one by an
appropriate renormalization of the energy. There is no loss of
generality, since only the ferromagnetic case $J>0$ will be
considered. On the contrary, the second term
of~(\ref{hamilkardarnagel}) corresponds to an interaction between
nearest neighbors along a one-dimensional lattice (periodic boundary
conditions are chosen). The coupling constant $K$ might be either
positive or negative.

The ferromagnetic state with all spins up ($S_i=1, \forall i$) or
down ($S_i=-1, \forall i$) has a negative energy $E_F=-N/2$. For the
antiferromagnetic state with alternate signs of nearest neighbour
spins, the first term of (\ref{hamilkardarnagel}) gives a vanishing
contribution to the energy and the energy is $E_A=KN$. At $T=0$, one
can determine if the model has a phase transition at some value of
$K$ by comparing the energy of the ferromagnetic state with that of
the antiferromagnetic one, since only the energy term of the free
energy matters. Hence, by imposing $E_A=E_F$ one gets the phase
transition value $K_t=-1/2$ at which a discontinuity of the order
parameter is found, from $m=0$ to $m=1$. Therefore the transition is
first order.

{\setlength{\unitlength}{0.5pt}
\begin{picture}(100,380)(-100,50)
\put(60,90){\framebox(230,40){Antiferromagnetic state}}
\put(60,240){\framebox(230,40){Paramagnetic state}}
\put(360,90){\framebox(200,40){Ferromagnetic state}}
\put(250,30){ $1^{st}$ order PT} %\put(235,10){Phase Transition}
%\put(-60,245){$2^{nd}$ order} \put(-40,225){PT}
\put(600,150){${K}{}$}
\put(295,150){$-\frac{1}{2}$}\put(435,150){$0$}\put(435,300){$1$}\put(455,380){$T$}
\put(450,300){\vector(1,0){50}}\put(505,300){ $2^{nd}$ order PT}
%\put(40,130){\scriptsize 0}
%\put(30,280){$T$}\put(60,235){\scriptsize $J$}
\put(50,135){\vector(1,0){550}} \put(450,135){\vector(0,1){250}}
\put(310,135){\vector(0,-1){80}}
%\put(50,240){\vector(-1,0){30}}
\end{picture}}
\begin{figure}[htbp]
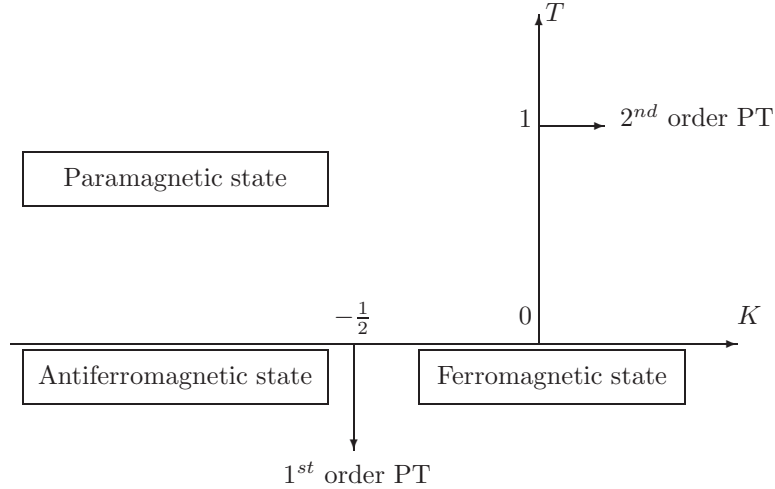

\caption{Elementary features of the phase diagram of the short plus
long-range Ising model showing the phase transitions on the
temperature $T$ and local coupling $K$ axis, respectively.}
\label{qualIsingLS}.
\end{figure}

For non zero temperature, one has to take into account the entropic
term of the free energy, which measures disorder. When the coupling
constant $K$ vanishes, one fully recovers the Curie-Weiss
Hamiltonian~(\ref{hamiladditi}), which exhibits a {\em second} order
phase transition at $T=1$. One therefore expects that the $(T,K)$
phase diagram displays a transition line which is first order at low
$T$ and second order at high $T$ (see Fig.~\ref{qualIsingLS}). The
transition line separates a
ferromagnetic from a paramagnetic state, although exactly at $T=0$
and $K< -1/2$ the state is antiferromagnetic. Let us now determine
analytically this transition line in both the canonical and the
microcanonical ensemble.

\paragraph{The solution in the canonical ensemble}

\begin{figure}[htb]
\begin{center}
\includegraphics[width=.45\textwidth]{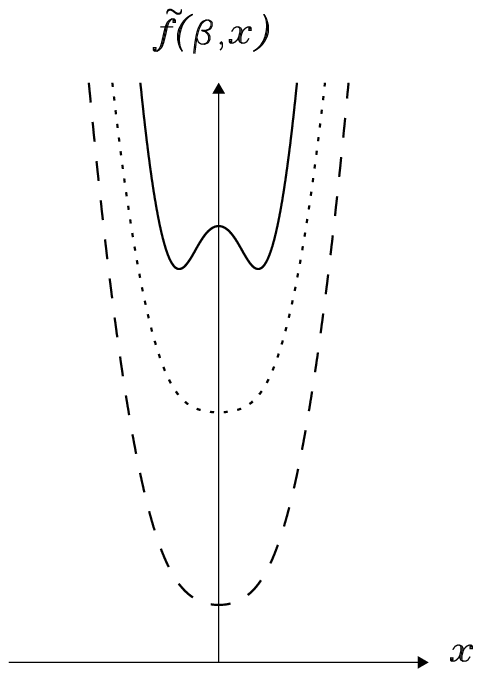}
\includegraphics[width=.45\textwidth]{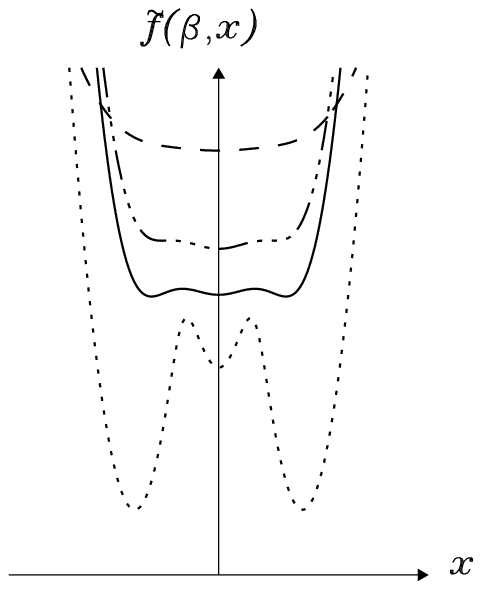}
\end{center}
\vskip -1truecm \caption{$\tilde{f}(\beta,x)$ for different values
of the inverse temperature. Left panel presents $\beta=1.1$ (dashed
line), $\beta_c \simeq 1.4$ (dotted), $\beta=2.5$ (solid) when
$K=-0.25$: a {\em second} order phase transition. Right panel shows
the case $K=-0.4$ when $\beta=10$ (dotted), $\beta_t \simeq 2.4$
(solid), $\beta=2.35$ (dash-triple dot), $\beta=2$ (dashed): a {\em
first} order phase transition. Note that the different curves have
been vertically shifted for readability purposes.}
\label{ftildelongplushort}
\end{figure}

The phase diagram has been studied in the canonical ensemble by
Nagle~\cite{Nagle} and Khardar~\cite{Khardar,Khardar83}. The
partition function is
\begin{eqnarray}
Z(\beta,N)&=& \sum_{\{S_1,\ldots,S_N\}} e^{-\beta H}
 =\sum_{\{S_1,\ldots,S_N\}}\exp\left[{\displaystyle \frac{\beta}{2N}
 \left(\sum_{i=1}^NS_i\right)^2+\frac{\beta K}{2}\sum_{i=1}^N\left(S_iS_{i+1}-1\right)}\right]~.
 \label{partfunctionNagle}
\end{eqnarray}
To get rid of the quadratic term, we use the Hubbard-Stratonovich
transformation
\begin{eqnarray}
e^{\displaystyle
\frac{\beta}{2N}\left(\sum_{i=1}^NS_i\right)^2}&=&\sqrt{\frac{\beta
N}{2\pi}}\int_{-\infty}^{+\infty}\dd x \ e^{\displaystyle
-\frac{\beta N}{2}x^2 +\beta x\sum_{i=1}^NS_i}~,
\end{eqnarray}
so that the partition function~(\ref{partfunctionNagle}) can be
rewritten as
\begin{eqnarray}
Z(\beta,N) &=&\sqrt{\frac{\beta N}{2\pi}}\int_{-\infty}^{+\infty}\dd
x \ \,e^{\displaystyle -\frac{\beta N}{2}x^2}\,
\sum_{\{S_1,\ldots,S_N\}}\Biggl[e^{\displaystyle\beta x\sum_{i=1}^N
S_i+ \frac{\beta
K}{2}\sum_{i=1}^N\left(S_iS_{i+1}-1\right)}\Biggr]\\
&=&\sqrt{\frac{\beta N}{2\pi}}\int_{-\infty}^{+\infty}\dd x\
\,e^{\displaystyle  -N\beta \tilde{f}\left(\beta,x\right)}~.
\label{partfunctionNaglebis}
\end{eqnarray}
The free energy can be written as
\begin{eqnarray}
\tilde{f}(\beta,x)=\frac{1}{2}x^2 + f_0(\beta,x),
\end{eqnarray}
where $f_0(\beta,x)$ is  the free energy of the nearest-neighbor
Ising model with an external field $x$. Such an expression can be
easily derived using the transfer
matrix~\cite{KramersWannier,Huang,Khardar}. By an easy calculation,
we find $f_0(\beta,x)=-\ln(\lambda_+^N+\lambda_-^N)/(\beta N)$ where
the two eigenvalues of the transfer matrix are
\begin{equation}
\lambda_\pm=e^{\beta K/2}\cosh(\beta x)\pm\sqrt{e^{\beta
K}\sinh^2(\beta x)+e^{-\beta K}}.
\end{equation}
As $\lambda_+>\lambda_-$ for all values of $x$, only the larger
eigenvalue $\lambda_+$ is relevant in the limit
$N\rightarrow\infty$. One thus finally gets
\begin{eqnarray}
\tilde{\phi}(\beta,x)=\beta\tilde{f}(\beta,x)= \frac{\beta }{2}x^2
-\ln \left[e^{\beta K/2}\cosh(\beta x)+\sqrt{e^{\beta
K}\sinh^2(\beta x)+e^{-\beta K}}\right]~, \label{philongplusshort}
\end{eqnarray}
which is shown in Fig.~\ref{ftildelongplushort} for different values
of the inverse temperature $\beta$ and for two values of the
nearest-neighbor coupling~$K$.
In the large $N$-limit, the application of the saddle point method
to Eq.~(\ref{partfunctionNaglebis}) finally leads to the free
energy, which is obtained by taking the value of $x$ which minimizes
$\tilde{\phi}(\beta,x)$ in formula (\ref{philongplusshort}).

\begin{figure}[htbp]
\begin{center}
\resizebox{6truecm}{!}{\includegraphics{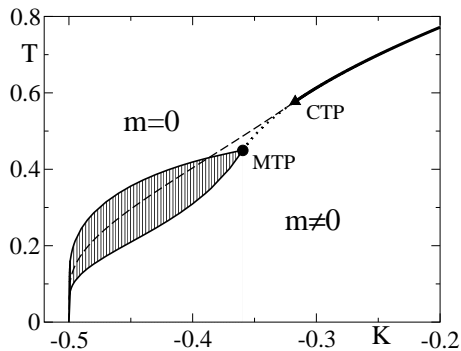}}
\end{center}
\caption{The canonical and microcanonical $(K,T)$ phase diagram. In
the canonical ensemble, the large $K$ transition is continuous (bold
solid line) down to the tricritical point CTP where it becomes first
order (dashed line). In the microcanonical ensemble the continuous
transition coincides with the canonical one at large $K$ (bold
line). It persists at lower $K$ (dotted line) down to the
tricritical point MTP where it turns first order, with a branching
of the transition line (solid lines). The shaded area is not
accessible in the microcanonical ensemble.}
\label{PhaseDiagrammLongplusshort}
\end{figure}

From the knowledge of the free energy, as anticipated, one gets
either a second or a first order phase transition depending on the
value of the coupling constant~$K$. As usual, the expansion of
$\tilde{f}(\beta,x)$ in power of $x$ is the appropriate procedure to
define the critical lines and points. One gets here
\begin{equation}
\tilde{f}(\beta,x)= -\ln 2\cosh\frac{\beta K}{2}+ \frac{\beta
}{2}x^2\left(1-\beta  e^{\beta K}\right)+ \frac{\beta^4}{24}
e^{\beta K}\left(3e^{2\beta K}-1\right)x^4+{\cal
O}(x^6).\label{expansionlongshort}
\end{equation}
The critical point of the second order transition is obtained for
each $K$ by computing the value $\beta_c$ at which the quadratic
term of the expansion~(\ref{expansionlongshort}) vanishes provided
the coefficient of the fourth order term is positive, obtaining
$\beta_c =\exp{(-\beta_c K)}$. When also the fourth order
coefficient vanishes, $3\exp(2\beta K)=1$ one gets the canonical
tricritical point (CTP) $K_{CTP}=-\ln3/(2\sqrt{3})\simeq -0.317$.
The first order line is obtained numerically by requiring that
$f(\beta,0)=f(\beta,x^*)$, where $x^*$ is the further local minimum
of $f$. Figure~\ref{PhaseDiagrammLongplusshort} represents the phase
diagram in both the canonical and the microcanonical ensembles. The
features of this phase diagram are very close to those of the BEG
model (see Fig.~\ref{schematic}) and of the generalized HMF (see
Fig.\ref{phasediagrammquatre}). We will comment them below, after
having obtained the solution in the microcanonical
ensemble~\cite{schreiber}.

\paragraph{The solution in the microcanonical ensemble}

In the microcanonical ensemble, a simple counting method has been
recently proposed~\cite{schreiber}. The magnetization
$M=\sum_{i=1}^NS_i$ can be rewritten as $M=N_+-N_-$ by introducing
the number of spins up, $N_+$, and of spins down, $N_-$. The first
term of the Hamiltonian~(\ref{hamilkardarnagel}) can be
straightforwardly rewritten as $-M^2/(2N)$. As two identical
neighboring spins would not contribute to the second term of
Hamiltonian~(\ref{hamilkardarnagel}) ($S_iS_{i+1}-1$ being equal to
zero) while two different ones would give a contribution equal to
$K$, the total contribution of the second term is $KU$, where $U$ is
the number of ``kinks" in the chain, i.e. links between two
neighboring spins of opposite signs.

For a chain of $N$ spins, the number of microstates corresponding to
an energy $E$ can be written as
\begin{equation}
\label{omegaSL} \Omega(N_+,N_-,U)\simeq
\left(\begin{array}{ll}  N_+ \\
{\displaystyle U}/{\displaystyle 2}\end{array}\right)
\left(\begin{array}{ll}  N_- \\
{\displaystyle U}/{\displaystyle 2}\end{array}\right)~.
\end{equation}
The formula is derived by taking into account that we have to
distribute $N_+$ spins among $U/2$ groups and $N_-$ among the
remaining $U/2$. Each of these distributions gives a binomial term,
and, since they are independent, the total number of states is the
product of the two binomials. The expression is not exact because we
are on a ring, but corrections are however of order $N$ and do not
affect the entropy. A slight correction to formula (\ref{omegaSL})
is present for small $N_+$, $N_-$ and $U/2$, and all these numbers
should be indeed reduced by a unity.

\begin{figure}[htb]
\begin{center}
\includegraphics[width=.45\textwidth]{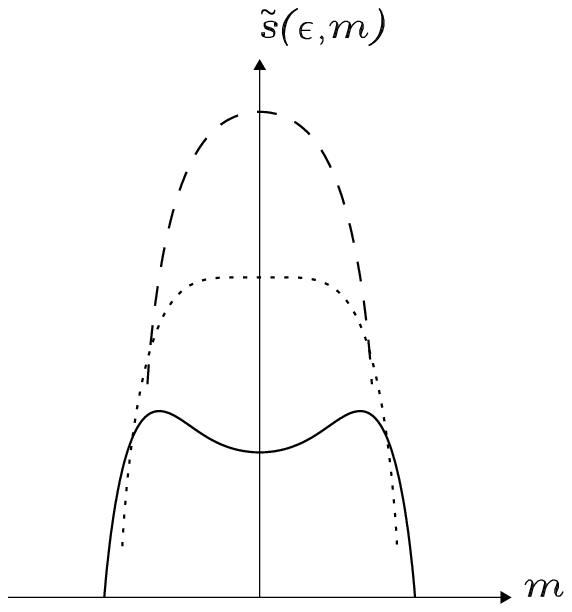}
\includegraphics[width=.45\textwidth]{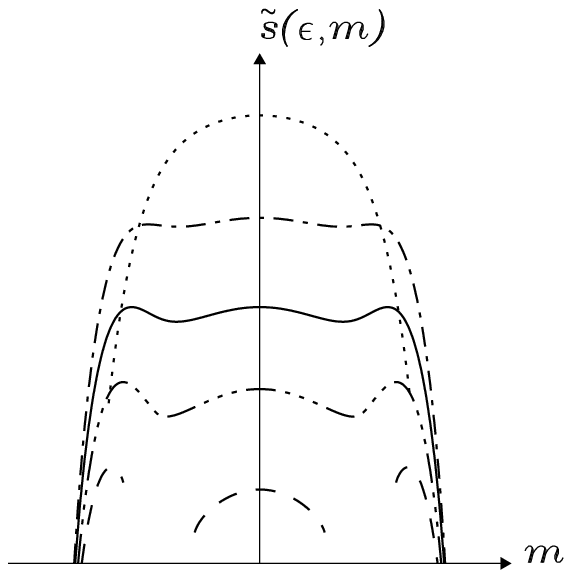}
\end{center}
\vskip-2truecm \caption{$\tilde{s}(\varepsilon,m)$ for different
values of the energy. Left panel presents $\varepsilon=-0.1$ (dashed
line), $\varepsilon_c \simeq -0.15$ (dotted), $\varepsilon=-0.2$
(solid) when $K=-0.25$: a {\em second} order phase transition. Right
panel shows the case $K=-0.4$ when $\varepsilon=-0.25$ (dotted),
$\varepsilon=-0.305$ (dash-dotted), $\varepsilon_t=-0.3138$ (solid),
$\varepsilon=-0.32$ (dash-triple dot), $\varepsilon=-0.33$ (dashed):
a {\em first} order phase transition. The gaps present in the lower
dashed curve are related to ergodicity breaking (see text). Note
that the different curves have been vertically shifted for
readability purposes.} \label{evolentropy}
\end{figure}

Introducing $m=M/N$, $u=U/N$ and $\varepsilon=E/N=-m^2/2+Ku$, one
thus finally gets the entropy
\begin{eqnarray}
\tilde{s}(\varepsilon,m)=\frac{1}{N}\, \ln \Omega%\\&=&
&=&\frac{1}{2}(1+m)\ln(1+m)+\frac{1}{2}(1-m)\ln(1-m)-u \ln
u\nonumber\\
&&\null\hskip
1truecm-\frac{1}{2}(1+m-u)\ln(1+m-u)-\frac{1}{2}(1-m-u)\ln(1-m-u)\ ,
\label{entropylongplusshort}
\end{eqnarray}
which is shown in Fig.~\ref{evolentropy} for different values of the
energy $\varepsilon$ and for two values of the nearest-neighbor
coupling~$K$.

In the large $N$-limit, the last step is to maximize the entropy
$\tilde{s}(\varepsilon, m)$ with respect to the magnetization $m$,
leading to the final entropy $s(\varepsilon)=\tilde{s}(\varepsilon,
m^*)$, where $m^*$ is the equilibrium value. As anticipated, one
gets either a second or a first order phase transition depending on
the value of the coupling constant~$K$. As usual, and analogously to
what has been done in the canonical ensemble, the expansion of
$\tilde{s}(\varepsilon,m)$ in power of $m$ is the appropriate
procedure to define the critical lines and points. One gets here
\begin{equation}
\tilde{s}(\varepsilon,m)=s_0(\varepsilon)+A\,m^2+B\,m^4+{\cal
O}(m^4),\label{expansionentropylong+short}
\end{equation}
with the paramagnetic zero magnetization entropy
\begin{eqnarray}
s_0(\varepsilon)&=&
-\frac{\varepsilon}{K}\ln\frac{\varepsilon}{K}-\left(1-\frac{\varepsilon}
{K}\right)\ln\left(1-\frac{\varepsilon}{K}\right)
\end{eqnarray}
and the expansion coefficients
\begin{eqnarray}
A_{mc}&=&
\frac{1}{2}\left[\frac{1}{K}\ln\frac{K-\varepsilon}{\varepsilon}
-\frac{\varepsilon}{K-\varepsilon}\right]\\
B_{mc}&=& \frac{\varepsilon^3}{12(\varepsilon-K)^3}-\frac{K^2+K}
{4(\varepsilon-K)^2}+\frac{1}{8K\varepsilon}.
\end{eqnarray}
Using these expression, it is straightforward to find the second
order phase transition line by requiring that $A_{mc}=0$
($B_{mc}<0$), finding $\beta_c=\exp(-\beta_cK)$, which is the same
equation found for the canonical ensemble. Again, as far as second
order phase transitions are concerned, the two ensembles are
equivalent. The tricritical point is obtained by the condition
$A_{mc}=B_{mc}=0$), which gives  $K_{MTP}\simeq-0.359$ and
$\beta_{MTP} \simeq 2.21$, which is close but definitely different
from $K_{CTP}\simeq -0.317$ and $\beta_{CTP}=\sqrt{3}$. The
microcanonical first order phase transition line is obtained
numerically by equating the entropies of the ferromagnetic and
paramagnetic phases. At a given transition energy, there are two
temperatures, thus giving a {it temperature jump} as for the BEG and
the generalized HMF model. Similarly to mean-field models, also this
model exhibits a region of negative specific heat when the phase
transition is first order in the canonical ensemble.

\paragraph{Equilibrium dynamics: breaking of ergodicity}

Model (\ref{hamilkardarnagel}) exhibits breaking of ergodicity, as
shown in Ref.~\cite{schreiber}. In order to reveal the dynamical
consequences of this effect, one has to define a {\it microcanonical
dynamics}. An appropriate one is given by the Creutz
algorithm~\cite{creutz}, which probes the microstates of the system
with energy lower or equal to the energy~$E$. Indeed, there are two
definitions of the microcanonical ensemble, that become equivalent
in the thermodynamic limit. In the first one, only states contained
on an energy shell $(E,E+\dd E)$ are counted (this is the definition
we used throughout this review); in the second one all the
phase-space volume contained within the energy $E$ hypersurface is
considered \cite{Huang}. Creutz algorithm is based on this second
definition.

The algorithm is implemented by adding an auxiliary variable, called
a {\it demon}, which has the following properties. One initiates the
procedure with the demon at zero energy, while the system has an
energy $E$. One then attempts to flip a spin. The move is accepted
if it corresponds to an energy decrease and the excess energy is
given to the demon. One then attempts to flip another spin. If the
flip decreases the energy, it is accepted, while if it increases the
total energy, it is accepted only if the needed energy can be
withdrawn from the demon energy. Demon serves really as a bank with
deposit and withdrawn: however, the total energy of the ``bank" is
always non negative (clearly different from typical banks in the
21st century!).

Creutz dynamics can be used to test the predictions on breaking of
ergodicity discussed in Sec.\ref{generalizedHMF}. Indeed, as it is
already apparent in Fig.~\ref{evolentropy}, not all magnetization
values are accessible in certain regions of the $(K,\varepsilon)$
plane, which manifests itself in the gaps of the entropy curves
shown in Fig.\ref{evolentropy}. This property is in turn a
consequence of the non convexity of the space of thermodynamic
parameters (see Sec.~\ref{Convexity})

\begin{figure}[htb]
\begin{center}
\resizebox{9truecm}{!}{\includegraphics{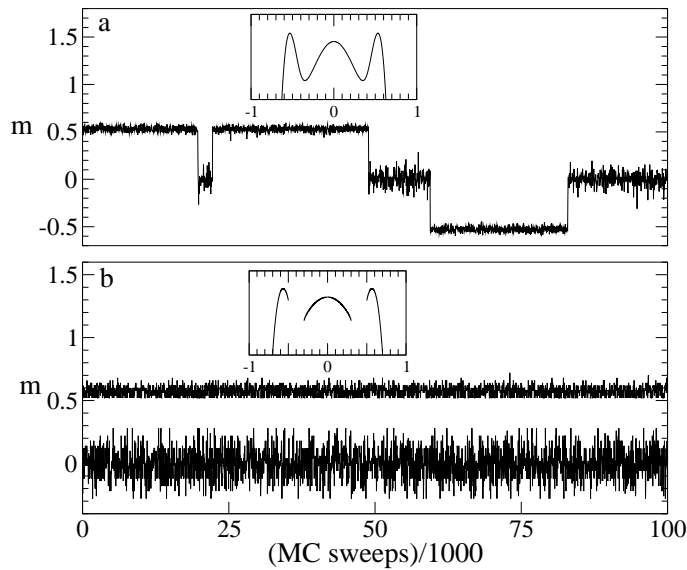}}
\end{center}
\caption{Time evolution of the magnetization for $K=-0.4$ (a) in the
ergodic region $(\varepsilon=-0.318$) and (b) in the non ergodic
region ($\varepsilon=-0.325$). Two different initial conditions are
plotted simultaneously. The corresponding entropy curves are shown
in the insets.} \label{dynamicsplusshort}
\end{figure}

To demonstrate ergodicity breaking, we show in
Fig.~\ref{dynamicsplusshort} the time evolution of the magnetization
in two cases: in the first case, where the whole magnetization
interval is accessible, one clearly see switches between the
metastable state $m^*=0$ and two stable symmetric magnetized states.
On the contrary, in the second example, the metastable state belongs
to an interval which is disconnected from the stable one (see the
inset for the corresponding entropy curve). The dynamics maintains
the system either in the stable or in the metastable interval,
depending only on the initial condition. The system is unable to
jump to the other interval even when the latter is more stable.

Similar results had been obtained using Hamiltonian dynamics for the
generalized HMF model, see Fig.~\ref{fig:tracem}.

\subsubsection{XY model}
\label{xylongplusshort}

\paragraph{Introduction}

In this subsection we study a generalization of the HMF model, in
which the Hamiltonian, besides the mean-field interaction, has a
nearest neighbour interaction between rotators. The presence of such
a term requires that we specify the properties of the lattice where
the model is defined. As in subsection~\ref{Isinglongplusshort}, we
study a one-dimensional lattice with a ring geometry; the first
property allows the use of the transfer integral technique, while
the second property, equivalent to the introduction of periodic
boundary conditions, is convenient for the calculations, but it is
irrelevant in the thermodynamic limit.

The model has been introduced in Ref.~\cite{campaphysica2006}, and
its Hamiltonian is given by
\begin{equation}
\label{hamilxy} H_N=\sum_{i=1}^N \frac{p_i^2}{2} +
\frac{1}{2N}\sum_{i,j=1}^N \left[1-\cos \left( \theta_i -\theta_j
\right) \right] -K \sum_{i=1}^N \cos \left( \theta_{i+1}-\theta_i
\right) \,,
\end{equation}
with $\theta_{N+1}=\theta_1$. The parameter $K$ is the coupling
constant of the nearest-neighbour interaction. For $K=0$, the
Hamiltonian reduces to the HMF model, that has a second order phase
transition at $T_c=0.5$. The similarities of this model with the
Ising model studied in the previous subsection are clear: there is a
mean-field interaction and a nearest neighbour interaction. However,
there is also a kinetic energy term, which allows to define a
Hamiltonian dynamics. Also in this case the interesting properties
are found for $K<0$. Let us first locate the first order transition
at $T=0$. This is done, as for the Ising case, comparing the energy
per particle of the fully magnetized $m=1$ state with that of the
staggered non magnetic $m=0$ state. The energy density of the former
is $\veps = -K$, while that of the latter is $\veps = 1/2 + K$. The
magnetized state is therefore favoured for $K \ge -1/4$.
Qualitatively, it is the same phase diagram as the one observed in
Fig.~\ref{qualIsingLS}.

\paragraph{Solutions in the canonical and microcanonical ensembles}

We adopt the min-max procedure, described in
subsection~\ref{minmaxsect}. We remind that, in this procedure, once
an expression for the canonical partition function has been obtained
in the form of Eq.~(\ref{zminmax}), not only can we compute the
canonical free energy $\phi(\beta)$ by the minimization defined in
Eq.~(\ref{freeminmax}), but we can also obtain the microcanonical
entropy $s(\veps)$ by Eq. (\ref{entrminmax1}).

Let us first derive the canonical partition function. As for the HMF
model, we use the Hubbard-Stratonovich transformation. The model
shares with the HMF model the invariance under global rotations,
therefore the spontaneous magnetization is defined only in modulus,
while there is degeneracy with respect to its direction. We can
exploit this fact to simplify slightly the computation; namely, we
assume from the beginning that the saddle point value of the
variable $x_2$ in Eq.~(\ref{hmfhubb1}) is $0$. The spontaneous
magnetization is in the direction of the $x_1$-axis, but this will
not cause any loss of generality. We therefore obtain
\begin{eqnarray}
\label{xylongshortpartfun} Z(\beta,N) &=& \frac{N\beta}{2\pi} \exp
\left(-\frac{N\beta}{2}\right)\left(\frac{2\pi}{\beta}
\right)^{N/2}\nonumber \\
&&\int \dd \theta_1 \dots \dd \theta_N \dd x \exp \left[ -\frac{N
\beta x^2}{2} +\beta x \sum_{i=1}^N \cos \theta_i +\beta K
\sum_{i=1}^N \cos \left( \theta_{i+1}-\theta_i \right) \right] \, .
\end{eqnarray}
The integral over the $\theta$'s is performed applying the transfer
integral method. For our one-dimensional geometry this is given by
\begin{equation}
\label{transfergeneral} \int \dd \theta_1 \dots \dd \theta_N \exp
\left[ \beta x \sum_{i=1}^N \cos \theta_i +\beta K \sum_{i=1}^N \cos
\left( \theta_{i+1}-\theta_i \right)\right] =\sum_{j}
\lambda_j^N(\beta x,\beta K) \, ,
\end{equation}
where $\lambda_j(\gamma,\sigma)$ is the $j$-th eigenvalue of the
symmetric integral operator
\begin{equation}\label{tranferoperator}
({\cal T} \psi) (\theta) = \int \dd \alpha \exp \left[ \frac{1}{2}
\gamma (\cos \theta +\cos \alpha) + \sigma \cos (\theta - \alpha)
\right] \psi (\alpha) \, .
\end{equation}
In the thermodynamic limit, only the largest eigenvalue
$\lambda_{max}$ will contribute to the partition function. This
eigenvalue can be computed numerically by a suitable discretization
of the integral operator (\ref{tranferoperator}). The rescaled free
energy is
\begin{equation}
\label{freeenerxy} \phi(\beta)=\beta f(\beta)=\inf_x
\tilde{\phi}(\beta,x) \, ,
\end{equation}
with
\begin{equation}
\label{freeenerxy1} \tilde{\phi}(\beta,x) = -\frac{1}{2}\ln
\frac{2\pi}{\beta}+\frac{\beta}{2}(1+x^2) -\lambda_{max}(\beta
x,\beta K) \, ~,
\end{equation}
where we have not explicitly written the $K$ dependence of
$\tilde{\phi}(\beta,x)$ and $\phi(\beta)$.

Accordingly to the min-max procedure, the microcanonical entropy
$s(\veps)$ is given by
\begin{equation}
\label{entropyxyminmax} s(\veps) = \sup_x \tilde{s}(\veps,x) \, ,
\end{equation}
with
\begin{equation}\label{entropyxyminmax1}
\tilde{s}(\veps,x) = \inf_{\beta} \left[ \beta \varepsilon
-\tilde{\phi}(\beta,x) \right] = \inf_{\beta} \left[ \beta
\varepsilon + \frac{1}{2}\ln \frac{2 \pi}{\beta}
-\frac{\beta}{2}(1+x^2)+\ln \lambda_{max}(\beta x, \beta K) \right]
\, .
\end{equation}
Again the dependence of $\tilde{s}(\veps,x)$ and $s(\veps)$ on $K$
has not been explicitly written. The expressions (\ref{freeenerxy1})
and (\ref{entropyxyminmax1}) are used to study canonical and
microcanonical thermodynamic phase diagrams, respectively, and then
to check if there is ensemble inequivalence. As for the Ising case,
to obtain the critical lines on the $(T,K)$ plane, one has to expand
$\tilde{\phi}(\beta,x)$ and $\tilde{s}(\veps,x)$ in powers of $x$.
The power expansion of $\lambda_{max}(\beta x,\beta K)$ can be
obtained explicitly as a function of modified Bessel functions. We
will avoid here the details of the calculation leading to the final
result.

Taking into account that our system is invariant under the symmetry
$\theta \to -\theta$, the expansion of $\tilde{\phi}(\beta,x)$ will
have only even powers. Let us write explicitly the $K$ dependence of
the expansion coefficients
\begin{equation}
\label{expansionphi} \tilde{\phi}(\beta,x)= \phi_0(\beta,K) +
\phi_1(\beta,K)\,x^2 +\phi_2(\beta,K)\,x^4 +o(x^6)\, .
\end{equation}
The canonical second order transition line in the $(T,K)$ plane is
determined by
\begin{equation}
\label{criticalxycanonical} \phi_1(\beta,K) = 0, \, \, \, \, \, \,
\, {\rm with} \, \, \, \, \phi_2(\beta,K) >0 \, .
\end{equation}
Inserting this expansion in the first equality in
Eq.~(\ref{entropyxyminmax1}) and minimizing with respect to $\beta$,
one obtains
\begin{equation}
\label{betaforepsilon} \veps= \phi_0'(\beta,K)
+\phi_1'(\beta,K)\,x^2 +\phi_2'(\beta,K)\,x^4 +o(x^6) \, ,
\end{equation}
where the prime denotes derivation with respect to $\beta$. This
equation gives an expansion of the form
\begin{equation}
\label{betaepskm} \beta(\veps,K,x) = \beta_0(\veps,K) +
\beta_1(\veps,K)\,x^2 +\beta_2(\veps,K)\,x^4 + o(x^6) \, .
\end{equation}
Using this expression, we obtain the expansion of the form
\begin{equation}\label{expansionsepsmgen}
s(\veps,m)= s_0(\veps,K) + s_1(\veps,K)m^2 +s_2(\veps,K)m^4 +o(m^6)
\,,
\end{equation}
where
\begin{eqnarray}
s_0(\veps,K)&=&\beta_0(\veps,K)\veps -\phi_0(\beta_0(\veps,K))\\
s_1(\veps,K)&=& \beta_1(\veps,K)\veps -
\phi_0'(\beta_0(\veps,K))\beta_1(\veps,K)
-\phi_1(\beta_0(\veps,K)) \\
s_2(\veps,K)&=&  \beta_2(\veps,K)\veps
-\frac{1}{2}\phi_0''(\beta_0(\veps,K))
\beta_1^2(\veps,K)-\phi_0'(\beta_0(\veps,K))\beta_2(\veps,K)\nonumber\\
&&\hskip 2truecm
-\phi_1'(\beta_0(\veps,K))\beta_1(\veps,K)-\phi_2(\beta_0(\veps,K)).
\end{eqnarray}
The microcanonical second order transition line is determined by
\begin{equation}
\label{criticalxymicrocanonical} s_1(\veps,K) = 0, \, \, \, \, \, \,
\, {\rm with} \, \, \, \, s_2(\beta,K) <0 \, .
\end{equation}
Now we can use the following equalities, that are obtained by
inserting back (\ref{betaepskm}) into (\ref{betaforepsilon})
\begin{eqnarray}
\label{equalitiescriticala}
\phi_0'(\beta_0(\veps,K)) &=& \veps \\
\phi_1'(\beta_0(\veps,K)) &=& -\phi_0''(\beta_0(\veps,K))\,
\beta_1(\veps,K) \, . \label{equalitiescriticalb}
\end{eqnarray}
Equation (\ref{equalitiescriticala}) implies that $\beta_0(\veps,K)$
is the temperature on the second order transition line. One obtains
\begin{eqnarray}\label{expansionsepsm1}
s(\veps,x)&=&\beta_0(\veps,K)\,\veps -\phi_0(\beta_0(\veps,K))
-\phi_1(\beta_0(\veps,K)) \, x^2 \nonumber \\
&& - \left[ -\frac{1}{2}\phi_0''(\beta_0(\veps,K)) \,
\beta_1^2(\veps,K) + \phi_2(\beta_0(\veps,K))\right] x^4 +o(x^6) \,
.
\end{eqnarray}
Therefore, the microcanonical second order transition line is
determined by
\begin{equation}
\label{criticalxymicrocanonical1} \phi_1(\beta_0(\veps,K)) = 0, \,
\, \, \, \, \, \, {\rm with} \, \, \, \,
\frac{1}{2}\phi_0''(\beta_0(\veps,K)) \, \beta_1^2(\veps,K) -
\phi_2(\beta_0(\veps,K)) < 0 \, .
\end{equation}
The concavity of $\phi(\beta)$ implies that, again on the critical
line, $\phi_0''(\beta_0(\veps,K))<0$ (This is valid in models where
parity with respect to $x$ implies that $\tilde{\phi}_{\beta x}=0$
for $m=0$). Comparing Eqs.~(\ref{criticalxycanonical}) and
(\ref{criticalxymicrocanonical1}), and taking into account the last
observation, we see that, as we expected from the general discussion
about ensemble equivalence at the end of subsection \ref{posneg},
the points belonging to the canonical second order transition line
also belong to the microcanonical one. The microcanonical line,
however, goes beyond and includes more points, as we have already
seen in several models, since by continuity the inequality in
(\ref{criticalxymicrocanonical1}) is satisfied also for some points
where $\phi_2(\beta_0(\veps,K))<0$.

\begin{figure}[htbp]
%\begin{center}
\resizebox{0.5\textwidth}{!}{\includegraphics{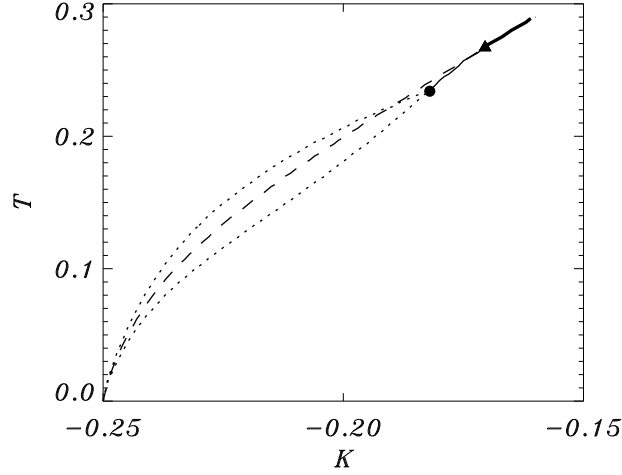}}
%\end{center}
%\includegraphics{phasediagxylongshort.eps}
\vskip -.5truecm \caption{Canonical and microcanonical $(K,T)$ phase
diagram of the one-dimensional $XY$ model with mean-field and
nearest-neighbour interactions. The canonical critical line (bold
solid line) ends at the tricritical point CTP indicated by a
triangle; then the transition becomes first order (dashed line). The
microcanonical second order transition line coincides with the
canonical one at large $K$ (bold solid line). It continues at lower
$K$ (light solid line) down to the tricritical point MTP, indicated
by a filled circle; then the transition becomes first order, with a
branching of the transition line (dotted lines), giving the two
extremes of the temperature jump.} \label{phasediagxylongshort}
\end{figure}

The study of the canonical and microcanonical critical lines
involves the analytic determination of $\lambda_{max}$, which, in
the limit $x \to 0$, can be performed using perturbation theory for
Hermitian operators \cite{TheseLeonardo}. For both ensembles, one
gets
\begin{equation}
\frac{I_1(\beta K)}{I_0(\beta K)}=\frac{4 -\beta}{2 \beta + 4}~,
\end{equation}
where $I_0$ and $I_1$ are modified Bessel functions. The canonical
tricritical point turns out to be $K_{CTP} \simeq -1.705$ and
$T_{CTP} \simeq 0.267$, while microcanonical tricritical point is at
$K_{MTP} \simeq -0.182$ and $T_{MTP} \simeq 0.234$. In
Fig.~\ref{phasediagxylongshort}, we plot the relevant part of the
phase diagram on the $(K,T)$ plane. As commented above, it is shown
how the microcanonical critical line (ending at the microcanonical
tricritical point MTP) extends beyond the canonical critical line,
which ends at the canonical tricritical point CTP.

\paragraph{Ergodicity breaking}

We here prove that model (\ref{hamilxy}) shows ergodicity breaking,
similarly to the Ising long- plus short-range model. Let us first
study separately the bounds of the short-range term
\begin{equation}
\label{potenkxy} \frac{1}{N}\sum_{i=1}^N \cos \left(
\theta_{i+1}-\theta_i \right) \, .
\end{equation}
The study of maxima and minima of this term can be reduced to those
configurations where the spins are aligned parallel or antiparallel
to a given direction. There is no loss of generality, due to
rotational symmetry, to then choose this direction as the $x$-axis.
Therefore, only the $x$ component of $m$ is non zero and in the
following we will restrict to $m=m_x \geq 0$. It turns out that, in
the thermodynamic limit, both extrema are attained with a fraction
$(1+m)/2$ of rotators parallel to the $x$-axis and a fraction
$(1-m)/2$ antiparallel to the $x$-axis, for all given values of $m$.
The maximum is $1$ and is achieved when the parallel rotators and
the antiparallel rotators are grouped in two separated blocks. The
minimum  $2m-1$, is instead attained when the antiparallel rotators
are all isolated, which is possible since we are considering $m \ge
0$.

\begin{figure}[!htbp]
\centering
\resizebox{0.9\textwidth}{!}{\includegraphics{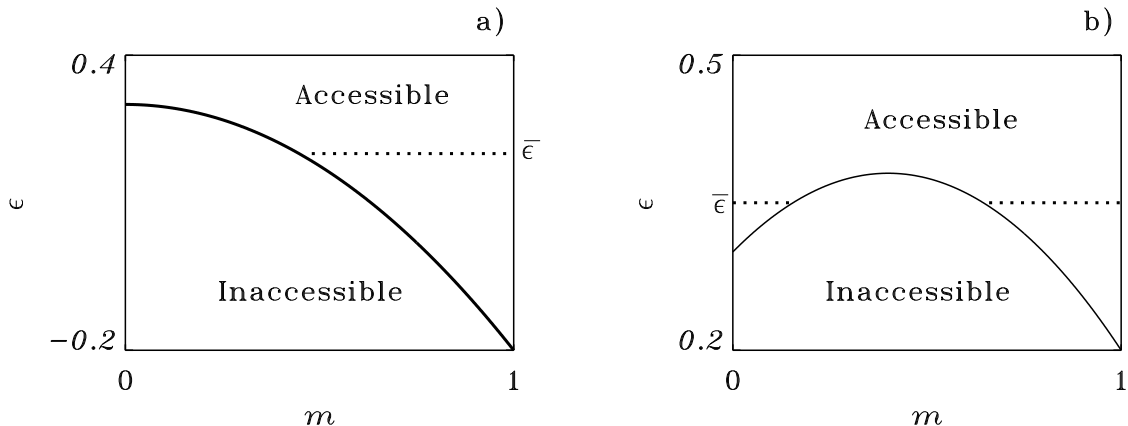}}
\caption{Accessible regions in the $(\veps,m)$ plane for the
one-dimensional XY model with both mean-field and nearest neighbour
interactions. Solid curves represent a) $\veps_{min}^{K>0}(m)$
(\ref{minimumppotxykpos}) and b) $\veps_{min}^{K<0}(m)$
(\ref{minimumppotxykneg}). The accessible and inaccessible regions
in parameter space are indicated. a) $K=0.2$: no breaking of
ergodicity is present since at fixed energy $\bar{\veps}$ the
accessible interval of $m$ is connected (dotted line); b) $K=-0.2$:
there is breaking of ergodicity for values of $\veps$ between $0.30$
and $0.38$. For instance for $\bar{\veps}=0.35$ the are two
disconnected intervals of accessible magnetizations (dotted lines).}
\label{minimaenerxy}
\end{figure}

Now we have to distinguish the two cases $K>0$ and $K<0$. In the
positive case, the minimum of this contribution to the energy
density is $-K$ (thus actually independent of $m$), while in the
negative case it is $-K(2m-1)$. We obtain the minimum of the total
potential energy per particle just by adding $(1-m^2)/2$. This is
actually the minimum of the total energy per particle $\veps$, since
the kinetic energy is positive definite. We finally obtain that the
minimum of the energy per particle is
\begin{equation}
\label{minimumppotxykpos} \veps_{min}^{K>0}(m)=\frac{1-m^2}{2} -K
\end{equation}
for $K >0$, and
\begin{equation}
\label{minimumppotxykneg} \veps_{min}^{K<0}(m)=\frac{1}{2}(1-m^2)
-K(2m-1)
\end{equation}
for $K<0$. Plotting the two functions $\veps_{min}^{K>0}$ and
$\veps_{min}^{K<0}$ we will find the accessible $m$ values for a
given energy $\veps$. The functions are plotted in
Fig.~\ref{minimaenerxy} for two representative values: $K=0.2$ and
$K=-0.2$. For positive $K$, we see that below the energy $-K + 1/2$
the system cannot have all magnetizations, and values of $m$ near
$0$ are inaccessible. We note however that the region of accessible
values of $m$, for any energy, is connected; therefore there is no
ergodicity breaking. Nevertheless, we remark that the accessible
region in the $(\veps,m)$ plane is not convex, as it can happen only
for long-range interactions. For negative $K$, we note instead, that
for a given energy interval ($K+1/2<\veps<2K^2+K+1/2$ for $K>-1/4$
and $-K<\veps<2K^2+K+1/2$ for $K<-1/4$), the attainable values of
$m$ are separated in two disconnected regions. Therefore for $K<0$
we have breaking of ergodicity.

\subsection{Weakly decaying interactions}
\label{decayinginteraction}

All the models that have been considered in previous sections have
an infinite range term in the Hamiltonian. This undoubtly limits the
applicability of our analysis to realistic physical systems, where
interactions decay with the distance. We present below four models
of this kind which are direct generalizations of models presented
above and represent a step forward in the study of long-range weakly
decaying interactions. Some of the models are exactly solvable and
reproduce features of the phenomenology observed for infinite range
systems (ensemble inequivalence, negative specific heat, ergodicity
breaking, etc.).

\subsubsection{$\alpha$-Ising model}
\label{alphaIsing}

Let us consider the one-dimensional $\alpha$-Ising  Hamiltonian
\begin{equation}
 H_N = \frac{J}{N^{1-\alpha}}\sum_{i>j=1}^N \frac{1-S_iS_j}{|i-j|^{\alpha}},
\label{Hlattice}
\end{equation}
where $J>0$ and spins $S_i=\pm 1$ sit on a one-dimensional lattice
with either free or periodic boundary conditions (in the latter
case, $|i-j|$ is the minimal distance along the ring). The
$N^{\alpha-1}$ prefactor is introduced in order to have an extensive
energy as explained in Sec.~\ref{additivity}. This model has been
first introduced by Dyson~\cite{dyson} and studied for the
``integrable'' case, $\alpha>1$, in the canonical ensemble without
the $N^{\alpha-1}$ prefactor. We show here that it is possible to
obtain an exact microcanonical solution using large deviation theory
when $0\leq \alpha<1$~\cite{junext01,MukamelHouches,bbdrjstatphys}.
Moreover, the study of this model gives also the opportunity to
emphasize the important role played by boundary conditions when the
interactions are long-range.

We adopt the same scheme described in section~\ref{Methode_generale}
to obtain the solution of model (\ref{Hlattice}). The method can be
generalized to lattices of higher dimension.

In the first step, the Hamiltonian $H_N$ is rewritten in terms of
global variables by introducing a {\it coarse-graining}. Let us
divide the lattice in $K$ boxes, each with $n=N/K$ sites, and let us
introduce the average magnetization in each box $m_k$, $k=1\dots K$.
In the limit $N\to \infty$, $K\to \infty$, $K/N \to 0$, the
magnetization becomes a continuous function $m(x)$ of the $[0,1]$
interval. After a long but straightforward calculation, described in
Ref.~\cite{bbdrjstatphys}, it is possible to express $H_N$ as a
functional of $m(x)$
\begin{eqnarray}
H_N =  N H[m(x)] +o(N)  , \label{approxH1}
\end{eqnarray}
where
\begin{eqnarray}
H[m(x)] = \frac{J}{2} \int_0^1\! \dd x \int_0^1 \!\dd y \frac{1-m(x)
m(y)}{|x-y|^{\alpha}}.\label{formulepourh}
\end{eqnarray}
The estimation is uniform on all configurations.

In the second step, we evaluate the probability to get a given
magnetization $m_k$ in the $k$-th box from all {\it a priori}
equiprobable microscopic configurations. This probability obeys a
local large deviation principle $P(m_k)\propto \exp[{n
\tilde{s}(m_k)}]$, with
\begin{equation}
\tilde{s}(m_k)=-\frac{1+m_k}{2}\ln{\frac{1+m_k}{2}}-
\frac{1-m_k}{2}\ln{\frac{1-m_k}{2}}. \label{formuleentropie}
\end{equation}
Since the microscopic random variables in the different boxes are
independent and no global constraints has been imposed, the
probability of the full global variable  $(m_1,\ldots,m_K)$ can be
expressed in a factorized form as
\begin{eqnarray}
P(m_1,m_2,\ldots,m_K) & = & \prod_{i=1}^KP(m_i)
 \simeq  \prod_{i=1}^Ke^{\displaystyle n \tilde{s}(m_i)}\\&=&\exp\left[{ nK\sum_{i=1}^K
 \frac{\tilde{s}(m_i)}{K}}\right] \label{defsbar}\\
 &\simeq& e^{\displaystyle N \bar{s}[m(x)]} , \label{largedev2}
\end{eqnarray}
where $\bar{s}[m(x)]=\int_0^1 \tilde{s}(m(x)) \, \dd x$ is the
entropy functional associated to the global variable $m(x)$. Large
deviation techniques rigorously justify these
calculations~\cite{refnewellis}, proving that entropy is
proportional to~$N$, also in the presence of long-range
interactions. This result is independent of the specific model
considered.

\begin{figure}[htbp]
\resizebox{0.45\textwidth}{!}{\includegraphics{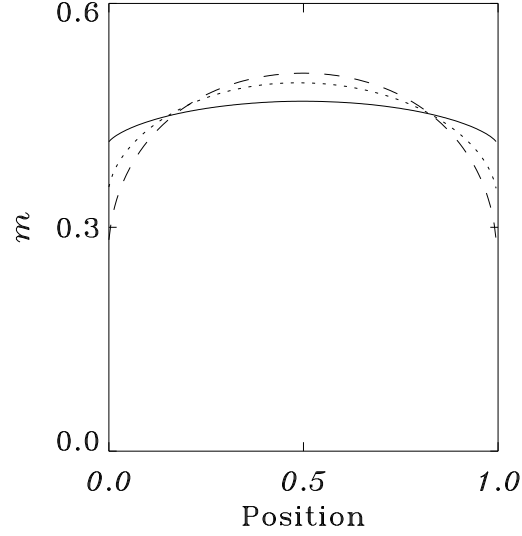}}
\caption{ Equilibrium magnetization profile  for the $\alpha$-Ising
model with free boundary conditions at an energy density
$\varepsilon=0.1$ for $\alpha=0.2$ (solid line), $\alpha=0.5$
(dotted line) and $\alpha=0.8$ (dashed line).}
\label{magnprofile_fig}
\end{figure}

In the third step, we formulate the variational problem in the
microcanonical ensemble to get the entropy
\begin{equation}
\label{isingmicro} s(\varepsilon)=\sup_{m(x)}\left( \bar{s}[m(x)]\:
\left| \varepsilon=H[m(x)]\right. \right).
\end{equation}
Let us remark that this optimization problem has to be solved in a
functional space. In general, this has to be done numerically,
taking into account boundary conditions. In the case of free
boundary conditions, the only available solutions are numerical. An
example of a maximal entropy magnetization profile obtained for free
boundary conditions is shown in Fig.~\ref{magnprofile_fig} for
different values of $\alpha$. The profile become more inhomogeneous
when increasing $\alpha$ (for $\alpha=0$ one recovers the mean-field
result with a homogeneous profile).

\begin{figure}[htbp]
\resizebox{0.8\textwidth}{!}{\includegraphics{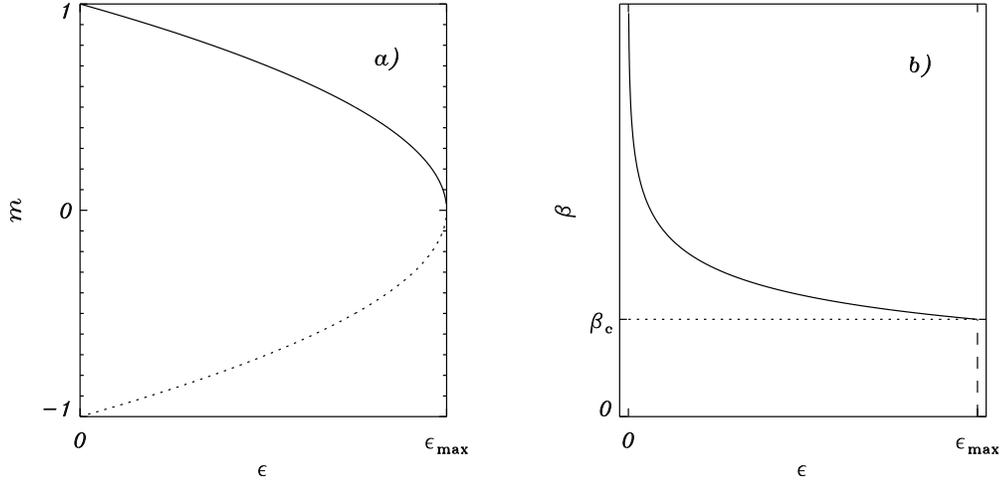}}
\caption{a) Equilibrium magnetization in the allowed energy range in
the microcanonical ensemble for the $\alpha$-Ising model with
$\alpha=0.5$; the negative branch is also reported with a dotted
line. b) Inverse temperature versus energy in the microcanonical
ensemble (solid line). The canonical ensemble result superposes to
the microcanonical one in the interval $[\beta_c,\infty)$ and is
represented by a dashed line for $\beta\in[0,\beta_c]$. $\beta_c$ is
then the inverse critical temperature in the canonical ensemble. In
the microcanonical ensemble, no phase transition is present.}
\label{mbetaising}
\end{figure}

In the following, we will treat the periodic boundary conditions
case, for which analytical result can be obtained. Both entropy and
free energy can be obtained in analytical form for homogeneous
magnetization profiles, which have been shown to be locally stable
in both the high-temperature and the low-temperature
phase~\cite{bbdrjstatphys}. It has indeed been proven that for
$\beta<\beta_c=(1-\alpha)/(J2^\alpha)$ there is a unique global
maximum of~$s(\veps)$, corresponding to a constant zero
magnetization profile. The variational problem (\ref{isingmicro}),
where $\bar{s}$ is defined in Eqs.~(\ref{formuleentropie}),
(\ref{defsbar}) and (\ref{largedev2}), leads to the consistency
equation
\begin{equation}
\label{isingmicro3} \tanh^{-1}\left(m(x)\right)={\beta\,
J}\int_0^1\frac{m(y)}{|x-y|^{\alpha}} \:\dd y,
\end{equation}
where $\beta$ is a Lagrange multiplier. For $\beta>\beta_c$, we
restrict ourselves to constant magnetization profiles, which are
locally stable, i.e. close non constant profiles have a smaller
entropy. In this case, using the relations $\int_0^1\dd x
|x-y|^{-\alpha} =2^\alpha/(1-\alpha)$ and
$\varepsilon_{\text{max}}=1/(2\beta_c)$ one can obtain the
magnetization vs. energy curve $m=\pm \sqrt{1-\veps/\veps_{max}}$,
see Fig.~\ref{mbetaising}(a). Moreover, fixing the energy implies
fixing the magnetization and, consequently, the Lagrange multiplier
$\beta$ in Eq.~(\ref{isingmicro3}). Expressing the magnetization in
terms of the energy in the entropy formula (\ref{isingmicro}) allows
us to derive the caloric curve, see Fig.~\ref{mbetaising}(b) (solid
line). The limit temperature $\beta_c$ (dotted line) is attained at
zero magnetization, which is a boundary point.

In the canonical ensemble, one has to solve the variational problem
(\ref{freeminsmu}). This leads to exactly the same consistency
equation (\ref{isingmicro3}), where the Lagrange multiplier is
replaced by the inverse temperature $\beta$. Solving this
consistency equation for $\beta>0$, one finds a zero magnetization
for $\beta<\beta_c$ and a non vanishing one for $\beta>\beta_c$. One
can also derive the canonical caloric curve, which is reported in
Fig.~\ref{mbetaising}(b) and superposes to the microcanonical
caloric curve from $\beta=\infty$ down to $\beta_c$ while it is
represented by the dashed line for $\beta<\beta_c$. It follows that
in the region $[0,\beta_c]$, the two ensembles are not equivalent.
In this case, a single microcanonical state at
$\varepsilon_{\text{max}}$ corresponds to many canonical states with
canonical inverse temperatures in the range $[0,\beta_c[$. Thus, in
such a case, the canonical inverse temperature is not equal to the
microcanonical one. In the microcanonical ensemble, the full high
temperature region is absent and, therefore, no phase transition is
present or, in other terms, the phase transition is at the boundary
of the accessible energy values. The entropy is always concave,
hence no inequivalence can be present in the allowed energy range,
apart from the boundaries. This situation is often called partial
equivalence~\cite{Ellis99,TouchettePhysRep,Casetti2007}. Partial
equivalence persists for all $\alpha$ values below one, and is
removed only for $\alpha=1$ when $\varepsilon_{\text{max}}\to\infty$
and $\beta_c\to0$: the phase transition is not present in both
ensembles and the system is always in its magnetized phase. The main
drawback of this analysis is the difficulty to obtain analytical
solutions of Eq.~(\ref{isingmicro3}) for non constant magnetization
profiles, which is the typical situation when boundary conditions
are not periodic.

\subsubsection{$\alpha$-HMF model}
\label{alphaxymodel}

Let us consider now a generalization of the HMF
model~(\ref{Ham_HMF}) which has been originally proposed in
Ref.~\cite{Anteneodo97,Anteneodo99}. The interaction between two
rotators has the same form as in the original model, but the
coupling constant is a weakly decaying function of the distance
between the two lattice sites where the rotators sit. Note that we
consider the lattice in general dimension $d$. In Ref.~\cite{CGM},
the generalized case of $n$-vector spin models has been analyzed,
including the HMF model (for $n=2$) and the Ising spins (for $n=1$).
Only periodic boundary conditions have been considered, as we will
do here.

%As emphasized in
%the previous subsection, boundary conditions are important in
%long-range systems, and they have a role in the determination of the
%equilibrium state.
%Therefore the results just presented for the
%$\alpha$-Ising model with free boundary conditions were not included
%in the analysis of Ref. \cite{CGM}.

%Anticipating that the $\alpha$-XY model, as the original HMF model,
%has only a second order phase transition, we can restrict our study
%to the canonical ensemble.

The Hamiltonian of the $\alpha$-HMF model is
\begin{equation}
\label{hamilalphaxy} H_N = \sum_{i=1}^N \frac{p_i^2}{2}
+\frac{1}{2\widetilde{N}} \sum_{i,j=1}^N \frac{1-\cos (\theta_i
-\theta_j)}{r_{ij}^\alpha} \, ,
\end{equation}
where $r_{ij}$ is the minimal distance between lattice sites $i$ and
$j$ (minimal distance convention in periodic boundary conditions).
We consider here cases for which the exponent $\alpha$, determining
the decay with distance of the coupling constant between rotators,
is less than the spatial dimension $d$, so that the system is
long-range as shown in Sec.~\ref{decayint}. The $\widetilde{N}$
prefactor is introduced in order to have an extensive energy as
explained in Sec.~\ref{additivity}. We show below that, in any
spatial dimension $d$ and for each $\alpha < d$, the thermodynamics
of this model is the same as that of the HMF model; in particular,
there is a second order phase transition at the inverse temperature
$\beta=2$.

Let us introduce the matrix of the coupling constants
$J_{ij}=1/r_{ij}^\alpha$. Since the diagonal elements of this matrix
are not defined, we assign to them the common arbitrary value $b$.
The prefactor is
\begin{equation}
\label{definitionntilde} \widetilde{N} = \sum_{j=1}^N J_{ij}
\end{equation}
that does not depend on $i$, because we adopt the minimal distance
convention for  $r_{ij}$. It is clear that in the thermodynamic
limit $N \to \infty$, the prefactor $\widetilde{N}$ behaves as
$N^{1-\alpha/d}$. The arbitrary parameter $b$, which appears only in
$\tilde{N}$, becomes negligible in thermodynamic limit. Its
introduction is however useful for the computation, as it will
become clear.

The most relevant steps of the solution in the canonical ensemble
(we refer the reader to Ref.~\cite{CGM} for details) can be obtained
starting from the partition function
\begin{eqnarray}
Z(\beta,N) &=& \left(\frac{2\pi}{\beta} \right)^{N/2}\int \dd
\theta_1 \dots \dd \theta_N \, \exp \left(
-\frac{\beta}{2\widetilde{N}} \sum_{i,j=1}^N J_{ij}\left[ 1- \cos
(\theta_i - \theta_j)\right]\right)\\
&=& \left(\frac{2\pi}{\beta} \right)^{N/2}\exp
\left(-\frac{N\beta}{2}\right) \int \dd \theta_1 \dots \dd \theta_N
\, \exp \left( \sum_{i,j=1}^N \left[ \cos \theta_i R_{ij} \cos
\theta_j +\sin \theta_i R_{ij} \sin \theta_j \right] \right) \, ,
\label{canonpartalphaxy}
\end{eqnarray}
where we have introduced the matrix $R_{ij} = \beta
J_{ij}/(2\widetilde{N})$ and used the property $\sum_{ij} J_{ij}=
N\widetilde{N}$. In order to use the Hubbard-Stratonovich
transformation, one diagonalizes the expression in the second
exponential. Denoting by $R$ the symmetric matrix with elements
$R_{ij}$, and by $V$ the unitary matrix that reduces $R$ to its
diagonal form $D=VRV^T$, one gets
\begin{equation}
\label{diagtranfscos} \sum_{i,j=1}^N \cos \theta_i R_{ij} \cos
\theta_j = \sum_{i=1}^N R_i a_i^2 \, ,
\end{equation}
where the quantities $R_i$ are the real eigenvalues of the matrix
$R$, and $a_i = \sum_{j=1}^N V_{ij}\cos \theta_j$. An analogous
expression holds for the sum with the sines in
Eq.~(\ref{canonpartalphaxy}). In order to express $\exp (R_i a_i^2)$
using the Hubbard-Stratonovich transformation, the eigenvalues $R_i$
must be positive. By choosing a sufficiently large value of $b$, but
in any case of order $1$, all eigenvalues $R_i$ are positive. One
can therefore write
\begin{eqnarray}
\label{hubbapplication} \exp \left( \sum_{i=1}^N R_i a_i^2 \right)
&=& \frac{1}{\sqrt{(4\pi)^N \det R}}\int \dd y_1 \dots \dd y_N \exp
\left[ \sum_{i=1}^N \left( -\frac{y_i^2}{4R_i}+a_i y_i
\right) \right]  \\
&=&\frac{1}{\sqrt{(4\pi)^N \det R}}\int \dd z_1 \dots \dd z_N \exp
\left[-\frac{1}{4} \sum_{i,j} z_i \left( R^{-1}\right)_{ij}z_j
+\sum_{i=1}^N z_i \cos\theta_i \right] \, .
\end{eqnarray}
In this $N$-dimensional Hubbard-Stratonovich transformation, we have
used that $\det R = \prod_{i=1}^N R_i$, we have introduced the
unitary change of integration variables defined by $y_i=\sum_{j=1}^N
V_{ij}z_j$ and we have indicated with $R^{-1}$ the inverse matrix of
$R$. Using the analogous expression for the term with the sines, we
arrive, after performing the integration over the angles $\theta$,
to the following expression for the partition function
\begin{eqnarray}
\label{partfinalalphaxy} Z(\beta,N)=\left(\frac{2\pi}{\beta}
\right)^{N/2}\frac{\exp \left(-{N\beta}/{2}\right)}{(4\pi)^{N} \det
R} \int \dd z_1 \dots \dd z_N \dd z_1' \dots \dd z_N' && \exp \left[
-\frac{1}{4} \sum_{i,j} z_i \left( R^{-1}\right)_{ij}z_j
-\frac{1}{4} \sum_{i,j} z_i' \left( R^{-1}\right)_{ij}z_j' \right.
\nonumber \\ &&\qquad\qquad\left. \phantom{aaaa} + \sum_{i=1}^N \ln
I_0\left(\sqrt{z_i^2+z_i'^2} \right)\right] \, .
\end{eqnarray}
This integral can be computed using the saddle point method, since
most eigenvalues $R_i$ vanish in the thermodynamic limit \cite{CGM}.
Introducing $\rho_i=\sqrt{z_i^2+z_i'^2}$, the stationary points are
solutions of the following consistency equations
\begin{eqnarray}
\label{statpointsalphaxy}
%\frac{1}{2}\, \sum_{j=1}^N
%\left(R^{-1}\right)_{ij} z_j &=&
%\frac{I_1\left(\sqrt{z_i^2+z_i'^2}\right)}{I_0\left(\sqrt{z_i^2+z_i'^2}\right)}
%\frac{z_i}{\sqrt{z_i^2+z_i'^2}}  \\
\frac{1}{2}\, \sum_{j=1}^N \left(R^{-1}\right)_{ij} z_j &=&
\frac{I_1\left(\rho_i\right)}{I_0\left(\rho_i\right)}
\frac{z_i}{\rho_i}  \\
\frac{1}{2}\, \sum_{j=1}^N \left(R^{-1}\right)_{ij} z_j' &=&
\frac{I_1\left(\rho_i\right)}{I_0\left(\rho_i\right)}
\frac{z_i'}{\rho_i}~,
%\\
%\frac{1}{2}\, \sum_{j=1}^N \left(R^{-1}\right)_{ij} z_j' &=&
%\frac{I_1\left(\sqrt{z_i^2+z_i'^2}\right)}{I_0\left(\sqrt{z_i^2+z_i'^2}\right)}
%\frac{z_i'}{\sqrt{z_i^2+z_i'^2}} \,\,\,\,\,\,\,\,\,\, i=1,\dots,N
\end{eqnarray}
which, after inversion with respect to $z_i,z_i'$, become
\begin{eqnarray}
\label{statpointsalphaxy1} z_i &=& 2 \, \sum_{j=1}^N R_{ij}
\frac{I_1\left(\rho_j\right)}{I_0\left(\rho_j\right)}
\frac{z_j}{\rho_j}  \\
z_i' &=& 2 \, \sum_{j=1}^N
R_{ij}\frac{I_1\left(\rho_j\right)}{I_0\left(\rho_j\right)}
\frac{z_j'}{\rho_j}~. \label{statpointsalphaxy1prime}
\end{eqnarray}
Let us first look for homogeneous solutions, i.e., for solutions in
$z_i,z_i'$ that do not depend on $i$. The previous system reduces to
the following pair of equations for $z\equiv z_i$ and $z'\equiv
z_i'$
\begin{eqnarray}
\label{statpointsalphaxy2} z = \beta
\frac{I_1\left(\rho\right)}{I_0\left(\rho\right)} \frac{z}{\rho}
\quad \mbox{and}\quad z' = \beta \frac{I_1\left(\rho
\right)}{I_0\left(\rho\right)} \frac{z'}{\rho} \, ,
\label{statpointsalphaxy2prime}
\end{eqnarray}
where we have used that $2\sum_{j=1}^N R_{ij}= \beta$ for each $i$.
By expressing $z=\rho \cos \gamma $ and $z'=\rho \sin \gamma$, we
see that these two equations determine only the modulus $\rho
\equiv\sqrt{z^2+z'^2}$, leaving the angle $\gamma$ undetermined.
They thus reduce to a single equation for the modulus
\begin{equation}
\label{statpointsalphaxy3} \rho = \beta
\frac{I_1\left(\rho\right)}{I_0\left(\rho \right)} \, .
\end{equation}
This equation, after the change of variable $\rho=\beta m$,
coincides with Eq.~(\ref{solhmfcan}) of the HMF model.

In order for this homogeneous stationary point to be the relevant
one for the evaluation of the integral in
Eq.~(\ref{partfinalalphaxy}), it is necessary that the exponential
has an absolute maximum at this stationary point. One can easily
show that the homogeneous stationary point is a local maximum
\cite{CGM}. Here, we will directly prove that it is also a global
maximum with respect to any other possible inhomogeneous stationary
point.

Let us then go back to Eqs.~(\ref{statpointsalphaxy1}) and
(\ref{statpointsalphaxy1prime}) and suppose that they have
inhomogeneous solution, considering as a first case a solution with
different values for the moduli $\rho_i\equiv\sqrt{z_i^2+z_i'^2}$.
If $k$ is the site with the largest modulus~$\rho_k$, let us choose
the $x$-axis of the XY rotators such that $z_k=\rho_k$ and $z_k'=0$.
Then one can write
\begin{eqnarray}
\label{estimatemoduli} z_k=\rho_k = 2 \, \sum_{j=1}^N R_{ij}
\frac{I_1\left(\rho_j\right)}{I_0\left(\rho_j\right)}
\frac{z_j}{\rho_j} &\le& 2 \, \sum_{j=1}^N R_{ij}
\frac{I_1\left(\rho_j\right)}{I_0\left(\rho_j\right)}
 \\
&<& 2 \, \sum_{j=1}^N R_{ij}
\frac{I_1\left(\rho_k\right)}{I_0\left(\rho_k\right)} =\beta
\frac{I_1\left(\rho_k\right)}{I_0\left(\rho_k\right)} \, ,
\label{estimatemoduliprime}
\end{eqnarray}
where use has been made of the monotonicity of the function
$I_1/I_0$. We emphasize that the last inequality is strict since, by
hypothesis, not all moduli $\rho_i$ are equal. From the properties
of the function $I_1/I_0$, it follows from the last inequality that
also the largest modulus $\rho_k$ of this inhomogeneous stationary
point is smaller than that of the homogeneous solution. One gets the
same conclusion for a possible stationary point with all equal
moduli but with different directions; in that case inequality
(\ref{estimatemoduliprime}) becomes an equality, but inequality
(\ref{estimatemoduliprime}) is strict. The proof is completed by
showing that the exponent in Eq.~(\ref{partfinalalphaxy}), if
computed at a stationary point, has a positive derivative with the
respect the moduli $\rho_i$ of the stationary point. In fact, it can
be easily shown that at a stationary point the exponent in Eq.
(\ref{partfinalalphaxy}), let us call it $A$, can be written as a
function of the moduli $\rho_i$ of the stationary point in the
following way
\begin{equation}
\label{exponentstationary} A= \sum_{i=1}^N \left[
-\frac{1}{2}\rho_i\frac{I_1\left(\rho_i\right)}{I_0\left(\rho_i\right)}
+ \ln I_0\left(\rho_i\right)\right] \, .
\end{equation}
Its derivative with respect to $\rho_i$ is
\begin{equation}
\label{exponentstationaryder} \frac{\partial A}{\partial \rho_i} =
\frac{1}{2}\left[
\frac{I_1\left(\rho_i\right)}{I_0\left(\rho_i\right)} - \rho_i
\frac{\partial}{\partial
\rho_i}\frac{I_1\left(\rho_i\right)}{I_0\left(\rho_i\right)}\right]
\, .
\end{equation}
The concavity of the function $I_1/I_0$, for positive values of the
argument, assures that the right hand side is positive. This
concludes the proof.

We conclude that for each dimension $d$ and for each decaying
exponent $\alpha < d$, the free energy of the $\alpha$-HMF model
with periodic boundary conditions is the same as that of the HMF
model. Maybe the derivation of the solution in the microcanonical
ensemble does not pose particular technical
problems~\cite{JulienBarrePrivate}.

\subsubsection{One-dimensional gravitational models}
\label{taka}

A direct study of the full three-dimensional $N$-body gravitational
dynamics is particularly heavy~\cite{Heggiehut} and special purpose
machines have been built to this aim~\cite{grape5taka}. Therefore,
lower dimensional models have been introduced to describe
gravitational systems with additional symmetries. For instance, the
gravitational sheet model, describing the motion of infinite planar
mass distributions perpendicularly to their surface, has been
considered~\cite{Hohl67}. Although this model shows interesting
behaviors~\cite{Tsuchiya94,Koyama01}, the specific heat is always
positive and no phase transition is present. Long ago,
H\'enon~\cite{Henon67} introduced a system of concentric spherical
mass shells that interact via gravitational forces. This model has
been recently shown to display phase transitions and ensemble
inequivalence in a mean-field limit~\cite{Youngkins}.

\begin{figure}[htbp]
\resizebox{140mm}{!}{ \includegraphics{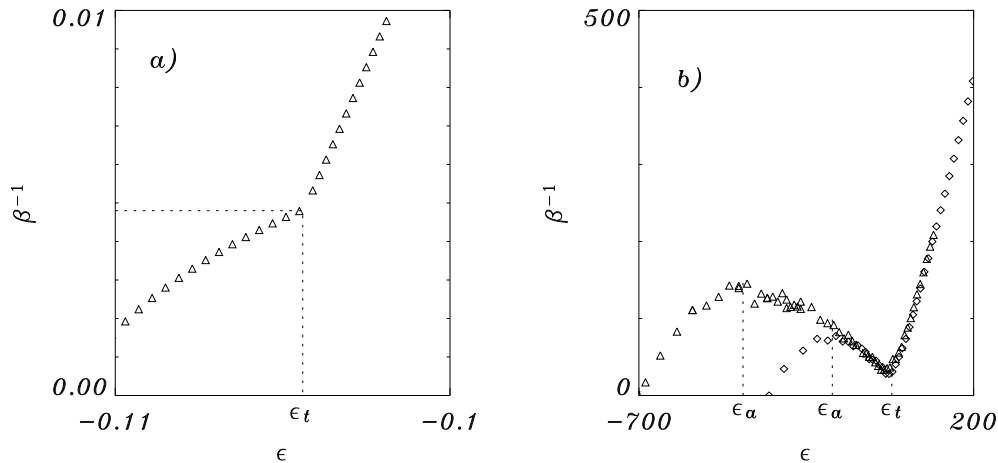}}
\caption{Caloric curves of the Self Gravitating Ring (SGR) model
obtained from numerical simulations of
Hamiltonian~(\ref{eqn:SGR-H}). Panel (a) refers to the softening
parameter value $\delta=10$, for which a second order phase
transition appears at $\veps_t$. No backbending of the caloric
curve, indicating a negative specific heat, is present. Simulations
were performed for $N=100$. Panel (b) shows the caloric curves for
two different values of the softening parameter, $\delta_1=1.0
\times 10^{-6}$ and $\delta_2= 2.5 \times 10^{-7}$, and $N=100$. The
transition is here first order in the microcanonical ensemble. A
negative specific  heat phase appears for $\veps_a< \veps< \veps_t$,
and expands as the softening parameter is reduced.}
\label{fig:SGR-caloric}
\end{figure}

In this section we discuss the phase diagram, in both the
microcanonical and the canonical ensemble, of the
Self-Gravitating-Ring (SGR)~\cite{SGR-model} model. In this model,
particle motion is constrained on a ring  and particles interact via
a true 3D Newtonian potential. The Hamiltonian of the SGR model is
\begin{eqnarray}
\label{eqn:SGR-H} H_N &=& \frac{1}{2} \sum_{i=1}^N p_i^2 +
\frac{1}{2N} \sum_{i,j}
 V_{\delta} (\theta_i - \theta_j) \,, \\
\mbox{with}\qquad V_{\delta} (\theta_i - \theta_j)& = &
-\frac{1}{\sqrt{2}}
 \frac{1}{\sqrt{1-\cos (\theta_i - \theta_j) + \delta}} \,,\label{eqn:SGR-Hbis}
\end{eqnarray}
where $\delta$ is the softening parameter, which is introduced, as
usual, in order to avoid the divergence of the potential at short
distances. Taking the  large $\delta$ limit, the potential becomes
\begin{equation}
\label{eqn:large-e-limit} V_{\delta}= \frac{1}{\sqrt{2 \delta} }
\left[\frac{1-\cos \left(\theta_i - \theta_j\right)}{2\delta}-1
\right]  + O(\delta^{-2}) \,,
\end{equation}
which is the one of the Hamiltonian Mean-Field (HMF) model
(\ref{Ham_HMF}). Since the interaction is regularized at short
distances, a global entropy maximum always exists, and
thermodynamics is well defined in the mean-field limit. In a
situation close to that of the HMF model e.g. for $\delta=10$, the
caloric curve determined from microcanonical numerical simulations
is reported in Fig.~\ref{fig:SGR-caloric}(a). In the homogeneous
phase $\veps>\veps_t(\delta)$, the caloric curve is almost linear,
while in the clustered phase $\veps<\veps_t(\delta)$, it is bent
downward.  Nonetheless, temperature always grows with energy and one
does not observe any negative specific heat energy range.
However, as it happens for 3D Newtonian gravity
simulations~\cite{Heggiehut}, when one reduces the softening
parameter, a negative specific heat phase develops. For instance, in
Fig.~\ref{fig:SGR-caloric}(b), we show two cases at small $\delta$
where three phases can be identified~\cite{SGR-model}:
\begin{itemize}
\item a low-energy clustered phase for $\veps<\veps_a$,
 where $\veps_a$ is defined as the energy at which the negative specific
 heat region begins.
\item an intermediate-energy phase, $\veps_a(\delta)<\veps<\veps_t(\delta)$,
with negative specific heat.
\item a high-energy gaseous phase for $\veps_t(\delta) <\veps$.
\end{itemize}
The clustered phase is created by the presence of softening, without
which the particles would fall into the zero distance singularity.
In the gas phase, the particles are hardly affected by the potential
and behave as almost free particles.  The intermediate phase is
expected to show the characters of gravity, persisting and even
widening in the $\delta\to0$ limit.

\begin{figure}[htbp]
\resizebox{70mm}{!}{\includegraphics{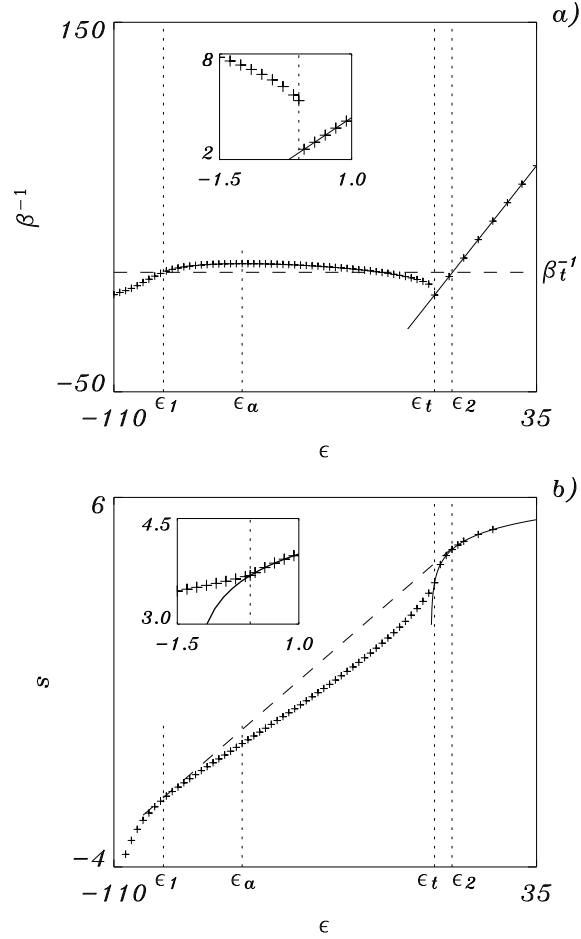}}\\
%\resizebox{80mm}{!}{\includegraphics{SGR-entropy-e-5.eps}}
\caption{Temperature (panel (a)) and entropy (panel (b)) versus
energy $\veps$ for the softening parameter value $\delta=10^{-5}$.
Four values of the energy, indicated by the short-dashed vertical
lines, can be identified from this picture: $\veps_1\simeq-93$
and~$\veps_2\simeq6$ bound from below and above the  region of
inequivalence of ensembles. $\veps_t\simeq0$ is the first order
phase transition energy in the microcanonical ensemble.
$\veps_a\simeq-66$ limits from below the negative specific heat
region.  $\beta_t^{-1}\simeq15$, represented with a dashed line in
panel (a), is the canonical transition temperature. The solid lines
represent the analytical solutions of temperature and entropy in the
high energy phase. They are extended slightly below $\veps_t$, in
the metastable phase. The insets in panels (a) and (b) show a zoom
of temperature and entropy around $\veps_t$, revealing a temperature
jump. } \label{fig:SGR-S-e-5}
\end{figure}

It is shown analytically in Ref.~\cite{TAKA} that ensembles are not
equivalent and a phase of negative specific heat and temperature
jumps in the microcanonical ensemble appears in a wide intermediate
energy region, if the softening parameter is small enough. In
Fig.~\ref{fig:SGR-S-e-5}, that is drawn for $\delta=10^{-5}$, we
show all the features that we had already commented in the context
of the study of the phase transition of the BEG model
(\ref{begmodel}). In particular, we would like to draw the attention
of the reader to the similarities of Fig.~\ref{fig:SGR-S-e-5}b with
Fig~\ref{entropyBEG}a.

\begin{figure}[htbp]
 \resizebox{80mm}{!}{ \includegraphics{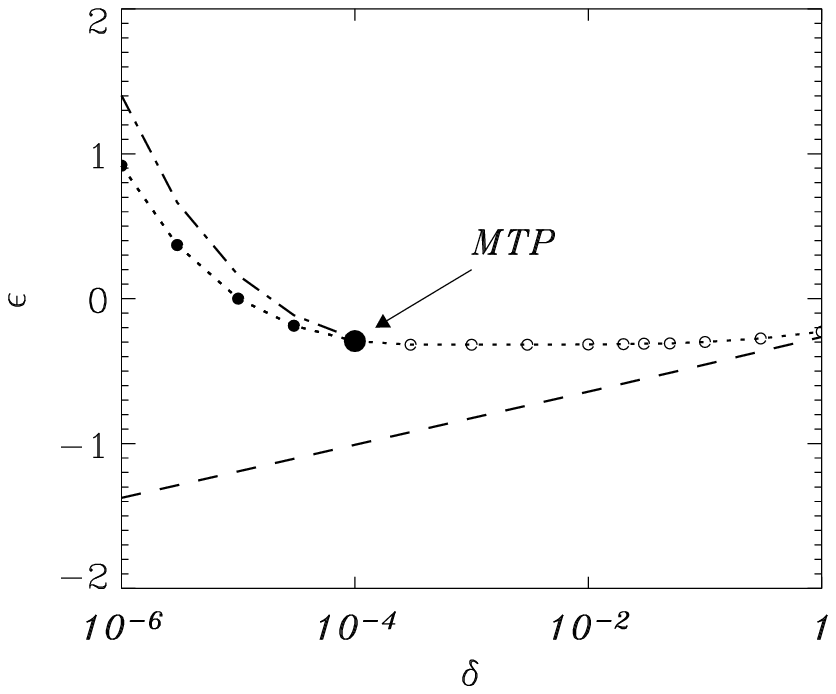}}
 \caption{Phase diagram for the Self-Gravitating Ring model, see
 Eqs.~(\ref{eqn:SGR-H}) and~(\ref{eqn:SGR-Hbis}).
 The dash-dotted line represents the upper energy of the metastable
  low energy phase, the dashed line is instead the lower energy of
  the metastable high energy phase. They bound the region of phase coexistence.
  The dotted line joining the points
  is the phase transition line, which changes order at the microcanonical
  tricritical point (MTP). The filled (open) circles belong to the first (second)
  order phase transition line.}
\label{fig:SGR-MCE}
\end{figure}

The phase transition changes from second to first order at a
tricritical point, whose location is not the same in the two
ensembles. In Fig.~\ref{fig:SGR-MCE}, we represent the
$\delta$-dependence of the critical energy $\veps_t$ and the
so-called coexistence region. The transition line separating the low
energy and high energy phases has a tricritical point: phase
transitions are first order at small $\delta$ and second order at
large $\delta$.

\subsubsection{Dipolar interactions in a ferromagnet}
\label{ramaz}

In Sec.~\ref{physicalexamples}, where we have discussed several
physical examples of long-range interacting systems, we have
remarked that dipolar interactions are marginal as far as the
long-range character of the interaction is concerned. Indeed, in
this case, the exponent governing the decay of the interaction
potential is $\alpha=d=3$. In spite of this, we will show here that
magnetic dipolar interactions offer a relatively simple way to
observe a direct manifestation of long-range interactions in samples
of suitable shape.

Generally, Heisenberg exchange interaction between electronic spins
is much higher (usually a few order of magnitudes) than magnetic
dipolar energy, and therefore the only role played by the dipolar
interaction is the introduction of anisotropies of the spontaneous
magnetization. On the other hand, magnetic ordering in nuclear
spins, governed by dipolar interaction alone, would require
experiments performed at nanokelvin temperatures \cite{ojamodphys}.
Here, we describe an example in which the dipolar interaction
between electronic spins acquires, through the careful preparation
of the sample, an effective strength in a suitable range of
temperatures, such that its long-range character gives rise to an
interesting dynamical effect related to a phase
transition~\cite{CampaPRB}.

Compounds of the type $({\rm C}_\nu {\rm H}_{2\nu+1}{\rm NH}_3)_2
{\rm CuCl}_4$ have been considered in
Refs.~\cite{sievers1,sievers2,sievers3}, where the interest was the
observation of intrinsic localized modes (discrete breathers). These
compounds, organized in a face centered orthorombic crystal, are
layered spin structures, in which the weak magnetic interlayer
interaction is antiferromagnetic for $\nu > 1$ and ferromagnetic for
$\nu = 1$. The relevant variables are the $s=1/2$ spins of the ${\rm
Cu}^{2+}$ ions, that are placed in two-dimensional layers. The hard
axis of the magnetic interaction is orthogonal to the layers, that
therefore are the easy planes. However, also the in-plane
interaction is anisotropic: an easy and a ``second easy" axis can be
determined. The cited works focused on the antiferromagnetic case
$\nu = 2$. Here, we are interested in the ferromagnetic case $\nu =
1$, i.e. to the compound $({\rm C}_1 {\rm H}_3{\rm NH}_3)_2{\rm
CuCl}_4$, called bis(Methylammonium) tetrachloro-copper
\cite{Chapuis}.

When writing the Hamiltonian of the system, we adopt a coordinate
frame in which the hard axis is the $x$-axis, while the $z$-axis and
the $y$-axis are the easy and ``second easy" axis belonging to the
two-dimensional layers, respectively. The Hamiltonian reads
therefore
\begin{equation}
H = -W \sum_{I,<i,j>} \left( s_{Ii}^z s_{Ij}^z + \eta s_{Ii}^y
s_{Ij}^y +\xi s_{Ii}^x s_{Ij}^x \right) -w \sum_{I,<i,j>}
\overrightarrow {s}_{Ii} \cdot \overrightarrow{s}_{I+1,j} +
\sum_{Ii\neq Jj} \frac{2\mu_B^2}{r^3}\left( \overrightarrow{s}_{Ii}
\cdot \overrightarrow{s}_{Jj} -3 \frac{(\overrightarrow{s}_{Ii}
\cdot \overrightarrow{r}) (\overrightarrow{s}_{Jj} \cdot
\overrightarrow{r})}{r^2}\right) \, , \label{hamculayers}
\end{equation}
where the first, second and third sum represents the intralayer
exchange interaction, the interlayer exchange interaction and the
dipolar interaction, respectively. The capital indices number the
layers, while the lowercase indices denote the spins. In the first
sum $<i,j>$ refers to nearest-neighbors within the same layer and in
the second sum to nearest-neighbors in adjacent layers; the last sum
is extended over all pairs of spins of the system. The intralayer
exchange constant $W$ is much larger than the interlayer one $w$,
for example $W \simeq 10^4 w$ \cite{dupas}. The parameters $\xi$ and
$\eta$, both smaller than $1$ are related to the out-of-plane and
in-plane anisotropies. In the dipolar interaction, $\mu_B$ is Bohr's
magneton and $r$ is the modulus of $\overrightarrow{r}$, the vector
between the sites of the spins $\overrightarrow{s}_{Ii}$ and
$\overrightarrow{s}_{Jj}$.

The large $W/w$ ratio determines the existence, for a given shape of
the sample, of a temperature range in which all the spins of a
single layer are ferromagnetically ordered, such that the spin
vector $\overrightarrow{s}_{Ii}$ can be considered as independent of
the index $i$, $\overrightarrow{s}_{I}\equiv
\overrightarrow{s}_{Ii}$, while full three-dimensional ordering has
not yet been reached. This situation arises if the temperature is
below the single layer ordering temperature, of the order of $W$,
and if in addition the number $n$ of spins in each single layer is
such that $nw < W$. Under such conditions the thermodynamic and
dynamical properties of the system will be determined by the
interlayer exchange constant $w$ and by the long-range dipolar
interaction. Exploiting the fact that all spins in a layer are
ordered, the Hamiltonian can be cast in a one-dimensional form. To
this purpose, one has to use a procedure employed in the treatment
of dipolar forces, in which the short-range contribution (here
including the interaction between nearest-neighbors) and the
long-range contribution are treated separately
\cite{LandauLifshitz}. The latter contribution gives rise to
shape-dependent terms. Here, we consider a rod-shaped sample, with
short sides along the out-of-plane $x$-axis and the in-plane
$y$-axis, and long sides along the in-plane $z$-axis. With this
shape, the long-range dipolar interaction produces an interaction
term in the Hamiltonian which is typical of an ellipsoidal shape
\cite{LandauLifshitz,akhiezer}. The magnetization in the sample
varies only along the $x$-axis. If we denote by $N$ the number of
layers of the sample, the Hamiltonian is then transformed to
\cite{englishprb,akhiezer}
\begin{equation}
\label{hamillayers1d} H_{N} =n\left[B_x\sum_{J=1}^{N}
\left(s_J^x\right)^2 + B_y\sum_{J=1}^{N} \left(s_J^y\right)^2 -
2\omega_{ex}\sum_{J=1}^{N-1} \left(s_J^y s_{J+1}^y + s_J^z
s_{J+1}^z\right) - \frac{\omega_M}{N} \left( \sum_{J=1}^{N} s_J^z
\right)^2 + \frac{\omega_M}{2N} \left(\sum_{J=1}^{N} s_J^y \right)^2
\right] \, .
\end{equation}
The first two sums come from the first sum in
Eq.~(\ref{hamculayers}), and describe the intralayer exchange
interaction. Here $B_x=4W(1-\xi)$ and $B_y=4W(1-\eta)$. Since for
this compound the out-of-plane anisotropy is much larger than the
in-plane one, we have that $B_x\gg B_y$; a constant additive term
involving $W$ has been neglected. The third sum in
Eq.~(\ref{hamillayers1d}) is a combination of the interlayer
exchange interaction (second sum in (\ref{hamculayers})) and the
nearest-neighbor interlayer dipolar interaction, while the
intralayer nearest-neighbor dipolar interaction produces a constant
term that has been neglected. In this sum, $\omega_{ex}$ is given by
\begin{equation}
\label{omegaeffective}
\omega_{ex}=2w-\frac{2\mu_B^2}{r_b^3}\left(2-\frac{3r_a^2}{2r_b^2}\right),
\end{equation}
where $r_a$ and $r_b$ are the distances between nearest-neighbor
spins within the same layers and adjacent layers, respectively. The
last two mean-field terms in Eq.~(\ref{hamillayers1d}) come from the
long-range part of the dipolar interaction. This contribution is
proportional to a combination of the square of the components of the
magnetization density vector, with the coefficients of the
combination depending on sample shape. In this case, where the
magnetization varies along the $x$-axis, one has to consider the
average magnetization density vector \cite{akhiezer}. In these
mean-field terms, $\omega_M$ is given by
$\omega_M=(4\pi/3)(2\mu_B^2/v_0)$, where $v_0$ is the volume of the
unit cell of the lattice. In the effective Hamiltonian
(\ref{hamillayers1d}), we have neglected all terms including the $x$
component of the spins which emerge from dipolar or exchange
interlayer forces, since $B_x$ is much larger than $B_y$, $\omega_M$
and $\omega_{ex}$.

Hamiltonian (\ref{hamillayers1d}) governs the dynamics of the vector
spin ${\overrightarrow{s}_J}$ through the torque equation
\begin{equation}
\hbar \frac{\dd {\overrightarrow{s}_J}}{\dd
t}={\overrightarrow{s}_J} \times {\overrightarrow{{\cal H}}_J} \, ,
\label{torqueequation}
\end{equation}
where ${\overrightarrow{{\cal H}}_J}=-{\partial
H_{N}}/{\partial{\overrightarrow{s}_J}}$ is an effective magnetic
field acting on the spin ${\overrightarrow{s}_J}$. The numerical
results that we are going to present shortly are based on the
integration of the last equation. To make a connection with the
analytical results, we introduce an approximation in the
Hamiltonian. We define $\overrightarrow{S}_J\equiv 2
\overrightarrow{s}_J$, $S_J^y=\sqrt{1-(S_J^x)^2}\sin\theta_J$ and
$S_J^z=\sqrt{1-(S_J^x)^2}\cos\theta_J$, which yields the equations
of motion in terms of the angular variables $\theta_J$ and $S_J^x$.
Noting that $B_x$ is much larger than all the other parameters,
which implies that $\sqrt{1-(S_J^x)^2}\approx 1$, the equations of
motion are simplified as follows
\begin{equation}
\hbar \frac{d \theta_J}{dt}=-nB_xS_J^x \,~.
\label{equationsmotionsx}
\end{equation}
This dynamics is consistent with the following effective
Hamiltonian, which includes only the angles $\theta_J$'s, and their
time derivatives
\begin{equation}
H_{N}^\prime=\frac{2 H_{N}}{n\omega_M}= \frac{1}{2}\sum_{J=1}^{N}
\left[ \frac{\dd \theta_J}{\dd
t}\right]^2-\frac{\omega_{ex}}{\omega_M} \sum_{J=1}^{N-1}
\cos(\theta_{J+1}-\theta_J) + \frac{B_y}{2\omega_M}
\sum_{J=1}^{N}\sin^2\theta_{J} -\frac{1}{2N}\left( \sum_{J=1}^{N}
\cos\theta_J\right)^2 + \frac{1}{4N}\left( \sum_{J=1}^{N}
\sin\theta_J\right)^2 \, , \label{hamilapproxlayers}
\end{equation}
where we have introduced a dimensionless time via the transformation
$t\rightarrow tn\sqrt{B_x\omega_M}/\hbar$. We note the similarity of
this Hamiltonian with the Hamiltonian (\ref{hamilxy}) of the chain
of rotators coupled by a nearest-neighbor interaction and a
mean-field one studied in subsection \ref{xylongplusshort}. With
respect to Hamiltonian (\ref{hamilxy}), the one given in
Eq.~(\ref{hamilapproxlayers}) is not invariant under global
rotations, because of the presence of the term proportional to $B_y$
and of the difference between the coefficients of the two mean-field
terms. However, if $B_y>0$, which is the case experimentally, these
differences do not change qualitatively  \cite{TheseLeonardo} the
phase diagram obtained in subsection \ref{xylongplusshort}.

Let us now discuss the dynamical effect that could in principle be
observed in experiments with this type of samples. We have shown
that the XY model with nearest-neighbor and mean-field interactions
has, in particular, a second order ferromagnetic phase transition
for all positive values of the nearest-neighbor coupling. We have
also shown that, below the critical energy, there is another energy
threshold, below which ergodicity is broken, with an inaccessible
range of values of the magnetization around $0$ (contrary to the
case $K > 0$ of subsection \ref{xylongplusshort}, the
inaccessibility of a magnetization range around $0$ has to be
considered a breaking of ergodicity, since now the lack of
invariance under global rotation can forbid the passage from $m_x>0$
to $m_x < 0$). For the particular values of the parameters in
Hamiltonian (\ref{hamilapproxlayers}), one finds \cite{CampaPRB}
that the critical energy is $\veps_c \simeq 0.376$, and the
threshold below which there is breaking of ergodicity is
$\veps=-0.3$. Besides that, because of the lack of global rotational
invariance, the spontaneous magnetization has only two possible
directions, i.e., $\theta_J=0$ and $\theta_J = \pi$. If we consider
the dynamics of such a system, at an energy between the threshold
for ergodicity breaking and the critical energy, and with a finite
number of degrees of freedom, we should therefore observe that the
modulus of the magnetization will fluctuate around that of the
spontaneous magnetization, and from time to time will flip between
the two directions of the magnetization. We expect that the flips
will be more and more rare as the system size increases and
approaching the ergodicity breaking threshold \cite{celardo}. The
sudden flips are due to the fact that, contrary to short-range
systems, the formation of domains with different directions of the
magnetization is not possible.

\begin{figure}[htbp]
\begin{center}
\resizebox{0.4\textwidth}{!}{\includegraphics{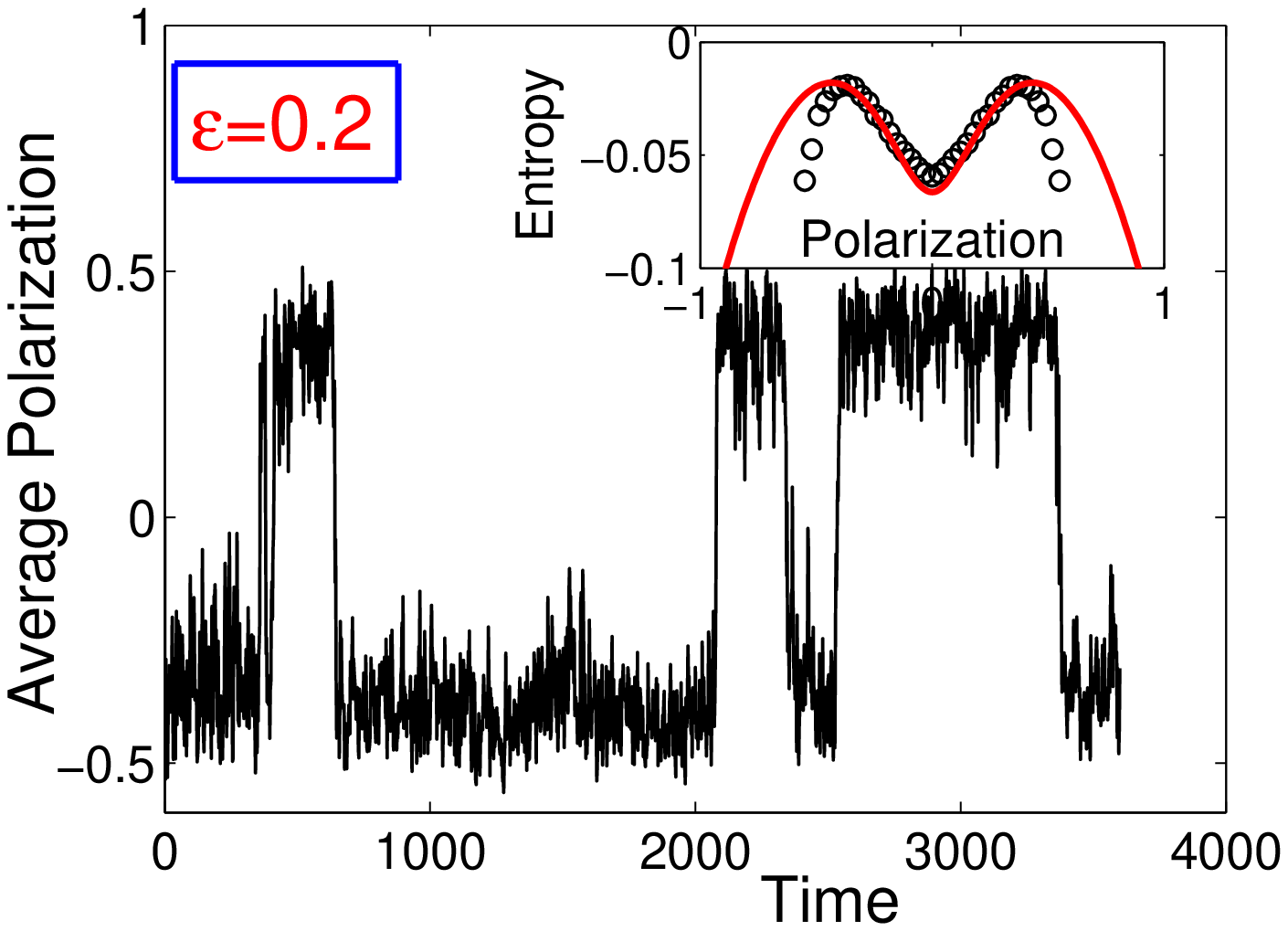}}
\resizebox{0.4\textwidth}{!}{\includegraphics{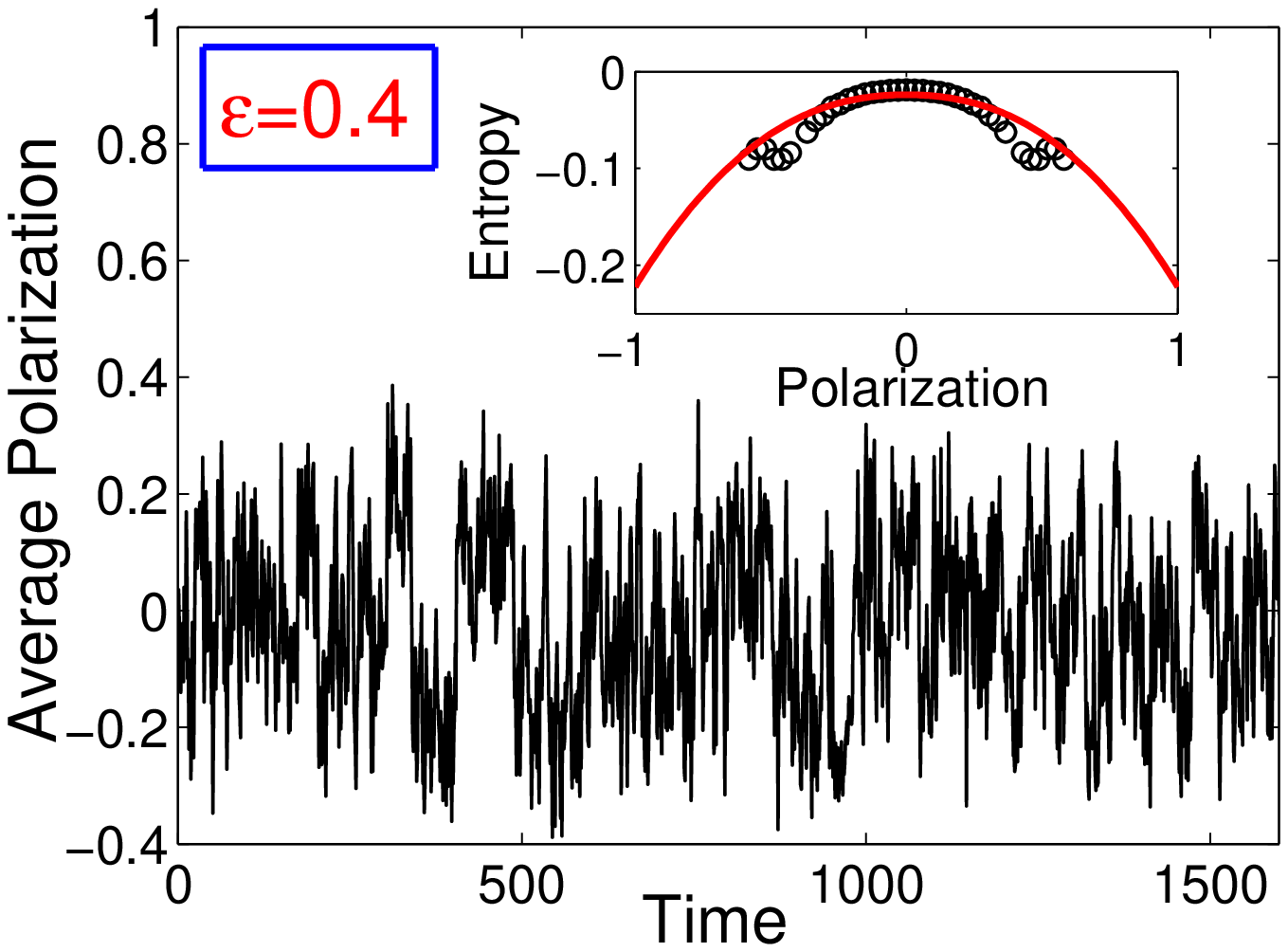}}
\end{center}
\caption{Numerical simulations of the one dimensional spin chain
model, Eqs.~(\ref{hamillayers1d}) and (\ref{torqueequation}), with
two different energies: $\veps=0.2$ (left panel) and $\veps=0.4$
(right panel), i.e. below and above the second order phase
transition energy $\veps_c \simeq 0.376$ predicted for Hamiltonian
(\ref{hamilapproxlayers}). The number of layers in the simulations
is $N=100$. The insets show the corresponding entropy curves (solid
lines) obtained for Hamiltonian (\ref{hamilapproxlayers}) and the
data obtained from the probability distribution function of the
magnetization (open circles) for Hamiltonian (\ref{hamillayers1d}).}
\label{resultslayers}
\end{figure}

Numerical experiments have been performed using Hamiltonian
(\ref{hamillayers1d}) and the torque equation (\ref{torqueequation})
\cite{CampaPRB}. We here present results concerning the two energies
$\veps = 0.2$ and $\veps = 0.4$ (in the adimensional values computed
after the transformation to Hamiltonian (\ref{hamilapproxlayers})).
The first value is below the critical energy but still far enough
from the ergodicity breaking threshold to observe magnetization
flips on a reasonable time scale. The second value is above the
critical energy, and the system should fluctuate around zero
magnetization. In Figs.~\ref{resultslayers}, we show the results of
the numerical simulations that confirm the expectations.

From the numerical data, it is also possible to obtain the
probability distribution function for the magnetization, shown in
the insets of Figs.~\ref{resultslayers} by the open circles. The
logarithm  of the probability distributions obtained numerically is
compared with that computed analytically at the corresponding energy
values on the basis of Hamiltonian (\ref{hamilapproxlayers}). The
comparison is then made between the numerical distribution functions
obtained by direct simulations of Hamiltonian (\ref{hamillayers1d})
and the expression $\tilde{s}(\veps,x)$ given by
Eq.~(\ref{entropyxyminmax1}), obtained for Hamiltonian
(\ref{hamilapproxlayers}). The agreement is good although not
perfect. The effect which leads to disagreement is twofold. First,
one should have obtained the expression of $\tilde{s}(\veps,x)$ for
Hamiltonian (\ref{hamillayers1d}), which is in principle possible.
%, which is work in progress \cite{TheseLeonardo}.
%There will be a slight difference in the integral operator (\ref{tranferoperator}), due to
%the presence of the $\sin^2 \theta_J$ term in Hamiltonian (\ref{hamilapproxlayers}),
%and therefore the numerical values of the maximum eigenvalue $\lambda$ will be
%slightly different.
Second, we have taken $x$ as playing the role of the spontaneous
magnetization $m$. However, we have already emphasized several times
that the parameter $x$ coming from the Hubbard-Stratonovich
transformation coincides with the spontaneous magnetization only at
equilibrium. The reasonably good agreement that we obtain justifies
a posteriori this identification, which is valid only close to
equilibrium.

In summary, our main purpose here was to see that a system which is
realizable in the laboratory can be used to observe the
peculiarities predicted for long-range interacting systems. We have
emphasized the importance of sample shape. For instance: for
spherical or approximately spherical, samples, spin flips will not
be observed, since the mean-field terms in the Hamiltonian will not
introduce any anisotropy.

\section{Conclusions}
\label{conclusions}

\subsection{Summary}

It should be now clear that the dynamics and thermodynamics of systems
with long-range interactions is a rich and fascinating field. It has
been shown how, changing the range of the interactions, deeply renew
several aspects of statistical mechanics. We have encountered many
unusual properties: ensemble inequivalence, negative specific heat and
negative susceptibility, ergodicity breaking, quasi-stationary states,
algebraic relaxation and anomalous diffusion. Rather than being
contradictory with standard statistical mechanics, they extend its
domain of validity.
It should be remarked that the analysis of statistical ensembles for
long-range systems privileges the use of microcanonical ensemble, since all
canonical macrostates can be realized in the microcanonical ensemble, but the
converse is not true: this is the essence of ensemble inequivalence.
These properties are of course not found in all
models and for any value of the thermodynamic
parameters. Table~\ref{table:models} summarizes the main features of
the models studied in this review.

\begin{table}[htbp]
\centering
\begin{tabular}
{|c|c|c|c|c|c|c|c|} \hline Model&Variable&Ensemble&Negative&
Negative&Ergodicity&Computable&Section\\
&&Inequivalence&specific heat&susceptibility&Breaking&
Entropy&\\\hline BEG &D&Y&Y&&Y&Y&\ref{begmodel}\\\hline 3 states
Potts&D&Y&Y&N&N&Y&\ref{Pottsmodel}\\\hline  Ising L+S
&D&Y&Y&&Y&Y&\ref{Isinglongplusshort}\\\hline  $\alpha$-Ising
&D&Y&N&&&Y&\ref{alphaIsing}\\\hline HMF
&C&N&N&&&Y&\ref{exemple_HMF}\\\hline XY L+S
&C&Y&Y&&Y&Y&\ref{xylongplusshort}\\\hline $\alpha$-HMF
&C&N&N&&&&\ref{alphaxymodel}\\\hline Generalized HMF
&C&Y&Y&&Y&Y&\ref{generalizedHMF}\\\hline Mean-Field $\phi^4$
&C&Y&N&Y&Y&Y&\ref{modelphi4}\\\hline Colson-Bonifacio
&C&N&N&N&N&Y&\ref{Colson-Bonifaciomodel}\\\hline SGR
&C&Y&Y&N&Y&Y&\ref{taka}\\\hline
\end{tabular}
\caption{Summary of the important properties of the different models
studied in this review. C for continuous and D for discrete. Y for
Yes, N for No, while the slot is left empty when the answer is not
known. L+S means long-range plus short-range
interactions.}\label{table:models}
\end{table}

We have mostly restricted our analysis to mean-field models. However,
we have shown that the properties that we have discussed extend also
to non mean-field interactions, as presented in
Sec.~\ref{perspectives}. Recently, new analytical methods have been developed to
solve models with long-range interactions at equilibrium: they all aim
at the direct derivation of the entropy. We have in particular
presented in full detail the large deviation method in
Sec.~\ref{Methode_generale}. Relaxation to equilibrium is a key topic
in systems with long-range interactions, which historically began with
the realization that self-gravitating systems cannot converge to
statistical equilibrium. By considering a much simpler model, that
nevertheless is representative of long-range interacting systems, we
have shown that relaxation to equilibrium proceeds in two steps. On a
fast time scale, one observes relaxation to quasi-stationary states
that are successfully described using a theoretical approach proposed
by Lynden-Bell. On a longer time scale, one indeed observes relaxation
to Boltzmann-Gibbs equilibrium. In connection with this complex
relaxation to equilibrium, we have extensively discussed kinetic
equations for systems with long-range interactions (Klimontovich,
Vlasov, Lenard-Balescu, Fokker-Planck). All this is presented in
Sec.~\ref{outofequilibrium}.

\subsection{Open problems}

Before closing this review, we would like to briefly comment on a few topics
that we have discussed but which would need a wider treatment
since it is clear that the present review opens a ``Pandora box''.

Another source of negative specific heat which has not discussed at
all in this review is thermodynamic instability. Even if a system is
short-range, attractive and the interaction is regularized at short
distances, energy density can be unbounded from below and both
potential energy and kinetic energy will increase as a consequence of
the formation of a cluster. This phenomenon has been discussed in a
series of papers beginning with
Refs.~\cite{Compagner89,PoschNarnhoferThirring1990}.

For two-dimensional and geophysical flows, Chavanis~\cite{chavanism},
as well as Bouchet and
Venaille~\cite{venaillebouchet}, obtain different equilibrium states
depending on the shape of the domain.  Following a classification of
phase transitions previously derived by Barr\'e and
Bouchet~\cite{julienfreddyjstat}, they exhibit the first example of
bicritical points and second order azeotropy for long-range
interacting systems. There is a close connection between
two-dimensional fluid dynamics and two-dimensional electron
plasmas. In this latter context, experiments have shown the relaxation
towards different quasi stationary
states~\cite{HuangDriscoll,Fine,Kiwamoto}, and recently theoretical
frameworks have been proposed~\cite{Kawahara, brands}. Electron plasmas could
also constitute experimental setups to observe in the future the new
type of phase transitions predicted in Ref.~\cite{julienfreddyjstat}.

While the initial violent relaxation has been successfully
characterized, the slow collisional process that leads to the final
equilibrium is still an open field of research. Chavanis has carefully
investigated this problem and has proposed a rich kinetic theory
approach~\cite{ChavanisAssisi}. As an application, collisional
relaxation has been numerically investigated for the HMF model in
Ref.~\cite{CampaChavanis}.

Concerning self-gravitating systems, one should quote the remarkable
observation of a sequence of transient states in the long-term stellar
dynamical evolution.  Density distributions of such states are well
fitted by stellar polytropes~\cite{TaruyaSakagamiPRL,TaruyaSakagami}.
Similar observations have been reported for the HMF model in
Ref.~\cite{CampaChavanis}. Recently, the problem of the
robustness of quasi-stationary states with respect to noise has been
investigated~\cite{chavanisn}.
Gravitational clustering in finite and infinite systems has been
numerically investigated by Joyce, Sylos Labini and
collaborators~\cite{SylosLabiniAssisi}. These authors have tried to
assess if simulations performed with a finite number of particles
reproduce the mean-field Vlasov limit and to understand the universal
features of halo structures.

An interesting direction of investigation is the one followed by
Barr\'e and Gon\c calves, that considers models on random graphs. Using
the large deviation cavity method, analytical solutions of the Potts
spins systems on a random $k$-regular graph have been recently derived
in both the canonical and microcanonical ensembles. The analytical
solution, confirmed by numerical Metropolis and Creutz simulations,
clearly demonstrate the presence of a region with negative specific
heat and, consequently, ensemble inequivalence between canonical and
microcanonical ensembles~\cite{BarreGoncalves}.

An alternative method has been used to describe quasi-stationary
states. These states, according to the method of non extensive
statistics proposed by Tsallis~\cite{Tsallisjsp}, should be
maximum entropy states of a generalized Boltzmann entropy defined as
$S_{q}=k_B\ ({1-\sum_ip_i^{q}})/({q-1})$, where $p_i$ is the
normalized probability of state~$i$. As correctly pointed out by
Tsallis and coworkers~\cite{Tsallis,lrt2002}, the presence of
quasi-stationary states is tightly linked to the non-commutability of
the infinite time limit with the thermodynamic limit for systems with
long-range interactions.  Velocity distributions that maximize Tsallis
entropy $S_q$ are $q$-Gaussian defined by $G_q(x) = A(1-(1 - q)\eta
x^2)^{1/1-q}$, where $A$ is a normalization constant and $\eta$
controls the width of the distribution. These distributions do not
convincingly fit numerical data of the HMF
model~\cite{Tsallis,lrt2002}. Moreover, no theoretical approach has
been proposed which leads to the derivation of such distributions for
a specific case, like the HMF model. For this reason, this theory
has not shown any predictive power in this field.

Although unsuccessful to explain quasi-stationary states, Tsallis
approach has led to the introduction of interesting models and
concepts. A relevant model, that we also discuss in
Sec.~\ref{alphaxymodel}, is the $\alpha$-HMF model, which has been
introduced in Ref.~\cite{Anteneodo97} to discuss the behavior of the
maximal Lyapunov exponent in systems with long-range interactions:
this is still an open and interesting question.  Algebraic decay of
time correlations and anomalous diffusion within the HMF model were
discovered by Rapisarda and
coworkers~\cite{lrt2001,Pluchino,Correlation,Correlationb}. However,
the approach to this problem developed in
Secs.~\ref{fokkerplancksubsection} and~\ref{temporalcorrelations} is
able to explain all numerical observations without resorting to non
extensive statistics.  Moreover, although motivated by non extensive
statistics, the work in Ref.~\cite{celiabis} concludes that
quasi-stationary states, originally looked at by Rapisarda and
coworkers, do not show anomalous diffusion, as also previously
reported in Ref.~\cite{yamaPRE}.

\subsection{Perspectives}

The study of long-range interactions has certainly overcome the
pioneering era where it has been important to become acquainted with
basic facts and concepts that looked odd initially, like negative
specific heat. Having passed to a more mature period, this field,
besides strengthening its foundations, should incorporate methods and
tools developed in other disciplines and look towards the realizations
of well devised experiments that could lead to the observations of the
effects predicted by the theory.

From the mathematical point of view, one is faced to interesting and
difficult problems for what concerns both equilibrium statistical
mechanics and out-of-equilibrium dynamics. It would be important to
develop large deviation methods in order to allow the treatment of
systems with slowly decaying interactions as was done for instance
for the $\alpha$-Ising model~\cite{bbdrjstatphys} and of systems
with both short and long-range interactions. For what concerns the
dynamical properties, several significant advances have been
recently obtained in kinetic theory concerning the study of the
solutions of Vlasov-Poisson and Vlasov-Maxwell systems. Both the
difficulties of incorporating in the theory the singularity of the
Coulomb or gravitational potential and the enlargement of the
treatable initial conditions have been addressed (see
Ref.~\cite{LesHouches2009}).

When going to systems where the interaction is long-range, but not
mean-field, the difficulty of performing numerical simulations is
related to the fact that algorithms are intrinsically of order $N^2$
for each time step, where $N$ is the number of particles.  Because
of this, for gravitational systems, dedicated computers have been
built (e.g. GRAPE in Japan \cite{grape5taka}) and sophisticated
numerical codes have been developed~\cite{aaserth}. The issue of
devising better algorithms is central for the improvements of the
simulations and has not yet been seriously faced.

We have contributed with this review to convey the idea that many of
the effects that were previously attributed to gravitational systems
appear also in other systems, as was early stressed in two-dimensional
hydrodynamics~\cite{Miller90,ChavanisSommeriaRobert}.
We have already mentioned in Sec.~\ref{physicalexamples}
two-dimensional hydrodynamics, two-dimensional elasticity, charged and
dipolar systems, small systems. We would like to concentrate here on
three physical examples that are more likely to lead in the near
future to experimental realizations: the free electron laser, cold
atoms and lattice dipolar systems.

The cooperative effect leading to coherent emission of radiation in
the Free Electron Laser (FEL) has been described in
Sec.~\ref{Colson-Bonifaciomodel}, using an analogy with phase
transitions. In a realistic setting, convergence to Boltzmann-Gibbs
equilibrium will never occur because of the extremely elevated number
of electrons, compared to the length of the wiggler. That's why it is
highly probable that FELs will always operate in regions of parameters
dominated by quasi-stationary states. As we know from the analysis of
such states in the HMF model, they show phase transitions from
homogeneous to inhomogeneous macroscopic
distributions (see Sec.~\ref{Lyndenbellentropy}). Such transitions, together with
the characterization of quasi-stationary states, could be the object of
an experimental study with linear FELs~\cite{FELPRE}.

Cold atomic systems offer a promising laboratory for testing the
predictions of the statistical mechanics of systems with long-range
interactions. These systems range from ionized atoms confined by an
external potential which interact via unscreened Coulomb repulsion, to
neutral atoms that interact by means of Van der Waals type
forces. Additionally, dipolar gases have been recently the subject of
extensive experimental studies~\cite{MorigiAssisi}. Initially, some
attention had been paid to the possible realization of true
gravitational-like classical interaction induced by external laser
fields~\cite{Kurizki}. However, this hope has failed until now due to
technical difficulties. When the wave-like behavior of matter can be
neglected, leading to a semi--classical approximation, and laser light
is far from resonance, a fully Hamiltonian (microcanonical) treatment
of Bose-Einstein condensates is viable. Moreover, ensemble
inequivalence for bosonic gases is reminiscent of the
ensemble inequivalence encountered for long-range interacting
systems~\cite{Holtauss,wilkens,Ziff}. However, it has been later realized that
this type of ensemble inequivalence is of a different nature and
should be analyzed in itself. A potentially fruitful seed has
recently emerged in the study of cold atoms in optical high-Q
cavities. For these systems, the interaction between two atoms is
independent of the distance, because they are coupled via the
back-scattering of photons within the cavity. The interaction can lead
to collective instabilities and self-organization phenomena, known as
CARL~\cite{CARL,Courteille,LesHouches2009}. Models of the CARL
phenomenon have strong similarities with those of FEL. Therefore, we
expect that the study of phase transitions and of quasi-stationary
states could be realized also for these laboratory systems.

We finally argue that dipolar condensed matter systems should be
seriously considered to check some features of long-range
interactions. Indeed, it has been remarked since long time that
composites like $HoRhB_4$, where dipolar interactions dominate over
Heisenberg exchange interactions, have a mean-field critical
behavior~\cite{Ott}. More recently, such results have been extended to diluted
models which simulate the effect of disorder, making the
Sherrington-Kirpatrick spin-glass behavior experimentally
accessible~\cite{Ocio}. The importance of long-range dipolar interactions
has been also recognized for pyrochlore lattice
structures, that show the strongly degenerate and frustrated ``spin
ice'' phase~\cite{Bramwell01}.

\medskip

In the final section of this review, we have briefly mentioned
experimental applications, because we think that theoretical results
obtained in the study of systems with long-range interactions have
nowadays reached a level of understanding that might allow, in the
near future, the experimental verifications of curious effects like
negative specific heat, ergodicity breaking, quasi-stationary states.

\section*{Acknowledgements}

We would like to warmly thank our collaborators Mickael Antoni,
Andrea Antoniazzi, Romain Bachelard, Julien Barr\'e, Fausto
Borgonovi, Freddy Bouchet, Francesco Califano, Gian-Luca Celardo,
Cristel Chandre, Pierre-Henri Chavanis, Giovanni De Ninno, Yves
Elskens, Duccio Fanelli, Marie-Christine Firpo, Andrea Giansanti,
Dieter Gross, Alessio Guarino, Haye Hinrichsen, Peter Holdsworth,
Ramaz Khomeriki, Tetsuro Konishi, Hiroko Koyama, Raman Johal, Vito
Latora, Xavier Leoncini, Stefano Lepri, Fran\c cois Leyvraz,
Leonardo Lori, Gianluca Morelli, Daniele Moroni, David Mukamel,
Andrea Rapisarda, Nir Schreiber, Luca Sguanci, Takayuki Tatekawa,
Alessandro Torcini, Hugo Touchette, Yoshi Yamaguchi. We also thank
Angel Alastuey for helpful discussions and for reading parts of the
manuscript. We thank for financial support the Galileo program of
the French-Italian University ``Study and control of models with a
large number of interacting particles'' and  the COFIN07-PRIN
program ``Statistical physics of strongly correlated systems at and
out of equilibrium'' of the Italian MIUR. Finally, SR and AC thank
the ENS Lyon for hospitality and financial support, and SR thanks
Universit\'e de Provence for hospitality and financial support.

\section*{Appendix A: Proof of min-max inequality}

Let us consider the function of two variables $f(x,y)$. Under quite
general conditions on $f$, the following inequality holds
\begin{equation}
\sup_x \inf_y f(x,y) \leq \inf_y \sup_x f(x,y)~.
\label{maxminrelation}
\end{equation}
Let us give a sketchy proof of this property, which is used in the
text to prove that microcanical entropy (\ref{entrminmax1}) is
always smaller or equal than canonical entropy
(\ref{canentrminmax}). Let us denote by $(x_1,y_1)$ the extremum
that satisfies the l.h.s. of (\ref{maxminrelation}), then
\begin{equation}
f(x_1,y_1) \leq f(x_1,y)~, \quad \quad \quad \quad \forall y~.
\end{equation}
Similarly, the extremum that satisfies the r.h.s., denoted
$(x_2,y_2)$ satisfies
\begin{equation}
f(x_2,y_2) \geq f(x,y_2)~, \quad \quad \quad \quad \forall x~.
\end{equation}
Then,
\begin{equation}
f(x_1,y_1) \leq f(x_1,y_2) \leq f(x_2,y_2)~,
\end{equation}
which proves inequality (\ref{maxminrelation}).

\section*{Appendix B: Evaluation of the Laplace integral outside the analyticity strip}

The microcanonical partition function in formula~(\ref{laplacea}) can
be expressed, using the Laplace representation of the Dirac $\delta$ function as follows
\begin{equation}\label{omeminmax}
\Omega (\veps,N) =\frac{1}{2\pi i} \int_{\beta -i\infty}^{\beta
+i\infty} \dd \lambda \, e^{N\lambda\veps}Z(\lambda,N) \, ,
\end{equation}
with $\beta>0$; this is the expression given in (\ref{laplacec}).
As explained in subsection \ref{minmaxsect},
we divide the integral in (\ref{omeminmax}) in three
intervals, defined by $\lambda_I < -\delta$, $-\delta < \lambda_I < \delta$
and $\lambda_I > \delta$, respectively, with $0 < \delta < \Delta$.
Here we show that the contribution to the integral
in $\lambda$ coming from values of $\lambda_I$ outside the
strip, i.e. for values of $\lambda_I$ with
$|\lambda_I| > \Delta$, is exponentially small in $N$.

Let us then consider first the value of $Z(\lambda,N)$ in the two external
intervals, i.e., for $|\lambda_I|> \Delta$. We have
\begin{equation}\label{zminmaxim}
Z(\beta +i\lambda_I,N) = \sum_{\{S_1,\ldots,S_N\}} \exp \left\{
-\beta H(\{ S_i \})\right\} \exp \left\{ -i\lambda_I H(\{S_i
\})\right\} \, .
\end{equation}
We see that this expression is proportional to the canonical
expectation value $\langle \exp \left( -i\lambda_I H\right)
\rangle$, that we expect to be exponentially small for
large $N$. We confirm this expectation rewriting the last expression
as
\begin{equation}\label{zminmaxim2}
Z(\beta +i\lambda_I,N) = N\int \dd \varepsilon \exp \left( -N\left[\beta \varepsilon
+i\lambda_I \varepsilon -s(\varepsilon)\right]\right) \, ,
\end{equation}
where we have used the expression of the canonical partition
in terms of the microcanonical entropy as in (\ref{addedforapp}).
This is an integral with a
large phase. For $N$ going to infinity, its value will be determined
by the value of the integrand for $\varepsilon$ equal to the integration
extremes and to the values of the possible nonanalyticities of $s(\varepsilon)$,
all denoted by $\veps_k$ (see, e.g., Ref. \cite{bender}). We then have
\begin{equation}\label{zminmaxim3}
Z(\beta +i\lambda_I,N) \sim N \sum_k c_k \exp \left( -N\left[\beta
\varepsilon_k +i\lambda_I \varepsilon_k -s(\varepsilon_k)\right]\right) \, ,
\end{equation}
where $c_k$ are coefficients that could in principle be determined. It
is clear that the successive integration over $\lambda_I$ in any one
of the two external intervals of integration will then give a
vanishing contribution, due to the very large oscillations. We are
then left with
\begin{equation}\label{omeminmax1}
\Omega (\veps,N) \stackrel{N\to+\infty}{\sim} \frac{1}{2\pi i}
\int_{\beta -i\delta}^{\beta +i\delta} \dd \lambda \,
e^{N\lambda\veps}Z(\lambda,N) \, ,
\end{equation}
which is the first equality given in Eq. (\ref{stripintegral}).

\section*{Appendix C: Differentiability of the function $\bar{\phi}(\lambda_1,\dots,\lambda_n)$}

We give a proof that the function
$\bar{\phi}(\lambda_1,\dots,\lambda_n)$, defined in
Eq.~(\ref{canpartmumin}), is differentiable, for real values of the
$\lambda_i$, when the functions $\mu_k(x)$ are given by sums of
one-particle functions
\begin{equation}
\label{defmuspecial} \mu_k(x)= \frac{1}{N}\sum_{i=1}^N g_k(q_i,p_i)
\,\,\,\,\,\,\,\,\, k=1,\dots,n \, .
\end{equation}
In this case, the canonical partition function is given by
\begin{equation}
\label{ztildespecial} \bar{Z}(\lambda_1,\dots,\lambda_n)  = \left(
\int \dd p \, \dd q \, \exp \left[ -\sum_{k=1}^n \lambda_k g_k(q,p)
\right] \right)^N \, .
\end{equation}
Therefore we have
\begin{equation}
\label{phitildespecial} \bar{\phi}(\lambda_1,\dots,\lambda_n) = -\ln
\left( \int \dd p \, \dd q \, \exp \left[ -\sum_{k=1}^n \lambda_k
g_k(q,p) \right] \right) \, .
\end{equation}
In the range of $\lambda_i$ where the integral in
Eq.~(\ref{phitildespecial}) is defined, $\bar{\phi}$ is continuous.
Besides, differentiating with respect to $\lambda_i$ we have:
\begin{equation}
\label{phiderspecial} \frac{\partial \tilde{\phi}}{\partial
\lambda_i} (\lambda_1,\dots,\lambda_n) = \frac{\displaystyle \int
\dd p \, \dd q \, g_i(q,p) \exp \left[ -\sum_{k=1}^n \lambda_k
g_k(q,p) \right]}{\displaystyle \int \dd p \, \dd q \, \exp \left[
-\sum_{k=1}^n \lambda_k g_k(q,p) \right]} \, .
\end{equation}
Also this function is continuous, under the only hypothesis that the
expectation value of the observable $g_i(q,p)$ is finite for all the
allowed values of the $\lambda_i$ in the canonical ensemble defined
by $\bar{Z}(\lambda_1,\dots,\lambda_n)$.

\section*{Appendix D: Autocorrelation of the fluctuations of the one-particle density}

Using the definition of the Fourier transform and
formula~(\ref{deffetdeltaf}), one gets
\begin{eqnarray}
\langle \delta f \left(k,p,0\right) \delta
f\left(k',p',0\right)\rangle &=& \int_0^{2\pi}\frac{\dd
\theta}{2\pi} \int_0^{2\pi}\frac{\dd \theta'}{2\pi}\
e^{-i(k\theta+k'\theta')} \langle \delta f \left(\theta,p,0\right)
\delta
f\left(\theta',p',0\right)\rangle  \\
&=& \int_0^{2\pi}\frac{\dd \theta}{2\pi} \int_0^{2\pi}\frac{\dd
\theta'}{2\pi}\ e^{-i(k\theta+k'\theta')} N \left[\langle   f_d
\left(\theta,p,0\right)f_d\left(\theta',p',0\right)\rangle
-f_0\left(p\right)f_0\left(p'\right)\right]\, . \label{aremplcaer}
\end{eqnarray}
The expression of the discrete density
function~(\ref{discretedensityfunction}) leads then to
\begin{eqnarray}
\langle f_d
\left(\theta,p,0\right)f_d\left(\theta',p',0\right)\rangle &=&
\frac{1}{N^2} \biggl\langle \displaystyle  \sum_{j=1}^N\delta
\left(\theta -\theta _{j}\right)\delta \left(p-p_{j} \right)\delta
\left(\theta
-\theta'\right)\delta \left(p-p' \right)\rangle \nonumber \\
&&\hskip 3truecm+ \sum_{i\neq j}\delta \left(\theta -\theta
_{j}\right)\delta \left(p-p_{j}\right)\delta \left(\theta' -\theta
_{i}\right)\delta \left(p'-p_{i}\right)\biggr\rangle  \\
&=& \frac{1}{N^2} \biggl[ \displaystyle  N \, \langle
f_d\left(\theta,p,0\right)\rangle\, \delta \left(\theta
-\theta'\right)\delta \left(p-p' \right)+
N(N-1)f_2(0,\theta,p,\theta',p')\biggr]  \\
&=& \frac{1}{N} f_0\left(p\right) \delta \left(\theta
-\theta'\right)\delta \left(p-p' \right)+
f_0\left(p\right)f_0\left(p'\right)+h_2(\theta,p,\theta',p',0),
\label{expressioncorrelation}
\end{eqnarray}
where we used the following definition of the correlation function
$h_2$
\begin{eqnarray}
f_2(\theta,p,\theta',p',0)&=&\langle \delta \left(\theta -\theta
_{j}\right)\delta \left(p-p_{j} \right) \delta \left(\theta' -\theta
_{i}\right)
\delta \left(p'-p_{i} \right)\rangle  \\
&=&\frac{N}{N-1}\left[
f_0\left(p\right)f_0\left(p'\right)+h_2(\theta,p,\theta',p')
\right] \, .
\end{eqnarray}
Substituting expression~(\ref{expressioncorrelation}) in
Eq.~(\ref{aremplcaer}), we find
\begin{eqnarray}\label{separf2theta}
\langle \delta f \left(k,p,0\right) \delta
f\left(k',p',0\right)\rangle&=& \int_0^{2\pi}\frac{\dd \theta}{2\pi}
\frac{f_0\left(p\right)}{2\pi}\ e^{-i(k+k')\theta} \delta \left(p-p'
\right)+\int_0^{2\pi}\frac{\dd \theta}{2\pi} \int_0^{2\pi}\frac{\dd
\theta'}{2\pi}\ N e^{-i(k\theta+k'\theta')}
h_2(\theta,p,\theta',p')  \\
&=&  \frac{f_0\left(p\right)}{2\pi}\ \delta_{k,-k'}
\delta \left(p-p' \right)+ \frac{1}{2\pi}\ \delta_{k,-k'}\, \mu(k,p,p') \label{separf2thetabis}\\
&=&\frac{\delta_{k,-k'}}{2\pi}\left[ {f_0(p)}\delta
(p-p')+\mu(k,p,p')\right] \, .
\end{eqnarray}
In the passage from (\ref{separf2theta}) to~(\ref{separf2thetabis}),
we have used the fact that $h_2$  depends only on the difference
$\theta-\theta'$. Besides, it decays rapidly to zero in a range
$(\theta-\theta')\sim 1/N$, so that $\mu(k,p,p')$ is of order $1$.

\section*{Appendix E: Derivation of the Fokker-Planck coefficients}

As mentioned in Sec.~\ref{fokkerplancksubsection}, the derivation of
the Fokker-Planck equation~(\ref{generalFokkerPlanckequation}) is
performed in the time range $1 \ll t \ll N$%, obtaining in this way
%the approximate expressions~(\ref{differentmoments1ici})
%and~(\ref{differentmoments2}), in which the coefficients actually do
%not depend on time
. Within this approximation, expressions
(\ref{momentsFokkerPlanck1}) and (\ref{momentsFokkerPlanck}) are
thus replaced by
\begin{eqnarray}
A(p,t)&=& \frac{1}{t}\langle \left( p(t)-p(0)\right)\rangle_{p(0)=p}
\label{appmomentsFokkerPlanck1} \\
B(p,t)&=& \frac{1}{t}\langle \left(
p(t)-p(0)\right)^2\rangle_{p(0)=p} \,
.\label{appmomentsFokkerPlanck}
\end{eqnarray}
Looking at Eq. (\ref{equation_p}), we see that, at the order $1/N$,
we need to compute the following averages
\begin{eqnarray}
&&\frac{1}{\sqrt{N}}\, \left\langle \frac{\partial \delta
v}{\partial \theta}(\theta(0)+p(0)t,t) \right\rangle
\label{averageformoment1a}
\end{eqnarray}
and\begin{eqnarray}\frac{1}{N}\, \left\langle \frac{\partial^2
\delta v} {\partial \theta^2}(\theta(0)+p(0)t_1,t_1) \frac{\partial
\delta v}{\partial \theta} (\theta(0)+p(0)t_2,t_2)\right\rangle
\label{averageformoment1b}
\end{eqnarray}
for the first moment, and
\begin{equation}
\frac{1}{N} \left\langle \frac{\partial \delta v}{\partial
\theta}(\theta(0)+p(0)t_1,t_1) \frac{\partial \delta v}{\partial
\theta}(\theta(0)+p(0)t_2,t_2)\right\rangle
\label{averageformoment2}
\end{equation}
for the second one. The fact that the position and the momentum of
the test-particle at time $0$ are given is specified inside the
dependence on $\theta$ of the derivatives of $\delta v$. The
condition of given initial angle and momentum of the test-particle
produces corrections of the averages in
Eqs.~(\ref{averageformoment1a}), (\ref{averageformoment1b}) and
(\ref{averageformoment2}), with respect to their unconditioned
values. These corrections being of order $1/\sqrt{N}$, one gets that
at order $1/N$ they have to be taken into account only for
Eq.~(\ref{averageformoment1a}). Note that the unconditioned value of
that equation vanishes.

Let us start with Eq. (\ref{averageformoment2}). To compute it, it
is necessary to rewrite Eq.~(\ref{correldeltaVfinal}) as
\begin{equation}
\left\langle  {\delta v}(k,\omega){\delta
v}(k',\omega')\right\rangle =2\pi^3
\delta_{k,-k'}\left(\delta_{k,1}+\delta_{k,-1}\right)\,
\frac{\delta(\omega+\omega')} {\left|\tilde
D(\omega,k)\right|^2}\,\int\dd p'\, f_0(p')\delta(\omega-p'k) \,
.\label{correldeltaVfinalbis}
\end{equation}
By the inverse Fourier-Laplace transform, we then have
\begin{eqnarray}
\left\langle \frac{\partial \delta v}{\partial
\theta}(\theta(0)+pt_1,t_1)\frac{\partial \delta v}{\partial
\theta}(\theta(0)+pt_2,t_2)\right\rangle
&=&\frac{1}{\left(2\pi\right)^2}\sum_{k=-\infty}^{+\infty}
\sum_{k'=-\infty}^{+\infty}\, \int_{-\infty}^{+\infty} \dd \omega \,
\int_{-\infty}^{+\infty} \dd \omega' \, (-kk')\
e^{ik(\theta(0)+pt_1) -i \omega t_1} \nonumber\\
&&\hskip 3truecm \times e^{ik'(\theta(0)+pt_2) - i\omega't_2}\,
\left\langle {\delta v}(k,\omega){\delta
v}(k',\omega')\right\rangle  \\
&=&\frac{\pi}{2}\sum_{k=-\infty}^{+\infty}\,
\!\!\!\!\int_{-\infty}^{+\infty}\!\!\!\!\!\!\!\!  \dd \omega \, k^2
\, e^{i(kp-\omega)(t_1 - t_2)}
\frac{\delta_{k,1}+\delta_{k,-1}}{\left|\tilde
D(\omega,k)\right|^2}\,\int\dd p'\,
f_0(p')\delta(\omega-p'k)  \phantom{\hskip 1truecm}\\
&=&\frac{\pi}{2}\int_{-\infty}^{+\infty} \dd \omega \,
\frac{f_0(\omega)}{\left|\tilde D(\omega,1)\right|^2}\left[
e^{i(p-\omega)(t_1-t_2)} + e^{-i(p-\omega)(t_1-t_2)} \right] \, .
\label{autocorr1t1t2}
\end{eqnarray}
Equation~(\ref{equation_p}) shows that the second moment appearing
in the right hand side of Eq.~(\ref{appmomentsFokkerPlanck}) is
determined by the integral of the last expression in $t_1$ and $t_2$
from $0$ to $t$. More precisely
\begin{eqnarray}
B(p,t)&=& \frac{1}{t}\frac{1}{N}\frac{\pi}{2} \int_0^t \dd t_1 \,
\int_0^t \dd t_2 \, \int_{-\infty}^{+\infty} \dd \omega \,
\frac{f_0(\omega)}{\left|\tilde D(\omega,1)\right|^2}\left[
e^{i(p-\omega)(t_1-t_2)} + e^{-i(p-\omega)(t_1-t_2)} \right] \\
&=&\frac{1}{t}\frac{1}{N}\pi\, \int_{-\infty}^{+\infty} \dd \omega
\, \frac{f_0(\omega)}{\left|\tilde D(\omega,1)\right|^2} \int_0^t
\dd s \, \int_0^{t-s} \dd t_2 \, \left[
e^{i(p-\omega)s} + e^{-i(p-\omega)s} \right]  \\
&=&\frac{1}{t}\frac{1}{N}\pi\, \int_{-\infty}^{+\infty} \dd \omega
\, \frac{f_0(\omega)}{\left|\tilde D(\omega,1)\right|^2} \int_0^t
\dd s \, (t-s) \left[
e^{i(p-\omega)s} + e^{-i(p-\omega)s} \right]  \\
&\stackrel{t \to +\infty}{\sim}& \frac{1}{N}2 \pi\,
\int_{-\infty}^{+\infty} \dd \omega \,
\frac{f_0(\omega)}{\left|\tilde D(\omega,1)\right|^2}\
 \pi \delta \left(p-\omega \right)   \\
&=&\frac{1}{N}2 \pi^2\, \frac{f_0(p)}{\left|\tilde D(p,1)\right|^2}
=\frac{2}{N} D(p) \, , \label{appmomentsFokkerPlanckfin}
\end{eqnarray}
which is expression (\ref{differentmoments2}).

For what concerns Eq.~(\ref{averageformoment1b}), starting again
from Eq.~(\ref{correldeltaVfinalbis}), we obtain
\begin{eqnarray}
\left\langle \frac{\partial^2 \delta v}{\partial
\theta^2}(\theta(0)+pt_1,t_1) \frac{\partial \delta v}{\partial
\theta}(\theta(0)+pt_2,t_2)\right\rangle
&=&\frac{1}{\left(2\pi\right)^2}\sum_{k=-\infty}^{+\infty}
\sum_{k'=-\infty}^{+\infty}\, \int_{-\infty}^{+\infty} \dd \omega \,
\int_{-\infty}^{+\infty} \dd \omega' \, (-ik^2k')\,
e^{ik(\theta(0)+pt_1) -i \omega t_1}\nonumber \\
&& \hskip 4truecm \times\, e^{ik'(\theta(0)+pt_2) - i\omega't_2}\,
\left\langle {\delta v}(k,\omega){\delta
v}(k',\omega')\right\rangle  \\
&=&i \frac{\pi}{2}\sum_{k=-\infty}^{+\infty}\!\!
\!\!\int_{-\infty}^{+\infty}\!\!\!\!\!\! \dd \omega \, k^3 \,
e^{i(kp-\omega)(t_1 - t_2)}
\frac{\delta_{k,1}+\delta_{k,-1}}{\left|\tilde
D(\omega,k)\right|^2}\,\int\dd p'\,
f_0(p')\delta(\omega-p'k) \phantom{\hskip1truecm} \\
&=&i\frac{\pi}{2}\int_{-\infty}^{+\infty} \dd \omega \,
\frac{f_0(\omega)}{\left|\tilde D(\omega,1)\right|^2}\left[
e^{i(p-\omega)(t_1-t_2)} - e^{-i(p-\omega)(t_1-t_2)} \right] \, .
\label{autocorr2t1t2}
\end{eqnarray}
Equation~(\ref{equation_p}) shows that the contribution $A_1(p,t)$
to the first moment appearing in the right hand side of
Eq.~(\ref{appmomentsFokkerPlanck1}) is given by
\begin{eqnarray}
A_1(p,t)&=& \frac{1}{t} \int_0^t \dd u \, \int_0^u \dd u_1 \,
\int_0^{u_1} \dd u_2 \, \left\langle \frac{\partial^2 \delta
v}{\partial \theta^2}(\theta(0)+pu,u) \frac{\partial \delta
v}{\partial \theta}(\theta(0)+pu_2,u_2)
\right\rangle \\
&=& \frac{1}{t} \frac{1}{N}\frac{i\pi}{2} \int_0^t \dd u \, \int_0^u
\dd u_1 \, \int_0^{u_1} \dd u_2 \, \int_{-\infty}^{+\infty} \dd
\omega \, \frac{f_0(\omega)}{\left|\tilde
D(\omega,1)\right|^2}\left[
e^{i(p-\omega)(u-u_2)} - e^{-i(p-\omega)(u-u_2)} \right]\phantom{\hskip1truecm} \\
&=& -\frac{1}{t} \frac{1}{N} \pi {\rm Im} \int_{-\infty}^{+\infty}
\dd \omega \, \frac{f_0(\omega)}{\left|\tilde D(\omega,1)\right|^2}
\int_0^t \dd u \, \int_0^u \dd u_2 \,
(u-u_2)\, e^{i(p-\omega)(u-u_2)} \\
&=& -\frac{1}{t} \frac{1}{N} \pi {\rm Re} \int_{-\infty}^{+\infty}
\dd \omega \, \frac{f_0(\omega)}{\left|\tilde D(\omega,1)\right|^2}
\frac{\partial}{\partial \omega} \int_0^t \dd u \, \int_0^u \dd u_2
\,
e^{i(p-\omega)(u-u_2)} \\
&=& \frac{1}{t} \frac{1}{N} \pi {\rm Re} \int_{-\infty}^{+\infty}
\dd \omega \, \left(\frac{\partial}{\partial \omega}
\frac{f_0(\omega)}{\left|\tilde D(\omega,1)\right|^2} \right)
\int_0^t \dd u \, \frac{i}{p-\omega}
\left[ 1 - e^{i(p-\omega)u-u_2}\right] \\
&=& \frac{1}{t} \frac{1}{N} \pi \int_{-\infty}^{+\infty} \dd \omega
\, \left(\frac{\partial}{\partial \omega}
\frac{f_0(\omega)}{\left|\tilde D(\omega,1)\right|^2} \right)
\int_0^t \dd u \, \frac{\sin (p-\omega)u}{p-\omega} \\
&=& \frac{1}{t} \frac{1}{N} 2 \pi \int_{-\infty}^{+\infty} \dd
\omega \, \left(\frac{\partial}{\partial \omega}
\frac{f_0(\omega)}{\left|\tilde D(\omega,1)\right|^2} \right)
\frac{1}{\left(p-\omega\right)^2} \sin^2
\frac{(p-\omega)}{2} \\
&\stackrel{t \to +\infty}{\sim}& \frac{1}{t} \frac{1}{N} 2 \pi
\left(\frac{\partial}{\partial p} \frac{f_0(p)}{\left|\tilde
D(p,1)\right|^2} \right) \int_{-\infty}^{+\infty} \dd \omega \,
\frac{1}{\left(p-\omega\right)^2} \sin^2
\frac{(p-\omega)}{2} \\
&=& \frac{1}{N} \pi^2 \left(\frac{\partial}{\partial p}
\frac{f_0(p)}{\left|\tilde D(p,1)\right|^2} \right) =\frac{1}{N}
\frac{\dd}{\dd p} D(p) \, .\label{appmomentsFokkerPlanck1fin1}
\end{eqnarray}

For what concerns Eq. (\ref{averageformoment1a}), let us rewrite
Eq.~(\ref{fouriertransformofdeltaVsuiteert}) as
\begin{equation}
\label{fouriertransformofdeltaVsuiteertbis} \widetilde{\delta
v}(k,\omega)=-\frac{\pi \left(\delta_{k,1}+\delta_{k,-1}\right)}
{\tilde D(\omega,k)}\int_{-\infty}^{+\infty}\!\!\dd p'\
\frac{\widehat{\delta f}(k,p',0)}{i(p'k-\omega)}.
\end{equation}
where we have neglected the $k=0$, that will not contribute. The
condition on given initial angle and momentum of the test particle
implies that
\begin{equation}
\left\langle\widehat{\delta f}(k,p',0)\right\rangle =
\frac{1}{\sqrt{N}}\frac{1}{2\pi}e^{-ik\theta(0)}\delta (p-p').
\label{conditiondeltafkp0}
\end{equation}
We therefore have
\begin{eqnarray}
\left\langle \frac{\partial \delta v}{\partial
\theta}(\theta(0)+pu,u) \right\rangle
&=&-\frac{1}{\sqrt{N}}\frac{1}{\left(2\pi\right)^2}\sum_{k=-\infty}^{+\infty}
\, \int_{\cal C} \dd \omega \, (ik)\, e^{ik(\theta(0)+pu) -i \omega
u} \, \frac{\pi \left(\delta_{k,1}+\delta_{k,-1}\right)}{\tilde
D(\omega,k)}
\frac{e^{-ik\theta(0)}}{i(pk-\omega)} \\
&=&-\frac{1}{\sqrt{N}}\frac{1}{4\pi} \, \int_{\cal C} \dd \omega \,
\left[ \frac{e^{i(p - \omega) u}}{(p-\omega)\tilde D(\omega,1)}
+\frac{e^{-i(p+\omega)u}}{(p+\omega)\tilde D(\omega,-1)}\right] \\
&=&-\frac{1}{\sqrt{N}}\frac{1}{4\pi}\left( {\cal P}\,
\int_{-\infty}^{+\infty} \dd \omega \, \frac{1}{p-\omega} \left[
\frac{e^{i(p - \omega) u}}{\tilde D(\omega,1)}
+\frac{e^{-i(p-\omega)u}}{\tilde D(-\omega,-1)}\right]
+i\pi \left[ \frac{1}{\tilde D(p,1)} -\frac{1}{\tilde D(-p,-1)}\right] \right)\phantom{\hskip1truecm} \\
&=&-\frac{1}{\sqrt{N}}\frac{1}{4\pi}\left( 2 \, {\rm Re}\, {\cal
P}\, \int_{-\infty}^{+\infty} \dd \omega \,
\frac{1}{p-\omega}\frac{1}{\left|\tilde D(\omega,1)\right|^2} \left[
\tilde D^*(\omega,1)\, e^{i(p - \omega) u)} \right] +2 \pi^3
\frac{f_0'(p)}{\left|\tilde D(p,1)\right|^2} \right) \,.
\label{autocorr3t1t2}
\end{eqnarray}
From Eq. (\ref{equation_p}), one sees that the contribution
$A_2(p,t)$ of this term to the first moment appearing in the right
hand side of Eq. (\ref{appmomentsFokkerPlanck1}) is given by
\begin{eqnarray}
A_2(p,t)&=& \frac{1}{t} \frac{1}{N}\frac{1}{4\pi} \int_0^t \dd u \
\left( 2 \, {\rm Re}\, {\cal P}\, \int_{-\infty}^{+\infty} \dd
\omega \, \frac{1}{p-\omega}\frac{1}{\left|\tilde
D(\omega,1)\right|^2} \left[ \tilde D^*(\omega,1)\, e^{i(p - \omega)
u)} \right]
+2 \pi^3 \frac{f_0'(p)}{\left|\tilde D(p,1)\right|^2} \right) \\
&=& \frac{1}{N}\frac{1}{2}\pi^2 \frac{f_0'(p)}{\left|\tilde
D(p,1)\right|^2} +\frac{1}{t} \frac{1}{N}\frac{1}{4\pi} 2 \, {\rm
Re}\, {\cal P}\, \int_{-\infty}^{+\infty} \dd \omega \,
\frac{1}{\left(p-\omega\right)^2}\frac{\tilde
D^*(\omega,1)}{\left|\tilde D(\omega,1)\right|^2}
\left[ 2 i \sin^2 \frac{p-\omega}{2} +\sin (p-\omega)t \right] \\
&\stackrel{t \to +\infty}{\sim}& \frac{1}{N}\frac{1}{2}\pi^2
\frac{f_0'(p)}{\left|\tilde D(p,1)\right|^2} +\frac{1}{N}\frac{1}{2}
\, {\rm Re}\,
\frac{i \tilde D^*(\omega,1)}{\left|\tilde D(\omega,1)\right|^2} \\
&=&\frac{1}{N}\frac{1}{2}\pi^2 \frac{f_0'(p)}{\left|\tilde
D(p,1)\right|^2} +\frac{1}{N}\frac{1}{2} \pi^2
\frac{f_0'(p)}{\left|\tilde D(p,1)\right|^2} = \frac{1}{N}\pi^2
\frac{f_0'(p)}{\left|\tilde D(p,1)\right|^2}
=\frac{1}{N}\frac{1}{f_0}\frac{\partial f_0}{\partial p}D(p) \, .
\label{appmomentsFokkerPlanck1fin2}
\end{eqnarray}
Adding Eqs. (\ref{appmomentsFokkerPlanck1fin1}) for $A_1(p,t)$ and
(\ref{appmomentsFokkerPlanck1fin2}) for $A_2(p,t)$
we obtain Eq. (\ref{differentmoments1ici}) for $A(p,t)$.%
%\begin{equation}
%\label{differentmoments1icibis}
%A(p,t) \sim\!\!
%\frac{1}{N}\left(\frac{\dd D}{\dd p}(p)
%+\frac{1}{f_0}\frac{\partial f_0}{\partial p}D(p)\right) \, .
%\end{equation}


\begin{thebibliography}{9999}


\bibitem{aaserth}
S. J. Aarseth, {\em Gravitational N-Body Simulations: Tools and
Algorithms}, Cambridge University Press, Cambridge (2003).

\bibitem{akhiezer}
A. I. Akhiezer, V. G. Baryakhtar, S. V. Peletminskij,
{\it Spin Waves}, Amsterdam, North Holland, (1968).

\bibitem{Alastueybook}
A. Alastuey, M. Magro, P. Pujol, {\em Physique et Outils
Math\'ematiques: M\'ethodes et Exemples}, CNRS Editions and EDP
Sciences (2008).

\bibitem{Angelani2003}
L. Angelani, L. Casetti, M. Pettini, G. Ruocco, F. Zamponi,
Europhysics Letters 62, 775 (2003).

\bibitem{Angelani2005}
L. Angelani, L. Casetti, M. Pettini, G. Ruocco, F. Zamponi, Physical
Review E 71, 036152 (2005).

\bibitem{Angelani2007}
L. Angelani, G. Ruocco, Physical Review E 76, 051119 (2007).

\bibitem{celia}
C. Anteneodo, Physica A {342}, 112 (2004).

\bibitem{Anteneodo97}
C. Anteneodo, C. Tsallis, %``Breakdown of exponential
%sensitivity to initial conditions: Role of the range of interactions'',
{Physical Review Letters} {80}, 5313 (1998).

\bibitem{Vallejos}
C. Anteneodo, R. Vallejos,  Physica A {344}, 383 (2004).

%\bibitem{Antoni98}M. Antoni, Y. Elskens, D. Escande, %``Explicit reduction of
%N-body dynamics to self-consistent particle-wave interaction'',
%{Physics of Plasmas} \textbf{5}, 841-852 (1998).

\bibitem{antoniHinrichsenruffo}
M. Antoni, H. Hinrichsen, S. Ruffo, {Chaos, Solitons and Fractals}
{13}, 393 (2002).

\bibitem{Antoni95}
M. Antoni, S. Ruffo,
%``Clustering and relaxation in long-range Hamiltonian dynamics'',
{Physical Review} {E} {52}, 2361 (1995).

\bibitem{Antoni4}
M. Antoni, S. Ruffo,  A. Torcini, %"First and second order
%clustering transitions for a system with infinite-range attractive
%interaction" ,
Physical Review E  66, 025103 (2002) .

\bibitem{Antoni5}
M. Antoni, S. Ruffo, A. Torcini, %" First-order microcanonical
%transitions in finite mean-field models ", submitted to
Europhysics Letters 66, 645 (2004).

\bibitem{Antoni1}
M. Antoni, A. Torcini, %"Anomalous Diffusion as a Signature of a
%Collapsing Phase in 2-d Self-Gravitating Systems", 2D HMF
Phys. Rev. E  57, R6233 (1998).

\bibitem{Antoniazzi07_1}
A. Antoniazzi, F. Califano,  D. Fanelli, S. Ruffo,
Physical Review Letters 98, 150602 (2007).

\bibitem{Antoniazzi06}
A. Antoniazzi, Y. Elskens, D. Fanelli, S. Ruffo,
European Physical Journal B 50, 603 (2006).

\bibitem{Antoniazi}
A. Antoniazzi, D. Fanelli, J. Barr\'e, P. H. Chavanis, T. Dauxois, S.
Ruffo, Physical Review E 75, 011112 (2007).

\bibitem{Antoniazzi07_2}
A. Antoniazzi, D. Fanelli, S. Ruffo, Y. Y. Yamaguchi, Physical
Review Letters 99, 040601 (2007).

\bibitem{Antonov}
V. A. Antonov, Vest. Leningrad Gros. Univ. {7}, 135 (1962).

\bibitem{Aronson}
E. B. Aronson, C. J. Hansen, Astrophysical Journal  {177}, 145
(1972).

\bibitem{Ayuela}
A. Ayuela, N. H. March, Physics Letters A 372, 5617 (2008).

\bibitem{Bachelard08}
R. Bachelard, C. Chandre, D. Fanelli, X. Leoncini, S. Ruffo,
Physical Review Letters (2008).

\bibitem{chavanisn}
F. Baldovin, P. H. Chavanis, E. Orlandini, Physical Review E 79,
011102 (2009).

\bibitem{BaldovinOrlandini}
F. Baldovin, E. Orlandini, Physical Review Letters {96}, 240602
(2006).

\bibitem{BaldovinOrlandinibis}
F. Baldovin, E. Orlandini, Physical Review Letters {97}, 100601
(2006).

%\bibitem{baldoreblo}
%F. Baldovin, A. Robledo, [cond-mat/0205356].

\bibitem{Balescu}
{R. Balescu}, {Physics of Fluids} {3,} {52} (1960).

\bibitem{Balescubook}
R. Balescu, {\em Statistical Mechanics of Charged Particles},
Interscience, New York, (1963).

\bibitem{Balescubook1}
R. Balescu, {\em Equilibrium and nonequilibrium statistical
mechanics}, Wiley, New York, (1975).

\bibitem{Balian}
R. Balian, {\em From Microphysics to Macrophysics: Methods and
Applications of Statistical Physics}, Springer-Verlag, Berlin
(1992).

\bibitem{Griffiths00}
S. Banerjee, R. B. Griffiths, M. Widom, Journal of
Statistical Physics, 93, 109 (2004).

\bibitem{Barbara}
B. Barbara, private communication (2008).

\bibitem{junext01}
J.  Barr{\'e}, %``Microcanonical solution of lattice models
%with long-range interaction'',
{Physica A} {305}, 172 (2002).

\bibitem{barrethesis}
J. Barr{\'e}, {M{\'e}canique statistique et
dynamique hors {\'e}quilibre de syst{\`e}mes avec int{\'e}ractions
{\`a} longues port{\'e}es}, PhD Thesis, ENS Lyon (2003).

%\bibitem{barrebouchet}  J. Barr\'e, F. Bouchet,
%%" Mean-Field justified by large deviations results in long-range interacting systems"
%Proceedings of the conference Dynamics and thermodynamics of systems
%with long-range interactions, Les Houches, France, February 18-22
%2002, Eds. T.~Dauxois, E.~Arimondo, S.~Ruffo, M.~Wilkens; published
%on http://www.ens-lyon.fr/~tdauxois/procs02/.

%\bibitem{barrenbouchetCRAS}
%J. Barr\'e, F. Bouchet, Compte Rendus de l'Acad\'emie des Sciences:
%Physique 7, 414 (2006).
%%Statistical mechanics and long-range interactions

\bibitem{JulienBarrePrivate}
J. Barr\'e, Private communication (2009).

\bibitem{BBDR2002}
J. Barr\'e, F. Bouchet, T. Dauxois, S. Ruffo, Physical Review Letters 89, 110601 (2002).

\bibitem{BBDR2002b}
J. Barr\'e, F. Bouchet, T. Dauxois, S. Ruffo, European Physical Journal B 29, 577 (2002).

\bibitem{bbdrjstatphys}
J. Barr{\'e}, F. Bouchet, T. Dauxois, S. Ruffo,
Journal of Statistical Physics {119}, 677 (2005).

\bibitem{BBDRY2006}
J. Barr\'e, F. Bouchet, T. Dauxois, S. Ruffo, Y. Y. Yamaguchi,
Physica A 365, 177 (2006).

\bibitem{FELPRE}
J. Barr\'e, T. Dauxois, G. De Ninno, D. Fanelli, S. Ruffo, Physical
Review E, Rapid Communication 69, 045501 (R) (2004).

\bibitem{BDR2001}
J. Barr\'e, T. Dauxois, S. Ruffo, Physica A 295, 254 (2001).

\bibitem{BarreGoncalves}
J. Barr\'e, B. Gon\c calves,  Physica A 386, 212 %-218
 (2007).

\bibitem{BMR}
J. Barr{\'e}, D. Mukamel, S. Ruffo,
%``Inequivalence of ensembles in a system with long-range interactions'',
{Physical Review Letters} {87}, 030601 (2001).

\bibitem{MukamelHouches}
J. Barr{\'e}, D. Mukamel, S. Ruffo, ``Ensemble
inequivalence in mean-field models of magnetisms'', in Ref.
\cite{leshouches}.

\bibitem{bonasera}
M. Belkacem, V. Latora, A. Bonasera, Physical Review C {52}, 271
(1995).

\bibitem{bender}
C. M. Bender, S. A. Orszag, {\em Advanced Mathematical Methods for
Scientists and Engineers}, McGrawHill, New-York (1978), Chapter 6.

%\bibitem{Berry88}
%R. S. Berry, T. L. Beck, H. L. Davis, J. Jellineck, {Advances in
%Chemical Physics} {90}, 75 (1988).

\bibitem{Binder03}
K. Binder,
%"Theory of the evaporation/condensation transition of equilibrium droplets in finite volumes"
Physica A {319}, 99 %-114
 (2003).

%\bibitem{beckcohen} C. Beck, E.G.D. Cohen, Physica A {322},
%267-275 (2003).

\bibitem{binneytremaine}
J. Binney, S. Tremaine, {\em Galactic Dynamics}, Princeton Series in
Astrophysics (1987).

\bibitem{Blume0}
M.~Blume, Physical Review {141}, 517 (1966).

\bibitem{Blume}
M. Blume, V. J. Emery, R. B. Griffiths, Physical Review A
{4}, 1071 (1971).

\bibitem{Holtauss} D. Boers, M. Holthaus, ``Canonical statistics of occupation numbers
for ideal and weakly interacting Bose-Einstien condensates'', in Ref.~\cite{leshouches}.
%\bibitem{Bogoliubov}{N. N. Bogoliubov}, {Problems of a Dynamical Theory in Statistical Physics},
%{State Technical Press} {(1946)}.

\bibitem{Boltzmann}
L. Boltzmann, Wien, Ber. {66}, 275 (1872).

\bibitem{Bonifacio90}
R. Bonifacio, F. Casagrande, G. Cerchioni, L. De
Salvo Souza, P. Pierini, N. Piovella, {Rivista del Nuovo Cimento}
{\bf 13}, 1 (1990).

\bibitem{CARL}
R. Bonifacio, L. De Salvo,  Nuclear Instruments and Methods A 341, 360 (1994).

\bibitem{Bonifacio84}
R. Bonifacio, C. Pellegrini, L. M. Narducci, Optical Communications,
50, 373 (1984).

\bibitem{borgonovi1} F. Borgonovi, G. L. Celardo,  M. Maianti, E. Pedersoli, Journal
of Statistical Physics  116, 1435 (2004).

\bibitem{borgonovi2} F. Borgonovi, G. L. Celardo, A. Musesti, R. Trasarti-Battistoni, P.
Vachal, Physical Review E 73, 039903 (2006).

%\bibitem{Bonifacio94} R. Bonifacio, L. de~Salvo, L. M. Narducci,
%E. J. d'Angelo,
%%``Exponential gain and self-bunching in a collective atomic recoil laser'',
% {Physical Review A} {50}, 1716 (1994).

%\bibitem{Borges02} E. P. Borges, C. Tsallis, %``Negative specific heat in a
%%%Lennard-Jones like gas with long-range interactions'',
%{Physica A} {305}, 148 (2002).

%\bibitem{borgo}
%F. Borgonovi, G. L. Celardo, M. Maianti, E. Pedersoli, Journal of
%Statistical Physics {116}, 1435 (2004).
%
%\bibitem{borgobis}F. Borgonovi, G. L. Celardo, A. Musesti, R.
%Trasarti-Battistoni, P. Vachal, Physical Review E, {73}, 026116
%(2006).

%\bibitem{BG}{M. Born, H. S. Green}, {A General Kinetic Theory of
%Liquids}, {Cambridge University Press} ({1949}).

\bibitem{refnewellis}
C. Boucher, R. S. Ellis, B. Turkington, Annals of Probability {\bf
27}, 297 (1999). Erratum, Annals of Probability {30}, 2113 (2002).

\bibitem{Bouchet01}
F. Bouchet, {M{\'e}canique Statistique pour des
{\'E}coulements G{\'e}ophysiques}, PhD thesis, Universit{\'e} Joseph
Fourier, Grenoble (2001).

\bibitem{BouchetPRE}
F. Bouchet, Physical Review E {70}, 036113 (2004).

\bibitem{julienfreddyjstat}
F. Bouchet, J. Barr\'e, %"Classification of phase transitions and ensemble inequivalence, in systems with long-range interactions",
Journal of Statistical Physics 118, 1073 (2005).

\bibitem{fredthierPRE}
F. Bouchet, T. Dauxois, Physical Review E 72, 045103(R) (2005).

\bibitem{bdmrPRE}
F. Bouchet, T. Dauxois, D. Mukamel, S. Ruffo, Physical Review E 77, 011125 (2008).

\bibitem{Bouchet02}
F. Bouchet, J. Sommeria,
%``Emergence of intense jets and Jupiter's Great Red Spot as maximum-entropy structures'',
{Journal of Fluid Mechanics} {464}, 165 (2002).

\bibitem{Bramwell01}
S. T. Bramwell, M. J. P. Gingras, Science, 294 1495 (2001).

%\bibitem{StevenPeter}S. T. Bramwell, M. J.
%P. Gingras, P. C. W. Holdsworth, ``Spin Ice'', Chapter 7, in
%"Frustrated Spin Systems", H. T. Diep (Ed.), World Scientific
%(2004).

\bibitem{brands}
H. Brands, P. H. Chavanis, R. Pasmanter, J. Sommeria, Physics of
Fluids 11, 365 (1999).

\bibitem{BraunHepp}
W. Braun, K. Hepp, Communications in  Mathematical Physics {56}, 101
(1977).

\bibitem{Martin}
D. C. Brydges, P. A.  Martin,
%    Coulomb systems at low density: A review
{Journal of Statistical Physics } {96}, 1163 (1999).

\bibitem{debuyl}
P. de Buyl, D. Mukamel, S. Ruffo, AIP Conferences Proceedings, 800,
533 (2005).
%
%\bibitem{cagliotti}  E. Caglioti. P. L. Lions, C. Marchioro, M. Pulvirenti,
%{Communications in Mathematical Physics} {143},
% 501 (1992).
%
\bibitem{Caglioti_95}  E. Caglioti, P. L. Lions, C. Marchioro, M. Pulvirenti,
{Communications in Mathematical Physics} {174}, 229 (1995).

\bibitem{CagliottiRousset}
E. Caglioti, F. Rousset,
Archive for Rational Mechanics and Analysis 190, 517 (2008).

\bibitem{CampaChavanis}
A. Campa, P. H. Chavanis, A. Giansanti, G. Morelli, {Physical Review
E} {78}, 040102(R) (2008).

\bibitem{campaorder}
A. Campa, A. Giansanti, {Physica A} {340}, 170 (2004).

\bibitem{Assisi}
A. Campa, A. Giansanti, G. Morigi, F. Sylos Labini, {\em Dynamics
and Thermodynamics of systems with long-range interactions: Theory
and Experiment}, AIP Conference Proceedings {970} (2008).

%\bibitem{Campa2000}
%A. Campa, A. Giansanti, D. Moroni, {Physical Review E} {62}, 303
%(2000).

\bibitem{CGM2007}
A. Campa, A. Giansanti, G. Morelli, Physical Review E 76, 041117 (2007).

\bibitem{CGM}
A. Campa, A. Giansanti, D. Moroni, {Journal of Physics} A {36}, 6897
(2003).

\bibitem{campaphysica2006}
A. Campa, A. Giansanti, D. Mukamel, S. Ruffo, {Physica A} {365}, 120
(2006).

\bibitem{CampaPRB}
A. Campa, R. Khomeriki, D. Mukamel, S. Ruffo,
Physical Review B, {76}, 064415 (2007).

\bibitem{campa2006}
A. Campa, S. Ruffo, {Physica A} {385}, 233 (2007).

\bibitem{campa2007}
A. Campa, S. Ruffo, H. Touchette, {Physica A} {369}, 517 (2007).

\bibitem{Capel}
H. W. Capel, Physica {32}, 966 (1966).

\bibitem{Casetti1999}
L. Casetti, E. G. D. Cohen, M. Pettini,  Physical Review Letters 82,
4160 (1999).

\bibitem{Casetti2006}
L. Casetti, M. Kastner, Physical Review Lett 97, 100602 (2006).

\bibitem{Casetti2007}
L. Casetti, M. Kastner, Physica A 384, 318 (2007).

\bibitem{Casetti2000}
L. Casetti, M. Pettini, E. G. D. Cohen, Physics Reports 337, 237
(2000).

\bibitem{Casetti2003}
L. Casetti, M. Pettini, E. G. D. Cohen, Journal of Statistical
Physics 111, 1091 (2003).

\bibitem{Castaing}
B. Castaing, {An Introduction to hydrodynamics}, in {\em
Hydrodynamics and nonlinear instabilities}, Eds C. Godr\`eche, P.
Manneville, Cambridge University Press (1995).

\bibitem{celardo}
G. L. Celardo, J. Barr\'e, F. Borgonovi, S. Ruffo, Physical Review
E, {73}, 011108 (2006).

%\bibitem{Pettini02} M. Cerruti-Sola , P. Cipriani, M. Pettini, % ``On the
%%clustering phase transition in self-gravitating N-body systems'',
%{Monthly Notices of the Royal Astronomical Society} {328}, 338-352
%(2001).

\bibitem{Chabanol}
L.-L Chabanol, F. Corson, Y. Pomeau, Europhysics Letters {50}, 148 (2000).

\bibitem{Chaikin95}
P. M. Chaikin, T. C. Lubensky, {\em Principles of Condensed Matter
Physics}, Cambridge University Press (1995).

\bibitem{chandrasekhar}
S. Chandrasekhar, {\em Principles of Stellar Dynamics}, University
of Chicago Press (1942).

\bibitem{Chapuis}
G. Chapuis, G. Brunisholz, C. Javet, R. Roulet, Inorganic Chemistry
22, 455 (1983).

%\bibitem{Chavanis96} P. H. Chavanis, {Contributions \`a la m{\'e}canique
%statistique des tourbillons bidimensionnels. Analogie
%avec la relaxation violente des systèmes stellaires}, PhD thesis,
%Ecole Normale Sup\'erieure de Lyon (1996).

%\bibitem{Chavanis98} P. H. Chavanis, %``From Jupiter's great red spot to the
%%structure of galaxies: statistical mechanics of two-dimensional
%%vortices and stellar systems'',
%{Annals of New-York Academay of Sciences} {867}, 120(1998).

\bibitem{Chavanisnewaddi}
P. H. Chavanis, Physical Review E 64, 026309 (2001).

\bibitem{chavanisaddit}
P. H. Chavanis, Physical Review E 65, 056123
(2002).

\bibitem{ChavanisHouches}
P. H. Chavanis, {Statistical mechanics
of two-dimensional vortices and stellar systems}, in
\cite{leshouches}.

\bibitem{ChavanisAssisi}
P. H. Chavanis, {Dynamics and thermodynamics of systems
with long-range interactions: interpretation of the different
functionals}, in \cite{Assisi}.

\bibitem{chavanisexpN}
P. H. Chavanis, Astronomy and Astrophysics 432, 117 (2005).

\bibitem{Chavanisreview2006}
P. H. Chavanis, International Journal of Modern Physics B {20}, 3113
(2006).

\bibitem{chavanisPhysicaI2006}
P. H. Chavanis, Physica A {361}, 55 (2006).

\bibitem{chavanisPhysicaII2006}
P. H. Chavanis, Physica A {361}, 81 (2006).
%
%\bibitem{chavaniscondmat} P. H. Chavanis,  Physica A {365}, 102
%(2006).

\bibitem{chavanisEPJB2006}
P. H. Chavanis, European Physical Journal B 52, 61 (2006).

\bibitem{Chavanis_phasetrans}
P. H. Chavanis, European Physical Journal B 53, 487 (2006).

\bibitem{chavanisPhysicaIII2008}
P. H. Chavanis, Physica A {387},
787 (2008).

\bibitem{chavanisPhysicaIV2008}
P. H. Chavanis, Physica A {387},
1504 (2008).

\bibitem{Chavanis_pharo}
P. H. Chavanis, G. De Ninno, D. Fanelli, S. Ruffo, ``Out of
equilibrium phase transitions in mean-field Hamiltonian dynamics",
in {\it Chaos, complexity and transport}, C. Chandre, X. Leoncini
and G. Zaslavsky (Eds.), World Scientific, p. 3(2008).

\bibitem{ChavanisLemou2005}
P. H. Chavanis, M. Lemou, Physical Review E {72} 061106 (2005).

\bibitem{chavanisEPJB2007}
P. H. Chavanis, M. Lemou, European Physical Journal B {59}, 217 (2007).

%\bibitem{chavanissire} P. H. Chavanis, C. Sire, Physica A {356}, 419
%(2005).
%
%\bibitem{Chavanis02shallow}P. H. Chavanis, J. Sommeria,
%%``Statistical mechanics of the Shallow Water system'',
%{Physical Review E,} {65}, (2002).

\bibitem{chavanism}
P. H. Chavanis, J. Sommeria, Journal of Fluid Mechanics
 314, 267 (1996).

\bibitem{chavanise}
P. H. Chavanis, J. Sommeria, Physical Review E 65, 026302 (2002).

\bibitem{ChavanisSommeriaRobert}
P. H. Chavanis, J. Sommeria, R. Robert, %``Statistical
%mechanics of two-dimensional vortices and collisionless stellar systems'',
{The Astrophysical Journal} {471}, 385 (1996).

\bibitem{Judith}
P. H. Chavanis, J. Vatteville, F. Bouchet,
European Physical Journal B 46, 61 (2005).

\bibitem{ChoiChoi1}
M. Y. Choi, J. Choi, Physical Review Letters {91}, 124101
(2003).

\bibitem{ChoiChoi2}
J. Choi, M. Y. Choi, Journal of Physics A, {38}, 5659
(2005).

\bibitem{chomazassisi}
Ph. Chomaz, {Phase Transitions in finite systems using information
theory}, in \cite{Assisi}.

\bibitem{chomazdd}
Ph. Chomaz, F. Gulminelli,
{Phase transitions in finite systems} in~\cite{leshouches}.

\bibitem{chonucphys1999}
Ph. Chomaz, F. Gulminelli, Nuclear Physics A 153, 647 (1999).

\bibitem{choepj2006}
Ph. Chomaz, F. Gulminelli, European Physical Journal A 30, 317
(2006).

\bibitem{chopre2001}
Ph. Chomaz, F. Gulminelli, V. Duflot, Physical Review Letters 64,
046114 (2001).

%\bibitem{cohen} E. G. D.  Cohen, I. Ispolatov,
%{Phase transitions in systems with $1/r^\alpha$ attractive
%interactions} in~\cite{leshouches}.

\bibitem{chavanish}
H. G. H. Clercx, G. J. F. van Heijst, Applied Mechanics Reviews 62,
020802 (2009).

\bibitem{Lecar}
L. Cohen, M. Lecar, Bulletin d'Astronomie, 3\`eme s\'erie, Tome III.
fasc. 2, 213 (1968).

\bibitem{Colson1976}
W. B. Colson, Physics Letters A, 59, 187 (1976).

\bibitem{Compagner89}
A. Compagner, C. Bruin, A. Roelse, Physical Review A 39, 5989
(1989).

%\bibitem{costeniuc}
%M. Costeniuc, R. S. Ellis, H. Touchette, B. Turkington, {Physical
%Review E} {73}, 026105 (2006).

\bibitem{creutz}
M. Creutz, Physical Review Letters {50}, 1411 (1983).

\bibitem{dagostino}
M. D'Agostino, F. Gulminelli, P. Chomaz, M. Bruno, F. Cannata, R.
Bougault, F. Gramegna, I. Iori, N. Le Neindre, G. V. Margagliotti,
A. Moroni, G. Vannini, Physics Letters B {473}, 219 (2000).

\bibitem{DHR2000}
T. Dauxois, P. Holdsworth, S. Ruffo,
European Physical Journal B, 16, 659 (2000).

\bibitem{Dauxois02}
T. Dauxois, V. Latora, A. Rapisarda, S. Ruffo, A. Torcini,
``The Hamiltonian Mean Field Model: from Dynamics to Statistical
Mechanics and back'', in Ref. \cite{leshouches}.

\bibitem{dauxois2003}
T. Dauxois, S. Lepri, S. Ruffo, {Communications in  Nonlinear Science and  Numerical Simulations}
{8}, 375 (2003).

\bibitem{dauxoispeyrard}
T. Dauxois, M. Peyrard, {\em Physics of Solitons}, Cambridge
University Press (2006)

\bibitem{leshouches}
T. Dauxois, S. Ruffo, E. Arimondo, M. Wilkens (Eds.), {\em Dynamics
and Thermodynamics of Systems with Long-Range Interactions},
{Lecture Notes in Physics} {602}, Springer (2002).

\bibitem{LesHouches2009}
T. Dauxois, S. Ruffo, L. Cugliandolo (Eds.), {\em Long-Range
Interacting Systems}, Oxford University Press (2009).
%
%\bibitem{diego} D. Del Castillo-Negrete,
%{Dynamics and self-consistent chaos  in a mean field Hamiltonian
%model} in~\cite{leshouches}.

\bibitem{diego98a}
D. Del Castillo-Negrete, Physics Letters A 241, 99 (1998).

\bibitem{diego98b}
D. Del Castillo-Negrete, Physics of Plasmas 5, 3886 (1998).

\bibitem{Dembo}
A. Dembo, O. Zeitouni, {\em Large Deviations Techniques and their
Applications}, Springer-Verlag, New York, (1998).

\bibitem{desai1978}
R. Desai, R. Zwanzig, {Journal of Statistical Physics} {19}, 1
(1978).

\bibitem{DubinOneil99}
D. H. Dubin, T. M. O'Neil, {Review of Modern Physics} {71}, 87 (1999).

\bibitem{DubinJin}
D. H. Dubin, D. Z. Jin, {Physics Letters A} {284}, 112 (2001).

\bibitem{Dubin03}
D. H. Dubin,  {Physics of Plasmas} {10}, 1338 (2003).

\bibitem{dupas}
A. Dupas, K. Le Dang, J.-P. Renard, P. Veillet, J. Phys. C:
Solid State Physics  10, 3399, (1977).

\bibitem{dyson}
F. J. Dyson, {Communications in  Mathematical Physics} {12}, 91 (1969).

\bibitem{eddigton}
A. S. Eddington, {\em The internal constitution of stars}, Cambridge
University Press (1926).

\bibitem{Ellis85}
R. S. Ellis, {\em Entropy, Large Deviations, and Statistical
Mechanics}, Springer-Verlag, New-York (1985).

\bibitem{Ellis99}
R. S. Ellis,
%``The theory of large deviations: from Boltzmann's 1877
%calculation to equilibrium macrostates in 2D turbulence'',
{Physica D} {133}, 106 (1999).

\bibitem{ellisdd}
R. S. Ellis, K. Haven, B. Turkington, %{The large
%    deviation principle and complete equivalence and nonequivalence
%    results for pure and mixed ensembles},
Journal of Statistical  Physics {101}, 999 (2000).

\bibitem{Ellis02}
 R. S. Ellis, K. Haven, B. Turkington, {Nonlinearity}
{15}, 239 (2002).

\bibitem{Touchette2003}
R. S. Ellis, H. Touchette, B. Turkington, %``Thermodynamic
%versus statistical nonequivalence of ensembles for the mean-field
%Blume-Emery-Griffiths model'',
{Physica A} {335}, 518 (2004).

%\bibitem{Elskens02houches} Y. Elskens,
%{Kinetic theory for plasmas and wave-particle hamiltonian dynamics}
%in~\cite{leshouches}.

\bibitem{Escande}
Y. Elskens, D. Escande, {\em Microscopic Dynamics of Plasmas and
Chaos}, IOP Publishing, Bristol (2002).

\bibitem{Emden}
R. Emden, {\em Gaskugeln}, Teubner, Leipzig (1907).

\bibitem{englishprb}
L. Q. English, M. Sato, A. J. Sievers,
Physical Review B 67, 024403 (2003).

\bibitem{EyingSreenivasan}
G. L. Eyink, K. R. Sreenivasan, %{Onsager and the theory of hydrodynamic turbulence},
Review of Modern Physics {78}, 87 (2006).

\bibitem{farago}J. Farago,
{Europhysics Letters} {52},  {379} ({2000}).

\bibitem{Farina} D. Farina, F. Casagrande, U. Colombo, R. Pozzoli, Physical Review E, 49, 1603 (1994).

\bibitem{farizonassisi}
M. Farizon, B. Farizon, S. Ouaskit, T. D. M\"ark, Fragment size
distributions and caloric curve in collision induced cluster
fragmentation, in \cite{Assisi}.

\bibitem{Fauve}
S. Fauve, Pattern forming instabilities, in {\em Hydrodynamics and
nonlinear instabilities}, Eds C. Godr\`eche, P. Manneville, Cambridge
University Press (1995).

%\bibitem{feldman}
%E. B. Feldman, S. Lacelle, Journal of Chemical Physics {108}, 4709
%(1998).

\bibitem{FeixBertrand}
M. R. Feix, P. Bertrand, {Transport Theory and Statistical Physics}
{34},  {7} ({2005})

\bibitem{Fine}
K. S. Fine, A. C. Cass, W. G. Flynn, C. F. Driscoll, {Physical
Review Letters} {75}, 3277 (1995).

\bibitem{FirpoThese}M. C. Firpo, Etude dynamique et statistique de l'interaction
onde-particule, PhD Thesis, Universit\'e de Provence (1999).

\bibitem{Firpo00}M. C. Firpo, Y. Elskens, %``Phase transition in the
%collisionless damping regime for wave-particle interaction'',
{Physical Review Letters} {84}, 3318 (2000).

\bibitem{franzosipettini1}
R. Franzosi,  M. Pettini, {Physical Review Letters} {92}, 060601
(2004).

\bibitem{franzosipettini2}
R. Franzosi,  M. Pettini, {Nuclear Physics B} {782}, 189 (2007).

\bibitem{franzosipettini3}
R. Franzosi, M. Pettini, {Nuclear Physics B} {782}, 219 (2007).

\bibitem{FrenkelSmith}
D. Frenkel, B. Smit, {\em Understanding Molecular Simulation: From
Algorythm to Applications}, Academic Press (1996).

\bibitem{gallavotti}
G. Gallavotti, {\em Statistical Mechanics: A Short Treatise},
Springer, Berlin (1999).

%\bibitem{gaspard} P. Gaspard, {Physical Review E} {68}, 056209 (2003).

\bibitem{Tsallis} M. Gell-Mann, C. Tsallis, {\em Nonextensive
Entropy-Interdisciplinary Applications}, Oxford University Press
(2004).

\bibitem{gobet}
F. Gobet, B. Farizon, M. Farizon, M. J. Gaillard, J. P. Buchet, M. Carr\'e, T. D. M\"ark,
{Physical Review Letters} {87}, 203401 (2001).

%\bibitem{Gozzi}
%E. Gozzi, Physics Letters B  {233}, 383 (1989).

%\bibitem{Gozzibis} E. Gozzi, International Journal of Modern Physics A {16}, 2709 (2001).

\bibitem{Griffiths68}
R. B. Griffiths,  Physical Review {176}, 655 (1968).

\bibitem{Dieter}
D. H. E. Gross, {\em Microcanonical Thermodynamics}, World
Scientific, Singapore (2001).

\bibitem{grossdd}
D. H. E. Gross, {Thermo-Statistics or Topology of the Microcanonical Entropy
Surface} in~\cite{leshouches}.

\bibitem{grossphysrep}
D. H. E. Gross, Physics Reports 279, 119 (1997).

\bibitem{Gross00}
D. H. E. Gross, E. V. Votyakov, {European Physical Journal B} {15}, 115 (2000).

%\bibitem{Tarjus01} M. Grousson, G. Tarjus, P. Viot,
%%``Phase diagram of an Ising model with long-range frustrating interactions: A
%%theoretical analysis'',
%{Physical Review E} {62}, 7781 (2000).

\bibitem{chopre2002}
F. Gulminelli, Ph. Chomaz, Physical Review E 66, 046108 (2002).

\bibitem{hahn2005} I. Hahn, M. Kastner, {Physical Review E}
{72}, 056134 (2005).

\bibitem{hahn2006}
I. Hahn, M. Kastner, {European Physical Journal B} {50}, 311 (2006).

\bibitem{haurayjabin}
M. Hauray, P. E. Jabin,  Archive for Rational Mechanics and Analysis, 183, 489 (2007).

%\bibitem{hawking} S. W. Hawking, Nature {248}, 30 (1974).

%\bibitem{Maltehenkel}
%M. Henkel, %{Sur la solution de Sundman du probleme des trois corps},
%Philosophia Scientiae {5}, 161-184 (2001).

\bibitem{Heggiehut}
D. Heggie, P. Hut,{\em The Gravitational Million-Body Problem},
Cambridge University Press (2003).

\bibitem{Henon64}
M. H\'enon, {Annales d'Astrophysique} {27}, 83 (1964).

\bibitem{Henon67}
M. H\'enon, M\'emoire de la Soci\'et\'e Royale des Sciences de
Liege, 55, 243 (1967).

\bibitem{Henon68}
M. H\'enon, Bulletin Astronomie, 3\`eme s\'erie, Tome III. fasc. 2,
213 (1968).

\bibitem{Thirring}
P. Hertel, W. Thirring, {Annals of Physics} {63},
520 (1971).

\bibitem{HertelThirringcommmathphys}
P. Hertel, W. Thirring, Communications in Mathematical Physics 24,
22 (1971).

\bibitem{Hohl67}
F. Hohl, M. R. Feix, {Astrophysical Journal} {147}, 1164 (1967).

\bibitem{Hohl}
F. Hohl, J. W. Campbell, Astronomical Journal, 73, 611 (1968).

\bibitem{Horwitz77}G. Horwitz, J. Katz, %``Steepest-descent technique and
%stellar equilibrium statistical mechanics 1. Newtonian clusters in
%a box'',
{Astrophysical Journal} {211}, 226 (1977).

\bibitem{Horwitz78}
G. Horwitz, J. Katz,
{Astrophysical Journal} {222}, 941 (1978).

\bibitem{Huang}
K. Huang, {\em Statistical Mechanics}, John Wiley and Sons Ed.
(1987).

\bibitem{HuangDriscoll}
X.-P. Huang, C. F. Driscoll, {Physical Review Letters} {72}, 2187
(1994).

\bibitem{ichimaru}
S. Ichimaru, {\em Basic Principles of Plasma Physics}, W. A.
Benjamin, Inc. Reading, Mass. (1973).

\bibitem{Inagaki93a}
S. Inagaki, Progress in Theoretical Physics {90}, 577 (1993).%, {``Dynamical
%stability of a simple model similar to self-gravitating systems"}

\bibitem{Inagaki93b}
S. Inagaki, T. Konishi, Publications of the Astronomical Society of Japan
 {45}, 733 (1993).%, {``Dynamical
%stability of a simple model similar to self-gravitating systems"}

\bibitem{Ispolatov01a}
I. Ispolatov, E. G. D. Cohen, %`` On first-order phase
%transitions in microcanonical and canonical non-extensive systems'',
{Physica A} {295}, 475 (2001).

%\bibitem{ISPOCohenPRL} I. Ispolatov, E. G. D. Cohen {Physical Review
%Letters} {87}, 210601 (2000).

%\bibitem{ISPOCohenPRE} I. Ispolatov, E. G. D. Cohen {Physical Review E}  {64}, 056103 (2000).

%\bibitem{Ispolatov03} I. Ispolatov, M. Karttunen,
%``Collapses and explosions in self-gravitating systems'',
%cond-mat/0302590.

\bibitem{JeongChoi2006}
D. Jeong, J. Choi, M. Y. Choi, Physical Review E 74, 0556106 (2006).

\bibitem{Joyce73}
G. Joyce, D. Montgomery, %``Negative temperature states for
%the two-dimensional guiding center plasma'',
{Journal of Plasma Physics} {10}, 107-121 (1973).

\bibitem{KacUhlenbeck}
M. Kac, G. E. Uhlenbeck, P. C. Hemmer, Journal of Mathematical
Physics {4},  216 (1963).
%``On Van der Waals Theory of vapor-liquid equilibrium. 1.Discussion of a
%1-dimensionnal model'',

\bibitem{KadomtsebPogutse}
B. B. Kadomtsev, O. P. Pogutse, Physical Review Letters {25}, 1155 (1970).

\bibitem{Kandrup}
H.E. Kandrup, Astrophysical Journal 244, 316 (1981).

\bibitem{Khardar}
M. Kardar, {Physical Review B} {28}, 244 (1983).

\bibitem{Khardar83}
M. Kardar, {Physical Review Letters} {51}, 523 (1983).

\bibitem{Kastner2006}
M. Kastner, Physica A {359}, 447 (2006).

\bibitem{kastnerrevmodphys}
M. Kastner, {Review of Modern Physics} {80}, 167 (2008).

\bibitem{KastnerSchnetz2006}
M. Kastner, O. Schnetz, Journal of Statistical Physics 122, 1195
(2006).

\bibitem{kastnerprl2008}
M. Kastner, O. Schnetz, {Physical Review Letters} {100}, 160601
(2008).

%\bibitem{Kastner2008}
%M. Kastner, O. Schnetz, S. Schreiber {Journal of Statistical
%Mechanics: Theory and Experiment} P04025 (2008).

\bibitem{kastnerprl2007}
M. Kastner, S. Schreiber, O. Schnetz, {Physical Review Letters}
{99}, 050601 (2007).

\bibitem{Katz}
J. Katz, Monthly Notices of the Royal Astronomical Society {183}, 765 (1978).

\bibitem{Kawahara}
R. Kawahara, H. Nakanishi, Journal of Physical Society of Japan, 75, 054001 (2006).

\bibitem{grape5taka}
A. Kawai, T. Fukushige, J. Makino, M. Taiji, Publications of the
Astronomical Society of Japan {52}, 659 (2000).

%\bibitem{Kiessling}
%M. Kiessling, Journal of Statistical Physics, {55}, 1572-9613
%(1989).

\bibitem{kiesling89}
M. K. H. Kiessling, {Journal of Statistical Physics,} {55}, 203 (1989).

%\bibitem{kiesling} M. K. H. Kiessling, {Communications in Pure and Applied Mathematics}
%{46}, 27 (1993).

%\bibitem{Kiessling97} M. K. H. Kiessling, J.L. Lebowitz, %``The
%%Micro-Canonical Point Vortex Ensemble: Beyond Equivalence'',
%{Letters Math. Phys.} {42}, 43 (1997).

\bibitem{KiesslingLebowitz97}
M. Kiessling, J. L. Lebowitz, Letters in Mathematical Physics {42},
43 (1997).

\bibitem{Kiessling03}
M. K. H. Kiessling, T. Neukirch, %``Negative Specific
%Heat of a Magnetically Self-Confined Plasma Torus'',
The Proceedings of the National Academy of Sciences (USA) {100}, 1510 (2003).

\bibitem{Kittel51}
C. Kittel, Physical Review, 82, 965 (1951).

\bibitem{Kiwamoto} Y. Kiwamoto, N. Hashizume, Y. Soga, J. Aoki, Y. Kawai, Physical Review Letters 99, 115002 (2007).

\bibitem{Klimontovich}
{Yu. L. Klimontovich}, {\em The Statistical Theory of
Non-equilibrium Processes in a Plasma}, {MIT Press} (1967).
%
%\bibitem{Kirkwood} {J. G. Kirkwood}, {Journal of Chemical Physics} {14,} {180} {(1946}).
%
%\bibitem{Kirkwoodb} {J. G. Kirkwood}, {Journal of Chemical Physics} {15,} {72} {(1947)}.

\bibitem{Koyama01}
H. Koyama, T. Konishi, Physics Letters A {279}, 226 (2001).

\bibitem{KramersWannier}
H. A. Kramers, G. H. Wannier, Physical Review {60}, 252 (1941).

\bibitem{konishi-kaneko}
T. Konishi, K. Kaneko,  Journal of Physics A {25}, 6283, (1992).

\bibitem{LabastieWhetten}
P. Labastie, R. L. Whetten, {Physical Review Letters} {65}, 1567
(1990).
%
%\bibitem{LaiSievers} R. Lai, A.. J. Sievers, Physical Review Letters {81}, 1937 (1998).
%
%\bibitem{Laskar}
%J. Laskar

\bibitem{Landau}
L. D. Landau, Phys. Z. Sowjetunion {10}, 154 (1936).

\bibitem{Landaucontour}
L. D. Landau, Journal of Physics (USSR) {10}, 25 (1946).

%\bibitem{LandauLifshitzstat} L. D. Landau, E. M. Lifshitz, {Course of
%Theoretical Physics. T. 5: Statistical Physics} (1984).

\bibitem{LandauLifshitz}
L. D. Landau, E. M. Lifshitz, {\em Course of Theoretical Physics. T.
8: Electrodynamics of continuous media} (1984).

\bibitem{lrr}
V. Latora, A. Rapisarda, S. Ruffo, {Physical Review Letters} {83}, 2104 (1999).

\bibitem{lrt2001}
V. Latora, A. Rapisarda, C. Tsallis, Physical Review E
{64}, 056134 (2001).

\bibitem{lrt2002}
V. Latora, A. Rapisarda, C. Tsallis, Physica A 305, 129 (2002).

\bibitem{LebowitzLieb69}
J. L. Lebowitz, E. H. Lieb, {Physical Review Letters}  22, 613
(1969).

%\bibitem{Lebowitz66} J. L. Lebowitz, O. Penrose, {Journal of Mathematical Physics}
%{7}, 98 (1966).

\bibitem{Lebwohl}
P. A. Lebwohl, G. Lasher, Physical Review A 6, 426 (1972).

\bibitem{Lenard}
{A. Lenard}, {Annals of Physics} {10,} {390} (1960).

\bibitem{Levin1}
Y. Levin, R. Pakter, T. Teles, Physical Review Letters, 100, 040604 (2008).

\bibitem{Levin2}
Y. Levin, R. Pakter, F. B. Rizzato, Physical Review E, 78, 021130 (2008).

\bibitem{leyvruf}
F. Leyvraz, S. Ruffo, {Journal of Physics A: Mathematical and
General} {35}, 285 (2002).

\bibitem{LiebLebowitz72}
E. H. Lieb, J. L. Lebowitz, {Advances in Mathematics} {9}, 316 (1972).

\bibitem{landau}
{E. M. Lifshitz, L. P. Pitaevskij}, {\em Physical Kinetics},
{Pergamon Press, Oxford} (1981).

\bibitem{TheseLeonardo}
L. Lori, Modelli con interazione a lungo e corto raggio con
variabili microscopiche continue, Tesi di Laurea, Universit\`a di
Firenze (2008)

\bibitem{Lundgren} T. S. Lundgren, Y. B. Pointin, Journal of
Statistical Physics 17, 232 (1977).

\bibitem{Lyndenbell67}
D. Lynden-Bell, Monthly Notices of the Royal Astronomical Society {136}, 101 (1967).

\bibitem{LyndenPhysA}
D. Lynden-Bell, {Physica} A 263, 29 (1999).

\bibitem{LyndenMrMME}
D. Lynden-Bell, R. M. Lynden-Bell, Monthly Notices of the Royal
Astronomical Society {181}, 405 (1977).

\bibitem{Lyndenbell_epl2008} D. Lynden-Bell, R.M. Lynden-Bell, Europhysics Letters 42, 83001
(2008).

\bibitem{Lyndenwood68}
D. Lynden-Bell, R. Wood, Monthly Notices of the Royal Astronomical
Society {138}, 495 (1968).

\bibitem{MadameLyndenBellFirst}
R. M. Lynden-Bell,   Molecular Physics 86, 1353 %-1374
 (1995).

\bibitem{MadameLyndenBell}
R. M. Lynden-Bell, in {Gravitational Dynamics}, Eds. O. Lahav, E.
Terlevich, R.J. Terlevich (Cambridge University Presss), Cambridge
(1996).

\bibitem{Binder04}
L. G. MacDowell, P. Virnau, M. M\"uller, K. Binder,
% "The Evaporation/Condensation Transition of Liquid Droplets"
Journal of Chemical Physics {120}, 5293 (2004).

%\bibitem{maggs} A. C. Maggs, V. Rossetto, Physical Review Letters
%  {88}, 196402 (2002).

\bibitem{MarchioroPulvirenti}
C. Marchioro, M. Pulvirenti,  {\em Mathematical Theory of
Incompressible Nonviscous Fluids}, Springer-Verlag, New York -
Heidelberg (1994).

\bibitem{marksteiner}
S. Marksteiner, K. Ellinger, P. Zoller, {Physical Review A} {53}, 3409 (1996).

\bibitem{Maxwell}
J. C. Maxwell, %{\it On Boltzmann's theorem on the
%average distribution of energy in a system of material points},
Cambridge Philosophical Society's Transactions, XII, p. 90 (1876).

\bibitem{spohn}
J. Messer, H. Spohn, {Journal of Statistical Physics} {29}, 561 (1982).

\bibitem{micciche}
S. Miccich\`e, Modeling long-range memory with stationary
markovian processes,  cond-mat/08060722 (2008).

\bibitem{MichelRobert94}
J. Michel, R. Robert, %``Large Deviations for Young
%measures and statistical mechanics of infinite dimensional
%dynamical systems with conservation law'',
{Communications in Mathematical Physics} {159}, 195 (1994).

\bibitem{Milanovic98}
Lj. Milanovic, H. A. Posch, W. Thirring, Physical Review E, 57, 2763 (1998).

\bibitem{Miller90}
J. Miller, %``Statistical mechanics of Euler's equation in
%two dimensions'',
{Physical Review Letters} {65}, 2137 (1990).

\bibitem{MorigiAssisi}
G. Morigi, Long-range interactions in cold atomic systems: a foreword, in \cite{Assisi}.

\bibitem{celiabis}
L. G. Moyano, C. Anteneodo, Physical Review E {74}, 021118 (2006).

\bibitem{schreiber}
D. Mukamel, S. Ruffo, N. Schreiber, Physical Review Letters {95},
240604 (2005).
%Breaking of Ergodicity and Long Relaxation Times in Systems with Long-Range Interactions

\bibitem{muskhelishvili}
N. I. Muskhelishvili, in {\em Some Basic Problems of the
Mathematical Theory of Elasticity}, P. Noordhoff, Groningen (1953).

\bibitem{Nagle}
J. F. Nagle, Physical Review A {2}, 2124 (1970).

\bibitem{Neunzert}
H. Neunzert, Neuere qualitative und numerische Methoden in der
Plasmaphysik, Skriptum einer Gastvorlesung an der Gesamthochschule
Paderborn, 1975

\bibitem{Neunzertbis}
H. Neunzert, Fluid Dynamics Transactions {8} (1978).

\bibitem{nicholson}
D. R. Nicholson, {\em Introduction to Plasma Theory}, {John Wiley}
({1983}).

%\bibitem{heap} A. Noullez, D. Fanelli, E. Aurell,
%``Heap base algorithm'', cond-mat/0101336 (2001).

\bibitem{Kurizki}
D. O'Dell, S. Giovanazzi, G. Kurizki, V. M.
Akulin, Physical Review Letters {84}, 5687 (2000).

\bibitem{Ocio}
M. Ocio, D. Herisson, ``Fluctuation dissipation relation in a non
stationary system: experimental investigation in a spin glass'', p.
605  in ``Les Houches Session LXXVII'', J.L. Barrat, M. Feigelman,
J. Kurchan, J. Dalibard (Eds.) (2003).

\bibitem{ojamodphys}
A. S. Oja, V. Lounasmaa, Review of Modern Physics 69, 1 (1997).

\bibitem{Onsager49}
L. Onsager, %``Statistical hydrodynamics'',
{Nuovo Cimento Supplement} {6}, 279 (1949).

\bibitem{Ott}
H. R. Ott,  G. Keller, W. Odoni, L. D. Woolf, M. B. Maple, D. C.
Johnston, H. A. Mook, Physical Review B 25, 477 (1982).

\bibitem{Padmanabhan90}
T. Padmanabhan, %``Statistical mechanics of gravitating %systems'',
{Physics Reports} {188}, 285 (1990).

\bibitem{paddy}
T. Padmanabhan, {Statistical mechanics of
gravitating systems in static and expanding backgrounds}
in~\cite{leshouches}.

%\bibitem{Perrin01} M. Perrin, G. L. Lippi, A. Politi, ``Phase transition in
%%a radiation-matter interaction with recoil and collisions'',
%{Physical Review Letters} {86}, 4520 (2001).

\bibitem{Pettinibook}
M. Pettini, {\em Geometry and Topology in Hamiltonian Dynamics and
Statistical Mechanics}, Springer, Berlin, (2007).

\bibitem{Pichon}
C. Pichon, Dynamics of Self-gravitating Disks, PhD thesis, Cambridge
(1994).

\bibitem{Pluchino}
{A. Pluchino, V. Latora, A. Rapisarda}, {Physica A} {69}, {056113} ({2004}).

\bibitem{Correlation}
A. Pluchino, V. Latora, A. Rapisarda, {Physica A} {340}, 187 (2004).

\bibitem{Correlationb}
A. Pluchino, V. Latora, A. Rapisarda, {Physica D} {193}, 315-328 (2004).

%\bibitem{PostonStewart}
%T. Poston, J. Stewart, {Catastrophe Theory and its Application},
%Pitman, London (1978).

\bibitem{PoschNarnhoferThirring1990}
H. A. Posch, H. Narnhofer, W. Thirring, Physical Review A 42, 1880
(1990).

\bibitem{PoschThirringpre2006}
H. A. Posch, W. Thirring, Physical Review E 74, 051103 (2006).

\bibitem{Potapenko1997}
I. F. Potapenko, A. V. Bobylev, C. A. de Azevedo, A. S. de Assis,
Physical Review E {56}, 7159 (1997).

\bibitem{PurcellPound51}
E. M. Purcell, R. V. Pound, Physical Review {81}, 279 (1951).

\bibitem{RamirezHernandezPRL}
A. Ramirez-Hernandez, H. Larralde, and F. Leyvraz, Physical Review
Letters 100, 120601 (2008).

\bibitem{RamirezHernandezPRE}
A. Ramirez-Hernandez, H. Larralde, and F. Leyvraz, Physical Review E
78, 061133 (2008).

%\bibitem{Rapisarda}
%A. Rapisarda, A. Pluchino, Europhysics News {\bf 36}, 202 (2005).

\bibitem{Ana2004}
A. C. Ribeiro Teixeira, D. A. Stariolo, Physical Review E 70, 016113
(2004).

\bibitem{Ana2005}
S. Risau-Gusman, A. C. Ribeiro-Teixeira, D. A. Stariolo, Physical
Review Letters 95, 145702 (2005).

\bibitem{Robert90}
R. Robert, Comptes  Rendus  de l'Acad\'emie des
Sciences: S\'erie I Maths {311}, 575  (1990).

\bibitem{Robert91}
R. Robert, Journal of Statistical Physics 65, 531 (1991).

%\bibitem{robert2000} R. Robert,
%%``On the statistical mechanics of 2D Euler equation'',
%{Communcations in Mathematical Physics} {212}, 245 (2000).

\bibitem{RobertSommeria91}
R. Robert, J. Sommeria, Journal of Fluid Mechanics, 229, 291 (1991).

\bibitem{RobertSommeria92}
R. Robert, J. Sommeria, Physical Review Letters, 69, 2776 (1992).

\bibitem{ruelle}
D. Ruelle, {\em Statistical Mechanics: Rigorous Results}, Benjamin,
New York, (1969).

\bibitem{RuffoMarseille}
S. Ruffo, ``Hamiltonian dynamics and phase transitions'', in
Transport and Plasma Physics, eds S. Benkadda, Y. Elskens, F.
Doveil, World Scientific Singapore (1994).

\bibitem{RUGH} H. H. Rugh, Phys Rev Lett, 78,  772 (1997).

\bibitem{Rybicki71}
G. B. Rybicki, Astrophysical and Space Sciences, 14, 56 (1971).

%\bibitem{Salazar02}R. Salazar, R.  Toral, A.R.  Plastino, %``Numerical
%%determination of the distribution of energies for the XY-model'',
%{Physica A} {305}, 144 (2002).

\bibitem{Saslaw85} W. C. Saslaw, {\em Gravitational Physics of Stellar
and Galactic Systems}, Cambridge University Press (1985).


\bibitem{sievers1}
M. Sato, A.J. Sievers, Nature  432, 486 (2004).

\bibitem{sievers2}
M. Sato, A. J. Sievers, Physical Review B  71, 214306, (2005).

\bibitem{haberland}
M. Schmidt, R. Kusche, T. Hippler, J. Donges, W. Kronm\"uller, B.
von Issendorff, H. Haberland, Physical Review Letters {86}, 1191
(2001).

\bibitem{Courteille}
S. Slama, G. Krenz, S. Bux, C. Zimmermann, P. W. Courteille, Scattering in a High-Q Ring
Cavity, in \cite{Assisi}.

%\bibitem{Smith90} R. A.  Smith, T. M.  O'Neil, %``Nonaxisymmetric thermal
%%equilibria of a cylindrically bounded guiding center plasma or
%%discrete vortex system'',
%{Physics of Fluids B} {2,} 2961-2975 (1990).

%\bibitem{smirnov}
%V. I. Smirnov, {Corso di Matematica Superiore}, Vol. 3, part II,
%Editori Riuniti, Roma, Italy (1978).

\bibitem{chavanisf}
J. Sommeria, Two-dimensional Turbulence
 in {\em New Trends in Turbulence}, Eds. M. Lesieur, A. Yaglom, F.
David, Vol. 74, Les Houches Summer School, p. 385 (2001).

\bibitem{SGR-model}
Y. Sota, O. Iguchi, M. Morikawa, T. Tatekawa, K. Maeda,
Physical Review E {64}, 056133 (2001).

\bibitem{Spohn_Livre}
H. Spohn, {\em Large Scale Dynamics of Interacting Particles},
Springer (1991).

\bibitem{stahkkieslinschindler}
B. Stahl, M. K. H. Kiessling, K.
Schindler, {Planetary and Space Sciences} {43}, 271-282 (1995).

\bibitem{SylosLabiniAssisi}
F. Sylos Labini, {Gravitational clustering: an overview}, in \cite{Assisi}.

\bibitem{chavanisg}
P. Tabeling, Physics Reports 362, 1 (2002).

\bibitem{Anteneodo99}
F. Tamarit, C. Anteneodo, %``Rotators with long-range
%interactions: Connection with the mean-field approximation'',
{Physical Review Letters} {84}, 208 (2000).

\bibitem{TaruyaSakagamiPRL}
A. Taruya, M. Sakagami, Physical Review Letters 90, 181101 (2003).

\bibitem{TaruyaSakagami}
A. Taruya, M. Sakagami, Monthly Notices of the Royal Astronomical Society  364, 990 (2005).

\bibitem{TAKA}
T. Tatekawa, F. Bouchet, T. Dauxois, S. Ruffo, Physical Review E 71,
056111 (2005).

\bibitem{Thirringdd}
W. Thirring, Zeitschrift f\"ur Physik {235}, 339 (1970).

\bibitem{toral}
R. Toral, {Journal of Statistical Physics} {114}, 1393 (2004).

%\bibitem{Torcini} A. Torcini, M. Antoni, Physical Review E
%   {59}, 2746 (1999).


\bibitem{Antoni2}
A. Torcini, M. Antoni, %"Equilibrium and dynamical properties of
%two-dimensional N-body systems with long-range attractive
%interactions" ,
Phys. Rev. E  59, 2746 (1999) .

\bibitem{TouchettePhysRep}
H. Touchette, {Physics Reports}, ``The large deviations approach to
statistical mechanics", arXiv 08040327v1 (2008).

\bibitem{Tsallisjsp}
C. Tsallis, Journal of Statistical Physics {52}, 479 (1988).

%\bibitem{RT} C. Tsallis, A. Rapisarda, V. Latora, F.  Baldovin,
%{Nonextensivity: from low-dimensional maps to Hamiltonian systems}
%in~\cite{leshouches}.

\bibitem{Tsuchiya94}
T. Tsuchiya, T. Konishi, N. Gouda, Physical Review E {50}, 2607 (1994).

\bibitem{Turkington}
B. Turkington, A. Majda, K. Haven, M. DiBattista,  The Proceedings
of the National Academy of Sciences (USA), {98}, 12346 (2001).

\bibitem{vanhove}
L. Van Hove, Physica {15}, 951 (1949).

\bibitem{VanKampen}
N. G. Van Kampen, {\em Stochastic Processes in Physics and
Chemistry}, North-Holland, Elsevier (1992).

%\bibitem{Devega01}H. J. de Vega, N. Sanchez,
%%``Statistical mechanics of the self-gravitating gas: I.
%%Thermodynamic limit and phase diagrams'',
%{Nuclear Physics B} {625}, 409 (2002).

\bibitem{VelazquezCurilef}
L. Velazquez, S. Curilef, Journal of Physics A  42, 095006 (2009).

\bibitem{velasquez}
L. Velazquez, R. Sospedra, J. C. Castro, F. Guzman, ``On the
dynamical anomalies in the Hamiltonian Mean Field model'',
{cond-mat/0302456}.

\bibitem{venaillebouchet}
A. Venaille, F. Bouchet, Physical Review Letters 102, 104501 (2009).
%``Statistical ensemble inequivalence and
%bicritical points for 2D and Fofonoff flows'', cond-mat/0710.5606
%(2007).

\bibitem{Villain}
J. Villain, Reflets de la Physique, 7, 10 (2008).

\bibitem{originalVlasovpaper}
A. A. Vlasov, Journal of Physics (USSR) {9}, 25 (1945).

%\bibitem{Vollmayr01}B. P. Vollmayr-Lee, E. Luijten,
%%``A Kac-potential treatment of non integrable interactions'',
%{Physical Review E} {63}, 031108 (2001).

\bibitem{wilkens}
M. Wilkens, F. Illuminati, M. Kr\"amer, Journal of Physics B 33, L779 (2000).

\bibitem{WillisPicard}
C. R. Willis, R. H. Picard, Physical Review A 9, 1343 (1974).

\bibitem{sievers3}
J. P. Wrubel, M. Sato, A. J. Sievers, Physical Review Letters  95, 264101 (2005).

\bibitem{yamaPRE}
{Y. Y. Yamaguchi}, {Physical Review E} {68}, 066210 (2003).

\bibitem{yama}
Y. Y. Yamaguchi, Physical Review E 78, 041114 (2008).

\bibitem{yoshi}
Y. Y. Yamaguchi, J Barr{\'e}, F. Bouchet, T. Dauxois, S. Ruffo,
{Physica A} {337}, 36 (2004).

\bibitem{yoshiJSM}
Y. Y. Yamaguchi, F. Bouchet, T. Dauxois,  {Journal of
Statistical Mechanics: Theory and Experiment} P01020 (2007).

\bibitem{SakagamiGouda}
T. Yamashiro, N. Gouda, M. Sakagami, Progress in Theoretical Physics, 88, 269 (1992).

\bibitem{Youngkins}
V. P. Youngkins, B. N. Miller, {Physical Review E}, 62, 4583 (2000).

%\bibitem{Yvon} {J. Yvon}, {La Th{\'e}orie des Fluides et l'{\'E}quation
%d'{\'E}tat}, {Hermann} {(1935)}.
%
%\bibitem{zanette}D. Zanette, M. Montemurro, {Physical Review E} {67}, {031105}
%({2003}).

\bibitem{Zaslavsky_77} G. M. Zaslavsky, V. F. Shabanov, K. S. Aleksandrov and
I. P. Aleksandrova, Soviet Physics JETP, 45, 315 (1977).

\bibitem{Ziff}
R. M. Ziff, G. E. Uhlenbeck, M. Kac, Physics Reports 32, 169 (1977). %-248



\end{thebibliography}
\end{document}